\documentclass[a4paper,oneside,11pt]{article}

\pdfoutput=1
\usepackage{cprform}
\usepackage{amsmath,amstext,amsfonts,amsbsy,amssymb,amscd,slashed}
\usepackage{pdflscape,color,svg,epsfig,cite,hyperref,bbm,multirow}
\usepackage{graphicx}
\usepackage{tabularx}
\usepackage{booktabs}
\usepackage{floatpag}
\usepackage{changepage}
\usepackage{caption}
\usepackage{subcaption}
\usepackage{longtable}
\usepackage{rotating}
\usepackage{makecell}
\usepackage{afterpage}
\usepackage{siunitx}

\usepackage{tikz}
\usetikzlibrary{shapes.misc}
\usetikzlibrary{arrows.meta}
\usetikzlibrary{arrows,shapes}
\usetikzlibrary{trees}
\usetikzlibrary{matrix,arrows}                          
\usetikzlibrary{positioning}                            
\usetikzlibrary{calc,through}                           
\usetikzlibrary{decorations.pathreplacing}              
\usetikzlibrary{decorations.pathmorphing}               
\usetikzlibrary{decorations.markings}
\tikzset{
    vector/.style={decorate, decoration={snake}, draw},
        provector/.style={decorate, decoration={snake,amplitude=2.5pt}, draw},
        antivector/.style={decorate, decoration={snake,amplitude=-2.5pt}, draw},
    fermion/.style={draw=black, postaction={decorate},
        decoration={markings,mark=at position .55 with {\arrow[draw=black]{>}}}},
    fermionbar/.style={draw=black, postaction={decorate},
        decoration={markings,mark=at position .55 with {\arrow[draw=black]{<}}}},
    fermionnoarrow/.style={draw=black},
    gluon/.style={decorate, draw=black,
        decoration={coil,amplitude=4pt, segment length=5pt}},                           
    scalar/.style={dashed,draw=black, postaction={decorate},
        decoration={markings,mark=at position .55 with {\arrow[draw=black]{>}}}},
    scalarbar/.style={dashed,draw=black, postaction={decorate},
        decoration={markings,mark=at position .55 with {\arrow[draw=black]{<}}}},
    scalarnoarrow/.style={dashed,draw=black},
    electron/.style={draw=black, postaction={decorate},
        decoration={markings,mark=at position .55 with {\arrow[draw=black]{>}}}},
        bigvector/.style={decorate, decoration={snake,amplitude=4pt}, draw},
}
	
\newcommand{\ba}{\begin{array}}
\newcommand{\ea}{\end{array}}

\newcommand{\req}[1]{Eq.~(\ref{#1})}

\newcommand{\reapp}[1]{Appendix~\ref{#1}}

\newcommand{\dif}{{\rm d}}

\newcommand{\tr}{{\mathrm{tr}}}

\newcommand{\Dslash}{\relax{\kern+.25em / \kern-.70em D}}

\newcommand{\alphas}{\alpha_{\rm\scriptscriptstyle s}}
\newcommand{\dd}{\mathrm{d}}

\newcommand{\MeV}{{\mathrm{MeV}}}

\newcommand{\Real}{\relax{\mathsf{\Gamma\kern-.35em R}}}
\newcommand{\Int}{\relax{\mathsf{Z\kern-.40em Z}}}

\newcommand{\UO}{\mbox{U}(1)}
\newcommand{\SUT}{\mbox{SU}(2)}

\newcommand{\half}{{\scriptstyle{{1\over 2}}}}

\newcommand{\tthird}{{\scriptstyle{{2\over 3}}}}

\newcommand{\ihalf}{{\scriptstyle{{i\over 2}}}}
\newcommand{\iquart}{{\scriptstyle{{i\over 4}}}}

\newcommand{\NF}{N_\mathrm{\scriptstyle f}}

\newcommand{\gbar}{\kern1pt\overline{\kern-1pt g\kern-0pt}\kern1pt}
\newcommand{\mbar}{\kern2pt\overline{\kern-1pt m\kern-1pt}\kern1pt}
\newcommand{\obar}[1]{\kern3pt\overline{\kern-2pt #1\kern-0pt}\kern1pt}

\newcommand{\mcrit}{m_{\rm cr}}

\newcommand{\fX}{f_{\mathrm{\scriptscriptstyle X}}}
\newcommand{\fP}{f_{\mathrm{\scriptscriptstyle P}}}

\newcommand{\fA}{f_{\mathrm{\scriptscriptstyle A}}}

\newcommand{\Zm}{Z_{\mathrm{m}}}
\newcommand{\ZP}{Z_{\mathrm{\scriptscriptstyle P}}}

\newcommand{\ZA}{Z_{\mathrm{\scriptscriptstyle A}}}

\newcommand{\Oa}{\mbox{O}(a)}
\newcommand{\Oasq}{\mbox{O}(a^2)}

\newcommand{\icsw}{c_{\mathrm{sw}}}

\newcommand{\icA}{c_{\mathrm{\scriptscriptstyle A}}}
\newcommand{\icT}{c_{\mathrm{\scriptscriptstyle T}}}
\newcommand{\icV}{c_{\mathrm{\scriptscriptstyle V}}}

\newcommand{\ibG}{b_{\scriptscriptstyle \Gamma}}
\newcommand{\ibS}{b_{\mathrm{\scriptscriptstyle S}}}
\newcommand{\ibP}{b_{\mathrm{\scriptscriptstyle P}}}
\newcommand{\ibA}{b_{\mathrm{\scriptscriptstyle A}}}
\newcommand{\ibT}{b_{\mathrm{\scriptscriptstyle T}}}
\newcommand{\ibV}{b_{\mathrm{\scriptscriptstyle V}}}

\newcommand{\ibGb}{\bar{b}_{\scriptscriptstyle \Gamma}}
\newcommand{\ibSb}{\bar{b}_{\mathrm{\scriptscriptstyle S}}}
\newcommand{\ibPb}{\bar{b}_{\mathrm{\scriptscriptstyle P}}}
\newcommand{\ibAb}{\bar{b}_{\mathrm{\scriptscriptstyle A}}}
\newcommand{\ibTb}{\bar{b}_{\mathrm{\scriptscriptstyle T}}}
\newcommand{\ibVb}{\bar{b}_{\mathrm{\scriptscriptstyle V}}}

\newcommand{\ibGt}{\tilde{b}_{\scriptscriptstyle \Gamma}}

\newcommand{\ibPt}{\tilde{b}_{\mathrm{\scriptscriptstyle P}}}
\newcommand{\ibAt}{\tilde{b}_{\mathrm{\scriptscriptstyle A}}}

\newcommand{\ibGc}{\check{b}_{\scriptscriptstyle \Gamma}}
\newcommand{\ibSc}{\check{b}_{\mathrm{\scriptscriptstyle S}}}
\newcommand{\ibPc}{\check{b}_{\mathrm{\scriptscriptstyle P}}}
\newcommand{\ibAc}{\check{b}_{\mathrm{\scriptscriptstyle A}}}
\newcommand{\ibTc}{\check{b}_{\mathrm{\scriptscriptstyle T}}}
\newcommand{\ibVc}{\check{b}_{\mathrm{\scriptscriptstyle V}}}
\newcommand{\ibGh}{\hat{b}_{\scriptscriptstyle \Gamma}}
\newcommand{\ibSh}{\hat{b}_{\mathrm{\scriptscriptstyle S}}}
\newcommand{\ibPh}{\hat{b}_{\mathrm{\scriptscriptstyle P}}}

\newcommand{\ibTh}{\hat{b}_{\mathrm{\scriptscriptstyle T}}}

\newcommand{\ival}{\mathrm{\scriptscriptstyle (val)}}
\newcommand{\isea}{\mathrm{\scriptscriptstyle (sea)}}
\newcommand{\trmsea}{\mathrm{tr}\left(\mathbf{m}^{\mathrm{\scriptscriptstyle (sea)}}\right)}

\newcommand{\abar}{\kern1pt\overline{\kern-1pt a\kern-0.5pt}\kern1pt}

\newcommand{\cL}{{\cal L}}

\newcommand{\cO}{{\cal O}}

\newcommand{\vx}{\mathbf{x}}

\newcolumntype{C}{>{$}c<{$}}
\newcolumntype{R}{>{$}r<{$}}
\newcolumntype{L}{>{$}l<{$}}

\begin{document}



\begin{titlepage}


  \vspace*{-30truemm}
  \begin{flushright}
    IFT-UAM/CSIC-25-108\\
    {\large \today}
  \end{flushright}
  \vspace{15truemm}


  \centerline{\bigbf Hadronic physics from a Wilson fermion mixed-action approach:}
  \centerline{\bigbf Setup and scale setting}
  \vskip 10 true mm
  \begin{center}
    \includegraphics[width=0.16\textwidth]{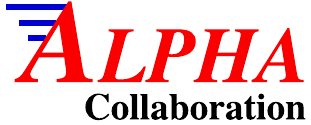}\\
  \end{center}

  \vskip -2 true mm
  \centerline{\elfrm 
    Andrea~Bussone$\,^{a}$,
    Alessandro Conigli$\,^{b,c}$,
    Julien~Frison$\,^{d}$,
    Gregorio~Herdo\'{\i}za$\,^{e,f}$,
  }
  \centerline{\elfrm 
    Carlos~Pena$\,^{e,f}$,
    David~Preti$\,^g$,
    Jos\'e~\'Angel~Romero$\,^e$,
    Alejandro S\'aez$\,^{e,f,h}$,
    Javier~Ugarrio$\,^{e,f}$}

  \vskip 4 true mm
  \centerline{\tenit $^a$Humboldt Universit\"at zu Berlin, Institut f\"ur Physik and IRIS Adlershof, Berlin, Germany}
  \vskip 1 true mm
  \centerline{\tenit $^b$Helmholtz Institute Mainz, Johannes Gutenberg University, Mainz, Germany}
  \vskip 1 true mm
  \centerline{\tenit $^c$GSI Helmholtz Centre for Heavy Ion Research, Darmstadt, Germany}
  \vskip 1 true mm
  \centerline{\tenit $^d$John von Neumann-Institut f\"ur Computing NIC,}
  \centerline{\tenit Deutsches Elektronen-Synchrotron DESY, Platanenallee 6, 15738 Zeuthen, Germany}
  \vskip 1 true mm
  \centerline{\tenit $^e$Instituto de F\'{\i}sica Te\'orica UAM-CSIC, Madrid, Spain}
  \vskip 1 true mm
  \centerline{\tenit $^f$Dpto. de F\'{\i}sica Te\'orica, Universidad Aut\'onoma de Madrid, Madrid, Spain}
  \vskip 1 true mm
  \centerline{\tenit $^g$INFN, Sezione di Torino, Turin, Italy}
  \vskip 1 true mm
  \centerline{\tenit $^h$Instituto de F\'isica Corpuscular (IFIC), CSIC-Universitat de Val\`encia, 46071, Valencia, Spain}
  \vskip 15 true mm


  %
  \noindent{\tenbf Abstract:}
  We introduce a lattice QCD mixed action approach that employs Wilson-type quarks in the sea and valence sectors.
  The sea sector is based on gauge ensembles with $\NF=2+1$ flavours of non-perturbatively O($a$)-improved Wilson fermions generated by the Coordinated Lattice Simulations (CLS) initiative.
  The parameter space of the considered ensembles encompasses five values of the lattice spacing, a range of pion masses extending down to the physical point, and large physical volumes.
  In the valence sector, we employ Wilson twisted-mass fermions at maximal twist, using the same massless Wilson-Dirac operator in both the sea and valence sectors.
  We describe the strategy applied for the required matching of the sea and valence quark masses along the target renormalised chiral trajectory.
  A precise universality test is then conducted by comparing the continuum-limit results of the mixed-action approach and of the unitary setup, in which the same Wilson fermion regularisation is employed in the sea and in the valence. 
  As a key application, we conduct a scale setting procedure based on lattice determinations of the masses and decay constants of the pion and kaon, as well as the gradient flow scale $t_0$.
  The scale setting can consequently be performed in three distinct ways, utilising the unitary setup, the mixed action approach, and their combination.
  We observe that the latter combination results in enhanced control of the systematic uncertainties, thereby yielding a precise determination of the physical value of $t_0$.
  \vspace{10truemm}

  \eject
\end{titlepage}

\tableofcontents

\cleardoublepage


\section{Introduction}
\label{sec:intro}

In the framework of the Standard Model (SM) of particle physics, lattice QCD calculations serve as an instrumental tool in the determination of its fundamental parameters and in the prediction of processes that are the subject of extensive experimental investigation. 
In particular, the quark flavour sector comprises processes where high-precision lattice QCD determinations are required to elucidate the degree of consistency between the SM predictions and the corresponding experimental measurements.
In this context, the availability of multiple independent determinations of the same quantity is crucial for the consolidation of the SM predictions.
The increased precision of lattice QCD calculations experienced in recent years~\cite{FlavourLatticeAveragingGroupFLAG:2021npn,FlavourLatticeAveragingGroupFLAG:2024oxs,Aoyama:2020ynm,Aliberti:2025beg} stems from a wide range of theoretical and computational advancements.
These encompass the development of improved lattice formulations that enable more precise control of systematic effects, along with numerical strategies aimed at reducing statistical uncertainties.
However, in this endeavour to further reduce the uncertainty of the lattice results, a number of challenges remain to be addressed.
For instance, in the heavy-quark sector, the study of leptonic and semileptonic decays of heavy-light mesons necessitates the use of exceedingly small values of the lattice spacing --- to account for the discretisation effects introduced by the heavy-quark mass --- in conjunction with large lattice volumes.
Ensuring reliable control over the continuum-limit extrapolation is essential, and represents a demanding task that requires the use of an improved lattice formulation, along with a sufficiently large set of fine lattice spacings.
In addition, the regime of small lattice spacings presents a computational challenge in itself, due to the difficulty of efficiently sampling the space of topological sectors~\cite{Luscher:2010iy,Schaefer:2010hu,Luscher:2011kk}.
In this work, we introduce a lattice QCD mixed-action setup with the aim of contributing to the refinement of the control of the lattice results.
Our approach is heavily reliant on the gauge-field configuration ensembles with $\NF=2+1$ flavours of non-perturbatively $\Oa$-improved Wilson fermions~\cite{Wilson:1974sk,Sheikholeslami:1985ij} generated by the CLS (Coordinated Lattice Simulations) initiative~\cite{Bruno:2014jqa,Bali:2016umi}
which we will henceforth refer to as the {\it sea sector} of the mixed-action approach.
A distinctive feature of the CLS programme is the use of open boundary conditions for gauge fields in the time direction, which has permitted the simulations of lattice spacings below $0.05\,{\mathrm{fm}}$ with adequate sampling of topological sectors.
This work will thus consider five values of lattice spacing, spanning from approximately 0.085\,fm to 0.039\,fm, together with a range of pion masses reaching down to the physical point.
Quark masses lie along a chiral trajectory in which the sum of the mass-degenerate (up, down) and strange quark masses --- corresponding to the trace of the sea-quark mass matrix, $\trmsea$ --- is kept constant.
In the valence sector, we employ a Wilson twisted mass formulation~\cite{Frezzotti:2000nk,Frezzotti:2003ni}, at maximal twist.
The corresponding Dirac operator includes the Sheikholeslami-Wohlert term~\cite{Sheikholeslami:1985ij}, with the same improvement coefficient $\icsw$ used in the sea sector.
This ensures that the sea and valence Dirac operators coincide in the chiral limit, and implies that the same renormalisation factors determined in the sea sector can be employed in the valence sector when a mass-independent renormalisation scheme is considered.
Another interesting feature of this setup is that, at maximal twist, physical observables are free from $\Oa$ lattice artefacts, save for sea-quark effects that induce residual lattice artefacts proportional to the trace of the sea-quark mass matrix.
In perturbation theory, these manifest as loop-suppressed lattice artefacts that first emerge at $\mbox{O}\left( \alphas^2(1/a) \, a\trmsea\right)$.
On the other hand, physical quantities will be free from leading-order lattice artefacts induced by the valence quark masses.
This property is particularly relevant in the context of heavy quark physics, as reported in our recent study of the charm-quark sector~\cite{Bussone:2023kag}.
The flavour-breaking discretisation effects induced by the Wilson twisted mass Dirac operator are, by construction, confined to the valence sector.
This, for instance, implies the absence of the undesirable effects resulting from the mass splitting between the charged and neutral sea pions.
Moreover, operators differing by flavour-breaking discretisation effects can, in principle, be considered, allowing their relative cut-off effects to be analysed and used to constrain the continuum extrapolation.\footnote{
Other types of mixed-action setups involving Wilson-type sea quarks have been considered in the existing literature.
For example, overlap valence quarks have been employed in conjunction with non-perturbatively $\Oa$-improved Wilson sea quarks~\cite{Bernardoni:2010nf} or with Wilson twisted mass sea quarks~\cite{Cichy:2010ta,Cichy:2012vg}.
Furthermore, Osterwalder-Seiler valence quarks have been employed in combination with dynamical simulations utilising Wilson twisted mass fermions at maximal twist~\cite{EuropeanTwistedMass:2014osg}.
For a detailed discussion of the mixed-action approach in lattice QCD, we refer Refs.~\cite{Bar:2002nr,Bar:2003mh,Golterman:2005xa,Hasenfratz:2006bq,Chen:2007ug,Aubin:2008wk,Chen:2009su}.
}

Given that the discretisations of the sea and valence Dirac operators differ in a mixed-action approach, a matching procedure is necessary to guarantee that the target QCD observables are reached in the continuum theory.
In particular, the equality of the quark masses in the sea and valence sectors has to be enforced.
A key validation test of a mixed-action approach is to monitor that, after a precise matching of sea and valence quark masses and a controlled continuum-limit extrapolation, the resulting mixed-action observables agree with those obtained from the unitary setup, in which the same action is used in both the sea and valence sectors.
The validation of such a universality test of the lattice discretisations thus paves the way for the combined use of the unitary and the mixed-action setups to improve the control of continuum-limit extrapolations.
In this work, we will pursue such a path. We will present a comprehensive account of the mixed-action properties along with a description of the matching of the sea and valence sectors.
We will furthermore present precise numerical evidence to support the universality of the considered discretisations. 
Once the construction and validation of the setup has been completed, we proceed to carry out a scale setting procedure, which considers three formulations: (i) the unitary setup based on $\Oa$-improved Wilson fermions in both the sea and valence sectors; (ii) the mixed-action approach with Wilson twisted mass fermions in the valence sector; and (iii) a combination of the previous discretisations, where a common continuum-limit result is enforced.
The precision achieved during the lattice scale setting has a direct impact on the precision of virtually all other lattice results; for example, it has been observed~\cite{DellaMorte:2017dyu} that the uncertainty in the scale setting contributes significantly to the uncertainty in the hadronic vacuum polarisation contribution to the muon anomalous magnetic moment (see, e.g., Ref.~\cite{Aliberti:2025beg} and references therein).
The lattice scale setting is the procedure by which the parameters of the theory are fixed, using the minimal set of necessary external physical input, thus leading to a determination of the values of the lattice spacing $a$ in physical units.
In the case of the isospin symmetric limit of QCD (isoQCD) with $\NF=2+1$ flavours --- as considered in this work --- two external inputs are required to fix the quark masses, and a third one is needed to fix the overall scale.
While the kaon and pion masses are natural choices for setting the quark masses, the choice of the observable used to set the scale is more intricate.
Ultimately, the choice is driven by the ability to accurately control the uncertainty of the lattice calculation for the selected quantity; the precision of the physical input also has a relevant role.
However, its relevance will become most important when considering the more complete theory incorporating both QCD and QED effects.
In the context of the isoQCD limit, the choice of the external input can be interpreted as a prescription for the separation of QCD and QED contributions (for a review including appropriate references, we refer to Refs.~\cite{FlavourLatticeAveragingGroupFLAG:2024oxs,Tantalo:2023onv}).
In this work, we will follow the proposal of Ref.~\cite{Bruno:2016plf} to consider the flavour-averaged linear combination of the pion and kaon decay constants, $f_{\pi K} = \frac{2}{3} \left(f_K + \frac{1}{2}f_{\pi}\right)$ to fix the scale.
This quantity can be determined with high precision on the lattice, and it exhibits a mild light-quark mass-dependence --- in comparison to the behaviour of, say, $f_\pi$ alone --- along the chiral trajectory considered in this work.
This is expected from the fact that it is only logarithmically dependent on the pion mass at next-to-leading order (NLO) in SU(3) Chiral Perturbation Theory~\cite{Gasser:1984gg}.
In practice, we will employ the gradient flow scale $t_0$~\cite{Luscher:2010iy} as an intermediate scale.
This enables the construction of suitable dimensionless quantities allowing to properly define the renormalised chiral trajectory and to perform the continuum-limit and chiral extrapolations.
While, as stated above, the CLS ensembles employed in this work are generated along a chiral trajectory with a constant value of the trace of the bare quark mass matrix that approximately crosses the physical point, for the sake of our analysis we will choose a slightly different line of constant physics.
The latter is given by a renormalised chiral trajectory on which the flavour-averaged combination of the pion and kaon masses in units of $t_0$, $\phi_4=8t_0\left(m_K^2+\frac{1}{2}m_{\pi}^2\right)$, is constrained to be constant and equal to its isoQCD physical value, $\phi_4=\phi_4^{\mathrm{ph}}$.\footnote{For the sake of simplicity, the inputs defining isoQCD will be referred to as physical inputs.}
Subsequently, the physical value of $t_0$ can be determined by employing the isoQCD physical input for $f_{\pi K}$.
This paper is structured as follows. In Section~\ref{sec:setup}, we introduce the discretisations of the sea and valence sectors that define the mixed-action setup, and describe their salient properties.
In Section~\ref{sec:match}, we present the Wilson fermion CLS gauge-field configuration ensembles that have been employed in the sea sector.
The relevant lattice observables are defined, along with the improvement and renormalisation procedure.
Furthermore, the matching procedure of the sea and valence quark masses is described, during which the tuning to maximal twist of the Wilson twisted mass valence quarks is simultaneously achieved.
In Section~\ref{sec:results}, we employ the determination of the pseudoscalar decay constants of the pion and kaon, based on both the sea and  the valence sectors --- i.e. in the unitary and in the mixed-action setup, respectively --- in conjunction with the gradient flow scale $t_0$, to carry out the continuum-limit and chiral extrapolations.
This procedure yields a determination of the physical value of $t_0$, which is then used to set the lattice spacing in physical units.
A detailed account of the associated systematic uncertainties is presented.
The conclusions are collected in Section~\ref{sec:concl}.
Additional detailed information regarding the analysis and field-theoretical aspects of the mixed-action formulation is provided in several accompanying appendices.
Appendix~\ref{app:twist} presents the correspondence between quark bilinears in the physical and twisted quark bases.
In Appendix~\ref{app:Oa}, we describe the $\Oa$-improvement of a generic regularisation scheme employing Wilson-type fermions in both the sea and valence sectors.
Further details on the application of the reweighting factor in the light (up, down) and strange quark sectors are given in Appendix~\ref{app:flagged_s}.
Additionally, Appendix~\ref{app:obc} provides an analysis of boundary effects arising from the use of open boundary conditions in the temporal direction.
In Appendix~\ref{app:analysis}, we present a concise account of the statistical error analysis based on the $\Gamma$-method and outline the methodology employed to estimate systematic uncertainties arising from model variations, including data cuts.
Appendix~\ref{app:derivatives} reports the combined continuum and chiral fits of mass derivatives used to perform mass shifts of the observables, while Appendix~\ref{app:FVE} addresses finite-size effects.
Appendix~\ref{app:chiral_w_wtm} contains further information on independent continuum and chiral fits of $\sqrt{8 t_0}\, f_{\pi K}$ for both the unitary and mixed-action regularisations.
The influence of varying the information criteria and the weighting of the chi-squared function in the continuum-limit and chiral fits involved in the determination of $t_0^{\mathrm{ph}}$ is examined in Appendix~\ref{app_variations}.
In Appendix~\ref{app:prescription}, we present the determinations of $t_0^{\mathrm{ph}}$ based on various prescriptions for the definition of isoQCD.
Finally, Appendices~\ref{apex_model_av_t0} and \ref{apex_model_av_t0_fpi} collect the results for $t_0^{\mathrm{ph}}$ obtained from each of the models considered in the analyses based on $f_{\pi K}^{\mathrm{isoQCD}}$ and $f_{\pi}^{\mathrm{isoQCD}}$, respectively, while Appendix~\ref{app:Tables} provides tables with results for the relevant lattice observables.
%



\section{Lattice QCD mixed-action setup}
\label{sec:setup}

We consider a lattice QCD setup in which two distinct discretisations of the Dirac operator, both constructed from the same massless Wilson--Dirac operator, are employed in the sea and valence sectors.
In this context, the term {\it sea sector} refers to the lattice action that is employed at the level of the generation of the gauge field configuration ensembles.
In contrast, the {\it valence sector} pertains to the discretisation employed in the construction of the observables.
Furthermore, we will refer to the standard approach, in which the same lattice regularisation is employed in both the sea and valence sectors, as the {\it unitary setup}.
This section presents the lattice regularisations that define the sea and valence sectors of the mixed action, and reviews some of their main features.

\subsection{Sea and valence lattice actions}

The sea sector of the mixed action employs the tree-level $\mbox{O}(a^2)$-improved L\"uscher-Weisz gauge action~\cite{Luscher:1984xn,Luscher:1985zq} and the non-perturbatively $\Oa$-improved Wilson fermion action~\cite{Wilson:1974sk,Sheikholeslami:1985ij} with $\NF=2+1$ flavours.
The corresponding gauge-field configuration ensembles have been generated by the Coordinated Lattice Simulation (CLS) consortium~\cite{Bruno:2014jqa,Bruno:2016plf,Bali:2016umi,Mohler:2017wnb,Mohler:2020txx}.
The massless Wilson-Dirac operator~\cite{Wilson:1974sk,Sheikholeslami:1985ij} has the form
\begin{gather}
  \label{eq:Dphys}
  D=\frac{1}{2}\gamma_\mu(\nabla_\mu^*+\nabla_\mu)-
  \frac{a}{2}\nabla_\mu^*\nabla_\mu 
  +\frac{i}{4}\,a\icsw\sigma_{\mu\nu}\hat F_{\mu\nu}\,,
\end{gather}
where $\nabla_\mu$ and $\nabla_\mu^*$ are, respectively, the forward and backward covariant difference operators, $\sigma_{\mu\nu}=\ihalf[\gamma_\mu,\gamma_\nu]$, and $\hat F_{\mu\nu}$ is the clover-leaf definition of the field strength tensor as given in Ref.~\cite{Luscher:1996sc}.
The non-perturbative value of the Sheikholeslami-Wohlert coefficient $\icsw$, as determined in Ref.~\cite{Bulava:2013cta}, is used.
In the sea sector, the quark mass term takes the standard form
\begin{gather}
  \label{eq:msea}
  \bar\psi^{\isea} \mathbf{m}^{\isea} \psi^{\isea}\,;
  \\
  \label{eq:mmatsea}
  \mathbf{m}^{\isea} = {\mathrm{diag}}(m_u^{\isea},m_d^{\isea},m_s^{\isea})\,,
\end{gather}
where $\psi^{\isea}$ is a vector in flavour space. The superscript ``$\mathrm{(sea)}$'' refers to the sea sector, while the flavour indices $u,d,s$ label the mass-degenerate (up, down) quarks and the strange quark, respectively.

In the mixed-action valence sector, we use a Wilson twisted mass fermion action~\cite{Frezzotti:2000nk}, which in the massless limit is identical to the Wilson-Dirac operator in \req{eq:Dphys} used in the sea sector.
The difference between sea and valence Dirac operators thus appears exclusively in the mass terms.
In the former case, it adopts the standard form in \req{eq:msea}, whereas in the valence sector, we adopt the four-flavour formulation of twisted-mass QCD introduced in Ref.~\cite{Pena:2004gb}
\begin{gather}
  \label{eq:mval}
  \bar\psi^{\ival} \left\{ \mathbf{m}^{\ival} + i \gamma_5 \boldsymbol{\mu}  \right\} \psi^{\ival}\,;
  \\
  \label{eq:mval2}
  \mathbf{m}^{\ival} = {\mathrm{diag}}(m_u^{\ival},m_d^{\ival},m_s^{\ival},m_c^{\ival})\,, \qquad
  \boldsymbol{\mu} = {\mathrm{diag}}(\mu_u,-\mu_d,-\mu_s,\mu_c)\,,
\end{gather}
where the superscript ``$\mathrm{(val)}$'' refers to the mixed-action valence sector; for simplicity, it is omitted for the twisted mass parameters in the twisted mass matrix $\boldsymbol{\mu}$ in \req{eq:mval2}, since there is no equivalent of the latter in the sea sector and therefore no ambiguity arises.
In \req{eq:mval2}, the parameters $\mu_i$ are non-negative, and can thus be directly interpreted as quark masses.
Alternative, inequivalent sign choices for the twisted mass parameters in $\boldsymbol{\mu}$, while possible, would correspond to so-called Osterwalder–Seiler quarks~\cite{Frezzotti:2004wz}; such variations do not significantly affect the discussion that follows, and we therefore adopt the convention specified in \req{eq:mval2} for definiteness and without loss of generality.
For brevity, in the following we will refer to our choice of regularisation in the valence sector as Wilson twisted mass, and use the acronym Wtm. Similarly, we will also denote the sea quark Wilson action by W.
It is implicit in Eqs.~(\ref{eq:Dphys})~and~(\ref{eq:mval}) that in the valence sector we are using the so-called twisted basis for quark fields.
This is technically beneficial, since, the regularised massless theory then being the same as with standard Wilson fermions,  all the relevant renormalisation constants from the latter can be carried over if a massless renormalisation scheme is employed.
The procedure for obtaining the correct continuum theory from this formulation is discussed in Refs.~\cite{Frezzotti:2000nk,Pena:2004gb}; here it is sufficient to recall that the formulation with physical (standard) quark fields is obtained by the following change of variables
\begin{gather}
  \label{eq:tmrot}
  \psi^{\ival}~\to~\psi^{'\ival} = e^{i\boldsymbol{\alpha}\gamma_5 \frac{\boldsymbol{T}}{2}}\psi^{\ival}\,, \quad \quad
  \bar\psi^{\ival}~\to~\bar\psi^{'\ival} = \bar\psi^{\ival} e^{i\boldsymbol{\alpha} \gamma_5 \frac{\boldsymbol{T}}{2}}\,,
\end{gather}
where $\boldsymbol{T} = {\mathrm{diag}}(\eta_u,\eta_d,\eta_s,\eta_c)$ with $\eta_u=\eta_c=-\eta_s=-\eta_d=1$, and $\boldsymbol{\alpha} = {\mathrm{diag}}(\alpha_u,\alpha_d,\alphas,\alpha_c)$. The twist angle $\alpha_i$ for a specific quark flavour is expressed in terms of the ratio of renormalised masses
\begin{gather}
  \label{eq:bare_twist_angle}
  \cot\alpha_i = \frac{m_{i,\mathrm{\scriptscriptstyle R}}^{\ival}}{\mu_{i,\mathrm{\scriptscriptstyle R}}}\,.
\end{gather}
When this chiral rotation is performed, the mass term in \req{eq:mval} acquires the standard form, while the Wilson term in \req{eq:Dphys} gets rotated.
We will only be concerned with the fully-twisted setup, $\alpha_i= \pi/2$, which can be obtained by setting the renormalised standard masses $m_{i,\mathrm{\scriptscriptstyle R}}^{\ival}$ to zero.
The precise procedure for this tuning to maximal twist will be discussed below.

\subsection{Symmetries and currents in the valence sector}

The sea sector has the standard symmetries of the Wilson fermion regularisation.
In particular, the Wilson term explicitly breaks the symmetry under axial flavour transformations.
In the valence sector, the pattern of global symmetry breaking is determined by the twist angles.
This is most easily understood by considering for a moment the light sector only, with mass-degenerate $u$ and $d$ quarks, and working in the physical basis.\,\footnote{
From now on, unless otherwise required by context, we omit the superscripts $\mathrm{(val)}$ on valence fields and mass parameters to streamline the notation.}
Out of the six generators of $\SUT_V \times \SUT_A$, three are always left unbroken by the Wilson term: at zero twist angle that is the standard $\SUT_V$, while at full twist one is left with $\UO_A^1 \times \UO_A^2 \times \UO_V^3$, where the superscript indicates the index of the Pauli matrix involved in the transformation, viz.
\begin{gather}
  \delta\psi'(x) = i\epsilon\gamma_5\tau^1\psi'(x)\,,
\end{gather}
under $\UO_A^1$, etc., where here ${\psi'}^T = (u,d)$.
As a result, in the fully twisted case there will be two exactly conserved axial currents and one exactly conserved vector current.
When the twisted basis is adopted, the change of variables in \req{eq:tmrot} results in a non-trivial mapping between composite operators, similar to the change in the form of the mass and Wilson terms discussed when the action was introduced above.
Let us spell this out explicitly for quark currents, for which we adopt the (otherwise standard) notation
\begin{gather}
  \label{eq:current_notation}
  \begin{split}
    V_\mu^{ij}=\bar\psi_i\,\gamma_\mu\psi_j\,,\qquad
    A_\mu^{ij}=\bar\psi_i\,\gamma_\mu\gamma_5\psi_j\,,\qquad
    P^{ij}=\bar\psi_i\,\gamma_5\psi_j\,,\\
    S^{ij}=\bar\psi_i\,\psi_j\,,\qquad
    T_{\mu\nu}^{ij}=\bar\psi_i\, i\sigma_{\mu\nu}\psi_j\,,\qquad
    \tilde T_{\mu\nu}^{ij}=\bar\psi_i\, i\tilde\sigma_{\mu\nu}\psi_j\,,
  \end{split}
\end{gather}
where $i,j$ are flavour indices and $\tilde\sigma_{\mu\nu}=\half\varepsilon_{\mu\nu\alpha\beta}\sigma_{\alpha\beta}=\iquart\varepsilon_{\mu\nu\alpha\beta}[\gamma_\alpha,\gamma_\beta]$.
In the channels corresponding to conserved currents, and at full twist, Dirac structures are swapped as $V \leftrightarrow A$, $P \leftrightarrow S$, and $T \leftrightarrow \tilde T$, up to trivial phases.
In the other channels the Dirac structure of the current is preserved.
A full dictionary of the rotations is provided in \reapp{app:twist}.
It is also worth reminding that, at non-vanishing values of the twist angles, the Wtm regularisation breaks parity and time-reversal by $\Oa$ effects.
Charge conjugation is, on the other hand, always preserved exactly.
The consequences of the $\Oa$ breaking of discrete symmetries are mostly relevant for the renormalisation and $\Oa$-improvement of the theory, as discussed in~\reapp{app:Oa}.
An important role is played by the Ward-Takahashi identities (WTI), that will be used to fix the value of the twist angle and to determine quark masses.
In the twisted basis, the generic form of the non-singlet ($i \neq j$) WTI for vector and axial currents is
\begin{align}
  \label{eq:axialWTI}
  \partial_\mu A_\mu^{ij} &= 2m_{ij} \, P^{ij} + i (\eta_i\mu_i+\eta_j\mu_j)\, S^{ij}\,,\\
  \label{eq:vectorWTI}
  \partial_\mu V_\mu^{ij} &= 2\tilde{m}_{ij} \, S^{ij} + i (\eta_i\mu_i-\eta_j\mu_j)\, P^{ij}\,,
\end{align}
where $m_{ij}$ and $\tilde{m}_{ij}$ denote the half-sum and the half-difference of the standard current quark masses, respectively.
Note that the physical interpretation of each of these relationships changes with the value of the twist angles.
As explained above, we are specifically interested in setting $\alpha_i = \pi/2~\forall i$, which implies vanishing renormalised standard mass $m_{i,\mathrm{\scriptscriptstyle R}}$.
This in turn implies that the current mass $m_{ij}$ in \req{eq:axialWTI} vanish, possibly up to $\Oa$ effects.
We will actually use this in our strategy to set full twist: bare parameters will be tuned such that the values of standard current masses obtained from WTI vanish, subject to the
constraint that matching conditions for physical observables in the valence and sea sectors are met, as discussed in Sec.~\ref{sec:match}.
For the time being we will simply observe that, after $\alpha_i=\pi/2$ has been set, the form of the WTI identities for each distinct choice of quark flavours is
\begin{align}
  \partial_\mu V_\mu^{ud} &= i (\mu_u+\mu_d)P^{ud}\,,\\
  \partial_\mu V_\mu^{us} &= i (\mu_u+\mu_s)P^{us}\,,\\
  \partial_\mu A_\mu^{uc} &= i (\mu_u+\mu_c)S^{uc}\,,\\
  \partial_\mu A_\mu^{ds} &= -i (\mu_d+\mu_s)S^{ds}\,,\\
  \partial_\mu V_\mu^{dc} &= i (\mu_d+\mu_c)P^{dc}\,,\\
  \partial_\mu V_\mu^{sc} &= i (\mu_s+\mu_c)P^{sc}\,,
\end{align}
for {\em physical} axial currents (cf. the relation between currents in the physical and twisted bases provided in \reapp{app:twist}), and
\begin{align}
  \partial_\mu A_\mu^{ud} &= i (\mu_u-\mu_d)S^{ud}\,,\\
  \partial_\mu A_\mu^{us} &= i (\mu_u-\mu_s)S^{us}\,,\\
  \partial_\mu V_\mu^{uc} &= i (\mu_u-\mu_c)P^{uc}\,,\\
  \partial_\mu V_\mu^{ds} &= -i (\mu_d-\mu_s)P^{ds}\,,\\
  \partial_\mu A_\mu^{dc} &= -i (\mu_d-\mu_c)S^{dc}\,,\\
  \partial_\mu A_\mu^{sc} &= -i (\mu_s-\mu_c)S^{sc}\,,
\end{align}
for {\em physical} vector currents.
The above expressions of course require proper current normalisation, and in practice will hold up to $\Oa$ mixing with higher-dimensional operators.
One key final observation is that, because of the preserved physical axial symmetries, exactly conserved currents will exist in the corresponding flavour channels.
Given our choice of twist, this implies the existence of exactly conserved $ud$, $us$, $dc$, and $sc$ axial currents.
This property will be used for doing away with the need of current normalisation in the determination of some matrix elements.

\subsection{Renormalisation and $\Oa$-improvement}

This far, we limited ourselves to introduce the Wilson twisted mass term as a distinct QCD discretisation on the lattice, without discussing the properties of this regularisation.
Among the original reasons behind the formulation of twisted mass QCD there was the need for an infrared cutoff that could remove instabilities arising from exceptional configurations.
However, it was soon realised that Wilson twisted-mass theories can simplify the renormalisation of certain operators with respect to standard Wilson fermions.
According to the symmetries of the action, it is straightforward to demonstrate that standard quark masses and couplings renormalise as in the Wilson formulation.
In addition, we have to consider the renormalisation pattern of the twisted mass term.
However, if we work in the full twist regime, i.e. the twisted angle  set to $\alpha_i=\pi/2$ for all $i$, standard quark masses vanish and all the physical information is encoded in the twisted mass term.
In this setup, the charged axial symmetry $\left[U_A(1)_{\frac{\pi}{2}}\right]_{1,2}$ ensures that the twisted mass term does not mix with other relevant operators under renormalisation, and consequently the twisted mass $\mu_i$ renormalises multiplicatively as
\begin{equation}
  \mu_{i,\mathrm{\scriptscriptstyle R}}(\mu_{\mathrm{ren}}) = Z_\mu(g_0^2, a\mu_{\mathrm{ren}})\mu_i\,,
\end{equation}
with $Z_\mu$ the associated renormalisation constant and $\mu_{\mathrm{ren}}$ the renormalisation scale.
Moreover, the exact flavour symmetry of massless Wilson fermions implies the existence of a point-split current $\tilde{V}_\mu^{ij}(x)$ such that the Ward identity holds exactly on the lattice~\cite{Frezzotti:2000nk,Shindler:2007vp}
\begin{equation}
  \langle \partial_\mu^* \tilde{V}^{ij}_\mu(x) \mathcal{O}(0) \rangle = 
  i(\eta_i\mu_i - \eta_j\mu_j) \langle P^{ij}(x) \mathcal{O}(0)\rangle\,,
  \label{eq:vector_ward_id}
\end{equation}
where $\partial_\mu^*$ is the lattice backward derivative.
Here  $\tilde{V}_\mu^{ij}(x)$ is defined in terms of the fermion fields in the twisted basis as 
\begin{equation}
  \tilde{V}_\mu^{ij}(x) = \frac{1}{2}
  \bigg[
    \bar{\psi}_{i}(x) (\gamma_\mu-1) U_\mu(x)\psi_j(x+a\hat{\mu}) 
    + 
    \bar{\psi}_{i}(x+a\hat{\mu}) (\gamma_\mu+1) U_\mu^\dagger(x)\psi_j(x)
    \bigg]\,,
  \label{eq:vector_pointsplit_current}
\end{equation}
where $\hat{\mu}$ denotes the unit vector in direction $\mu$.
The conservation of the vector Ward identity \req{eq:vector_ward_id} on the lattice implies that the point-split current $\tilde{V}_\mu^{ij}(x)$ renormalises trivially with $Z_{\tilde{V}}=1$ as in the continuum.
Therefore this entails  that  the renormalisation constants $\ZP$  for the pseudoscalar density is the same for all flavours and satisfy the following identity
\begin{equation}
  Z_\mu = \ZP^{-1}\,.
  \label{eq:zp_zmu_relation}
\end{equation}
At full twist, the quark masses can be readily determined, since the twisted mass $\mu_i$ is a parameter in our computation and renormalises multiplicatively with the pseudoscalar renormalisation constant.
Additionally, the underlying symmetries of Wilson twisted-mass fermions permit the extraction of the matrix element required for computing the pseudoscalar decay constant from the pseudoscalar correlator alone, without the need for renormalisation, as will be discussed in the following sections.
The introduction of the twisted mass term in principle requires additional improvement coefficients~\cite{Frezzotti:2001ea} to those required for a full improvement of the standard Wilson theory~\cite{Luscher:1996sc}.
In particular, for the case of non-degenerate quarks, as in the $\NF=2+1$ flavour theory considered in this work, the benefits of this formulation are thus not immediately apparent due to the considerable number of $\Oa$-improvement coefficients involved~\cite{Bhattacharya:2005rb}.
When considering Wilson twisted-mass fermions in both the sea and valence sectors, Frezzotti and Rossi~\cite{Frezzotti:2003ni} demonstrated the automatic $\Oa$-improvement of physical observables at maximal twist, where no improvement coefficients are required.
More specifically, parity-even correlation functions are free from $\Oa$ lattice artefacts, provided that the renormalised standard quark masses vanish in the continuum limit.
It is worth noticing that at full twist the theory is not improved, but the appropriate correlation functions and, consequently, physical observables are.
We hereby summarise the key points that lead to automatic $\Oa$-improvement of the two-flavour theory, and we refer to the original work~\cite{Frezzotti:2003ni} and to the reviews~\cite{Sint:2007ug,Shindler:2007vp} for a complete proof.
We begin from the Symanzik effective action where the leading order term $S_0$ is the action in the target continuum theory.
At full twist, the  action $S_0$ reads
\begin{equation}
  S_0[U,\psi,\bar{\psi}]  = \int\dd^4 x \, \bar{\psi}(x)
  \left[
    \gamma_\mu D_\mu + i\mu_R \gamma_5 \tau^3
    \right]\psi(x)\,,
\end{equation}
with the physical mass given by the twisted mass term only.
An essential element of the derivation is that the above continuum action is invariant under the discrete chiral symmetry
\begin{equation}
  \label{eq:R5_12}
  \mathcal{R}_5^{1,2} : 
  \begin{cases}
    \psi(x) \rightarrow i\gamma_5 \tau^{1,2}\psi(x)
    \\
    \bar{\psi}(x) \rightarrow \bar{\psi}(x)i\gamma_5 \tau^{1,2}
  \end{cases}\,,
\end{equation}
or, in other words, the continuum action $S_0$ is chirally invariant.
However, the dimension-5 term $\mathcal{L}_{1}$ of the Symanzik expansion for the action (see \req{eq:Oa}) is odd under the discrete chiral symmetry $\mathcal{R}_{5}^{1,2}$.
Recalling the Symanzik effective field theory (SymEFT) expression for a generic connected $n$-point correlation function $G^{(n)}$ of multilocal fields (see Sect.~\ref{sec:SymEFT} for further details, including the explicit case of a two-point function),
\begin{equation}
  G^{(n)}  = G^{(n)}_0 -a \int\dd^4 y \left\langle \phi_0(x_1) \ldots \phi_0(x_n) \mathcal{L}_1(y) \right\rangle_0 + a G^{(n)}_1  + \mbox{O}(a^2)\,,
  \label{eq:sym_tm_aux}
\end{equation}
if $G^{(n)}_0= \left\langle\phi_0(x_1) \ldots \phi_0(x_n) \right\rangle_0$ is chirally even, then the integrand in the right hand side of the above expression vanishes.
The $\Oa$ counterterm $G^{(n)}_{1}$ for $G^{(n)}_0$ includes insertions of higher-dimensional local fields with the same transformation properties as $\phi_{0}$ under the symmetries of the lattice theory.
It can be shown that $G^{(n)}_{1}$ has opposite transformation properties under \req{eq:R5_12} with respect to $G^{(n)}_{0}$ and, therefore, also the third term on the right-hand side of \req{eq:sym_tm_aux} vanishes.
For chirally-even operators, the SymEFT expansion thus coincides with the expectation value in the continuum theory up to $\Oasq$.
Consequently, for Wilson twisted-mass theories at full twist, no improvement coefficients are necessary, and physical observables are automatically $\Oa$-improved.
The only tuning required is that to maximal twist, which is achieved by setting the standard quark masses to vanish up to $\Oa$ effects.
In Appendix~\ref{app:Oa}, we present an account of the $\Oa$-improvement analysis of a generic regularisation employing Wilson-type fermions with non-degenerate quarks.
In the case of the considered mixed action with $\Oa$-improved Wilson sea quarks and Wilson twisted mass valence quarks, it is observed that when the valence sector is set to maximal twist, only residual  mass-dependent lattice artefacts proportional to the trace of the sea-quark mass matrix $\mathbf{m}^{\isea}$, introduced in \req{eq:msea}, survive at $\Oa$ for physical quantities.
In perturbation theory these effect arise from sea-quark loops with a leading contribution of $\mbox{O}(\alphas^2)$.
Consequently these terms are of $\mbox{O}\left( \alphas^2 \, a\trmsea\right)$ and are therefore expected to be very small in the $\NF=2+1$ theory, where the up, down, and strange quark masses in lattice units remain small.
On the other hand, an analysis along the lines of the aforementioned argument about automatic $\Oa$-improvement reveals that the symmetries of the lattice theory ensure that the inclusion of neither mass-independent nor valence-quark mass-dependent improvement coefficients is necessary at maximal twist.
We therefore expect that the considered mixed-action setup will effectively result in leading lattice artefacts of $\Oasq$ in physical quantities, with the exception of residual  $\mbox{O}\left( \alphas^2 \, a\trmsea\right)$  cutoff effects.
This type of lattice artefacts proportional to $\trmsea$ also arises in the context of the unitary setup.
In this study, we will neglect these residual contributions and perform accurate continuum-limit scaling tests that will allow to assess the validity of this approximation based on the precision of the current data.
%



\section{Computational setup and matching of sea and valence sectors}
\label{sec:match}

The use of different discretisations of the Dirac operator in the sea and valence is often motivated by the possibility to exploit a favourable property of the fermionic valence action on pre-existing gauge field configuration ensembles.
As such, it has been a common approach in lattice QCD.
An archetypal example is perhaps the quenched approximation, whereby the sea quarks can be interpreted as being infinitely heavy.
Partial quenching setups, in which additional valence quark flavours that are absent in the sea, or where different masses of a given quark flavour, are employed in the sea and valence sectors, are also being considered.\footnote{We refer to Refs.~\cite{FlavourLatticeAveragingGroupFLAG:2024oxs,Aliberti:2025beg} for comprehensive reviews.
}

A mismatch between the sea and valence quark masses can result in unphysical effects, such as the enhancement of chiral logarithms when the valence pion mass is smaller than the corresponding sea mass~\cite{Sharpe:1997by}.
In this context, differences in the discretisation of the fermionic actions can give rise to additional sources of lattice artefacts, which in turn may result in distortions of the fermionic correlation functions~\cite{Golterman:2005xa}.
Furthermore, disparities between the spectrum of the sea and the valence discretisations of the Dirac operator can, in certain regions of parameter space, result in significant fluctuations that are not effectively mitigated by the fermionic determinant~\cite{Cichy:2010ta,Cichy:2012vg}.
The restoration of unitarity in the continuum limit thus necessitates a careful procedure to match the sea and valence quark masses.
A critical validation of a mixed-action setup consists of performing a universality test, whereby continuum-limit results for well-controlled observables are compared with the corresponding results obtained using a unitary setup.
We will now proceed to the description of the gauge ensembles and the target chiral trajectory to the physical point.
As part of this discussion, we will describe how the physical point is approached through a line of constant physics involving a constraint on quark masses.
In order to impose the latter with high precision, a procedure to correct for small differences between target and simulated masses is applied.
This topic is covered in Sec.~\ref{subsec:chiral-traj}.
We will then discuss in Sec.~\ref{subsec:obs} the lattice observables considered in this work.
Some remarks about effects related to boundary conditions will be provided in Sec.~\ref{subsec:obc}.
Finally, the details of the matching strategy imposing the equality of sea and valence quark masses are discussed in Sec.~\ref{subsec:matching}.

\subsection{Gauge ensembles and chiral trajectory}
\label{subsec:chiral-traj}

\begin{table}[ht!]
  \begin{center}
    \begin{tabular}{c c @{\hspace{2em}} c c c c c c c c c}
      \toprule
      $\beta$ & \makecell{$a$ \\ $[\mathrm{fm}]$} & id & $L/a$ & $T/a$ & \makecell{$m_{\pi}$ \\ $[\mathrm{MeV}]$} & \makecell{$m_{K}$ \\ $[\mathrm{MeV}]$} & $m_{\pi}L$ & LMD & MDU & $N_\mathrm{cnfg}$ \\
      \toprule
      3.40 & 0.085 & H101r000 & 32 & 96 & 426 & 426 & 5.8 & no & 4004 & 1001 \\ 
      & & H101r001 & 32 & 96 & 426 & 425 & 5.8 & no & 4036 & 1009 \\
      \cline{3-11}
      & & H102r001 & 32 & 96 & 360 & 446 & 4.9 & yes & 3988 & 997 \\
      \cline{3-11}
      & & H102r002 & 32 & 96 & 360 & 446 & 4.9 & yes & 4032 & 1008 \\
      \cline{3-11}
      & & H105r001 & 32 & 96 & 286 & 470 & 3.9 & yes & 3788 & 947 \\
      & & H105r002 & 32 & 96 & 286 & 470 & 3.9 & yes & 4168 & 1042 \\
      \cline{3-11}
      & & H105r005 & 32 & 96 & 286 & 470 & 3.9 & no & 3348 & 837 \\
      \midrule
      3.46 & 0.075 & H400r001 & 32 & 96 & 429 & 429 & 5.2 & no & 2020 & 505 \\
      &  & H400r002 & 32 & 96 & 429 & 429 & 5.2 & no & 2160 & 540 \\
      \cline{3-11}
      &  & D450r011 & 64 & 128 & 221 & 483 & 5.4 & yes & 4000 & 250 \\
      \midrule
      3.55 & 0.063 & N202r001 & 48 & 128 & 418 & 418 & 6.5 & no & 3596 & 899 \\
      \cline{3-11}
      & & N203r000 & 48 & 128 & 350 & 448 & 5.4 & yes & 3024 & 756 \\
      & & N203r001 & 48 & 128 & 350 & 448 & 5.4 & yes & 3148 & 787 \\
      \cline{3-11}
      & & N200r000 & 48 & 128 & 288 & 470 & 4.4 & yes & 3424 & 856 \\
      & & N200r001 & 48 & 128 & 288 & 470 & 4.4 & yes & 3424 & 856 \\
      \cline{3-11}
      & & D200r000 & 64 & 128 & 204 & 488 & 4.2 & yes & 8004 & 2001 \\
      \cline{3-11}
      & & E250r001 & 96 & 192 & 131 & 497 & 4.0 & yes & 4000 & 100 \\
      \midrule
      3.70 & 0.049 & N300r002 & 48 & 128 & 427 & 427 & 5.1 & no & 6084 & 1521 \\
      \cline{3-11}
      & & N302r001 & 48 & 128 & 350 & 457 & 4.2 & no & 8804 & 2201 \\
      \cline{3-11}
      & & J303r003 & 64 & 192 & 261 & 481 & 4.1 & yes & 8584 & 1073 \\
      \cline{3-11}
      & & E300r001 & 96 & 192 & 177 & 499 & 4.2 & no & 4540 & 227 \\
      \midrule
      3.85 & 0.039 & J500r004 & 64 & 192 & 418 & 418 & 5.2 & no & 6288 & 198 \\
      & & J500r005 & 64 & 192 & 418 & 418 & 5.2 & no & 5232 & 130 \\
      & & J500r006 & 64 & 192 & 418 & 418 & 5.2 & no & 3096 & 94 \\
      \cline{3-11}
      & & J501r001 & 64 & 192 & 339 & 453 & 4.3 & no & 6536 & 371 \\
      & & J501r002 & 64 & 192 & 339 & 453 & 4.3 & no & 4560 & 248 \\
      & & J501r003 & 64 & 192 & 339 & 453 & 4.3 & no & 4296 & 168 \\
      \bottomrule
    \end{tabular}
  \end{center}
  \caption{
    Set of $\NF=2+1$ CLS gauge ensembles employed in this work.
    The table reports the values of the bare coupling $\beta = 6/g_{0}^{2}$, approximate values of the lattice spacings $a$, the ensemble name, the spatial and temporal lattice extents ($L/a$ and $T/a$, respectively), as well as the approximate pion and kaon masses, $m_{\pi}$ and $m_{K}$ (prior to mass shifting, cf.~\req{eqn:mass_shift}), and the values of $m_{\pi}L$.
    Additionally, the column LMD indicates whether low-mode deflation was employed in the computation of the reweighting factors~\cite{Kuberski:2023zky}.
    The length of the Monte Carlo chain in Molecular Dynamics Units (MDU) and the corresponding number of configurations $N_\mathrm{cnfg}$ employed in the computation of the mesonic correlation functions are also provided.
    For all ensembles, the number of stochastic sources $N_\mathrm{s}$ employed is $\mbox{O}(10)$.
    Open boundary conditions in the time direction are employed for the gauge fields, with the exception of ensembles E250 and D450, which have (anti-)periodic boundary conditions in time.
  }
  \label{tab:CLS_ens}
\end{table}
The set of CLS gauge ensembles considered in this work is listed in Table~\ref{tab:CLS_ens}.
They were generated with $\NF=2+1$ flavours of non-perturbatively $\Oa$-improved Wilson fermions  and the tree-level $\mbox{O}(a^2)$-improved L\"uscher-Weisz gauge action.
The considered ensembles employ open boundary conditions (OBC) in the temporal direction --- to mitigate the slowdown in the sampling of topological sectors at fine values of the lattice spacing --- with the exception of two ensembles, E250 and D450, that employ (anti-)periodic boundary conditions in time (which we will denote by PBC).
Further details regarding the role of the boundary conditions are reported in Sec.~\ref{subsec:obc} and in Appendix~\ref{app:obc}.
A reweighting procedure is applied to physical observables to eliminate (i) the effect of a twisted mass regulator in the light-quark sector used to stabilise the dynamical simulations~\cite{Luscher:2012av}, (ii) the rational approximation employed for simulating the strange quark with the RHMC algorithm~\cite{Clark:2006fx}; and (iii) isolated cases where the fermionic determinant of the strange quark changes sign~\cite{Mohler:2020txx}.
As indicated in Table~\ref{tab:CLS_ens}, for various ensembles, the reweighting factors (i) and (ii) were computed using the exact low-mode deflation method described in Ref.~\cite{Kuberski:2023zky}.
Further details on the reweighting procedure can be found in Appendix~\ref{app:flagged_s}.
At each value of the bare coupling, the ensembles were generated at a constant value of the trace of the bare sea quark mass matrix $\mathbf{m}^{\isea}$ introduced in \req{eq:mmatsea},
\begin{equation}
  \label{eqn:TrMq}
        \trmsea=m_u^{\isea}+m_d^{\isea}+m_s^{\isea}=\mathrm{cnst}\,.
\end{equation}
This choice implies that the trace in \req{eqn:TrMq}, expressed in terms of the bare subtracted quark masses,
\begin{equation}
  \label{eqn:msubtracted}
  m_i^{\isea}= m_{0,i}^{\isea} - \mcrit  = \frac{1}{2a} \left( \frac{1}{\kappa_i^{\isea}}-\frac{1}{\kappa_\mathrm{cr}} \right)\,,
\end{equation}
is held constant at each value of the bare coupling, since in a mass-independent renormalisation scheme the critical value of the hopping parameter $\kappa_{\mathrm{cr}}$ is a function of $g_0^2$.
In \req{eqn:msubtracted}, the bare unsubtracted quark mass is given by $m_{0,i}^{\isea} = \frac{1}{2a} \left( 1/\kappa_i^{\isea} - 8 \right)$ in terms of the sea-sector hopping parameter $\kappa_i^{\isea}$.
The improved bare gauge coupling
\begin{equation}
  \label{eq:impcoupling}
  \tilde{g}_0=g_0\left[1+\frac{a b_g}{\NF} \trmsea\right]\,,
\end{equation}
is thus kept constant when varying the quark masses according to \req{eqn:TrMq}, at each value of the lattice spacing, irrespective of the knowledge of the improvement coefficient $b_g$.
The aim of this procedure is to adjust the quark masses via \req{eqn:TrMq} in such a way that the combination of pion and kaon masses $m_K^2 + \frac{1}{2} m_\pi^2$ remains approximately constant in the vicinity of its isoQCD physical value, as expected from leading order (LO) chiral perturbation theory ($\chi$PT).
In practice, however, dimensionless quantities expressed in units of the gradient flow scale $t_0$ are considered.
We will thus use the following quantities
\begin{align}
  \phi_2&=8t_0m_{\pi}^2\,, \label{eqn:phi2}\\
  \phi_4&=8t_0\left(m_K^2+\frac{1}{2}m_{\pi}^2\right)\,, \label{eqn:phi4}
\end{align}
to locate the masses in the quark mass plane.
Further details on the determination of the gradient flow scale $t_0$ can be found in Sec.~\ref{subsec:t0lat}.
The description of the computation of fermionic observables in the sea and valence sectors is provided in Sec.~\ref{subsec:hadrons}.

\begin{figure}[!t]
  \begin{center}
    \includegraphics[width=0.8\linewidth]{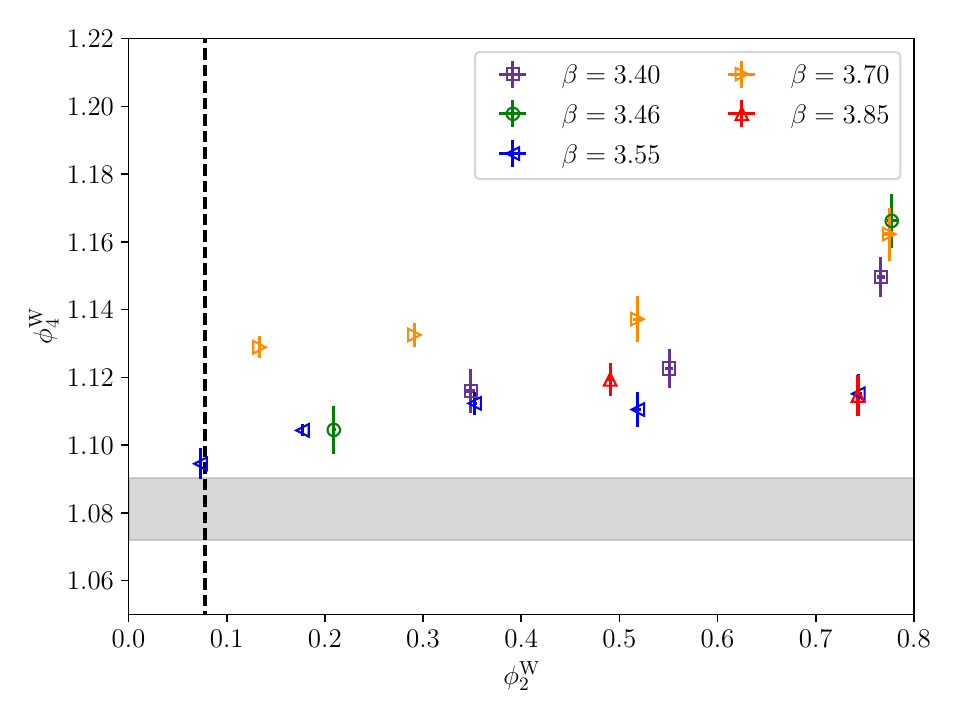}
  \end{center}
  \caption{
    The unshifted determinations of $\phi_4^{\mathrm{W}}$, as defined in \req{eqn:phi4}, for the ensembles considered in Table~\ref{tab:CLS_ens} are shown as a function of $\phi_2^{\mathrm{W}}$.
    The grey horizontal band represents the physical value, as determined from the scale setting analysis below (cf. \req{ch_ss:eq:phi4ph}), to which all observables will be mass-shifted, while the dashed vertical line marks the physical value of $\phi_{2}$ (cf. \req{ch_ss:eq:phi2ph}).
    These mass-corrections will also result in a slight displacement of the data points on the horizontal axis, i.e. on the light-quark mass proxied by $\phi_2$.
  }
  \label{fig:mass_shift_strange}
\end{figure}

\begin{figure}[!t]
  \begin{center}
    \includegraphics[width=.49\linewidth]{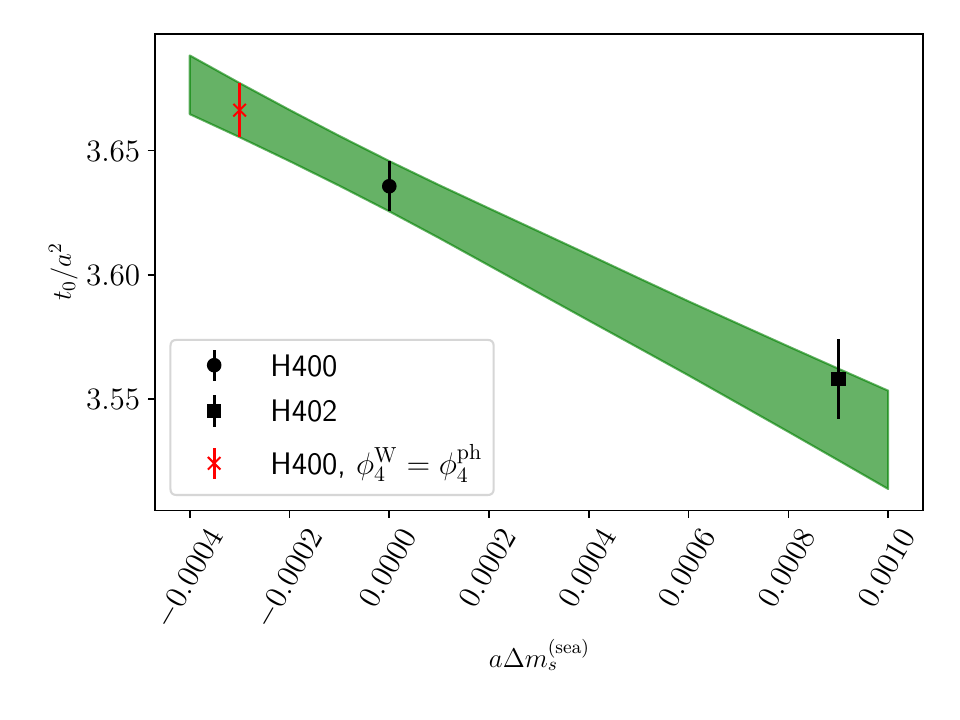}
    \includegraphics[width=.49\linewidth]{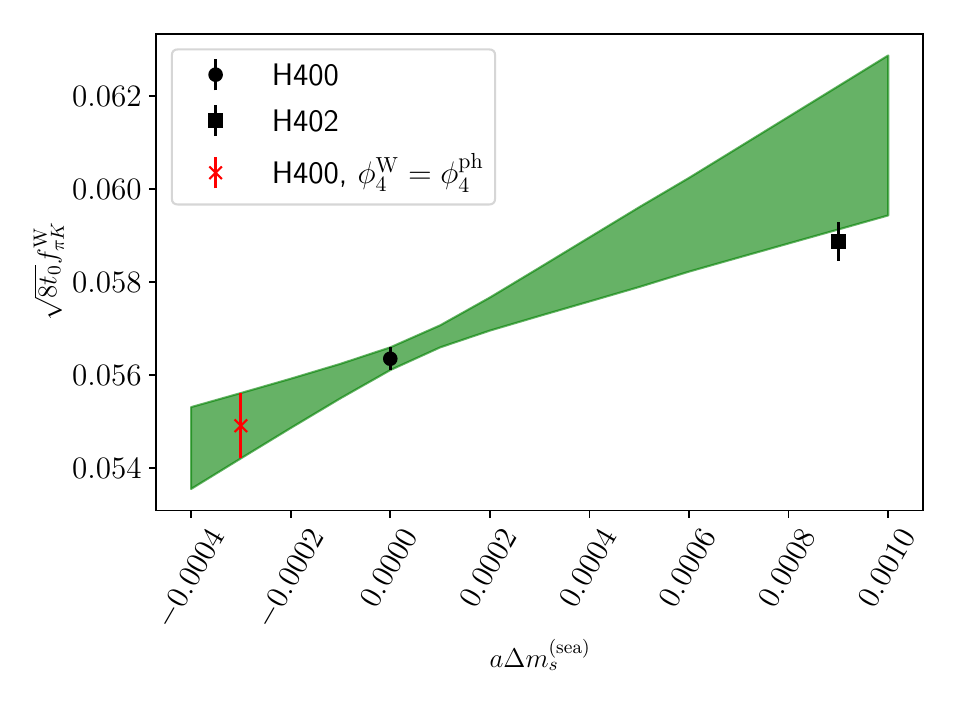}
  \end{center}
  \caption{
    Consistency checks of the mass-shifting procedure, in which direct measurements carried out on two CLS ensembles, H400 and H402, differing only by their sea quark masses, are compared to the Taylor expansion procedure in \req{eqn:mass_shift}.
    More specifically, we consider two observables, $t_0/a^2$ and $\sqrt{8t_0}f_{\pi K}$, from the unitary setup as a function of the quark mass-shift $\Delta m_i^{\isea}=m_i^{\prime\,{\isea}}-m_i^{\isea}$, starting from the H400 ensemble.
    The ensembles H400 and H402 share similar parameters, in particular $\beta=3.46,\;L/a=32,\;T/a=96$, and both lie on the SU(3) symmetric line where $m_\pi=m_K$.
    The H400 ensemble, indicated by the filled black circle, has a pion mass $m_\pi \approx 420\,\mathrm{MeV}$.
    In the case of the H402 ensemble (black square) with $m_\pi \approx 450\,\mathrm{MeV}$, the measurements reported in Ref.~\cite{Bruno:2016plf} are employed.
    The green band represents the mass-shift starting from the H400 point, summing over the three quarks flavours, $i=u,d,s$, in the last term on the right-hand side of \req{eqn:mass_shift}.
    The red cross symbol illustrates the position of the mass-shift from the H400 measurement to the physical value $\phi_4=\phi_4^{\mathrm{ph}}$, as determined in \req{ch_ss:eq:phi4ph}, required to fix the chiral trajectory. 
    In this regard, it is noteworthy that the H400 ensemble requires the largest shift among all the considered ensembles.
    We observe that a first order Taylor expansion allows to carry out these small mass-corrections at the expense of increased statistical uncertainty.
  }
  \label{fig:mass_shift}
\end{figure}

The simulation parameters used to impose \req{eqn:TrMq} may result in small deviations for the target value,
\begin{equation}\label{eqn:TrMqR}
  \phi_4^\mathrm{W} = \phi_4^{\mathrm{ph}} = \mathrm{cnst}\,,
\end{equation}
where $\phi_4^\mathrm{W}$ refers to the value of $\phi_4$ in the sea sector, i.e. computed in the unitary setup based on Wilson fermions, and $\phi_4^{\mathrm{ph}}$ denotes the isoQCD physical value of $\phi_4$, as determined from the scale setting analysis (cf. \req{ch_ss:eq:phi4ph}).
Fig.~\ref{fig:mass_shift_strange} indicates that the maximum deviations are at the 8\% level.
However, we aim to correct these small deviations so that the condition in \req{eqn:TrMqR}, which defines the renormalised chiral trajectory to the physical point, is enforced across all ensembles.
In order to impose \req{eqn:TrMqR}, it is thus necessary to perform small mass-shifts in the simulated sea quark masses $\left\{m_i^{\isea}\right\}_{i=u,d,s}$.
Following the procedure proposed in Ref.~\cite{Bruno:2016plf}, these mass-corrections will be implemented through a first-order Taylor expansion in the simulated sea quark masses.
For a given lattice observable $\mathcal{O}$, the first-order Taylor expansion reads
\begin{equation}
  \label{eqn:mass_shift}
  \mathcal{O} \left( \{ m_{i}^{\prime\,{\isea}} \}_{i=u,d,s} \right) = \mathcal{O} \left( \{ m_{i}^{\isea} \}_{i=u,d,s} \right) + \sum_{i=u,d,s} \left(m_i^{\prime\,{\isea}}-m_i^{\isea}\right)\frac{d\mathcal{O}}{dm_i^{\isea}}\,.
\end{equation}
In this context, we distinguish a \emph{primary} observable, whose estimator is computed directly as an average over configurations --- for example, the value of a two-point correlation function of local operators --- from a \emph{derived} observable, which is a function of the former --- for example, an effective mass or its average over a plateau to obtain a hadron mass.
In the case where $\mathcal{O}$ is a derived observable, $\mathcal{O}(\left< P_j \right>)$, depending on a set of primary observables $\left\{P_j\right\}_{j=1,2,...}$, the derivative term on the right-hand side of \req{eqn:mass_shift} can be expressed as
\begin{equation}
  \label{eqn:mass_shift2}
  \frac{d\mathcal{O}}{dm_i^{\isea}}=\sum_j\frac{\partial \mathcal{O}(\left< P_j \right>)}{\partial\left<P_j\right>}\left[\left<\frac{\partial P_j}{\partial m_i^{\isea}}\right>-\left<P_j\frac{\partial S}{\partial m_i^{\isea}}\right>+\left<P_j\right>\left<\frac{\partial S}{\partial m_i^{\isea}}\right>\right]\,,
\end{equation}
where $S$ is the lattice action employed in the sea sector. 
In practice, the set of considered primary observables $P_j$ will correspond to the gradient flow energy density and mesonic two-point correlation functions.
Derived observables will then correspond, e.g., to the gradient flow scale $t_0$, mesonic masses and decay constants, PCAC quark masses, and combinations thereof.
Of particular relevance will be the flavour-averaged combination $f_{\pi K}$ of the pion and kaon decay constants~\cite{Bruno:2016plf}
\begin{equation}
  \label{eq:fpik}
  f_{\pi K}=\frac{2}{3}\left(f_K+\frac{1}{2}f_{\pi}\right)\,,
\end{equation}
that will play a key role in scale setting.
A more detailed description of the considered observables is provided in Secs.~\ref{subsec:hadrons} and~\ref{subsec:t0lat}.
For the case of a gluonic quantity such as $t_0$, the quark mass-dependence only arises from sea quark effects, embodied in the derivative of the sea action in \req{eqn:mass_shift2}.
On the other hand, we emphasise that the sea-quark mass-corrections in \req{eqn:mass_shift} must also be applied to mixed-action observables.
As the derivatives on the right-hand side of \req{eqn:dOdphi4} are taken with respect to the sea-quark masses, a mixed-action observable $\mathcal{O}$ will only receive contributions from the derivatives involving the action $S$ in \req{eqn:mass_shift2}.
A more detailed examination of this point will be provided in Sec.~\ref{subsec:matching}.
In Fig.~\ref{fig:mass_shift} we illustrate the effect of the mass-shift on two derived observables from the unitary setup.
As a consistency check, a comparison to direct measurements on two CLS ensembles is also provided. These results corroborate the results from Ref.~\cite{Bruno:2016plf} where even larger mass-shifts were tested.
The checks demonstrate that a first-order Taylor expansion enables the application of these small mass-corrections, albeit at the cost of a slight decrease in statistical precision.
In \req{eqn:mass_shift}, the sum over the flavour index $i$ can be performed in various ways, depending on the direction in the quark mass plane that is selected in order to reach the target chiral trajectory in \req{eqn:TrMqR}.
The sum can for instance be taken over all three flavours $i=u,d,s$, with a uniform mass-shift applied to each, or it can be taken over a subset of the flavours.
Following the approach of Refs.~\cite{Strassberger:2021tsu,Strassberger:2023xnj}, we opt to apply a mass-shift exclusively to the strange quark, thereby restricting the sum to $i=s$, since this leads to a higher statistical precision on the target result.
In order to shift to a constant value of $\phi_4^{\mathrm{W}}$ across all ensembles and satisfy the condition in \req{eqn:TrMqR}, it is convenient to recast \req{eqn:mass_shift} in the following form~\cite{Strassberger:2023xnj}
\begin{equation}
  \label{eqn:derphi4}
  \mathcal{O}\left(\phi_4^{\mathrm{ph}}\right)=\mathcal{O}\left(\phi_4^\mathrm{W}\right)+\left(\phi_4^{\mathrm{ph}}-\phi_4^{W}\right)\frac{d\mathcal{O}}{d\phi_4^\mathrm{W}}\,,
\end{equation}
where
\begin{equation}
  \label{eqn:dOdphi4}
  \frac{d\mathcal{O}}{d\phi_4^\mathrm{W}}=\frac{d\mathcal{O}/dm_s^{\isea}}{d\phi_4^\mathrm{W}/dm_s^{\isea}}\,.
\end{equation}
The derivative in \req{eqn:dOdphi4} is an observable with a well-defined continuum limit.
Its physical value can be determined through a global fit of lattice data to carry out the continuum-chiral extrapolation. 
The information contained in such a global description of the lattice data can then be employed to improve the precision of the derivatives in \req{eqn:dOdphi4}, particularly for the noisier ensembles.
Additionally, it allows to incorporate in the analysis ensembles for which  direct measurements of the derivatives are unavailable, as it is  the case for ensembles D450, E250, E300, J500 and J501 in Table~\ref{tab:CLS_ens}.
The set of observables involved in the scale setting analysis --- and thus requiring mass-shifting --- are $\sqrt{8t_0}f_{\pi}$, $\sqrt{8t_0}f_{K}$, $\sqrt{8t_0}f_{\pi K}$, $\phi_2$, and $\phi_4$.
In the case of the latter, the derivative with respect to $\phi_4^{\mathrm{W}}$ is equal to one in the context of the unitary Wilson regularisation.
However, this will not be the case in the mixed action for the derivative of $\phi_4^{\mathrm{Wtm}}$ with respect to $\phi_4^{\mathrm{W}}$.
The generic definition of the PCAC quark mass $m_{ij}$ is given in Eqs.~(\ref{eqn:mpcac}) and (\ref{eqn:mPCACI}).\,\footnote{
In the following sections reporting numerical results, we find it convenient to employ the pairs of flavour indices $ij = 12, 13$ to refer to the non-singlet combinations up--down and up--strange, respectively.
Using this notation, for instance, the pseudoscalar meson masses of the pion and kaon correspond to $m_{\pi} = m_{\mathrm{\scriptscriptstyle PS}}^{12}$ and $m_{K} = m_{\mathrm{\scriptscriptstyle PS}}^{13}$, respectively.
Furthermore, the flavour indices $34$ will denote a non-singlet combination involving two distinct strange-quark flavours, taken to be mass-degenerate.
}
In the mixed-action regularisation, the valence Wtm light PCAC quark mass is employed in the tuning to maximal twist\,\footnote{
As previously indicated, the superscript ``$\mathrm{Wtm}$'' denotes observables computed in the mixed-action setup using Wilson twisted mass valence quarks, while ``$\mathrm{W}$'' refers to observables from the unitary setup with Wilson fermions.
},
consequently, the mass derivative of the quantity $\sqrt{8t_0}m_{12}^{\mathrm{Wtm,\,R}}$ must also be determined, where the superscript ``$\mathrm{R}$'' indicates that the renormalised quark mass is to be considered.
In the combined continuum–chiral fits of the derivatives $d\mathcal{O}/d\phi_4^{\mathrm{W}}$, we find that the lattice data are well described by the  functional form
\begin{equation}
  \label{eqn:md_1}
  F=\frac{d\mathcal{O}}{d\phi_4^\mathrm{W}}=A+B\phi_2+C\frac{a^2}{t_0}\,,
\end{equation}
for all observables $\mathcal{O}$ except for the light PCAC quark mass
$\sqrt{t_0}m_{12}^{\mathrm{Wtm\,,R}}$.
In this case, additional terms are required in order to describe the mass and lattice-spacing dependence adequately, viz.
\begin{equation}
  \label{eqn:md_2}
  F=\frac{d\mathcal{O}}{d\phi_4^\mathrm{W}}=A+B\phi_2+C\phi_2^2+(D+E\phi_2)\frac{a^2}{t_0}\,.
\end{equation}
The fit parameters corresponding to Eqs.~(\ref{eqn:md_1}) and (\ref{eqn:md_2}) are reported in Appendix~\ref{app:derivatives}, while Fig.~\ref{fig:dfpik_w} illustrates the mass and lattice-spacing dependence of some of the relevant observables.
\begin{figure}[!t]
  \centering
  \includegraphics[width=.49\textwidth]{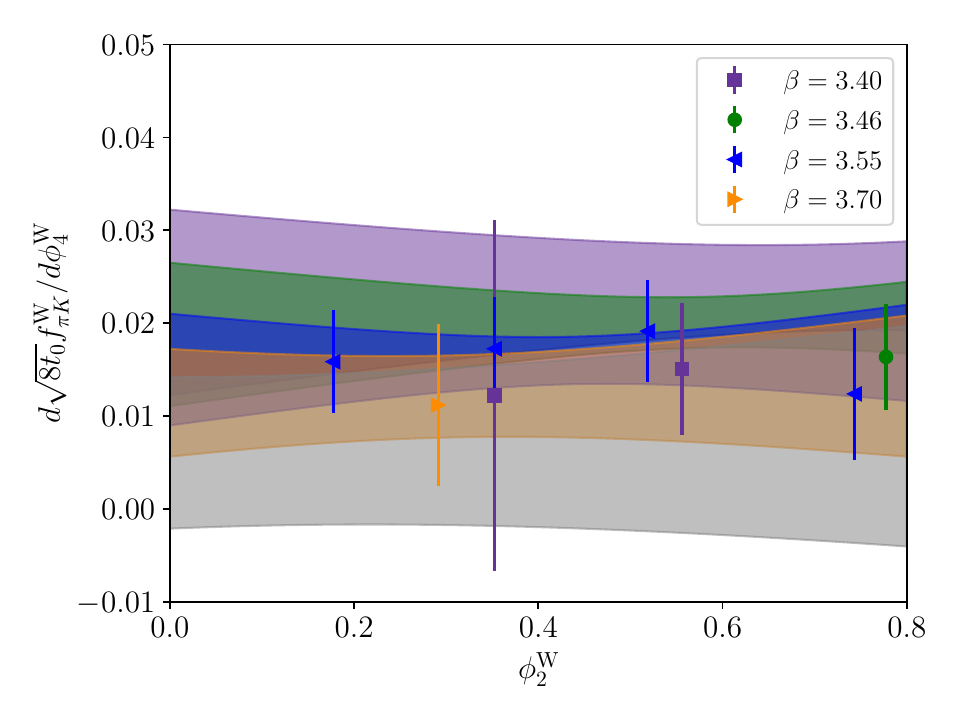}
  \includegraphics[width=.49\textwidth]{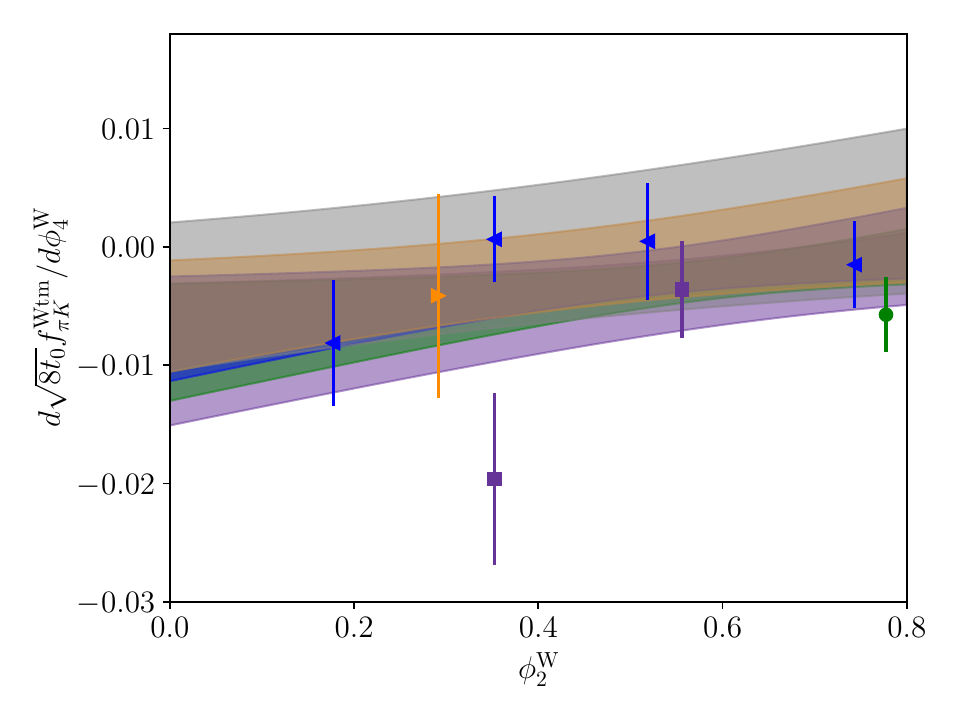}
  \includegraphics[width=.49\textwidth]{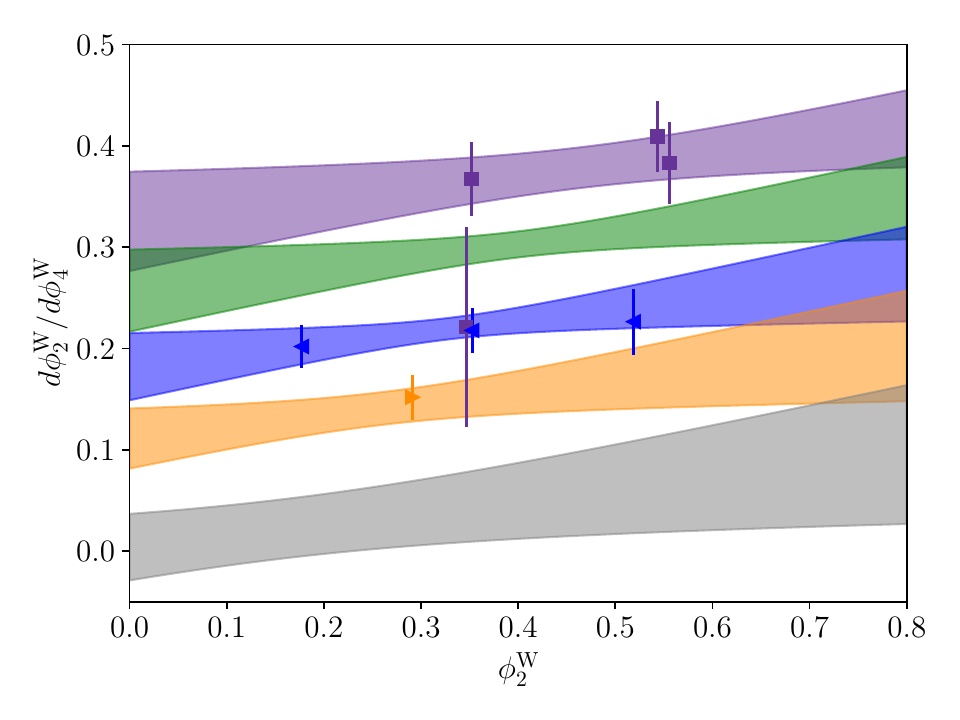}
  \includegraphics[width=.49\textwidth]{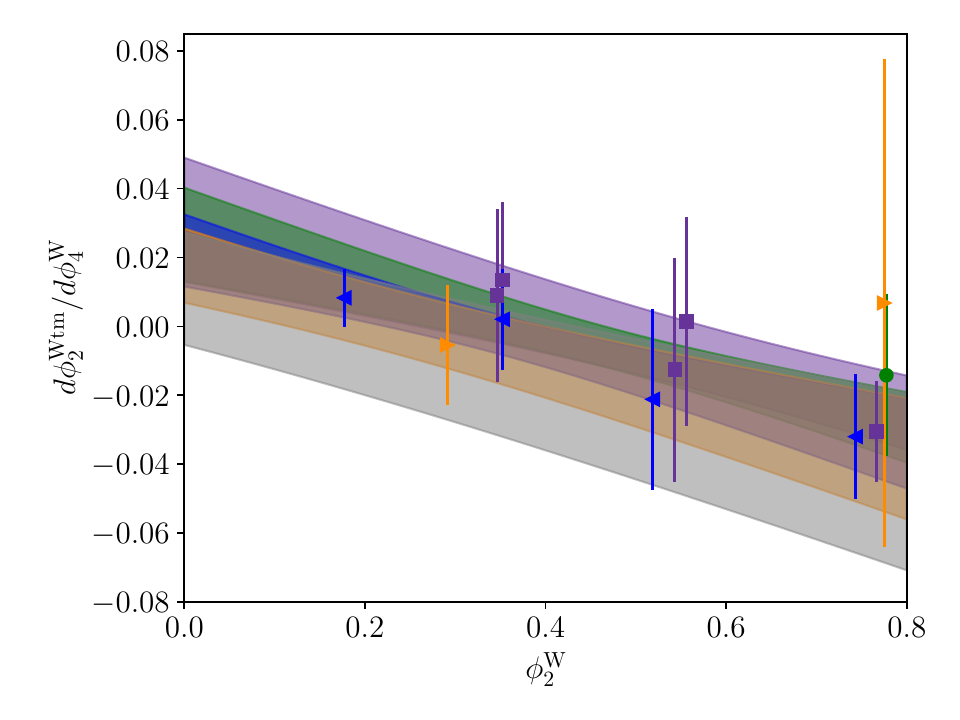}
  \caption{
    \textit{Top row}:
    Derivative $d\left(\sqrt{8t_0}f_{\pi K}\right)/d\phi_4^{\mathrm{W}}$, defined in \req{eqn:dOdphi4}, for the unitary ``W'' setup (left panel), and the mixed-action ``Wtm'' setup (right panel) as a function of $\phi_2^{\mathrm{W}}$.
    The coloured bands represent the result of the fit in \req{eqn:md_1} projected to each value of $\beta$.
    The grey bands correspond to the continuum-limit mass-dependence.
    The corresponding values of the fit parameters are collected in Table~\ref{tab:md}.
    Both p-values are approximately $0.2$.
    In the unitary setup, the derivative involves all the terms in the right-hand side of \req{eqn:mass_shift2}. In contrast, in the mixed-action regularisation, the derivative is a function of the sea contribution alone, which arises from the terms depending on the action $S$.
    In this latter case, the derivative is employed to impose the physical value of $\phi_{4}$ in the sea sector, whereas the corresponding condition in the valence sector is implemented through the matching procedure described in Sec.~\ref{sec:match}.
    \textit{Bottom row}:
    As with the top row, these panels display the corresponding derivatives of the observable $\phi_2$.
    Both p-values in this case are approximately $0.9$.
  }
  \label{fig:dfpik_w}
\end{figure}
At the $\mathrm{SU}(3)$ symmetric point, where up, down and strange quarks are mass-degenerate, the quantity $\phi_2^\mathrm{W}$ satisfies $\phi_2^\mathrm{W} = \tfrac{2}{3} \phi_4^\mathrm{W}$ by construction; thus $d\phi_2^{\mathrm{W,\,sym}}/d\phi_4^\mathrm{W} = 2/3$, an exact relation that we use instead of a fitted derivative.
As previously stated, the derivatives in \req{eqn:dOdphi4} are employed to apply a small mass-shift to $\phi_4^\mathrm{ph}$, the physical value of $\phi_4$, following \req{eqn:derphi4}.
However, the target value of $\phi_4^\mathrm{ph}$ can only be determined once the physical value of $t_0$ has been obtained as a result of the scale setting procedure.
For clarity, we summarise below the iterative procedure adopted to determine the target value of $\phi_{4}^{\mathrm{ph}}$ applied in the mass-shift calculations; a detailed description of the scale-setting procedure will be presented in Sec.~\ref{sec:results}.
The process begins with the selection of an educated guess, which is otherwise arbitrary, for the value of $t_0^{\mathrm{ph}}$.
This value may be selected without error, and it provides an initial estimate for the value of $\phi_4^{\mathrm{ph}}$.
The application of the full scale setting procedure results in a determination of $t_0^\mathrm{ph}$, which now includes an uncertainty that takes into account the correlations present in the lattice data.
This determination of $t_0^{\mathrm{ph}}$ serves to establish the value of $\phi_4^{\mathrm{ph}}$ to which the next iterative step of the mass-shift is applied.
After a few iterative steps, the procedure is stopped when convergence is achieved for the value of $t_0^{\mathrm{ph}}$.
The iterative procedure is illustrated in Fig.~\ref{ch_ss:fig:iter}.

\begin{figure}[!t]
  \begin{center}
    \includegraphics[width=0.8\linewidth]{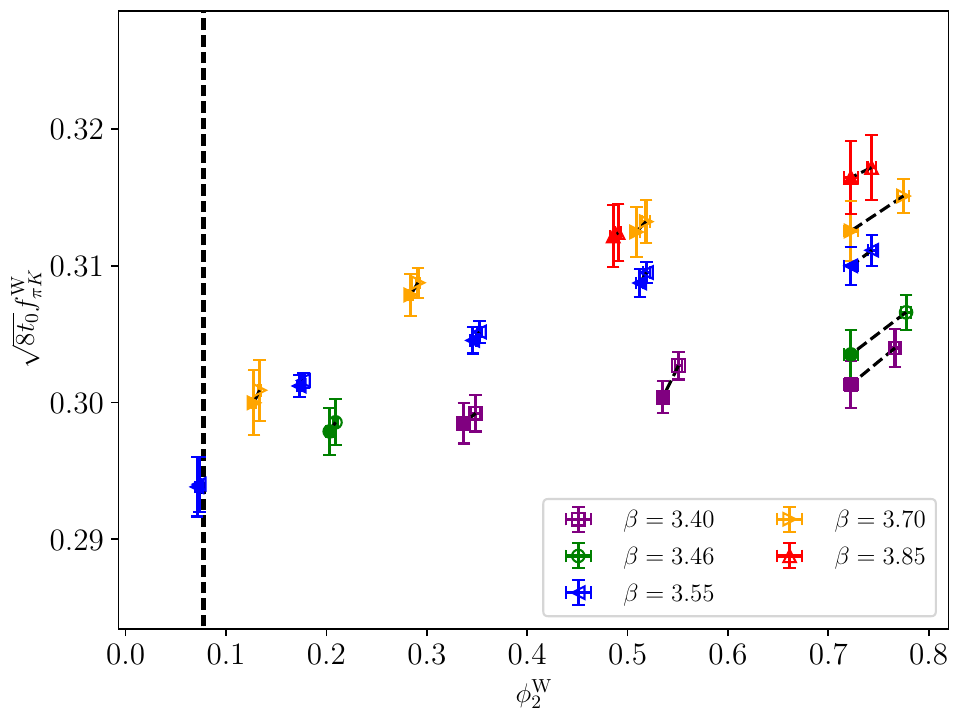}
  \end{center}
  \caption{
    Comparison of the unshifted and shifted determinations of $\sqrt{8t_0}f_{\pi K}$ for the unitary setup, where $f_{\pi K}$ is defined in \req{eq:fpik}.
    As discussed in the text, the mass-shifts to  $\phi_4^\mathrm{ph}$ based on \req{eqn:mass_shift} and \req{ch_ss:eq:phi4ph} are only applied to the strange quark mass.
    An empty symbol represents the unshifted determination, whereas the corresponding shifted result, indicated by a filled symbol, is connected by a black dashed line.
    The vertical dashed line represents the value of $\phi_2^{\mathrm{ph}}$ as determined from the scale-setting analysis (cf.~\req{ch_ss:eq:phi2ph}).
  }
  \label{fig:mass_shift_strangef}
\end{figure}

Figure~\ref{fig:mass_shift_strangef} displays the unshifted determination of $\sqrt{8t_0}f_{\pi K}$ as a function of $\phi_2$ in the unitary setup.
The impact of shifting $\phi_4$ to its physical value $\phi_4^\mathrm{ph}$, as defined in \req{ch_ss:eq:phi4ph}, is also illustrated.

\subsection{Computation of observables}
\label{subsec:obs}

The following section presents an account of the computation of lattice observables based on fermionic correlation functions in the context of both the unitary and mixed-action setups. Additionally, it provides information about the determination of the gradient flow scale $t_0$.

\subsubsection{Fermionic correlation functions}
\label{subsec:hadrons}

The fermionic observables considered in this work are the pion and kaon masses, $m_{\pi}$ and $m_K$, respectively, as well as the corresponding decay constants, $f_{\pi}$ and $f_K$.
These quantities are determined in both the unitary and mixed-action setups.
Furthermore, in the unitary setup, the PCAC quark mass $m_{ij}^{\mathrm{W}}$ is employed in the $\Oa$-improvement of the axial vector current and pseudoscalar density, while the light PCAC quark mass in the mixed-action setup, $m_{12}^{\mathrm{Wtm}}$, is required for the tuning of the valence Wilson twisted mass action to maximal twist.
As the same form of the correlation functions is applicable to both the unitary and mixed-action approaches, in the following, we will only distinguish between them, by adding the superscripts ``W'' and ``Wtm'', respectively, when the specific case needs to be identified.
In the context of the mixed-action setup, the quark fields that appear in the fermionic bilinears will be those of the twisted basis.
The relevant two-point correlation functions for extracting the aforementioned observables, projected to zero momentum, read
\begin{align}
  \label{eqn:fP}
  \fP^{ij}(x_0,y_0)&=\frac{a^6}{L^3}\sum_{\vec{x},\vec{y}}\left<P^{ij}(x_0,\vec{x})P^{ji}(y_0,\vec{y})\right>\,,\\
  \label{eqn:fA}
  \fA^{ij}(x_0,y_0)&=\frac{a^6}{L^3}\sum_{\vec{x},\vec{y}}\left<A_0^{ij}(x_0,\vec{x})P^{ji}(y_0,\vec{y})\right>\,,
\end{align}
defined in terms of the pseudoscalar density $P(x)$ and axial vector current $A_{\mu}(x)$ as follows\,\footnote{
To simplify the notation, in the following sections of the main text we depart from the definition of $A_{\mu}^{ij}(x)$ in \req{eq:current_notation} and adopt the useful definition given in \req{eqn:Amu}, which includes the mass-independent $\Oa$-improvement term proportional to the coefficient $c_{\mathrm{A}}$.
However, when discussing $\Oa$-improvement in Appendix~\ref{app:Oa}, we will return to the standard definition in \req{eq:current_notation}.}
\begin{align}
  \label{eqn:P}
  P^{ij}(x)&=\bar{\psi}_i(x) \gamma_5 \psi_j(x)\,,\\
  \label{eqn:Amu}
  A_{\mu}^{ij}(x)&=\bar{\psi}_i(x) \gamma_{\mu} \gamma_5 \psi_j(x) + a c_\mathrm{A} \tilde{\partial}_{\mu}P^{ij}(x)\,,
\end{align}
where the improvement coefficient $c_\mathrm{A}$ has  been  determined non-perturbatively in Ref.~\cite{Bulava:2015bxa}, and $\tilde{\partial}_{\mu}$ is the $\Oa$-improved lattice derivative, i.e. the symmetrised difference operator along direction $\mu$.
The renormalised pseudoscalar density and axial current in Eqs.~(\ref{eqn:P})~and~(\ref{eqn:Amu}), respectively, can be written as
\begin{align}
  \label{eqn:P_impr}
  P^{ij}_{\mathrm{\scriptscriptstyle R}} &=\ZP(g_0^2,a\mu_\mathrm{ren})\left(1+a\ibPb\trmsea+a\ibP m_{ij}^{\scriptscriptstyle (+)}\right)P^{ij}\,,\\
  \label{eqn:A_impr}
  A_{\mu, \mathrm{\scriptscriptstyle R}} ^{ij}&=\ZA(g_0^2)\left(1+a\ibAb\trmsea+a\ibA m_{ij}^{\scriptscriptstyle (+)}\right)A_{\mu}^{ij}\,,
\end{align}
where $m_{ij}^{\scriptscriptstyle (+)}$ is the subtracted mass for the flavour combination $i \neq j$, defined in \req{eq:avsubtm}.
In the unitary setup, the inclusion of the mass-dependent $\Oa$ counterterms, depending on the $b$-type coefficients in Eqs.~(\ref{eqn:P_impr})~and~(\ref{eqn:A_impr}), is sufficient to guarantee the $\Oa$-improvement of these fermionic bilinears.
As described in Appendix~\ref{app:Oa}, additional $\Oa$ terms depending on linear combinations of the valence twisted mass parameters $\mu_i$, corresponding to the coefficients $\ibPc$ in \req{eq:Pimp} and $\ibAc$ in \req{eq:Aimp}, arise when considering a valence Wilson twisted mass Dirac operator.
In the context of the mixed action, these terms induce the presence of the corresponding $\mu_i$-dependent $\Oa$ lattice artefacts in the tuning of the valence light PCAC quark mass to zero, i.e. when the maximal twist condition is imposed.
However, as described in Appendix~\ref{app:Oa}, the presence of these $\Oa$ effects in the tuning to maximal twist will not result in the emergence of associated $\Oa$ lattice artefacts in physical observables.
It is therefore unnecessary to explicitly include these $\Oa$-improvement coefficients proportional to the valence twisted-mass parameters in the mixed-action calculations.
As indicated in Table~\ref{tab:CLS_ens}, stochastic U(1) noise vectors~\cite{Michael:1998sg,Luscher:2010ae} are employed at the source position $y_0$ in Eqs.~(\ref{eqn:fP}) and (\ref{eqn:fA}) to compute the two-point correlation functions.
In the case of ensembles with open boundary conditions in the temporal direction, correlation functions are computed with sources at $y_0=a$ and $y_0=T-a$.
This allows for a symmetrisation with respect to the midpoint of the temporal extent $T$ through
\begin{equation}
  \label{eqn:sym}
  \fX^{ij}(x_0)=\frac{1}{2}\left[\fX^{ij}(x_0,a)\pm \fX^{ij}(x_0,T-a)\right]\,,
\end{equation}
where a positive sign is associated with $\mathrm{X}=\mathrm{P}$, while a negative sign is used for $\mathrm{X}=\mathrm{A}$.
In a slight abuse of notation, the omission of the argument $y_0$ in the two-point functions thus signifies that the symmetrisation, as defined in \req{eqn:sym}, has been taken.
The effects of the use of open boundary conditions in time will be examined in greater detail in Section~\ref{subsec:obc} and in Appendix~\ref{app:obc}.
The effective mass of the pseudoscalar meson with flavours $(i,j)$ is defined in terms of the pseudoscalar correlator $\fP^{ij}(x_0)$ in the standard way
\begin{equation}
  \label{eqn:meff}
  m_{\mathrm{\scriptscriptstyle PS, eff}}^{ij}(x_0)=-\frac{1}{a} \log\left(\frac{\fP^{ij}(x_0)}{\fP^{ij}(x_0+a)}\right)\,.
\end{equation}
The extraction of the corresponding ground-state pseudoscalar meson mass, $m_{\mathrm{\scriptscriptstyle PS}}^{ij}$, from a plateau interval in the effective mass requires a careful examination of the systematic effects due to the contamination from excited states and from boundary effects, when OBC are employed.
In the case of PBC, the pseudoscalar and axial correlators take the following generic form when restricting attention to the ground state and the first excited state,
\begin{align}
  \label{eqn:corrs_PBC}
  \fX^{ij}(x_0,y_0) &=a_{\mathrm{\scriptscriptstyle X}}^{ij} \left(e^{-m^{ij}_{\mathrm{\scriptscriptstyle PS}}(x_0-y_0)} \pm e^{-m^{ij}_{\mathrm{\scriptscriptstyle PS}}(T-x_0+y_0)}\right) \notag \\
  &+b_{\mathrm{\scriptscriptstyle X}}^{ij} \left(e^{-m^{\prime, ij}_{\mathrm{\scriptscriptstyle PS}}(x_0-y_0)}\pm e^{-m^{\prime, ij}_{\mathrm{\scriptscriptstyle PS}}(T-x_0+y_0)}\right),
\end{align}
where the $+$ sign corresponds to the pseudoscalar correlator ($\mathrm{X} = \mathrm{P}$), and the $-$ sign to the axial correlator ($\mathrm{X} = \mathrm{A}$), as defined in Eqs.~(\ref{eqn:fP}) and (\ref{eqn:fA}), respectively.
The meson masses can be extracted from a fit parameter $m_{\mathrm{\scriptscriptstyle PS}}^{ij}$ via the following expression
\begin{equation}
  \frac{\fP^{ij}(x_0,y_0)}{\fP^{ij}(x_0+a,y_0)}=\frac{{\mathrm{cosh}}\left(m^{ij}_{\mathrm{\scriptscriptstyle PS}}(x_0-y_0-T/2)\right)}{{\mathrm{cosh}}\left(m^{ij}_{\mathrm{\scriptscriptstyle PS}}(x_0+a-y_0-T/2)\right)}\,.
\end{equation}
To isolate the ground-state contribution, we perform a scan over multiple fit intervals in $x_0$, combined with a model averaging procedure (see below and Appendix~\ref{app:groundstate} for further details).
The pseudoscalar meson decay constant of the unitary Wilson regularisation, with flavours $(i,j)$, is determined via the following expression
\begin{align}
  \label{eqn:decay_const_W}
  f^{ij}_{\mathrm{\scriptscriptstyle PS, W}} = \ZA(g_0^2)\left[1+a\ibAb \trmsea+a\ibAt m_{ij}\right] f^{ij,\mathrm{bare}}_{\mathrm{\scriptscriptstyle PS, W}}\,,
\end{align}
where the presence of the index W in this and the following expressions is used to specify that the $\Oa$-improved Wilson-Dirac operator is employed in the valence sector.
The non-perturbative values of the renormalisation constant $\ZA$, determined in the chirally rotated Schr\"odinger functional~\cite{DallaBrida:2018tpn}, are reported in Table~\ref{tab:b-terms}.
In \req{eqn:decay_const_W}, $\ibAt$, denotes the improvement coefficient obtained from the coefficient $\ibA$ in \req{eqn:A_impr} when, in the corresponding $\Oa$ term, the subtracted mass $m_{ij}^{\scriptscriptstyle (+)}$, is replaced by the PCAC quark mass $m_{ij}$, as determined from \req{eqn:mpcac}.
More specifically, the substitution reads $\ibAt = \ibA Z^{-1}$, where $Z = \Zm \ZP / \ZA$.
This substitution holds when neglecting, in the expression for the renormalised subtracted quark mass, perturbative corrections of $\mbox{O}(\alphas^2)$ arising from $r_{m}$, the ratio of flavour-singlet to non-singlet scalar density renormalisation constants.
The one-loop perturbative values of $\ibAt$~\cite{Taniguchi:1998pf,Aoki:1998ar} employed in this work are listed in Table~\ref{tab:b-terms}.
The renormalisation constants and the mass-dependent improvement coefficients are to be evaluated at the improved coupling $\tilde{g}_0^2$ defined in \req{eq:impcoupling}.  
For $\ibA$, replacing $\tilde{g}_{0}^{2}$ by $g_{0}^{2}$ induces only an $\Oasq$ effect that can be neglected.
The one-loop perturbative expression from Ref.~\cite{Sint:1995ch}, $b_{g}(g_{0}^{2}) = 0.012 \NF\, g_{0}^{2} + \mbox{O}(g_{0}^{4})$, is employed to define $\tilde{g}_{0}^{2}$ in \req{eq:impcoupling} and used to evaluate the renormalisation factor $Z$ at the same perturbative order.
In practice, the one-loop improved coupling $\tilde{g}_{0}^{2}$ is numerically very close to $g_{0}^{2}$, and therefore we use $g_{0}^{2}$ when evaluating other renormalisation constants and improvement coefficients.
We note that the two-loop computation of $b_{g}$, recently reported in Ref.~\cite{Costa:2025xej}, induces at most a $0.5\%$ change in the value of the improved coupling compared to the one-loop result.
Furthermore, the available non-perturbative determinations of $b_{g}$~\cite{DallaBrida:2023fpl} do not cover the range of couplings relevant for this work.
For the contribution from $\ibAb$ multiplying $\Oa$ lattice artefacts depending on the sea quark masses, non-perturbative determinations based on coordinate-space methods~\cite{Korcyl:2016ugy,Korcyl:2016cmx} result in values compatible with zero with large uncertainties.
Given that $\ibAb$ first appears at the two-loop level in perturbation theory, the overall contribution of the corresponding term of  $\mbox{O}\left( \alphas^2 \, a\trmsea\right)$ is expected to be very small for up, down, and strange sea quarks, and is therefore omitted in the present analysis.
\begin{table}[!t]
  \begin{center}
    \begin{tabular}{c c c c}
      \toprule
      $\beta$ & $\ZA$ & $\ibAt$  \\
      \toprule
      3.40 & 0.75642(72) & 1.0832 \\
      3.46 & 0.76169(93) & 1.0818 \\
      3.55 & 0.76979(43) & 1.0797 \\
      3.70 & 0.78378(47) & 1.0765 \\
      3.85 & 0.79667(47) & 1.0735 \\
      \bottomrule
    \end{tabular}
  \end{center}
  \caption{
    The values of the renormalisation constant $\ZA$ and the improvement coefficient $\ibAt$ used in \req{eqn:decay_const_W} are collected for the bare couplings $\beta$ considered in this work.
    The non-perturbative value of $\ZA$ is taken from the chirally rotated Schr\"odinger functional calculation of Ref.~\cite{DallaBrida:2018tpn}, whereas the one-loop perturbative result for $\ibAt$ is based on Refs.~\cite{Taniguchi:1998pf,Aoki:1998ar}.
  }
  \label{tab:b-terms}
\end{table}
In the case of ensembles with OBC in the temporal direction, $f^{ij,\mathrm{bare}}_{\mathrm{\scriptscriptstyle PS, W}}$ is determined from a plateau average of an effective decay constant defined as follows
\begin{align}
  \label{eqn:decay_const_WR}
  f^{ij, \mathrm{bare}}_{\mathrm{eff, W}}(x_0)=\sqrt{\frac{2}{m^{ij}_{\mathrm{\scriptscriptstyle PS, W}}}}R^{ij}_{{\mathrm{\scriptscriptstyle A, W}}}(x_0)\,,
\end{align}
with the ratio $R^{ij}_{\mathrm{\scriptscriptstyle A, W}}(x_0)$ given by
\begin{equation}\label{eqn:R_W}
  R^{ij}_{{\mathrm{\scriptscriptstyle A, W}}}(x_0)=\left[\frac{f^{ij}_{{\mathrm{\scriptscriptstyle A, W}}}(x_0,a)f^{ij}_{{\mathrm{\scriptscriptstyle A, W}}}(x_0,T-a)}{f^{ij}_{{\mathrm{\scriptscriptstyle P, W}}}(T-a,a)}\right]^{1/2}\,,
\end{equation}
where the axial correlator $f^{ij}_{{\mathrm{\scriptscriptstyle A, W}}}$ in the unitary setup follows the generic definition of $f^{ij}_{\mathrm{\scriptscriptstyle A}}$ in \req{eqn:fA}, including the $c_\mathrm{A}$ improvement term appearing in \req{eqn:Amu}.
On the other hand, for ensembles with PBC in time, the expression
\begin{equation}
  f^{ij, \mathrm{bare}}_{\mathrm{\scriptscriptstyle PS, W}}=\frac{2}{m^{ij}_{\mathrm{\scriptscriptstyle PS, W}}}\frac{a^{ij}_{{\mathrm{\scriptscriptstyle A, W}}}}{\sqrt{a^{ij}_{{\mathrm{\scriptscriptstyle P, W}}}}}\,,
\end{equation}
is employed, where $a^{ij}_{{\mathrm{\scriptscriptstyle A, W}}}$ and $a^{ij}_{{\mathrm{\scriptscriptstyle P, W}}}$ are parameters extracted from a fit of the axial $f^{ij}_{{\mathrm{\scriptscriptstyle A, W}}}$ and pseudoscalar $f^{ij}_{{\mathrm{\scriptscriptstyle P, W}}}$ correlators, respectively, following the generic form in \req{eqn:corrs_PBC}.

We will now proceed to provide the corresponding definitions of the pseudoscalar meson decay constants in the mixed-action setup with Wtm valence quarks.
The definition of the decay constant involving a matrix element of the axial current in the physical basis can be mapped into a corresponding matrix element of the vector current in the twisted basis when working at maximal twist.\footnote{
We recall that all the numerical computations involving Wtm fermions are performed in the twisted basis.
}
The existence of a point-split vector current which is protected from renormalisation can be formally used to extract the Wtm decay constant, $f^{ij}_{\mathrm{\scriptscriptstyle PS}, \mathrm{Wtm}}$, without need to include any renormalisation constant.
More specifically, when employing OBC, the decay constant $f^{ij}_{\mathrm{\scriptscriptstyle PS}, \mathrm{Wtm}}$ is obtained from the plateau behaviour of the effective decay constant
\begin{align}
  \label{eqn:decay_const_Wtm}
  f^{ij}_{\mathrm{eff, Wtm}}(x_0)= \frac{\sqrt{2}}{\left(m^{ij}_{\mathrm{\scriptscriptstyle PS}, \mathrm{Wtm}}\right)^{3/2}}  \left( \eta_i\mu_i-\eta_j\mu_j \right) R^{ij}_{{\mathrm{\scriptscriptstyle P, Wtm}}}(x_0)\,,
\end{align}
where 
\begin{align}
  \label{eqn:R_Wtm}
  R^{ij}_{{\mathrm{\scriptscriptstyle P, Wtm}}}(x_0)=&\left[\frac{f_{{\mathrm{\scriptscriptstyle P, Wtm}}}^{ij}(x_0,a) f_{{\mathrm{\scriptscriptstyle P, Wtm}}}^{ij}(x_0,T-a)}{f_{{\mathrm{\scriptscriptstyle P, Wtm}}}^{ij}(T-a,a)}\right]^{1/2}\,.
\end{align}
Conversely, for the case of PBC ensembles, the decay constants can be extracted as
\begin{equation}
  f^{ij}_{\mathrm{\scriptscriptstyle PS}, \mathrm{Wtm}}=\left[ \frac{2}{\left(m^{ij}_{\mathrm{\scriptscriptstyle PS}, \mathrm{Wtm}}\right)^3}\, a^{ij}_{{\mathrm{\scriptscriptstyle P, Wtm}}} \right]^{1/2}\,,
\end{equation}
where, similarly to the Wilson case, the amplitude $a^{ij}_{{\mathrm{\scriptscriptstyle P, Wtm}}}$ is a parameter extracted from a fit of the Wtm pseudoscalar correlator following \req{eqn:corrs_PBC}.
In order to unburden the notation, in the following we shall refrain from including the labels W and Wtm, as well as in some cases the flavour indices $(i,j)$, in generic discussions of observables common to the mixed-action and unitary setups.
The lattice volumes considered in this work are such that $L>2.4\,{\mathrm{fm}}$ and $m_\pi L \geq 3.9$.
In order to correct for residual finite-volume effects in the pion and kaon masses and decay constants, we employ NLO $\chi$PT~\cite{Gasser:1986vb, Colangelo:2003hf, Colangelo:2010ba}
(we refer to Appendix~\ref{app:FVE} for further details).
We observe that they amount to corrections that are smaller than half the statistical uncertainty of the measured quantity.
The bare PCAC quark mass $m_{ij}$ is determined from the plateau behaviour of
\begin{equation}
  \label{eqn:mpcac}
  m_{ij}^{\mathrm{eff}}(x_0)=\frac{\tilde{\partial}_0\fA^{ij}(x_0)}{2\fP^{ij}(x_0)}\,,
\end{equation}
using the definitions in Eqs.~(\ref{eqn:fP})–(\ref{eqn:Amu}).
For the Wilson unitary setup, the $\Oa$-improved bare PCAC quark mass $m_{ij}^{\mathrm{\scriptscriptstyle I}}$ reads
\begin{equation}
  \label{eqn:mPCACI}
  m_{ij}^{\mathrm{\scriptscriptstyle I}}=\left[1+a\left(\ibAb-\ibPb\right)\trmsea+a\left(\ibAt-\ibPt\right) m_{ij}\right]m_{ij}\,,
\end{equation}
where, similarly to the case of $\ibAt$ in \req{eqn:decay_const_W}, the substitution $\ibPt = \ibP Z^{-1}$ is applied, implying in this case that higher-order perturbative corrections of $\mbox{O}(\alpha_{s}^{3})$ arising from $(\ibAt - \ibPt)(r_{m} - 1)$ are neglected.
The combination $\ibAt - \ibPt$ appearing in \req{eqn:mPCACI} has been non-perturbatively determined in Ref.~\cite{deDivitiis:2019xla}.
Similarly to the previously discussed case of $\ibAb$, the contribution of the term involving $\ibAb-\ibPb$ is expected to be very small.
It should be noted, however, that in the present work the PCAC quark mass of the Wilson unitary setup contributes only to the improvement of the decay constant in \req{eqn:decay_const_W}, where the unimproved mass can be employed.
In the case of the mixed action, a vanishing valence Wtm PCAC quark mass is used to define the maximal twist condition.
As previously discussed, the inclusion of additional $\Oa$ terms proportional to the twisted-mass parameters -- depending on the coefficients $\ibPc$ in \req{eq:Pimp} and $\ibAc$ in \req{eq:Aimp} -- is not required to enforce the maximal twist condition.
We therefore also employ the generic expressions in Eqs.~(\ref{eqn:mpcac}) and~(\ref{eqn:mPCACI}) when tuning the valence PCAC quark mass to zero.

\subsubsection{Gradient flow scale $t_0$}
\label{subsec:t0lat}

\begin{figure}[!t]
  \begin{center}
    \includegraphics[width=.49\linewidth]{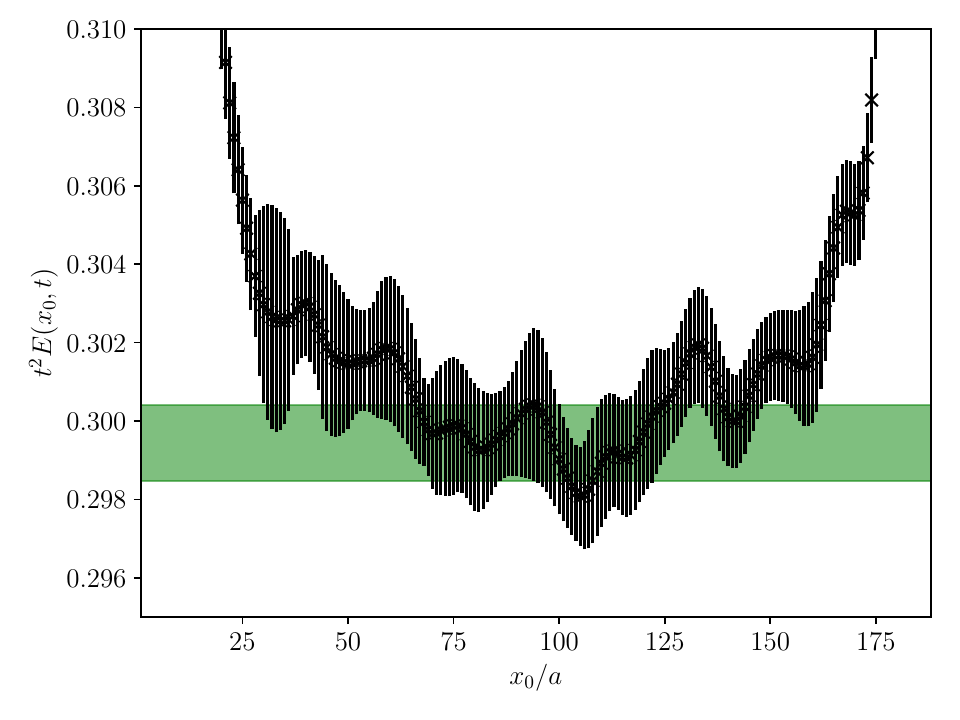}
    \includegraphics[width=.49\linewidth]{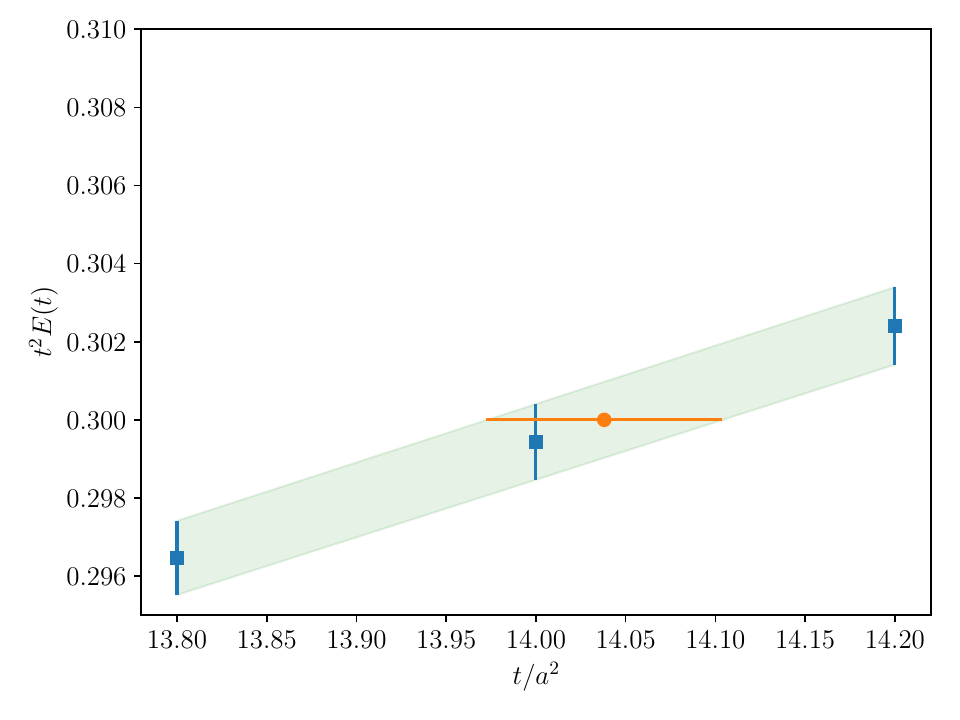}
  \end{center}
  \caption{
    Illustration of the determination of $t_0/a^2$ for the ensemble J501 with OBC.
    \textit{Left panel:} The Euclidean time dependence of $t^2E(x_0;t)$ is shown, with the time slice action density defined in \req{eqn:Ex0}, for a flow time $t$ in the neighbourhood of $t_0$.
    The green horizontal band corresponds to the model-average result for the extraction of the plateau in the bulk of the Euclidean time direction $x_0$.
    \textit{Right panel:} The blue squared points correspond to the determination of $t^2 E(t)$ for the three values of the flow time $t$ closest to $t_0$.
    From these three points, a small interpolation (blue band) in the flow time $t$ is performed in order to identify the value $t=t_0$ which satisfies the condition $t^2 E(t) = 0.3$ in \req{eqn:t0_def}.
    The interpolated value of $t_{0}/a^{2}$ is indicated by the amber circular data point.
  }
  \label{fig:t0}
\end{figure}

The gradient flow~\cite{Narayanan:2006rf,Luscher:2010iy} is a method that allows for the continuous smoothing of gauge fields according to a flow equation
\begin{equation}
  \label{eqn:Wflow}
  a^2\partial_tV_{\mu}(x,t)=-g_0^2\left\{\partial_{x,\mu}S_W[V]\right\} V_\mu(x,t)\,,
\end{equation}
where $t$ is the flow time and $\partial_{x,\mu}$ denotes the Lie-algebra valued derivative with respect to $V_\mu(x,t)$.
In practice, we use the Wilson flow --- corresponding to the choice where the gauge action on the right-hand side of \req{eqn:Wflow} is the Wilson gauge action, $S_W[V]$ --- which depends on the smooth gauge fields $V_\mu(x,t)$ subject to the initial condition
\begin{equation}
  V_{\mu}(x,t=0)=U_{\mu}(x)\,.
\end{equation}
The smooth field $V_{\mu}(x,t)$ is employed when defining the energy density $E(x_0,t)$~\cite{Luscher:2010iy} at the Euclidean time slice $x_0$ as
\begin{equation}
  \label{eqn:Ex0}
  E(x_0,t) =-\frac{a^3}{2L^3}\sum_{\vec x}  \tr \left\{ \hat G_{\mu\nu}(x,t) \,  \hat G_{\mu\nu}(x,t) \right\} \,.
\end{equation}
The field strength tensor $\hat{G}_{\mu\nu}$ defined via the clover discretisation reads
\begin{align}
  \label{eqn:Gmunuclover}
  \hat{G}_{\mu\nu}(x,t)&= \frac{1}{8a^2} \left\{ Q_{\mu\nu}(x,t) - Q_{\nu\mu}(x,t)  \right\}\,,\\
  Q_{\mu\nu}(x,t) &= P^{(1)}_{\mu\nu}(x,t)+ P^{(2)}_{\mu\nu}(x,t)
  +P^{(3)}_{\mu\nu}(x,t)+P^{(4)}_{\mu\nu}(x,t) \,,
\end{align}
where
\begin{align}
  \label{eqn:plaqp1}
  P^{(1)}_{\mu\nu}(x,t) &= V_\mu(x,t)V_\nu(x+a\hat\mu,t)V_\mu(x+a\hat\nu,t)^\dagger V_\nu(x,t)^\dagger\,,\\
  \label{eqn:plaqp2}
  P^{(2)}_{\mu\nu}(x,t) &= V_\nu(x,t)V_\mu(x-a\hat\mu+a\hat\nu,t)^\dagger V_\nu(x-a\hat\mu,t)^\dagger V_\mu(x-a\hat\mu,t)\,,\\
  \label{eqn:plaq3}
  P^{(3)}_{\mu\nu}(x,t) &= V_\mu(x-a\hat\mu,t)^\dagger V_\nu(x-a\hat\mu-a\hat\nu,t)^\dagger
  V_\mu(x-a\hat\mu-a\hat\nu,t) V_\nu(x-a\hat\nu,t)\,,\\
  \label{eqn:plaqp4}
  P^{(4)}_{\mu\nu}(x,t) &= V_\nu(x-a\hat\nu,t)^\dagger V_\mu(x-a\hat\nu,t) V_\nu(x+a\hat\mu-a\hat\nu,t)V_\mu(x,t)^\dagger\,.
\end{align}
The gradient flow scale $t_0$, introduced in Ref.~\cite{Luscher:2010iy}, is defined as the value of the flow time satisfying the condition
\begin{equation}
  \label{eqn:t0_def}
  \left. t^2 E(t) \right|_{t=t_0}=0.3\,,
\end{equation}
where the Euclidean time dependence of the action density in \req{eqn:Ex0} has been either summed over in the case of PBC, or isolated in the bulk of the lattice  when employing OBC. More specifically, in the case of OBC, a model variation over multiple plateau ranges is considered, following the procedure described in Sect.~\ref{subsec:obc} and in Appendix~\ref{app:MA}.
Small interpolations along the flow time $t$ are performed in order to set the value $t_0$ satisfying \req{eqn:t0_def}.
Fig.~\ref{fig:t0} illustrates the computation of the flow scale $t_0$ for ensemble J501.

\subsection{Boundary effects}
\label{subsec:obc}

In this work, the majority of CLS gauge ensembles under consideration employ open boundary conditions in the time direction~\cite{Luscher:2010we,Schaefer:2010hu,Luscher:2011kk,Luscher:2012av,Bruno:2014ova}.
In particular, OBC are employed for all ensembles at the two finest values of the lattice spacing, as indicated in Table~\ref{tab:CLS_ens}.
A more detailed specification of the boundary conditions is provided in Appendix~\ref{app:obc}.
The use of OBC results in a connected field space in the continuum theory, thereby enabling the topological charge to flow smoothly through the temporal boundaries, while preserving the transfer matrix and the space of physical states~\cite{Luscher:2011kk,Luscher:2012av}.
Consequently, all statistically relevant sectors of field space are expected to be explored during the ensemble generation, without the emergence of increasingly higher topology barriers as the continuum limit is approached.
In regions of the bulk sufficiently distant from the boundaries at $x_0 = 0$ and $x_0 = T$, local observables will approach their vacuum expectation values, with corrections that are exponentially suppressed.
The decay rate of these corrections is given by the lightest excitation with vacuum quantum numbers.
In order to isolate the vacuum expectation values, it is necessary to identify plateau regions where the contributions of the excited states can be safely neglected.
In practice, as discussed in Section~\ref{subsec:hadrons}, the quantities defined in Eqs.~(\ref{eqn:R_W}),~(\ref{eqn:R_Wtm})~and~(\ref{eqn:mpcac}) are considered, where matrix elements of operators situated in the proximity of the boundary drop out.
In addition to continuum effects, the time boundaries can also introduce lattice artefacts.
It should be noted that, although the coefficients of the boundary improvement terms are set to their tree-level values, an additional improvement is not required when considering correlation functions of fields situated far from the boundaries, where the effects of such terms are exponentially suppressed.
A more detailed account of these effects can be found in Appendix~\ref{app:obc}.
In the current study, the aim is therefore to identify the Euclidean time region where boundary effects can be considered to be negligible.
The identification of the plateau region can prove challenging due to the presence of long-range waves in the bulk, which are a consequence of the limited statistics in some of the ensembles.
Appendix~\ref{app:groundstate} reports the various approaches employed for the extraction of the ground state.
They include a two-stage procedure~\cite{Bruno:2014ova} aimed at identifying a regime where systematic errors from boundary effects and excited-state contamination are smaller than the statistical precision of the data, as well as model-variation techniques, described in Appendix~\ref{app:MA}, which explore multiple fit intervals in Euclidean time.
The statistical analysis method employed to address the autocorrelations present in Markov Chain Monte Carlo data is described in Appendix~\ref{app:analysis}.

\subsection{Matching procedure of sea and valence sectors}
\label{subsec:matching}

As previously discussed, in a mixed-action lattice regularisation, recovering unitarity in the continuum requires that the difference between the sea and valence quark masses for each flavour vanishes in the continuum limit.
One possible approach is to impose the matching procedure on the renormalised up, down and strange quark masses; more specifically, this procedure would involve fixing the renormalised twisted mass in the valence sector to equal the renormalised PCAC quark mass in the Wilson sea sector in the following way,
\begin{align}
  \mu_{12,\mathrm{\scriptscriptstyle R}}&=m_{12,\mathrm{\scriptscriptstyle R}}^{\mathrm{W}}\,,
  \label{eqn:matching_mq12}\\
  \mu_{13,\mathrm{\scriptscriptstyle R}}&=m_{13,\mathrm{\scriptscriptstyle R}}^{\mathrm{W}}\,,
  \label{eqn:matching_mq13}
\end{align}
where $\mu_{ij,\mathrm{\scriptscriptstyle R}} =\frac{1}{2}\left(\mu_{i,\mathrm{\scriptscriptstyle R}}+\mu_{j,\mathrm{\scriptscriptstyle R}}\right)$ and, as previously indicated, we use the indices $i = 1,2$ to label the mass-degenerate up and down flavours, and $i = 3$ for the strange quark.
In practice, this can be achieved on an ensemble-by-ensemble basis using the $\Oa$-improved quark masses, taking advantage of the fact that the $\ZP$ renormalisation factor on the two sides of Eqs.~(\ref{eqn:matching_mq12})~and~(\ref{eqn:matching_mq13}) cancels out, and only $\ZA$ and $\Oa$ improvement coefficients would be involved.
An alternative strategy is to match the masses of the pions and kaons rather than the sea and valence quark masses directly.
This is carried out in terms of the $\phi_2$ and $\phi_4$ quantities defined in Eqs.~(\ref{eqn:phi2})~and~(\ref{eqn:phi4}) as follows,
\begin{align}
  \phi_2^{\mathrm{Wtm}}&\equiv\phi_2^{\mathrm{W}}\,,
  \label{eqn:matching2} \\
  \phi_4^{\mathrm{Wtm}}&\equiv\phi_4^{\mathrm{W}}=\phi_4^{\mathrm{ph}}\,.
  \label{eqn:matching4}
\end{align}
The latter strategy will be pursued, since, in contrast to Eqs.~(\ref{eqn:matching_mq13})~and~(\ref{eqn:matching_mq13}), it does not require the use of current normalisations or improvement coefficients in either the unitary or the mixed-action setups.

\begin{figure}[!t]
  \begin{center}
    \includegraphics[width=1.\linewidth]{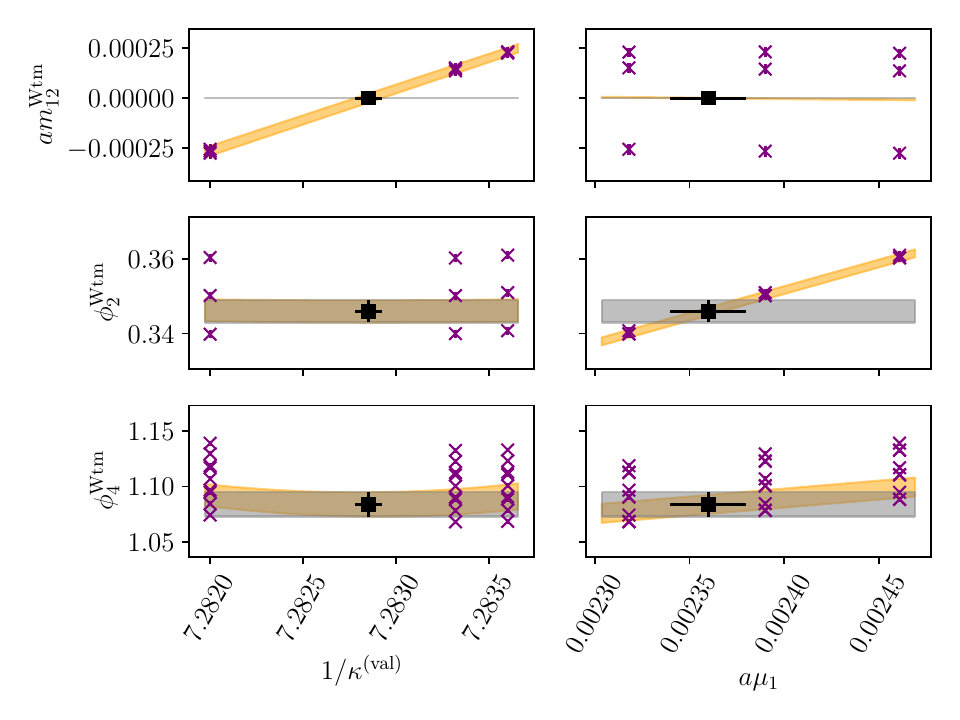}
  \end{center}
  \caption{
    Interpolations of $am_{12}^{\mathrm{Wtm}}$ (\textit{top row}), $\phi_2^{\mathrm{Wtm}}$ (\textit{middle row}), and $\phi_4^{\mathrm{Wtm}}$ (\textit{bottom row}) along the grid of valence mass parameters $\left(\kappa^{\ival},a\mu_1,a\mu_3\right)$ for the ensemble N200.
    The grid comprises three values of $\kappa^{\ival}$, three values of $a\mu_1$ and three values of $a\mu_3$.
    Consequently, it includes nine distinct values for $am_{12}^\mathrm{Wtm}$ and $\phi_{2}^\mathrm{Wtm}$, and 27 values for $\phi_4^\mathrm{Wtm}$, indicated by the cross symbols.
    The amber bands illustrate the fit result --- based on the functional forms in Eqs.~(\ref{eqn:matching_twist2})--(\ref{eqn:matching_twistm12}) --- projected onto the target matching parameters, with the exception of the parameter scanned along the horizontal axis of each panel.
    The p-value of this particular fit is approximately $0.4$.
    The horizontal grey bands indicate the target values of the interpolations in $am_{12}^{\mathrm{Wtm}}$, $\phi_2^{\mathrm{Wtm}}$, and $\phi_4^{\mathrm{Wtm}}$, respectively.
    Black square symbols indicate the valence parameter values at which the matching conditions in Eqs.~(\ref{eqn:matching2})--(\ref{eqn:maximal_twist}) are satisfied.
    In addition to the interpolations in the variables $\kappa^{\ival}$ and $a\mu_1$, a similar interpolation in $a\mu_3$, not displayed in these panels, is also carried out.
  }
  \label{fig:matching}
\end{figure}

\begin{figure} [!t]
  \begin{center}
    \includegraphics[width=.9\linewidth]{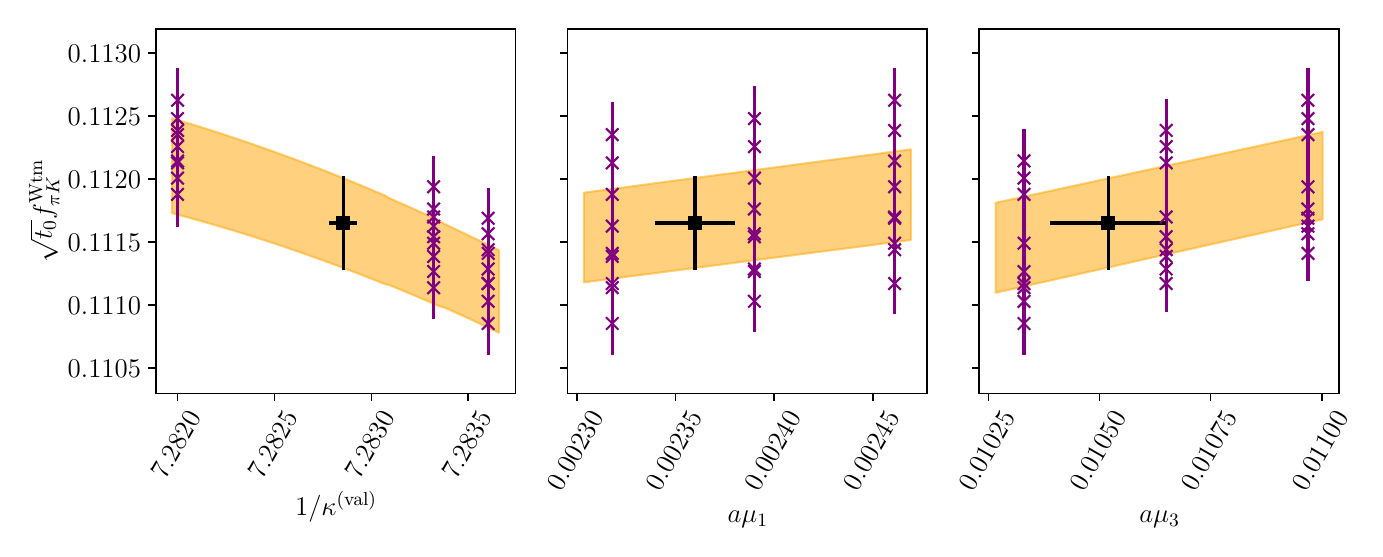}
  \end{center}
  \caption{
    Interpolations of $\sqrt{8t_0}f_{\pi K}^{\mathrm{Wtm}}$  along the grid of valence mass parameters $\left(\kappa^{\ival},a\mu_1,a\mu_3\right)$ for the ensemble N200.
    As the grid consists of three values for each parameter, namely $\kappa^{\ival}$, $a\mu_1$ and $a\mu_3$, there are 27 values for $\sqrt{8t_0}f_{\pi K}^{\mathrm{Wtm}}$, as indicated by the cross symbols.
    The amber bands illustrate the fit result, based on the functional forms in Eqs.~(\ref{eqn:match_fpi})--(\ref{eqn:match_fK}) from which the linear combination $f_{\pi K}$ defined in \req{eq:fpik} is built.
    The aforementioned bands are projected onto the target matching parameters, with the exception of the parameter scanned along the horizontal axis of each panel.
    The p-value of this fit is approximately $0.9$.
    Black square symbols indicate the result of the interpolation to the target valence parameters $\left(\kappa^{\ival*},a\mu_1^{*},a\mu_3^{*}\right)$ at which the matching conditions in Eqs.~(\ref{eqn:matching2})--(\ref{eqn:maximal_twist}) are satisfied.
  }
  \label{fig:decay_interp}
\end{figure}

In practice, to impose the matching conditions, we consider a set of input points --- referred to in the following as a {\it grid} ---  in the space of valence parameters ($\kappa^{\mathrm{( val)}}$, $a\mu_1$, $a\mu_3$).
This grid of points has been selected to ensure the coverage of a small region that enables the imposition of matching conditions in Eqs.~(\ref{eqn:phi2})~and~(\ref{eqn:phi4}), as well as the maximal twist condition, via small interpolations.
Along the grid, the hadron masses, $m_{\pi}^{\mathrm{Wtm}}$ and $m_K^{\mathrm{Wtm}}$, and the decay constants, $f_{\pi}^{\mathrm{Wtm}}$ and $f_K^{\mathrm{Wtm}}$, are computed for the mixed-action regularisation.
Hadron masses are used to determine $\phi_2^{\mathrm{Wtm}}$ and $\phi_4^{\mathrm{Wtm}}$, which are then interpolated to match the corresponding sea-sector values, $\phi_2^{\mathrm{W}}$ and $\phi_4^{\mathrm{W}}$, obtained from the analysis of the unitary Wilson setup.
The valence PCAC quark masses $m_{ij}^{\mathrm{Wtm}}$ are also determined.
The maximal twist condition is established by imposing a vanishing valence PCAC quark mass for the mass-degenerate light-quark flavours
\begin{equation}
  \label{eqn:maximal_twist}
  m_{12}^{\mathrm{Wtm}} \equiv 0\,.
\end{equation}
Also in this case, this condition can be realised through interpolations of the grid data points.
The selection of the maximal twist condition \req{eqn:maximal_twist} for the lightest quark flavours implies that the heavier quark flavours will be subject to deviations from the vanishing PCAC quark condition, stemming from lattice artefacts.
These lattice artefacts have been monitored in the strange-quark sector through $m_{34}^{\mathrm{Wtm}}$ and the corresponding twist angle $\alpha_{34}$, defined in \req{eqn:twist_34}, and are discussed in Appendix~\ref{app:twistangles}.
In particular, we observe that the full twist condition $\alpha_{34}=\pi/2$, is recovered in the continuum limit.
The interpolations of the valence observables to the matching conditions in Eqs.~(\ref{eqn:matching2})--(\ref{eqn:maximal_twist}) are carried on each ensemble using the following functional forms,
\begin{align}
  \phi_2^{\mathrm{Wtm}}\left(\kappa^{\ival},a\mu_1\right)&=\frac{p_1}{a\mu_1}\left(\frac{1}{\kappa^{\ival}}-\frac{1}{\kappa^{\ival*}}\right)^2+p_2(a\mu_1-a\mu_1^{*})+\phi_2^{\mathrm{W}}\,,
  \label{eqn:matching_twist2}\\
  \phi_4^{\mathrm{Wtm}}\left(\kappa^{\ival},a\mu_1,a\mu_3\right)&=\frac{p_3}{a\mu_1}\left(\frac{1}{\kappa^{\ival}}-\frac{1}{\kappa^{\ival*}}\right)^2+\frac{p_4}{a\mu_3}\left(\frac{1}{\kappa^{\ival}}-\frac{1}{\kappa^{\ival*}}\right)^2
  \nonumber\\
  &+p_5(a\mu_1-a\mu_1^{*})+p_6(a\mu_3-a\mu_3^{*})+\phi_4^{\mathrm{W}}\,,
  \label{eqn:matching_twist4}\\
  am_{12}^{\mathrm{Wtm}}\left(\kappa^{\ival},a\mu_1\right)&=p_7\left(\frac{1}{\kappa^{\ival}}-\frac{1}{\kappa^{\ival*}}\right)+p_8(a\mu_1-a\mu_1^{*})\,.\label{eqn:matching_twistm12}
\end{align}
These expressions correspond to polynomial approximations in the valence mass parameters ($\kappa^{\ival}$, $a\mu_1$, $a\mu_3$) around the matching point, denoted by the target parameters $\left(\kappa^{\ival*},a\mu_1^{*},a\mu_3^{*}\right)$.
A combined fit of $\phi_2^{\mathrm{Wtm}}$, $\phi_4^{\mathrm{Wtm}}$, and $m_{12}^\mathrm{Wtm}$ --- as given in Eqs.~(\ref{eqn:matching_twist2})–(\ref{eqn:matching_twistm12}) --- is performed for each ensemble.
This approach exploits the common dependence on the parameters $\kappa^{\ival*}$, $a\mu_{1}^{*}$, and $a\mu_{3}^{*}$, in addition to the individual dependence on the parameters $p_{i}$, for $i = 1, \dots, 8$.
Given that $\phi_4^{\mathrm{Wtm}}$ is minimised at maximal twist, its dependence on $\kappa^{\ival}$ in the neighbourhood of $\kappa^{\ival*}$ is weak; consequently, the small interpolations performed here make it difficult to determine the values of the nuisance parameters $p_3$ and $p_4$ in \req{eqn:matching_twist4} with high precision.
By relating the fit parameters $p_i$ ($i=1,\dots,6$) to the low‑energy constants (LECs) of SU(3) chiral perturbation theory ($\chi$PT), their signs can be constrained during the fit.
Imposing these constraints on $p_3$ and $p_4$ enforces the expected $\kappa^{\ival}$-dependence of $\phi_4^{\mathrm{Wtm}}$.
The matching and maximal‑twist tuning procedure is illustrated in Fig.~\ref{fig:matching}.\,\footnote{
In addition to the r.h.s. of Eqs.~(\ref{eqn:matching_twist2})--(\ref{eqn:matching_twistm12}), we have explored alternative functional forms inspired by SU(3) $\chi$PT as well as polynomial ans\"atze.
However, the target matching parameters remain stable under these variations, as only minor interpolations are required at this stage of the analysis.
}
Once the target parameter values $\left(\kappa^{\ival*},a\mu_1^{*},a\mu_3^{*}\right)$ have been obtained from the combined fit, the pion and kaon decay constants of the mixed-action regularisation must also be interpolated to these matched parameter values for each ensemble.
As before, low-order polynomials in deviations of input mass parameters with respect to their matched values are adequate for carrying out these interpolations of $f_{\pi}^{\mathrm{Wtm}}$ and $f_K^{\mathrm{Wtm}}$.
Specifically, the following functional forms were considered
\begin{align}
  af_{\pi}^{\mathrm{Wtm}}\left(\kappa^{\ival},a\mu_1\right)&=r_1+r_2a\mu_1+r_3 \left(\frac{1}{\kappa^{\ival}}-\frac{1}{\kappa^{\ival*}}\right)+r_4 \left(\frac{1}{\kappa^{\ival}}-\frac{1}{\kappa^{\ival*}}\right)^2\,,
  \label{eqn:match_fpi} \\
  af_{K}^{\mathrm{Wtm}}\left(\kappa^{\ival},a\mu_1,a\mu_3\right)&=r'_1+r'_2a\mu_1+r'_3a\mu_3+r'_4\left(\frac{1}{\kappa^{\ival}}-\frac{1}{\kappa^{\ival*}}\right) + r'_5 \left(\frac{1}{\kappa^{\ival}}-\frac{1}{\kappa^{\ival*}}\right)^2\,,
  \label{eqn:match_fK}
\end{align}
where $r_i$ and $r'_j$ are fit parameters.
An illustration of the interpolation of the decay constants is given in Fig.~\ref{fig:decay_interp}. 
%


\section{Scale setting}
\label{sec:results}

The lattice QCD setup under consideration is based on $\NF=2+1$ dynamical quarks, thereby corresponding to the isospin-symmetric limit of QCD (isoQCD) with mass-degenerate up and down quark flavours and absence of electromagnetic corrections.
The separation of a physical result into an isoQCD component and an isospin-breaking correction is a prescription-dependent procedure.
In this work, the so-called Edinburgh Consensus~\cite{FlavourLatticeAveragingGroupFLAG:2024oxs} is adopted for the definition of isoQCD
\begin{gather}
  \label{ch_ss:eq:isoQCD_mpi}
  m_{\pi}^{\mathrm{isoQCD}}=135.0\;{\mathrm{MeV}}\,, \\
  \label{ch_ss:eq:isoQCD_mk}
  m_{K}^{\mathrm{isoQCD}}=494.6\;{\mathrm{MeV}}\,, \\
  \label{ch_ss:eq:isoQCD_fpi}
  f_{\pi}^{\mathrm{isoQCD}}=130.5\;{\mathrm{MeV}}\,.
\end{gather}
These three inputs are sufficient to fix the quark masses and the QCD scale, required to define isoQCD with $\NF=2+1$ flavours.
However, an alternative to the use of solely $f_{\pi}$ in \req{ch_ss:eq:isoQCD_fpi} is the flavour-averaged combination $f_{\pi K}$
introduced in \req{eq:fpik}.
Accordingly, the inputs in Eqs.~(\ref{ch_ss:eq:isoQCD_mpi})--(\ref{ch_ss:eq:isoQCD_fpi}) are supplemented by an isoQCD value of $f_K$ from Ref.~\cite{FlavourLatticeAveragingGroupFLAG:2024oxs}
\begin{gather}
  \label{ch_ss:eq:isoQCD_fk}
  f_{K}^{\mathrm{isoQCD}}=157.4(2)_{\mathrm{exp}}(4)_{\mathrm{QED}}(4)_{|V_{us}|}\;{\mathrm{MeV}}\,.
\end{gather} 
The first error is due to the experimental measurement of the decay rate~\cite{ParticleDataGroup:2024cfk}. The second error is derived from an estimate of QED radiative corrections, and the third error originates from the input value of $|V_{us}|$.
In contrast to the case of Eqs.~(\ref{ch_ss:eq:isoQCD_mpi})--(\ref{ch_ss:eq:isoQCD_fpi}) corresponding to isoQCD definitions without errors, we have included in \req{ch_ss:eq:isoQCD_fk} the uncertainty of $f_{K}^{\mathrm{isoQCD}}$.
As indicated in \req{ch_ss:eq:isoQCD_fk}, this uncertainty depends on the input value of the CKM matrix element $|V_{us}|$ and on the estimates of the QED corrections.
The primary motivation to consider the linear combination $f_{\pi K}$ is that, given the current set of ensembles available in this study, the level of control of the continuum-limit and chiral extrapolations is higher than when $f_\pi$ alone is used.
Nonetheless, the results of the scale setting based exclusively on $f_\pi$ will also be reported in Section~\ref{subsec:t0fpi}.
In the scale setting procedure, the gradient flow scale $t_0$, described in Sect.~\ref{subsec:t0lat}, is employed as an intermediate reference scale, facilitating the construction of dimensionless quantities.
In particular, the dimensionless combination $\sqrt{8t_0}f_{\pi K}$ is used to perform a combined extrapolation to the continuum limit and to the physical values of the quark masses.
More specifically, the location of the physical values of the masses will be set through the dimensionless quantities $\phi_2^{\mathrm{ph}}$ and $\phi_4^{\mathrm{ph}}$, based on Eqs.~(\ref{eqn:phi2})--(\ref{eqn:phi4}), determined using the isoQCD input values for the pion and kaon masses in Eqs.~(\ref{ch_ss:eq:isoQCD_mpi})--(\ref{ch_ss:eq:isoQCD_mk}).
The mass-shifting procedure outlined in Sec.~\ref{subsec:chiral-traj} ensures the value of $\phi_4$ remains fixed at its physical value $\phi_4^{\mathrm{ph}}$ along the chiral trajectory in the quark mass plane.
Consequently, the chiral extrapolation only depends on $\phi_2$.
In practice, since the physical value of the gradient flow scale $t_0^{\mathrm{ph}}$ is initially unknown, the corresponding physical value $\phi_4^{\mathrm{ph}}$ must be determined via the iterative procedure described in Sec.~\ref{subsec:chiral-traj}.
The latter relies on the physical inputs in Eqs.~(\ref{ch_ss:eq:isoQCD_mpi})--(\ref{ch_ss:eq:isoQCD_mk})~and~(\ref{ch_ss:eq:isoQCD_fk}), and an educated guess for the physical value of $t_0$, which is updated at each iterative step with the new determination of $t_0^{\mathrm{ph}}$ coming from the scale setting analysis.
Therefore, at each iterative step, both the value of $\phi_2^{\mathrm{ph}}$ to which the chiral extrapolation is performed and the value of $\phi_4^{\mathrm{ph}}$ to which the observables are shifted are updated until convergence in the value of $t_0^{\mathrm{ph}}$ is obtained.
In practice, this occurs after only a few iterations.
After performing the continuum-chiral extrapolation of $\sqrt{8t_0}f_{\pi K}$, the physical value of the scale $t_0$ can be obtained through
\begin{equation}
  \label{eq:t0phinput}
  \sqrt{8t_0^{\mathrm{ph}}}=\frac{\left.\left(\sqrt{8t_0}f_{\pi K}\right)^{\mathrm{latt}}\right|_
    {a=0,\;\phi_2^{\mathrm{ph}}}}{f_{\pi K}^{\mathrm{isoQCD}}}\,.
\end{equation}
Since both the unitary Wilson and the valence Wtm mixed-action regularisations are considered, three types of scale setting analyses are performed: (i) analysis using only the Wilson regularisation, (ii) analysis using only the Wtm mixed-action setup, and (iii) a combined analysis that enforces a common continuum limit for both.
Each of these analyses will provide a different determination of $t_0^{\mathrm{ph}}$, though they will obviously be correlated.

\subsection{Determination of $t_0^{\mathrm{ph}}$ from $f_{\pi K}^{\mathrm{isoQCD}}$}
\label{subsec:t0}

We begin by describing the light-quark mass and lattice-spacing dependence of the quantities involved in the scale-setting procedure.
The expressions for $f_\pi$ and $f_K$ at next-to-leading order (NLO) in $\mathrm{SU(3)}$ $\chi$PT~\cite{Gasser:1984gg} read
\begin{align}
  \label{ch_ss:eq:fpifK_chiral}
  f_{\pi}&=f\left[1+\frac{16B_0L_5}{f^2}m_l+\frac{16B_0L_4}{f^2}(2m_l+m_s)-2L(m_{\pi}^2) -L(m_K^2)\right]\,, \\
  f_K&=f\left[1+\frac{8B_0L_5}{f^2}(m_l+m_s)+\frac{16B_0L_4}{f^2}(2m_l+m_s) -\frac{3}{4}L(m_{\pi}^2)-\frac{3}{2}L(m_K^2) \right. \notag \\
    &\left.-\frac{3}{4}L(m_{\eta}^2)\right]\,,
\end{align}
where $m_l=m_u=m_d$, and $f$, $B_0$, and $L_i$ are $\chi$PT LECs. The notation for the chiral logarithms $L(x)$ is
\begin{equation}
  L(x)=\frac{x}{(4\pi f)^2}\mathrm{log}\frac{x}{(4\pi f)^2}\,.
\end{equation}
At the order considered in the $\chi$PT expansion, the relation between quark masses and meson masses can be established by employing the LO expressions
\begin{align}
  m_{\pi}^2&=2B_0m_l\,, \\
  m_K^2&=B_0(m_l+m_s)\,, \\
  m_{\eta}^2&=\frac{4}{3}m_K^2-\frac{1}{3}m_{\pi}^2\,.
\end{align}
The NLO expression for $t_0$ in $\mathrm{SU(3)}$ $\chi$PT~\cite{Bar:2013ora} reads
\begin{align}
  \label{ch_ss:eq:t0_chiral}
  t_0&=t_{0,\mathrm{ch}}\left[1+k_1\frac{2m_K^2+m_{\pi}^2}{(4\pi f)^2}\right]\,,
\end{align}
depending of the LECs $t_{0,\mathrm{ch}}$ and $k_1$.

The choice of the flavour-averaged combination of decay constants $f_{\pi K}$ in \req{eq:fpik} is motivated by its chiral behaviour.
Indeed, for a fixed value of $\phi_4$, the NLO $\mathrm{SU(3)}$ $\chi$PT expression only depends on $\phi_2$ through chiral logarithms.
The light-quark mass-dependence of the combination $\Phi_{\pi K} \equiv \sqrt{8t_0}f_{\pi K}$ in the continuum thus reads
\begin{align}
  \label{ch_ss:eq:SU3ChPT}
  \Phi_{\pi K,\chi \mathrm{SU(3)}}^{\mathrm{cont}}(\phi_2)\equiv\left(\sqrt{8t_0}f_{\pi K}\right)^{\mathrm{cont}} &=\frac{A}{4\pi}\left[1-\frac{7}{6}\tilde{L}\left(\frac{\phi_2}{A^2}\right)-\frac{4}{3}\tilde{L}\left(\frac{\phi_4-\frac{1}{2}\phi_2}{A^2}\right)\right. \notag \\
    &\left.-\frac{1}{2}\tilde{L}\left(\frac{\frac{4}{3}\phi_4-\phi_2}{A^2}\right)+\frac{B}{A^2}\phi_4\right]\,,
\end{align}
where
\begin{equation}
  \label{ch_ss:eq:log}
  \tilde{L}(x)=x{\mathrm{log}}\left(x\right)\,,
\end{equation}
and the LECs are incorporated into the definition of the parameters $A$ and $B$ as 
\begin{align}
  A&=4\pi\sqrt{8t_{0,\mathrm{ch}}}f\,, \\
  B&=\frac{(16\pi)^2}{3}(L_5+3L_4)+k_1\,.
\end{align} 
The expression in \req{ch_ss:eq:SU3ChPT} will be employed to perform the combined continuum-limit and chiral fits of $\sqrt{8t_0}f_{\pi K}$ labelled as $[\chi \mathrm{SU(3)}]$.

In order to investigate the systematic effects associated with the approach to the physical pion mass, in addition to the $\mathrm{SU(3)}$ expressions, we also considered expressions inspired by NLO $\mathrm{SU(2)}$ $\chi$PT~\cite{Gasser:1983yg}, in which the mass-dependence of the strange quark is absorbed in the corresponding LECs,
\begin{align}
  \label{ch_ss:eq:SU2fpi}
  f_{\pi}&=f\left[1+\frac{8(2L_4+L_5)}{f^2}m_{\pi}^2-2L(m_{\pi}^2)\right]\,, \\
  \label{ch_ss:eq:SU2fk}
  f_K&=f^{(K)}(m_s)\left[1+\frac{c(m_s)}{f^2}m_{\pi}^2-\frac{3}{4}L(m_{\pi}^2)\right]\,.
\end{align}
As the expression of $f_{\pi}$ in \req{ch_ss:eq:SU2fpi} depends on $m_s$ only via sea quark loop effects, it is expected that the LECs $f$ and $L_{4,5}$ exhibit a mild dependence on $m_s$.
We thus assumed that they are independent of $m_s$.
On the other hand, the LECs $f^{(K)}(m_s)$ and $c(m_s)$ will be assumed to vary linearly with $m_s$.
The effect of keeping these parameters constant as a function of $m_{s}$ will also be investigated.
The corresponding expression for $\sqrt{8t_0}f_{\pi K}$, labelled $[\chi \mathrm{SU(2)}]$, may be expressed as follows
\begin{equation}
  \label{ch_ss:eq:SU2pik}
  \Phi_{\pi K,\chi \mathrm{SU(2)}}^{\mathrm{cont}}(\phi_2)=B+C\phi_2+D\phi_4-E\tilde{L}\left(\frac{\phi_2}{A^2}\right)\,,
\end{equation}
where the parameters $A,B,C,D,E$ are combinations of the LECs that appear in Eqs.~(\ref{ch_ss:eq:SU2fpi})--(\ref{ch_ss:eq:SU2fk}).
A term of type $D\phi_4$ may arise from the chiral expansion of $t_0$ in \req{ch_ss:eq:t0_chiral}, even when $f^{(K)}(m_s)$ and $c(m_s)$ are considered to be independent of $m_s$; but,
since a mass-shift to a constant value of $\phi_4$ is applied, the fit cannot distinguish between two terms $B$ and $D\phi_4$,
and we will therefore group these two terms into a single one to reduce the number of fit parameters.

An alternative approach to parameterising the trajectory towards the physical point is to use Taylor expansions in $\phi_{2}$ around its symmetric-point value, $\phi_{2}^{\mathrm{sym}}$, where $m_{\pi} = m_{K} \approx 420\,\mathrm{MeV}$.
Taylor expansions for $\sqrt{8t_0}f_{\pi K}$ up to second and fourth order have been considered as follows
\begin{equation}
  \label{ch_ss:eq:Tay2}
  \Phi_{\pi K, \mathrm{Tay2}}^{\mathrm{cont}}(\phi_2)=A+B\left(\phi_2-\phi_2^{\mathrm{sym}}\right)^2\,,
\end{equation}
and
\begin{equation}
  \label{ch_ss:eq:Tay4}
  \Phi_{\pi K, \mathrm{Tay4}}^{\mathrm{cont}}(\phi_2)=A+B\left(\phi_2-\phi_2^{\mathrm{sym}}\right)^2+C\left(\phi_2-\phi_2^{\mathrm{sym}}\right)^4\,.
\end{equation}
We will label these models as $[\mathrm{Tay2}]$ and $[\mathrm{Tay4}]$, respectively.
The flavour-averaged combination $f_{\pi K}$ is protected from the presence of terms with odd powers in $\phi_2-\phi_2^{\mathrm{sym}}$ due to symmetry constraints~\cite{Bietenholz:2011qq}.

\begin{figure}[!t]
  \centering
  \includegraphics[width=.7\textwidth]{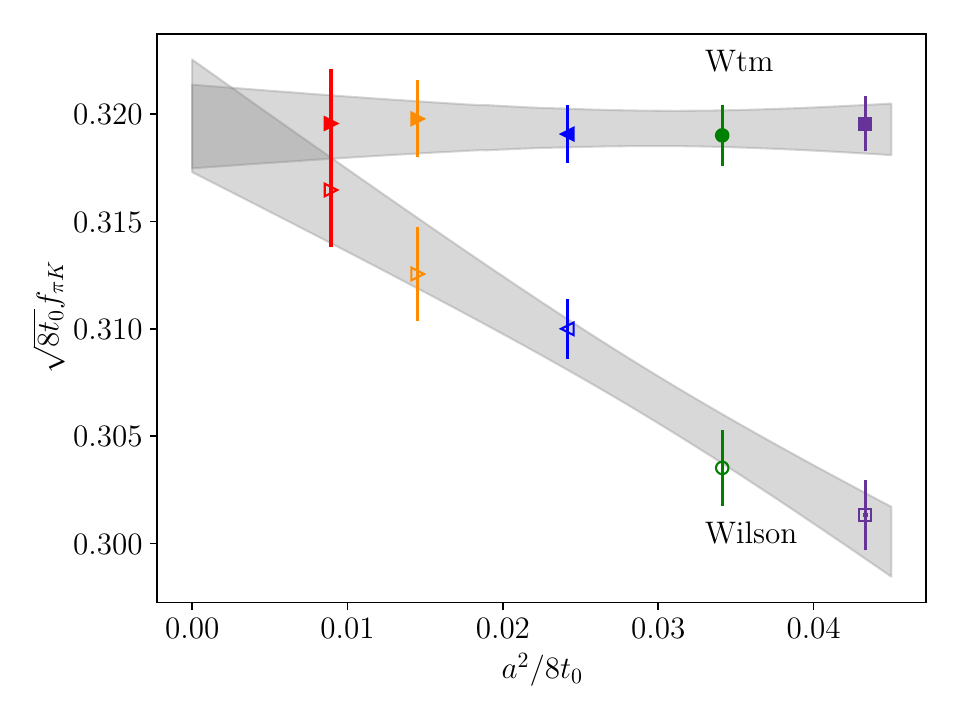}
  \caption{
    Continuum-limit extrapolation of $\sqrt{8t_0}f_{\pi K}$ for symmetric-point ensembles --- verifying $\phi_2^{\mathrm{sym}} = \frac{2}{3} \phi_4^{\mathrm{phys}}$ --- which provides a universality test for the regularisations under consideration.
    The results for the Wilson unitary setup are represented by empty symbols, while the results for the Wtm mixed action are represented by filled symbols.
    Lattice artefacts effects are modelled as independent coefficients of $\Oasq$ for each regularisation.
    No constraints are imposed on the continuum-limit values in the fit.
    The p-value of the fit is approximately $0.9$.
  }
  \label{ch_ss:fig:universality}
\end{figure}

In addition to the parameterisations of the continuum light-quark mass dependence described above, the functional forms governing the continuum-limit extrapolation must also be considered.
As discussed in Sect.~\ref{subsec:hadrons}, the inclusion of the relevant $\Oa$-improvement coefficients in the fermionic observables, in combination with the tuning to maximal twist of the mixed-action regularisation, imply that the leading lattice artefacts of both the unitary and the mixed-action setups are expected to start at $\Oasq$.
The lattice artefacts will therefore be parameterised according to the following three alternatives
\begin{align}
  \label{ch_ss:eq:a2}
  \Phi_{\pi K, a^2}^{\mathrm{latt}}(\phi_2, a/\sqrt{8t_0})&=\Phi_{\pi K}^{\mathrm{cont}}(\phi_2)+F\frac{a^2}{8t_0}\,,\\
  \label{ch_ss:eq:aas}
  \Phi_{\pi K,a^2\alphas}^{\mathrm{latt}}(\phi_2, a/\sqrt{8t_0})&=\Phi_{\pi K}^{\mathrm{cont}}(\phi_2)+F\frac{a^2}{8t_0}\alphas^{\hat{\Gamma}}(a)\,,\\
  \label{ch_ss:eq:a2phi2}
  \Phi_{\pi K,a^2+a^2\phi_2}^{\mathrm{latt}}(\phi_2, a/\sqrt{8t_0})&=\Phi_{\pi K}^{\mathrm{cont}}(\phi_2)+\left(F+G\phi_2\right)\frac{a^2}{8t_0}\,,
\end{align}
where $F$ and $G$ are fit parameters.
The labels $[a^2]$, $[a^2\alphas^{\hat{\Gamma}}]$ and $[a^2+a^2\phi_2]$ are assigned to characterise the lattice artefacts of the models in Eqs.~(\ref{ch_ss:eq:a2})--(\ref{ch_ss:eq:a2phi2}), respectively.
The form of \req{ch_ss:eq:aas} is motivated by the studies in Refs.~\cite{Husung:2022kvi,Husung:2024cgc}, which incorporate logarithmic corrections in the lattice spacing $a$, depending on a tower of possible values of the exponent $\hat{\Gamma}$ associated with the anomalous dimensions of the relevant operators appearing in the SymEFT.
The available data does not allow for multiple independent fitting parameters to be used to characterise these logarithmic corrections.
Therefore, only a single value of $\hat{\Gamma}$ has been considered in this study.
The impact of each choice of $\hat{\Gamma}$ on the quality of the fit and on the value of $t_0^{\mathrm{ph}}$ derived from the model average is monitored.
As $t_0^{\mathrm{ph}}$ is found to be largely unaffected by the choice of $\hat{\Gamma}$, we opted to consider only the smallest value, $\hat{\Gamma}=-0.111$, in the model variation.
We have also considered higher-order terms in the fit forms characterising the lattice-spacing dependence beyond those present in Eqs.~(\ref{ch_ss:eq:a2})--(\ref{ch_ss:eq:a2phi2}).
However, we observe that non-zero values of these coefficients cannot be accurately resolved given the current statistical precision of our data.
In addition to the previously described variations in the parameterisation of the continuum-limit and chiral extrapolations, we also analyse the impact of performing cuts in the data set.
This enables us to explore the systematic uncertainties arising from the removal of the coarsest ensembles, heaviest pion masses and smallest volumes.
More specifically, in addition to the ``no cut'' option, we are also considering the following set of cuts:
\begin{align}
  \label{ch_ss:eq:cuts1}
  \beta>3.40\,,  &~~\mathrm{i.e.}~ a < 0.08\,\mathrm{fm}\,, \\
  \label{ch_ss:eq:cuts2}
  \beta>3.46\,,  &~~\mathrm{i.e.}~ a < 0.07\,\mathrm{fm}\,, \\
  \label{ch_ss:eq:cuts3}
  m_{\pi}<420\,\mathrm{MeV}\,,  &~~\mathrm{i.e.}~ \phi_2<0.7\,, \\
  \label{ch_ss:eq:cuts4}
  m_{\pi}<350\,\mathrm{MeV}\,,  &~~\mathrm{i.e.}~ \phi_2<0.5\,, \\
  \label{ch_ss:eq:cuts5}
  \beta>3.40\; \& \; m_{\pi}<420\,\mathrm{MeV}\,,  &~~\mathrm{i.e.}~ a < 0.08\,\mathrm{fm} \;\mathrm{and}\; \phi_2<0.7\,, \\
  \label{ch_ss:eq:cuts6}
  m_{\pi}L>4.1\,.
\end{align}
For each of the cuts, the ensembles that satisfy the above condition are the ones that are kept in the analysis.
For instance, the cut in Eq.~(\ref{ch_ss:eq:cuts3}) corresponds to discarding the symmetric-point ensembles.
The systematic uncertainty in extracting $\sqrt{t_0^{\mathrm{ph}}}$ is assessed by a model variation procedure, as described in Appendix~\ref{app:MA}.
This model variation is followed by a model averaging (MA) in which a probability weight $W$ is assigned to each model based on the model selection criterion.
As described in Appendix~\ref{app:MA}, various model selection criteria will be considered to assess the stability of the results.
In general, we observe that the models included in the model average correspond to good fits in terms of their p-values, as defined in Ref.~\cite{Bruno:2022mfy}.
The p-values of the considered models are reported in Appendix~\ref{apex_model_av_t0}.
When evaluating the weight $W$ using \req{eqn:weight}, based on the Takeuchi Information Criterion (TIC)~\cite{Takeuchi76,Frison:2023lwb} as defined in \req{eqn:TICchiexp}, the combined fits of the continuum-limit and chiral extrapolations of $\sqrt{8t_{0}}\,f_{\pi K}$ that involve cuts in the dataset tend to exhibit markedly smaller weights compared to those without cuts.
On the other hand, the fits are strongly influenced by data from ensembles with the coarsest lattice spacings and/or heaviest pion masses, as these points typically have smaller uncertainties than those nearer to the continuum limit and physical masses.
The application of information criteria such as that in \req{eqn:weight} to models incorporating cuts can therefore lead to situations where the removal of ensembles with the coarsest lattice spacings and the heaviest masses cannot be effectively probed within the model-averaging procedure.
As a result, we observe that the systematic effects associated with this type of cuts are not always adequately explored within such a model averaging framework.
It is therefore desirable to consider a conservative procedure that incorporates the theoretical knowledge from effective theories into the fits and the model average, namely the fact that the Symanzik EFT is an expansion which has larger higher-order corrections when the lattice spacing grows, and $\chi$PT receives larger corrections as the pion mass grows.
Following Refs.~\cite{DallaBrida:2016kgh,Bruno:2022mfy}, when defining the chi-squared function of the fit (see \req{eqn:chisq} in Appendix~\ref{app:MA}) we consider a weight matrix $\mathcal{W}$ which, in addition to the statistical weight matrix $\left.\mathcal{W}\right|_{\mathrm{stat}}$, also includes a term accounting for systematic effects, $\left.\mathcal{W}\right|_{\mathrm{syst}}$, as follows
\begin{align}
  \label{ch_ss:eq:Wsyst}
  \mathcal{W}^{-1}=  \left. \mathcal{W}\right|_{\mathrm{stat}}^{-1} + \left. \mathcal{W}\right|_{\mathrm{syst}}^{-1} =  \left. \mathcal{W}\right|_{\mathrm{stat}}^{-1} \left( 1 +  \left. \mathcal{W}\right|_{\mathrm{stat}} \left. \mathcal{W}\right|_{\mathrm{syst}}^{-1} \right) \,.
\end{align}
We remark that the expectation value of the chi-squared allows to determine the p-value of a fit based on a generic positive weight matrix $\mathcal{W}$~\cite{Bruno:2022mfy} such as that in \req{ch_ss:eq:Wsyst}, including a term $\left.\mathcal{W}\right|_{\mathrm{syst}}$ which does not scale with the number of data points.
In the case of the continuum-limit and chiral extrapolations of $\sqrt{8t_{0}}\,f_{\pi K}$, the matrix $\left.\mathcal{W}\right|_{\mathrm{syst}}$ can be used to incorporate into the fit information about the regime in which SymEFT and $\chi$PT are expected to perform better.
When considering a correlated fit, $\left.\mathcal{W}\right|_{\mathrm{stat}}^{-1}$ corresponds to the covariance matrix $C$, although alternative weight matrices can be employed to account for the statistical fluctuations of the data.
More specifically, the elements of the weight matrix $\mathcal{W}$ employed in combined continuum-limit and chiral extrapolations of $\sqrt{8t_{0}}\,f_{\pi K}$, are written as follows
\begin{equation}
  \label{ch_ss:eq:Wpenal}
  \mathcal{W}_{ij}^{-1}=C_{ij}\times \left( 1+\frac{z_i}{C_{ii}} \right)^{\frac{1}{2}} \left( 1+\frac{z_j}{C_{jj}} \right)^{\frac{1}{2}}\,,
\end{equation}
where the elements of the covariance matrix $C_{ij}$ incorporate the statistical fluctuations of the data, while $z_{i}$ acts as an additional systematic factor encoding information about higher-order contributions in the Symanzik EFT and $\chi$PT as follows
\begin{equation}
  \label{ch_ss:eq:penal}
  z_i=\Biggl(\, p_{\beta}\, \left.\frac{a^4}{t_0^2}\right|_i\,\Biggr)^2 + \Biggl(\, p_{\phi_2}\, \Bigl.\phi_2^2\Bigr|_i\,\Biggr)^2\,.
\end{equation}
This is motivated by the fact that higher-order cutoff effects are most relevant for ensembles with coarser lattice spacings, whereas NNLO contributions of $\mbox{O}(m_{\pi}^{4})$ in $\chi$PT become increasingly significant for heavier pion masses.
The presence of the terms depending on $z_{i}$ effectively enlarges the uncertainties of ensembles with the coarsest lattice spacings and heaviest pion masses, and can only lead to a conservative increase in the error contribution to the weight of the chi-squared function.
We have explored various approaches to set the values of the coefficients $p_{\beta, \phi_2}$ in \req{ch_ss:eq:penal}, and to investigate how the results depend on these choices.
The values of $p_{\beta}$ and $p_{\phi_{2}}$ can be determined by directly assessing the magnitude of higher orders in the lattice spacing and the pion-mass dependence present in the lattice data, combined with the requirement that the statistical precision of ensembles at the coarsest lattice spacing or at the symmetric point does not dominate the fit.
This ensures a balanced exploration of systematic effects due to cuts in the data by appropriately weighting ensembles closer to the continuum limit and the physical pion mass.
Specifically, the effect introduced by $z_i$ will be applied to the coarsest ensembles corresponding to $\beta=3.40$, and to the symmetric point ensembles, corresponding to the heaviest pion mass, $m_\pi \approx 420$\, MeV.
Setting the coefficients $p_{\beta}$ or $p_{\phi_2}$ to infinity is thus equivalent to performing the cuts $\beta>3.40$ or $m_\pi< 420$\,MeV, respectively.
Conversely, setting them to zero corresponds to the absence of cuts.
As anticipated, the inclusion of the terms depending on $z_i$ in \req{ch_ss:eq:Wpenal} leads to a more balanced distribution of weights among well-fitting models --- based on their p-values --- both with and without data cuts.
This results in a more conservative estimate of the systematic uncertainty.
The impact on considering various ways to implement the modified weight in \req{ch_ss:eq:Wpenal} is discussed in Appendix~\ref{app_variations}.
Before proceeding to provide our results for the scale setting analysis, it is interesting to illustrate how the results coming from the Wilson unitary and the Wtm mixed-action setups not only are consistent, but allow for an excellent control of the continuum-limit systematics.
This can be seen through a universality test for the continuum extrapolations of $\sqrt{8t_0}f_{\pi K}$ based on a line of constant physics corresponding to the symmetric point ensembles, where $\phi_2^{\mathrm{sym}} = \frac{2}{3} \phi_4^{\mathrm{phys}}$.
As illustrated in Fig.~\ref{ch_ss:fig:universality}, the continuum-limit extrapolations provide evidence of a linear dependence in $a^2$ for both regularisations, and consistent results are observed in the continuum limit.
The validation provided by this universality test supports the approach of incorporating into the scale setting analysis a fit combining the two regularisations with a common continuum limit, in addition to conducting individual analyses of each of the two regularisations.
We observe that combining the two regularisations results in enhanced statistical precision and, most importantly, greater control over the continuum-limit extrapolation of $\sqrt{8t_0}f_{\pi K}$.

\begin{figure}[!t]
  \centering
  \includegraphics[width=.7\textwidth]{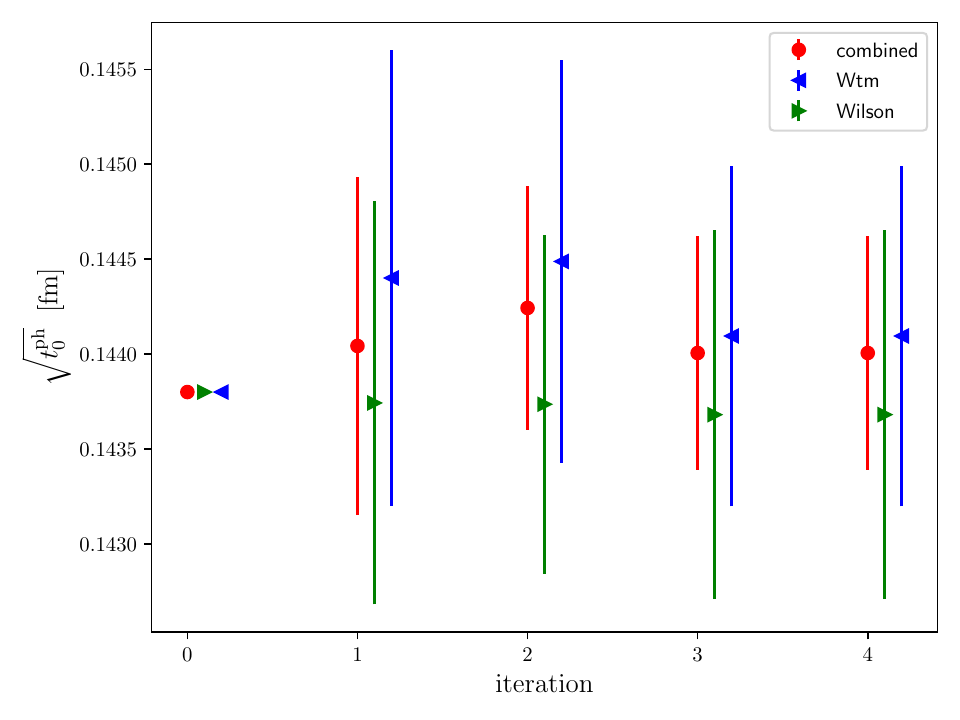}
  \caption{
    Convergence of the determination of $[t_0^{\mathrm{ph}}]^{1/2}$ employing the iterative procedure --- as explained in Sections~\ref{subsec:chiral-traj} and~\ref{subsec:t0} ---
    applied to the Wilson, Wtm and Combined analyses.
    The iterative process starts with an initial guess of $[t_0^{\mathrm{ph}}]^{1/2}$ that is free of error.
    This value is used to locate the target values of the physical masses, $\phi_2^{\mathrm{ph}}$ and $\phi_4^{\mathrm{ph}}$.
    The continuum and chiral extrapolations of $\sqrt{8t_0}f_{\pi K}$ lead to a determination of $[t_0^{\mathrm{ph}}]^{1/2}$ which is then used as input for the next iterative step.
    After a few iterative steps convergence is achieved for the three types of analysis --- Wilson, Wtm and Combined --- corresponding to the results in Eqs.~(\ref{ch_ss:eq:t0ph_w})--(\ref{ch_ss:eq:t0ph_c}).
  }
  \label{ch_ss:fig:iter}
\end{figure}

\begin{figure}[!t]
  \centering
  \includegraphics[width=.8\textwidth, angle=0]{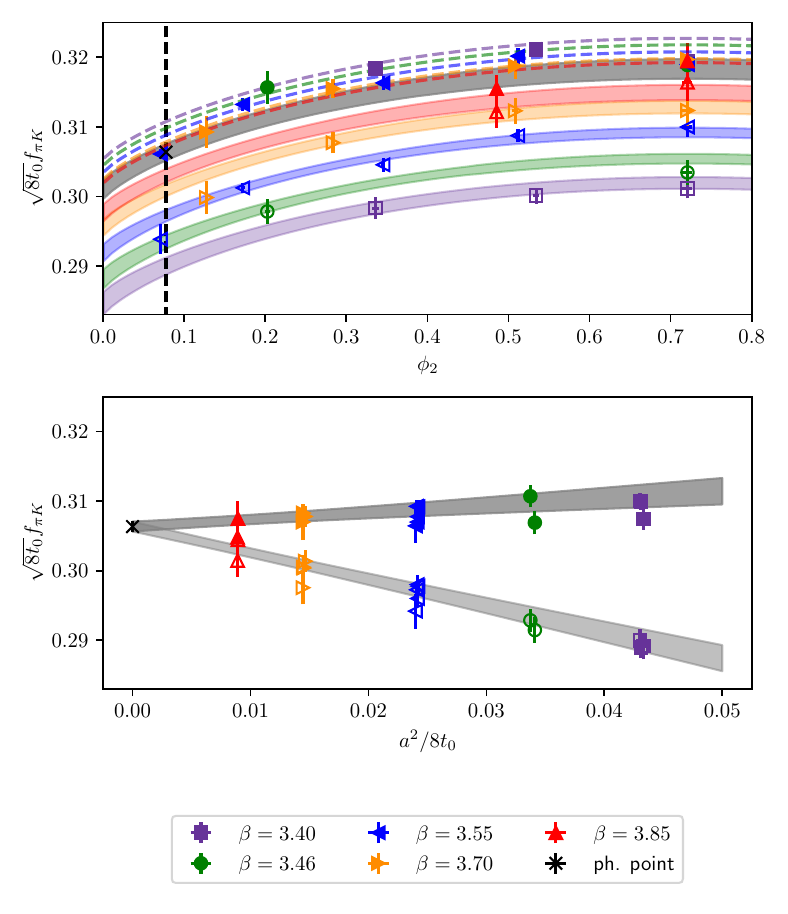}
  \caption{
    \textit{Top}:
    Light-quark mass-dependence of $\sqrt{8t_0}f_{\pi K}$ for the $\mathrm{SU(3)}$ $\chi$PT model with pure $\Oasq$ cutoff effects and absence of cuts in the data, corresponding to the label $[\chi \mathrm{SU(3)}][a^2][-]$, based on Eqs.~(\ref{ch_ss:eq:SU3ChPT}) and (\ref{ch_ss:eq:a2}).
    The combined fit of both Wilson (empty symbols) and Wtm mixed-action (filled symbols) results is considered.
    The modified weight in the chi-squared function, introduced in \req{ch_ss:eq:Wpenal}, is employed.
    The coloured bands display the pion mass-dependence at each value of the lattice spacing for the Wilson results.
    For a better visualisation, in the case of the mixed action, the dashed lines indicate the central values of the corresponding mass-dependence.
    The grey band indicates the common continuum-limit mass-dependence of the two regularisations.
    The position of the physical mass $\phi_2^{\mathrm{ph}}$ is indicated by the vertical dashed line.
    The value of $\sqrt{8t_0}f_{\pi K}$ at the physical point is indicated by the black cross.
    The p-value of this combined fit is $0.85$.
    \textit{Bottom}:
    Lattice spacing dependence of $\sqrt{8t_0}f_{\pi K}$ from the same model.
    The data points were projected to the physical pion mass $\phi_2^{\mathrm{ph}}$ using the fit result for the continuum mass-dependence $\Phi_{\pi K}^{\mathrm{cont}}(\phi_2)$.
  }
  \label{ch_ss:fig:SU3a2}
\end{figure}

\begin{figure}[!t]
  \centering
  \includegraphics[width=.6\textwidth]{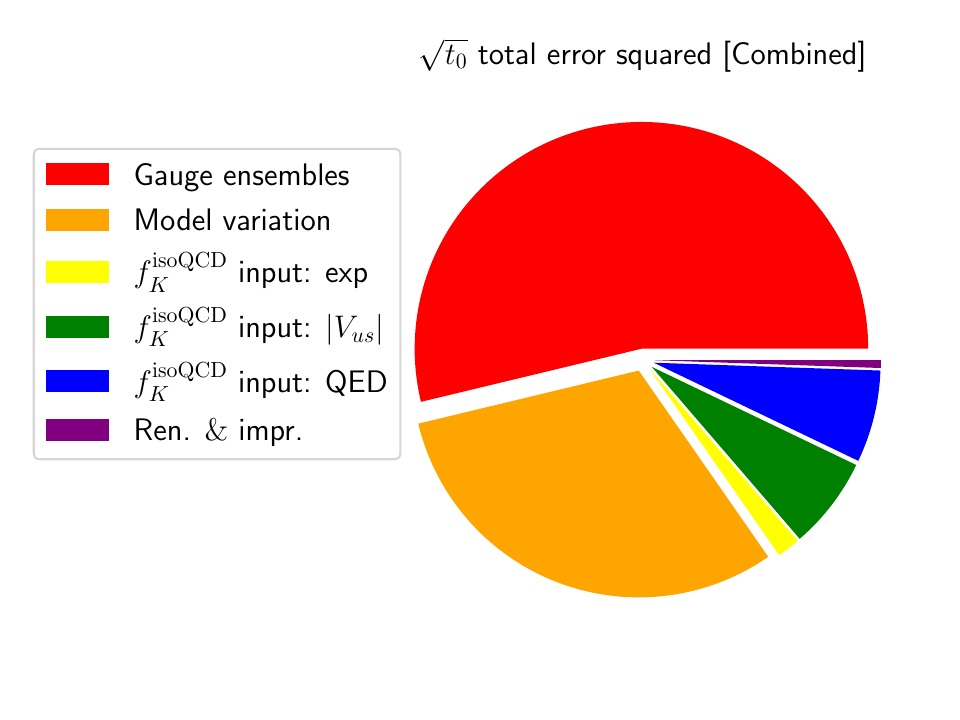}
  \caption{
    \label{fig:pie}
    Graphical representation of the various sources of error in the determination of $[t_0^{\mathrm{ph}}]^{1/2}$ in \req{ch_ss:eq:t0ph_c}, as collected in Table~\ref{ch_ss:tab:stat}.
  }
\end{figure}

\begin{figure}[!t]
  \centering
  \includegraphics[width=.7\textwidth]{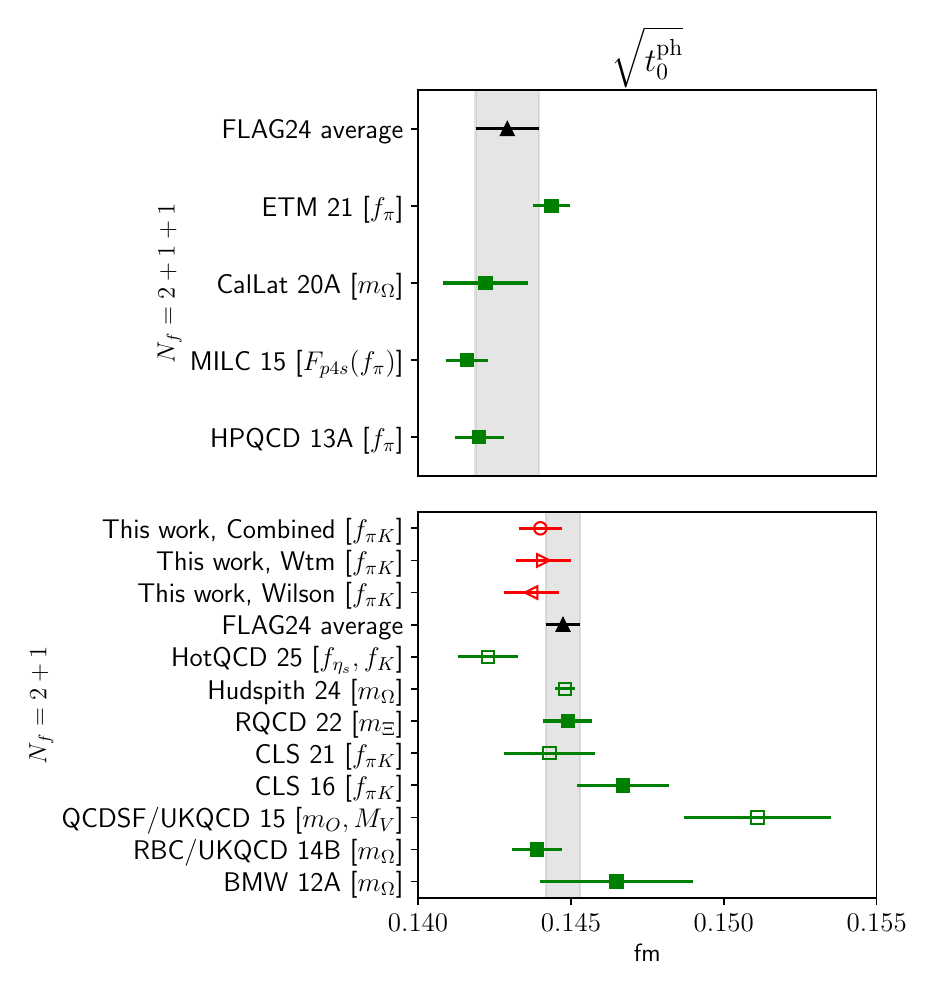}
  \caption{
    Comparison of the results of $[t_0^{\mathrm{ph}}]^{1/2}$ in Eqs.~(\ref{ch_ss:eq:t0ph_w})--(\ref{ch_ss:eq:t0ph_c}) with other lattice determinations that use $\NF=2+1$ or $2+1+1$ dynamical quarks.
    As for the present work, CLS 16~\cite{Bruno:2016plf}, CLS 21~\cite{Strassberger:2021tsu}, RQCD 22~\cite{RQCD:2022xux}, and Hudspith 24~\cite{Hudspith:2024kzk},
    are all based on $\NF=2+1$ CLS gauge ensembles.
    BMW 12A refers to~\cite{BMW:2012hcm}; RBC/UKQCD 14 to~\cite{RBC:2014ntl}; and QCDSF/UKQCD 14B to~\cite{Bornyakov:2015eaa}.
    HPQCD 13A is taken from~\cite{Dowdall:2013rya}, MILC 15 from~\cite{MILC:2015tqx}, CalLat 20A from~\cite{Miller:2020evg}, ETM 21 from~\cite{ExtendedTwistedMass:2021qui}, and HotQCD 25 from~\cite{Larsen:2025wvg}.
    The FLAG averages~\cite{FlavourLatticeAveragingGroupFLAG:2024oxs} are denoted by the vertical bands.
    Following the FLAG convention~\cite{FlavourLatticeAveragingGroupFLAG:2024oxs}, results that have not been included in the FLAG averages are shown with empty symbols.
    The external input used to set the scale is indicated in brackets.
  }
  \label{ch_ss:fig:t0_compar}
\end{figure}

Once the set of models employed to extrapolate $\sqrt{8t_0}f_{\pi K}$ to the continuum limit and the physical point have been considered, a determination of $t_0^{\mathrm{ph}}$ based in \req{eq:t0phinput} can be obtained from the weighted Model Average (MA) in \req{eq:OWMA}.
In the case of the TIC, the model weight $W$ is given in \req{eqn:weight}; moreover, alternative options have been considered to further assess the stability of the results when modifying the model-averaging procedure, as previously discussed and described in more detail in Appendix~\ref{app_variations}.
The systematic uncertainty associated with the model variations is determined using the weighted variance defined in \req{eq:WMsyst}.
Further details on the model average are collected in Appendix~\ref{app:MA}.
As outlined at the end of Sec.~\ref{subsec:chiral-traj} and at the beginning of the current section, the results for $[t_0^{\mathrm{ph}}]^{1/2}$ reported in Eqs.~(\ref{ch_ss:eq:t0ph_w})--(\ref{ch_ss:eq:t0ph_c}) are the outcome of an iterative process.
The convergence of the latter is illustrated in Fig.~\ref{ch_ss:fig:iter}.
The upper panel of Fig.~\ref{ch_ss:fig:SU3a2} illustrates the pion mass-dependence of $\sqrt{8t_0}f_{\pi K}$ based on the continuum-chiral extrapolation of model $[\chi \mathrm{SU(3)}][a^2]$, combining the complete Wilson and Wtm datasets (i.e. without cuts).
The corresponding figures for the cases of the fits based solely on the Wilson unitary and on the Wtm mixed-action data sets are collected in Appendix~\ref{app:chiral_w_wtm}.
Figures~\ref{ch_ss:fig:MA_w} to~\ref{ch_ss:fig:MA_comb} present the model-average results for the Wilson unitary setup, the Wtm mixed action and the combined analysis, respectively.
The full set of numerical results, including corresponding p-values and weights for each model considered, is compiled in Appendix~\ref{apex_model_av_t0}.
\begin{figure}[!t]
  \centering
  \includegraphics[width=1.\textwidth]{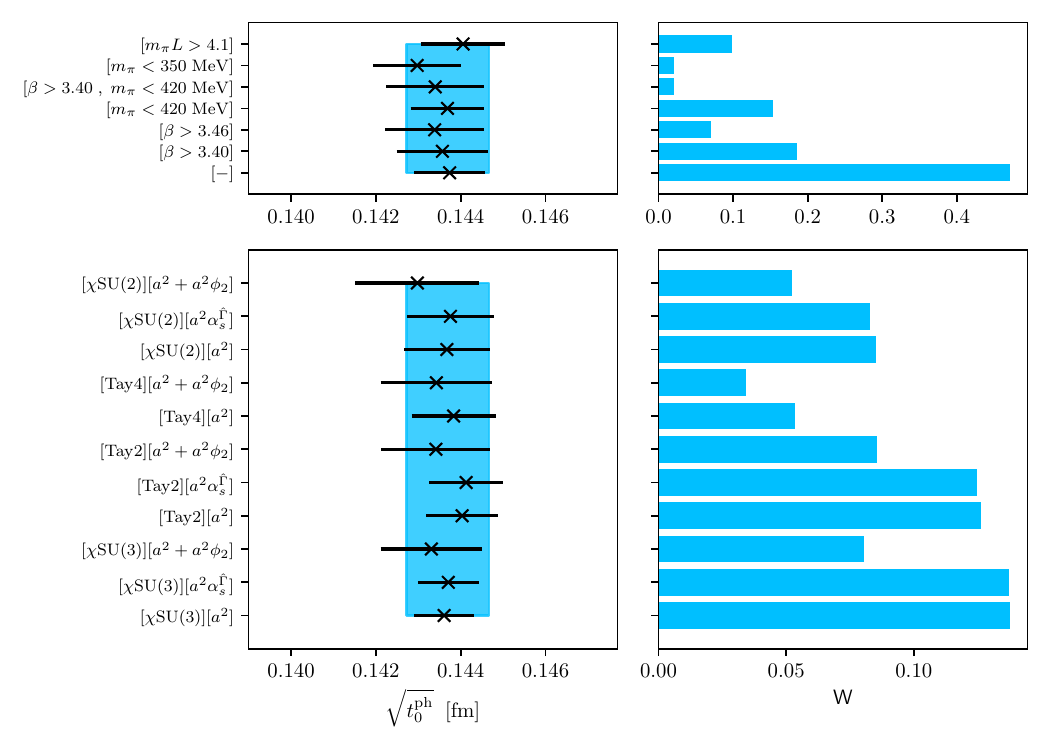}
  \caption{
    Model variations considered in the determination of $[t_0^{\mathrm{ph}}]^{1/2}$ reported in \req{ch_ss:eq:t0ph_w}, based solely on the Wilson unitary setup, together with physical input for $f_{\pi K}$.
    The model average, denoted by the vertical blue band in the left panels, employs the model weight $W$ in \req{eqn:weight} and includes the full set of models.
    The width of this band corresponds to the statistical and systematic uncertainties added in quadrature.
    \textit{Top}:
    The labels on the left correspond to the cuts applied to the datasets following Eqs.~(\ref{ch_ss:eq:cuts1})--(\ref{ch_ss:eq:cuts6}).
    They are supplemented by the label ``[--]'' to indicate cases where no cuts were applied.
    For each case, an average --- denoted by the black cross --- was taken according to the model weights of the complete set of fit forms employed to perform the continuum-limit and chiral extrapolations.
    The inclusion of the modification of the weight in the chi-squared function, as introduced in \req{ch_ss:eq:Wpenal}, implies that models incorporating cuts do induce a non negligible contribution to the systematic uncertainty.
    \textit{Bottom}:
    Model variations over the set of fit forms employed in the continuum-limit and chiral extrapolations.
    For each label, an average --- denoted by the black cross --- was taken according to the model weights of the complete set of cuts.
    The right panels display the corresponding weights $W$ of the models.
    For further details regarding the correspondence between each label and the respective fit models, as well as for tables collecting the results of $[t_0^{\mathrm{ph}}]^{1/2}$, the p-value, and the weight of each individual model, please refer to Appendix~\ref{apex_model_av_t0}.
    The decomposition of the model variations into sets of submodels -- e.g., exploring data cuts (\textit{top}) and lattice artefacts (\textit{bottom}) -- is described in Sect.~\ref{sec:decompsys}.
  }
  \label{ch_ss:fig:MA_w}
\end{figure}
\begin{figure}[!t]
  \centering
  \includegraphics[width=1.\textwidth]{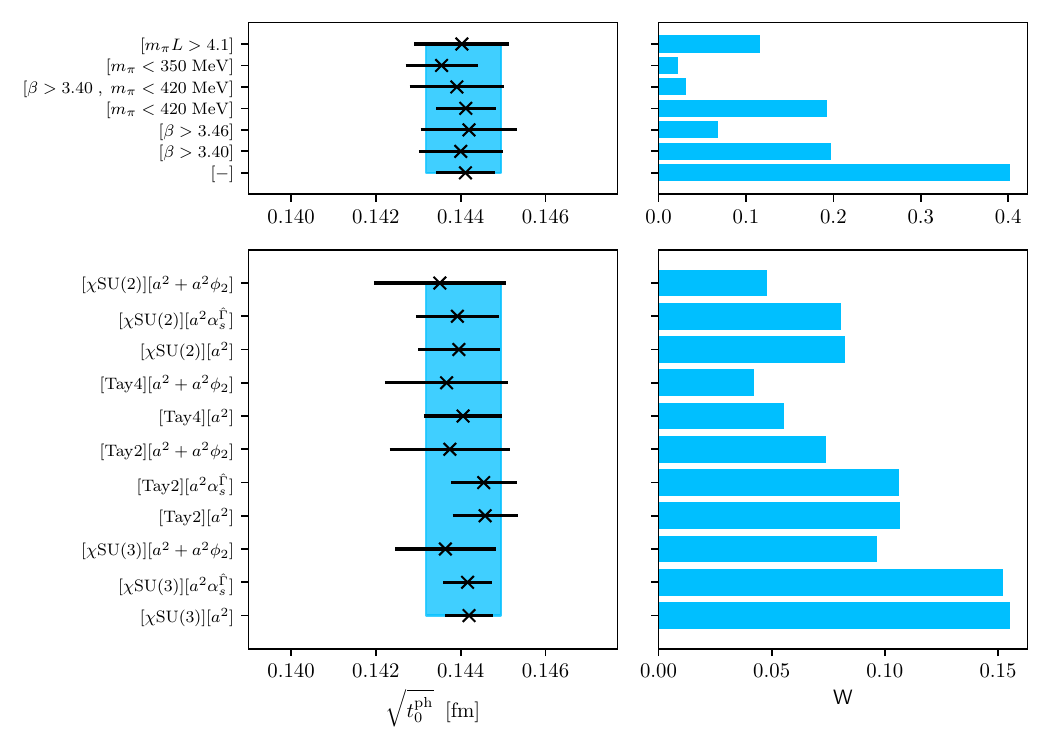}
  \caption{
    Model variations considered in the determination of $[t_0^{\mathrm{ph}}]^{1/2}$ reported in \req{ch_ss:eq:t0ph_tm}, based on Wtm mixed-action setup, together with physical input for $f_{\pi K}$.
    For further information we refer to the caption of Fig.~\ref{ch_ss:fig:MA_w}.
  }
  \label{ch_ss:fig:MA_tm}
\end{figure}
\begin{figure}[!t]
  \centering
  \includegraphics[width=1.\textwidth]{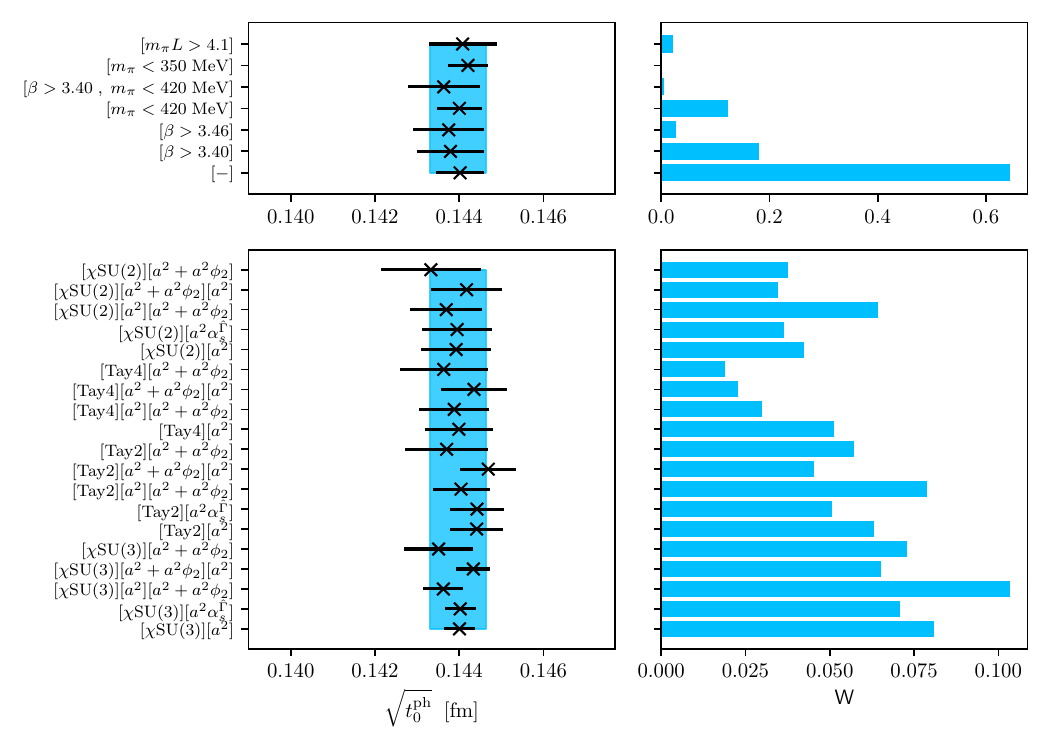}
  \caption{
    Model variations considered in the determination of $[t_0^{\mathrm{ph}}]^{1/2}$ reported in \req{ch_ss:eq:t0ph_tm}, based on the combination of the unitary Wilson and Wtm mixed-action regularisations, together with physical input for $f_{\pi K}$.
    For further information we refer to the caption of Fig.~\ref{ch_ss:fig:MA_w}.
  }
  \label{ch_ss:fig:MA_comb}
\end{figure}
The results for $\sqrt{t_0^{\mathrm{ph}}}$ in physical units --- using $f_{\pi K}^{\mathrm{isoQCD}}$ physical input --- as computed from the model average for the Wilson unitary setup, the Wtm mixed action, and the combination of both datasets, read
\begin{align}
  \label{ch_ss:eq:t0ph_w}
  \sqrt{t_0^{\mathrm{ph}}}&=0.1437(8)(4)\,\mathrm{fm}\,,&\mathrm{Wilson}\,, \\
  \label{ch_ss:eq:t0ph_tm}
  \sqrt{t_0^{\mathrm{ph}}}&=0.1441(8)(4)\,\mathrm{fm}\,,&\mathrm{Wtm}\,, \\
  \label{ch_ss:eq:t0ph_c}
  \sqrt{t_0^{\mathrm{ph}}}&=0.1440(6)(4)\,\mathrm{fm}\,,&\mathrm{Combined}\,.
\end{align}
The first error in Eqs.~(\ref{ch_ss:eq:t0ph_w})--(\ref{ch_ss:eq:t0ph_c}) originates from the gauge noise of the CLS ensembles and also includes the uncertainties in the renormalisation constants and improvement coefficients listed in Table~\ref{tab:b-terms}, as well as the physical input in \req{ch_ss:eq:isoQCD_fk}.
The second error corresponds to the systematic uncertainty resulting from the model variation, following \req{eq:WMsyst}.
The relative contribution of the various sources of error for the case of the combined analysis is displayed in Table~\ref{ch_ss:tab:stat} and Fig.~\ref{fig:pie}.
\begin{table}[!t]
  \centering
  \begin{tabular}{c c}
    Contributions to total error squared of $[t_0^{\mathrm{ph}}]^{1/2}$ [Combined] & \\
    \toprule
    Gauge ensembles & $53.7\%$ \\
    Model variation (systematic error) & $30.9\%$ \\
    $f_{K}^{\mathrm{isoQCD}}$ input: QED  & $6.6\%$ \\
    $f_{K}^{\mathrm{isoQCD}}$ input: $\left|V_{us}\right|$ & $6.6\%$ \\
    $f_{K}^{\mathrm{isoQCD}}$ input: exp & $1.6\%$ \\
    Renormalisation and improvement  & $0.6\%$ \\
    \bottomrule
  \end{tabular}
  \caption{
    Contributions to the total uncertainty of $[t_0^{\mathrm{ph}}]^{1/2}$ for the combined analysis of unitary Wilson and Wtm mixed-action regularisations corresponding to the result in \req{ch_ss:eq:t0ph_c}.
    The dominant source of uncertainty is related to the statistical gauge noise of the CLS ensembles.
    The model variation category comprises an estimate of the systematic error based on a model averaging procedure.
    This involves exploring a range of functional forms for the continuum-limit and chiral extrapolations, as well as applying cuts to the data set.
    The uncertainties associated with the input value of $f_{K}^{\mathrm{isoQCD}}$ are related to the three errors on \req{ch_ss:eq:isoQCD_fk}.
    The uncertainty in the renormalisation constants and improvement coefficients is derived from Table~\ref{tab:b-terms}.
  }
  \label{ch_ss:tab:stat}
\end{table}
The physical values of $\phi_{4}$ and $\phi_{2}$ derived from the the physical value of $t_0$ in \req{ch_ss:eq:t0ph_c} and the inputs for the pion and kaon masses in the Eqs.~(\ref{ch_ss:eq:isoQCD_mpi})--(\ref{ch_ss:eq:isoQCD_mk}) are given by
\begin{align}
  \label{ch_ss:eq:phi4ph}
  \phi_{4}^{\mathrm{ph}}&=1.081(9)(6)\,,  \\
  \label{ch_ss:eq:phi2ph}
  \phi_{2}^{\mathrm{ph}}&=0.0777(6)(4)\,,
\end{align}
where the two quoted uncertainties follow the same conventions explained after Eqs.~(\ref{ch_ss:eq:t0ph_w})--(\ref{ch_ss:eq:t0ph_c}).
The impact of varying the coefficients $p_{\beta}$ and $p_{\phi_2}$ in \req{ch_ss:eq:penal} is reported in Table~\ref{tab:variations_chi2W} of Appendix~\ref{app_variations}.
The corresponding values of $[t_0^{\mathrm{ph}}]^{1/2}$ are well compatible with those reported in Eqs.~(\ref{ch_ss:eq:t0ph_w})--(\ref{ch_ss:eq:t0ph_c}).
The choice that has been adopted, namely $p_{\beta}=1.5\times10^{-2}$ and $p_{\phi_2}= 5.8\times10^{-3}$, results in a more conservative error estimate when compared to the standard case, in which $p_{\beta}=p_{\phi_2}=0$.
We refer to Appendix~\ref{app:prescription} for an analysis of the impact on $[t_0^{\mathrm{ph}}]^{1/2}$ of considering alternative definitions of the isoQCD limit, as compared to the reference input values given in Eqs.~(\ref{ch_ss:eq:isoQCD_mpi})--(\ref{ch_ss:eq:isoQCD_fk}).
We also provide in Table~\ref{tab:ders} the values of the partial derivatives of $[t_0^{\mathrm{ph}}]^{1/2}$ with respect to the input quantities $m_{\pi,K}$ and $f_{\pi,K}$, enabling conversion of our results to a different scheme involving alternative values of the input quantities that define the isoQCD limit.
A comparison of the results of $[t_0^{\mathrm{ph}}]^{1/2}$ in Eqs.~(\ref{ch_ss:eq:t0ph_w})--(\ref{ch_ss:eq:t0ph_c}) with other lattice determinations based on $\NF=2+1$ and $2+1+1$ dynamical quarks is shown in Fig.~\ref{ch_ss:fig:t0_compar}.
The result in \req{ch_ss:eq:t0ph_c} amounts to an update with respect to the previous determination, $[t_0^{\mathrm{ph}}]^{1/2}=0.1445(6)$\,fm, reported in Ref.~\cite{Bussone:2023kag}.
The difference between these values is well within one sigma.
In comparison with the preceding determination, the present study incorporates additional ensembles with a finer lattice spacing, $a \approx 0.04$\,fm (ensembles J500 and J501), in addition to smaller pion masses (ensembles D450, N302 and E300, and the physical point ensemble E250).
Another improvement consists in the use of the reweighting factors computed from the low-mode deflation method for a subset of the ensembles (see Table~\ref{tab:CLS_ens}).

\subsection{Determination of $t_{0}^{\mathrm{sym}}$ and the lattice spacing $a$}

\begin{figure}[!t]
  \centering
  \includegraphics[width=.7\textwidth]{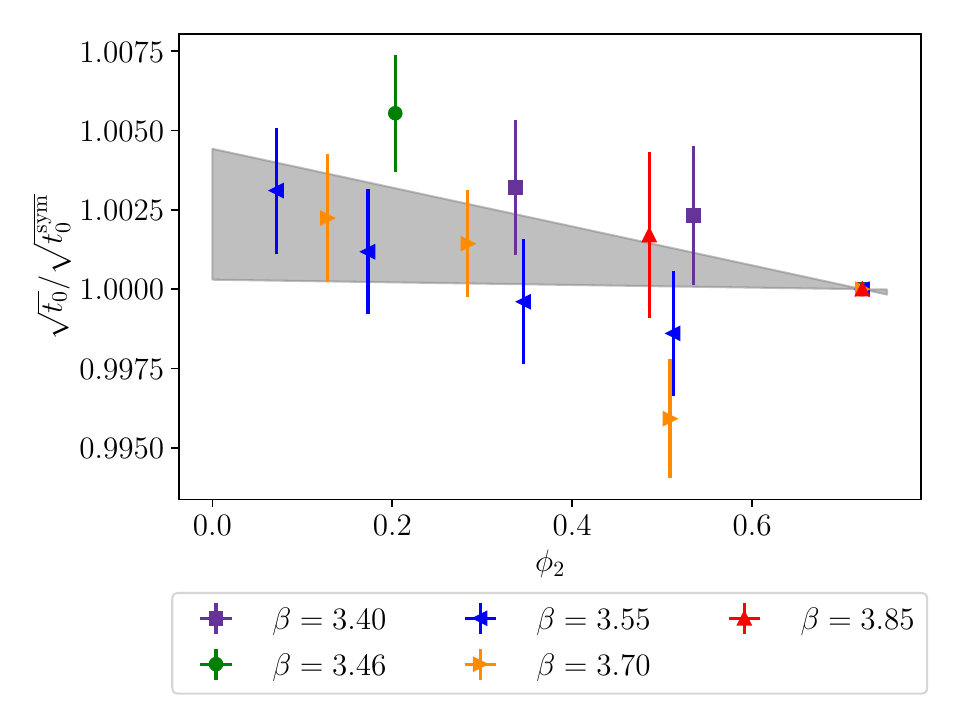}
  \caption{
    Light-quark mass-dependence of the ratio in \req{ch_ss:eq:ratio}, shown together with a fit to the functional form in \req{ch_ss:eq:fit_t0_sym}, which yields a p-value of approximately $0.2$.
  }
  \label{ch_ss:fig:t0_sym}
\end{figure}

As previously described, the symmetric point is defined as the intersection in the quark-mass plane of the symmetric line -- where the up, down, and strange quarks are mass-degenerate -- with the chiral trajectory specified in \req{eqn:TrMq}.
More specifically, in terms of the quantities $\phi_2$ and $\phi_4$ in Eqs.~(\ref{eqn:phi2})--(\ref{eqn:phi4}), the symmetric point satisfies
\begin{equation}
  \phi_2^{\mathrm{sym}}=\frac{2}{3}\phi_4^{\mathrm{ph}},
\end{equation}
where the physical value $\phi_4^{\mathrm{ph}}$, given in \req{ch_ss:eq:phi4ph}, was obtained by the iterative procedure used to determine $t_0^{\mathrm{ph}}$.
In order to extract the value of $t_0$ at the symmetric point, $t_0^{\mathrm{sym}}=t_0(\phi_2^{\mathrm{sym}},\phi_4^{\mathrm{ph}})$, we consider the ratio~\cite{Strassberger:2023xnj} 
\begin{equation}
  \label{ch_ss:eq:ratio}
  \frac{\sqrt{t_0/a^2}}{\sqrt{t_0^{\mathrm{sym}}/a^2}}\,,
\end{equation}
where $t_0/a^2$ is the gradient flow scale determined on a given ensemble, while the corresponding determination on the symmetric point ensemble at the same value of the inverse coupling $\beta$ is given by $\sqrt{t_0^{\mathrm{sym}}/a^2}$.
In accordance with the observation reported in Ref.~\cite{Strassberger:2023xnj}, the lattice determination of the ratio  in \req{ch_ss:eq:ratio} can be described by the following fit ansatz
\begin{equation}
  \label{ch_ss:eq:fit_t0_sym}
  F(\phi_2)=\sqrt{1+p(\phi_2-\phi_2^{\mathrm{sym}})}\,,
\end{equation}
where $p$ is a fit parameter.
Including additional terms in the fit form indicates that the available data do not provide sufficient sensitivity to reliably resolve lattice artefacts of $\Oasq$, $\mbox{O}(a^{2}\phi_2)$, or $\mbox{O}(a^{2}\alphas^{\hat{\Gamma}})$ in the ratio given in \req{ch_ss:eq:ratio}.
The fit based on \req{ch_ss:eq:fit_t0_sym} is illustrated in Fig.~\ref{ch_ss:fig:t0_sym}.
The evaluation of the functional form in \req{ch_ss:eq:fit_t0_sym} at $\phi_2^{\mathrm{ph}}$ leads to
\begin{align}
  \label{eq:R_sym}
  \frac{\sqrt{t_{0}^{\mathrm{ph}}}}{\sqrt{t_{0}^{\mathrm{sym}}}}=1.0021(18)\,.
\end{align}
Using the results in Eqs.~(\ref{ch_ss:eq:t0ph_w})--(\ref{ch_ss:eq:t0ph_c}), the following determinations of $t_0$ at symmetric point are then obtained,
\begin{align}
  \label{ch_ss:eq:t0_sym}
  \sqrt{t_0^{\mathrm{sym}}}&=0.1434(9)(4)\,\mathrm{fm}\,,&\mathrm{Wilson}\,, \\
  \sqrt{t_0^{\mathrm{sym}}}&=0.1438(10)(5)\,\mathrm{fm}\,,&\mathrm{Wtm}\,, \\
  \sqrt{t_0^{\mathrm{sym}}}&=0.1437(6)(4)\,\mathrm{fm}\,,&\mathrm{Combined}\,,
\end{align}
where, as before, the two quoted uncertainties follow the same conventions explained after Eqs.~(\ref{ch_ss:eq:t0ph_w})--(\ref{ch_ss:eq:t0ph_c}).
Furthermore, the fit to the ratio \req{ch_ss:eq:ratio} using the functional form of \req{ch_ss:eq:fit_t0_sym} can be used to determine
\begin{equation}
  \left(\sqrt{\frac{t_0}{a^2}}\right)^{\mathrm{ph}}=\sqrt{\frac{t_0^{\mathrm{sym}}}{a^2}}F(\phi_2^{\mathrm{ph}})\,.
\end{equation}
Using a determination of $t_{0}^{\mathrm{ph}}$, the values of the lattice spacing in physical units can thus be obtained from
\begin{equation}
  \label{ch_ss:eq:a}
  a=\frac{\sqrt{t_0^{\mathrm{ph}}}}{\left(\sqrt{\frac{t_0}{a^2}}\right)^{\mathrm{ph}}}.
\end{equation}
The results for the lattice spacing, as determined by the values of  $[t_0^{\mathrm{ph}}]^{1/2}$ in Eqs.~(\ref{ch_ss:eq:t0ph_w})--(\ref{ch_ss:eq:t0ph_c}), are presented in Table~\ref{ch_ss:tab:a}.
\begin{table}[!t]
  \centering
  \begin{tabular}{c c c c}
    \toprule
    $\beta$ & $a$\,[fm] Wilson & $a$\,[fm] Wtm & $a$\,[fm] Combined \\
    \toprule
    $3.40$ & $0.0844(6)(2)$ & $0.0846(6)(2)$ & $0.0846(4)(2)$ \\
    $3.46$ & $0.0749(5)(2)$ & $0.0751(5)(2)$ & $0.0750(4)(2)$ \\
    $3.55$ & $0.0630(4)(2)$ & $0.0632(4)(2)$ & $0.0631(3)(2)$ \\
    $3.70$ & $0.0489(3)(1)$ & $0.0490(3)(1)$ & $0.0490(2)(1)$ \\
    $3.85$ & $0.0383(2)(1)$ & $0.0384(3)(1)$ & $0.0384(2)(1)$ \\
    \bottomrule
  \end{tabular}
  \caption{
    Determination of the lattice spacing $a$ in physical units using \req{ch_ss:eq:a}.
    The values of the gradient flow scale $[t_0^{\mathrm{ph}}]^{1/2}$, determined in Eqs.~(\ref{ch_ss:eq:t0ph_w})–(\ref{ch_ss:eq:t0ph_c}) for the Wilson, Wtm mixed action, and Combined analyses, respectively, are employed.
  }
  \label{ch_ss:tab:a}
\end{table}

\subsection{Determination of $t_0^*$}

Following Refs.~\cite{Bruno:2016plf,Brida:2025gii}, we also consider the flow scale $t_0^{*}=t_0(\phi_2^{*},\;\phi_4^{*})$,
defined for degenerate up, down, and strange quark masses and the prescribed values
\begin{gather}
  \label{eq:phistar}
  \phi_4^{*}=1.11\,, \quad \phi_{2}^{*}=\frac{2}{3}\phi_4^{*}\,.
\end{gather}
Hence, the determination of $t_0^{*}$ follows a similar procedure as outlined in the previous section, but with the mass-shifting procedure from Sect.~\ref{subsec:chiral-traj} applied to the specific value $\phi_4^{*}$ given in \req{eq:phistar}, instead of the physical value $\phi_4^{\mathrm{ph}}$.
This analysis leads to the following determination of the ratio
\begin{equation}
  \label{eq:R_star}
  \frac{\sqrt{t_0^{\mathrm{ph}}}}{\sqrt{t_0^{*}}}=1.0016(18)\,.
\end{equation}
Combining this result with the determinations of $[t_0^{\mathrm{ph}}]^{1/2}$ in Eqs.~(\ref{ch_ss:eq:t0ph_w})–(\ref{ch_ss:eq:t0ph_c}) yields
\begin{align}
  \label{ch_ss:eq:t0*}
  \sqrt{t_0^*}&=0.1435(9)(4)\,\mathrm{fm}\,,&\mathrm{Wilson}\,, \\
  \sqrt{t_0^*}&=0.1438(9)(5)\,\mathrm{fm}\,,&\mathrm{Wtm}\,, \\
  \sqrt{t_0^*}&=0.1437(7)(4)\,\mathrm{fm}\,,&\mathrm{Combined}\,.
\end{align}
The results in Eqs.~(\ref{eq:R_star})--(\ref{ch_ss:eq:t0*}) are in good agreement with those reported in Refs.~\cite{RQCD:2022xux,Brida:2025gii}.

\subsection{Determination of $t_{0}^{\mathrm{ph}}$ from $f_{\pi}^{\mathrm{isoQCD}}$}
\label{subsec:t0fpi}

The determination of $t_{0}^{\mathrm{ph}}$ presented in Sec.~\ref{subsec:t0} was based on the isoQCD input value for the flavour averaged combination of the pion and kaon decay constants, $f_{\pi K}$, as defined in \req{eq:fpik}.
In this subsection, an alternative analysis for determining $t_{0}^{\mathrm{ph}}$ is considered, relying exclusively on physical external input from $f_{\pi}^{\mathrm{isoQCD}}$.
This choice is motivated by the fact that this analysis does not rely on the value on $f_{K}^{\mathrm{isoQCD}}$, whose determination requires prior knowledge of the CKM matrix element $|V_{us}|$.
Generally, it is preferable that this external input for $|V_{us}|$ is independent of lattice results; however, the input value of $|V_{us}|$ used to set $f_{K}^{\mathrm{isoQCD}}$ in Ref.~\cite{FlavourLatticeAveragingGroupFLAG:2024oxs} is derived from lattice determinations of the $K \to \pi \ell \nu$ semileptonic decay from factor $f_+(0)$.
In contrast, for the case of $f_{\pi}^{\mathrm{isoQCD}}$, the input value of $|V_{ud}|$ can be determined via super-allowed $\beta$ decays~\cite{ParticleDataGroup:2024cfk} without lattice inputs.
The CKM matrix element $|V_{us}|$ is instrumental in testing the unitarity of the first row of the CKM matrix, making it an important target for lattice QCD computations.
Moreover, compared to $f_{K}$, the current uncertainties associated with strong isospin-breaking and QED corrections are smaller for $f_{\pi}$~\cite{FlavourLatticeAveragingGroupFLAG:2024oxs}.
To set the scale using $f_{\pi}$, we perform continuum-limit and chiral extrapolations of $\sqrt{8t_{0}}f_{\pi}$, following a similar approach to that employed in the analysis of  $\sqrt{8t_{0}}f_{\pi K}$ in Sec.~\ref{subsec:t0}.
Employing Eqs.~(\ref{ch_ss:eq:t0_chiral})~and~(\ref{ch_ss:eq:SU2fpi}) to describe the mass-dependence, yields the following functional forms
\begin{align}
  \label{eq:SU3_fpi}
  \Phi_{\pi,\chi \mathrm{SU(3)}}^{\mathrm{cont}}(\phi_2)\equiv\left(\sqrt{8t_{0}}f_{\pi}\right)^{\mathrm{cont}}&=\frac{A}{4\pi}\left[1+B\phi_{2}+C\phi_{4}-2\tilde{L}\left(\frac{\phi_{2}}{A^{2}}\right)-\tilde{L}\left(\frac{\phi_{4}-\frac{1}{2}\phi_{2}}{A^{2}}\right)\right]\,,
  \\
  \label{eq:SU2_fpi}
  \Phi_{\pi,\chi \mathrm{SU(2)}}^{\mathrm{cont}}(\phi_2)\equiv\left(\sqrt{8t_{0}}f_{\pi}\right)^{\mathrm{cont}}&=\frac{A}{4\pi}\left[1+B\phi_{2}-2\tilde{L}\left(\frac{\phi_{2}}{A^{2}}\right)+C\phi_{4}\right]\,.
\end{align}
We label these parametrisations as $[\chi \mathrm{SU(3)}]$ and $[\chi \mathrm{SU(2)}]$, respectively.

\begin{figure}[!t]
  \centering
  \includegraphics[width=.8\textwidth]{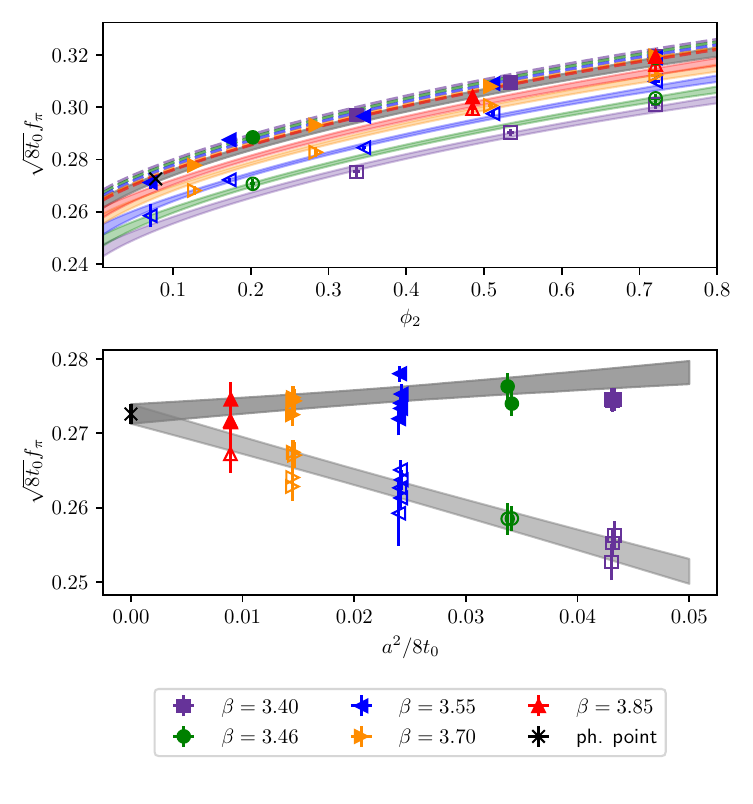}
  \caption{
    \textit{Top}:
    Light-quark mass-dependence of $\sqrt{8t_0}f_{\pi}$ based on \req{eq:SU2_fpi} with $\Oasq$ cutoff effects and no cuts applied to the dataset.
    This corresponds to the model label $[\chi \mathrm{SU(2)}][a^2][-]$.
    The combined fit of both Wilson (empty symbols) and Wtm mixed-action (filled symbols) results is considered.
    The modified weight in the chi-squared function, introduced in \req{ch_ss:eq:Wpenal}, is employed.
    The coloured bands indicate the pion mass-dependence at each lattice spacing for the Wilson data, while the dashed lines represent the dependence for the Wtm mixed-action data.
    For the latter, only the central values of the bands are shown, for clarity of the visualisation.
    The grey band denotes the common continuum-limit mass-dependence.
    The fit is performed simultaneously with that shown in Fig.~\ref{fig:t0fk_chiral_continuum}.
    The p-value of this combined fit is $0.15$.
    \textit{Bottom}:
    Lattice-spacing dependence of $\sqrt{8t_0}f_{\pi}$ based on the same model, with points projected to the physical pion mass $\phi_2^{\mathrm{ph}}$.
    \label{fig:t0fpi_chiral_continuum}}
\end{figure}

\begin{figure}[!t]
  \centering
  \includegraphics[width=.8\textwidth]{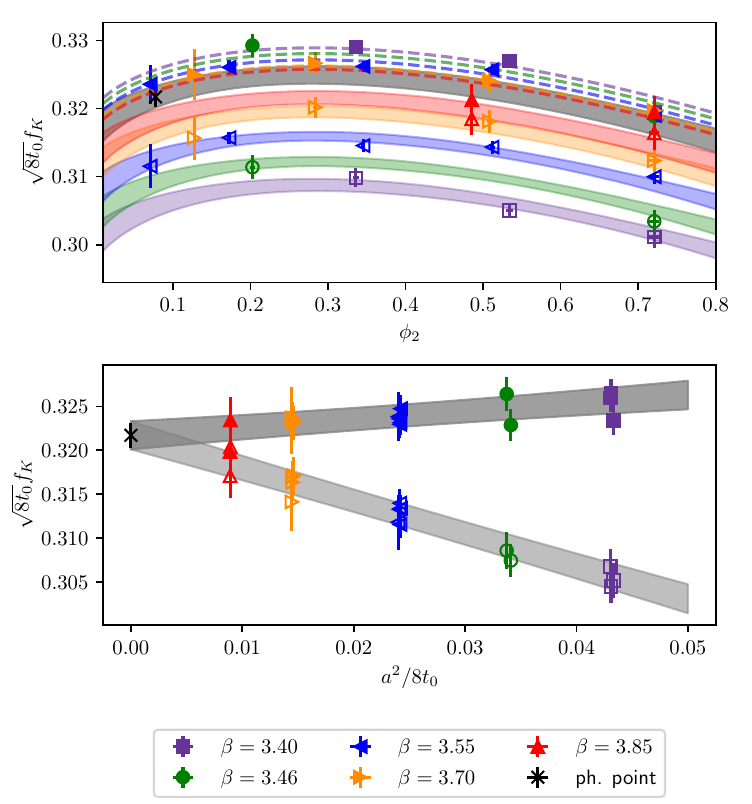}
  \caption{
    Light-quark mass-dependence and lattice-spacing dependence of $\sqrt{8t_0}f_{K}$, obtained from the same combined fit displayed in Fig.~\ref{fig:t0fpi_chiral_continuum}, to which we refer for further details.
    \label{fig:t0fk_chiral_continuum}
  }
\end{figure}

\begin{figure}[!t]
  \centering
  \includegraphics[width=1.\textwidth]{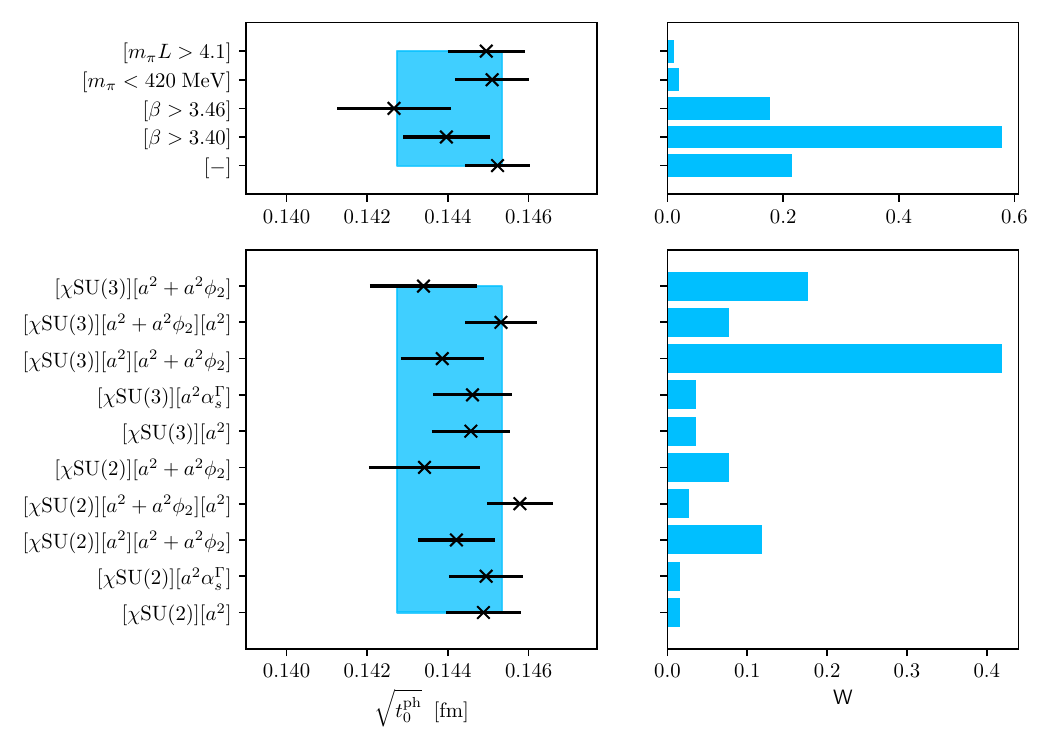}
  \caption{
    Model variations considered in the determination of $[t_0^{\mathrm{ph}}]^{1/2}$, as reported in \req{eq:t0ph_c_fpi}, using $f_{\pi}$ as physical input and combining unitary Wilson and Wtm mixed-action regularisations.
    The notation employed closely follows the one introduced in Fig.~\ref{ch_ss:fig:MA_w}.
    Appendix~\ref{apex_model_av_t0_fpi} contains tables detailing the correspondence between each label and its associated model, as well as the resulting values of $[t_0^{\mathrm{ph}}]^{1/2}$, the p-value, and the weight of each individual model.
  }
  \label{fig:MA_fpi}
\end{figure}

\begin{figure}[!t]
  \centering
  \includegraphics[width=.6\textwidth]{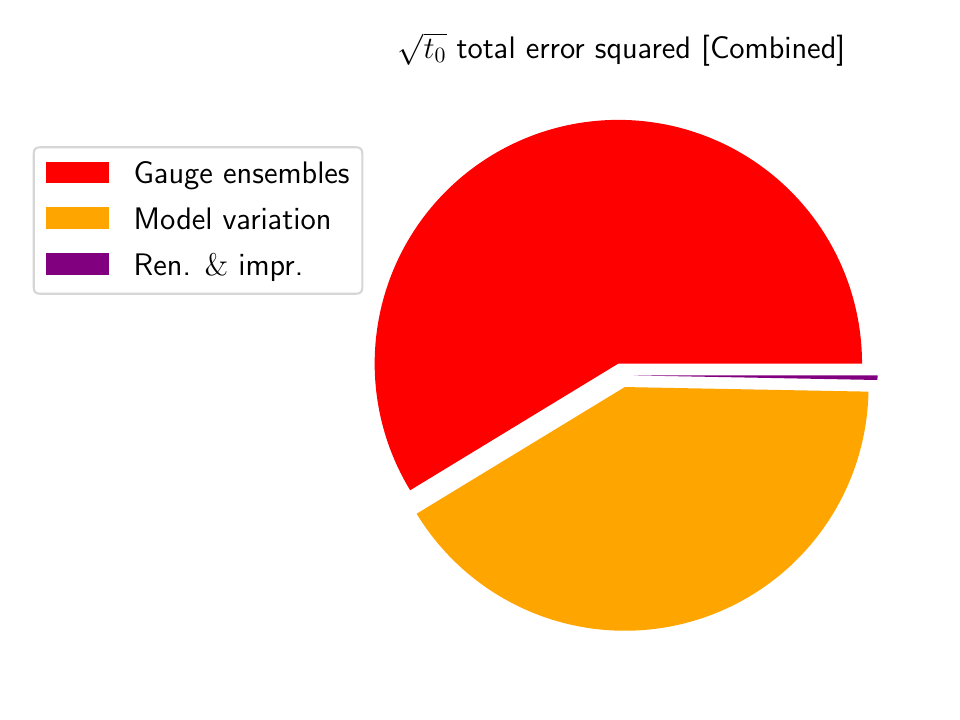}
  \caption{
    \label{fig:pie_fpi}
    Graphical representation of the contributions from different sources of uncertainty to the determination of $[t_{0}^{\mathrm{ph}}]^{1/2}$ in \req{eq:t0ph_c_fpi}, obtained using $f_{\pi}^{\mathrm{isoQCD}}$ as external input, as summarised in Table~\ref{tab:pie_t0_fpi}.
  }
\end{figure}

Although no physical input will be used for $f_{K}$, we will consider a combined fit of  $\sqrt{8t_{0}}f_{\pi}$ and $\sqrt{8t_{0}}f_{K}$, which enables an independent determination of $f_{K}^{\mathrm{isoQCD}}$.
The mass-dependence of $\sqrt{8t_{0}}f_{K}$ at NLO in $\chi$PT is given by
\begin{align}
  \label{eq:SU3_fk}
  \Phi_{K,\chi \mathrm{SU(3)}}^{\mathrm{cont}}(\phi_2)\equiv\left(\sqrt{8t_{0}}f_{K}\right)^{\mathrm{cont}}&=\frac{A}{4\pi}\left[1+B\left(\phi_{4}-\frac{1}{2}\phi_{2}\right)+C\phi_{4}\right. \notag \\
    &\left.-\frac{3}{4}\tilde{L}\left(\frac{\phi_{2}}{A^{2}}\right)-\frac{3}{2}\tilde{L}\left(\frac{\phi_{4}-\frac{1}{2}\phi_{2}}{A^{2}}\right)-\frac{3}{4}\tilde{L}\left(\frac{\frac{4}{3}\phi_{4}-\phi_{2}}{A^{2}}\right)\right]\,,
  \\
  \label{eq:SU2_fk}
  \Phi_{K,\chi \mathrm{SU(2)}}^{\mathrm{cont}}(\phi_2)\equiv\left(\sqrt{8t_{0}}f_{K}\right)^{\mathrm{cont}}&=\frac{D}{4\pi}\left[1+B\phi_{2}-\frac{3}{4}\tilde{L}\left(\frac{\phi_{2}}{A^{2}}\right)+C\phi_{4}\right]\,,
\end{align}
where the labelling $[\chi \mathrm{SU(3)}]$ and $[\chi \mathrm{SU(2)}]$, respectively, is also employed.
As certain fit parameters are shared between the expressions for the mass-dependence of $\sqrt{8t_{0}}f_{\pi}$ and $\sqrt{8t_{0}}f_{K}$ in Eqs.~(\ref{eq:SU3_fpi})--(\ref{eq:SU2_fpi}) and (\ref{eq:SU3_fk})--(\ref{eq:SU2_fk}), respectively, it is advantageous to fit both quantities simultaneously in order to improve the stability of the fits.
The expressions in Eqs.~(\ref{eq:SU3_fpi})--(\ref{eq:SU2_fk}) can be supplemented with terms that parameterise cutoff effects, as in Eqs.~(\ref{ch_ss:eq:a2})--(\ref{ch_ss:eq:a2phi2}), allowing for independent cutoff effects to contribute in the pion and kaon decay constants, as well as in the unitary Wilson and Wtm mixed-action regularisations.
In addition, the following data cuts are considered
\begin{align}
  \beta>3.40\,,  &~~\mathrm{i.e.}~ a < 0.08\,\mathrm{fm}\,, \\
  \beta>3.46\,,  &~~\mathrm{i.e.}~ a < 0.07\,\mathrm{fm}\,, \\
  m_{\pi}<420\,\mathrm{MeV}\,,  &~~\mathrm{i.e.}~ \phi_2<0.7\,, \\
  m_{\pi}L>4.1\,.&
\end{align}
To allow for a more conservative estimate of the systematic uncertainty --- by ensuring that fits with dataset cuts can contribute non-negligibly to the model average --- we employ the weighting scheme introduced in \req{ch_ss:eq:Wpenal}.
Fig.~\ref{fig:t0fpi_chiral_continuum} illustrates the continuum-limit and chiral extrapolations of $\sqrt{8t_{0}}f_{\pi}$, based on the mass-dependence in \req{eq:SU2_fpi}, including purely $\Oasq$ lattice artefacts and considering all data points (i.e. absence of cuts) from a combined fit to the Wilson unitary and Wtm mixed-action regularisations.
Fig.~\ref{fig:t0fk_chiral_continuum} shows the outcome of the same combined fit for the case of $\sqrt{8t_{0}}f_{K}$.
In Fig.~\ref{fig:MA_fpi}, we present the set of model variations for  $\sqrt{t_{0}^{\mathrm{ph}}}$, obtained from the continuum-limit and chiral extrapolations of $\sqrt{8t_{0}}f_{\pi}$.
Because of its more pronounced dependence on the light-quark mass, the analysis of $\sqrt{8t_{0}}f_{\pi}$ results in lower p-values and increased statistical and systematic uncertainties relative to the corresponding analysis of $\sqrt{8t_{0}}f_{\pi K}$.
Consequently, with the current dataset, the determination of ${t_{0}^{\mathrm{ph}}}$ from scale setting using $f_{\pi}$ is less precise than the one obtained in the analysis presented in Sec.~\ref{subsec:t0}.
Using physical input from $f_{\pi}$, and combining results from the Wilson unitary and Wtm mixed-action regularisations, the determination of $\sqrt{t_{0}^{\mathrm{ph}}}$ reads
\begin{equation}
  \label{eq:t0ph_c_fpi}
  \sqrt{t_{0}^{\mathrm{ph}}}=0.1439(10)(9)\,{\mathrm{fm}}\,,\quad\left[f_{\pi}^{\mathrm{isoQCD}}~\mathrm{input}\right]\,.
\end{equation}
The corresponding breakdown of uncertainties is provided in Table~\ref{tab:pie_t0_fpi} and Fig.~\ref{fig:pie_fpi}.
\begin{table}[!t]
  \centering
  \begin{tabular}{c c}
    \toprule
    Contributions to total error squared of $[t_0^{\mathrm{ph}}]^{1/2}$ [Combined, $f_{\pi}^{\mathrm{isoQCD}}$ input] & \\
    \toprule
    Gauge ensembles & $58.8\%$ \\
    Model variation (systematic error) & $41.0\%$ \\
    Renormalisation and improvement & $0.3\%$ \\
    \bottomrule
  \end{tabular}
  \caption{
    Breakdown of the total uncertainty in $[t_0^{\mathrm{ph}}]^{1/2}$, determined using $f_{\pi}^{\mathrm{isoQCD}}$ as external input, for the combined analysis of unitary Wilson and Wtm mixed-action regularisations corresponding to the result in \req{eq:t0ph_c_fpi}.
    The dominant source of uncertainty is related to the statistical gauge noise of the CLS ensembles.
    The model variation category comprises an estimate of the systematic error based on a model averaging procedure.
    This involves exploring a range of functional forms for the continuum-limit and chiral extrapolations, as well as applying cuts to the data set.
    No uncertainty is assigned to the physical input because the isoQCD limit~\cite{FlavourLatticeAveragingGroupFLAG:2024oxs} is defined using the values of $m_{\pi}$, $m_{K}$, and $f_{\pi}$ without errors, as specified in Eqs.~(\ref{ch_ss:eq:isoQCD_mpi})--(\ref{ch_ss:eq:isoQCD_fpi}).
  }
  \label{tab:pie_t0_fpi}
\end{table}
The determination of $t_{0}^{\mathrm{ph}}$ in \req{eq:t0ph_c_fpi}, together with the model average determination of the continuum-limit value of $\sqrt{8t_{0}}f_{K}$ with physical masses, allows for the extraction of the following result for the kaon decay constant
\begin{equation}
  \label{eq:fK_fpi_isoQCD}
  f_{K}^{\mathrm{isoQCD}}=156.8(1.1)(1.0)\,{\mathrm{MeV}}\,.
\end{equation}
This determination is in agreement with the value given in \req{ch_ss:eq:isoQCD_fk}, which served as input for the scale setting analysis employing $f_{\pi K}$.
%


\section{Conclusions}
\label{sec:concl}

In this work we have developed and implemented a partially quenched mixed-action setup, aimed at precision studies in lattice QCD.
The sea sector entails $\NF=2+1$ CLS ensembles \cite{Bruno:2014jqa,Bali:2016umi}, with non-perturbatively $\Oa$-improved Wilson quarks and the tree-level improved Symanzik gauge action.
Open boundary conditions are used for gauge fields in almost all ensembles, which allows to explore very fine values of the lattice spacing without suffering from the effects of topological freezing.
The ensembles used explore five values of the lattice spacing, and light-quark masses reach the physical point.
They lie along a chiral trajectory where the trace of the quark-mass matrix is kept approximately constant, which we exploit to obtain results along a line of constant physics, where $m_K^2+\half m_\pi^2$, expressed in units of $t_0$, remains strictly constant (within uncertainties).
In the valence sector, we employ a Wilson twisted-mass QCD action at maximal twist, which effectively eliminates the need to include $\Oa$-improvement counterterms.
A salient feature of this setup is the possibility of overconstraining the continuum limit through universality by combining the mixed-action results with those obtained using a fully unitary action, both exhibiting robust continuum-limit scaling properties.
The matching procedure of the two actions is carried out to very high precision, while respecting the prescribed line of constant physics and keeping light quarks fully twisted for each ensemble.
This ensures not only that universality is fully recovered in the continuum limit, which we demonstrate, but also that Symanzik effective theory applies strictly so that the continuum limit can be rigourously controlled.
One immediate byproduct of our approach is a precise scale setting, through the computation of the intermediate scale $t_0$, using the flavour-average combination of decay constants $f_{\pi K}=\tthird(f_K+\half f_\pi)$ as input to relate it to physical units.
We have computed the latter at the physical point and continuum limit, using SU(3) and SU(2) $\chi$PT formulae or a Taylor expansion in order to perform the probe the light-quark mass dependence.
In addition to this, different ways of parameterising cutoff effects were explored, as well as cuts in volume, pion masses and coupling $\beta$.
Our final result for $t_0^{\mathrm{ph}}$ is
\begin{equation}
  \sqrt{t_0^{\mathrm{ph}}}=0.1440(6)(4)\;\mathrm{fm}\;\left[f_{\pi K}\right]\,,
\end{equation}
where the first uncertainty originates from the statistical fluctuations of the gauge ensembles and also includes the uncertainties in the renormalisation constants and improvement coefficients, as well as those of the physical input in \req{ch_ss:eq:isoQCD_fk}.
The second error corresponds to the systematic uncertainty resulting from the model variation.
We emphasise that achieving high precision in scale setting with fully controlled systematics is a cornerstone of precision lattice QCD, as it is an essential ingredient in any computation and often plays a critical role in reducing the total uncertainty.
We have also computed $t_0$ at the symmetric point $m_\pi=m_K\approx 420~\MeV$ and at a fiducial reference point given by $8t_0(m_K^2+m_\pi^2/2)=1.11$, finding
\begin{align}
  \sqrt{t_0^{\mathrm{sym}}}&=0.1437(6)(4)\;\mathrm{fm}\;\left[f_{\pi K}\right]\,,\\
  \sqrt{t_0^{*}}&=0.1437(7)(4)\;\mathrm{fm}\;\left[f_{\pi K}\right]\,,
\end{align}
respectively.
From our result for $t_0^{\mathrm{ph}}$ in physical units, we have also determined the value of the lattice spacing at the different values of the coupling $\beta$.
Furthermore, we assessed the impact of the physical isoQCD input used for the pion and kaon pseudoscalar masses and decay constants, and provided the derivatives of $\sqrt{t_{0}^{\mathrm{ph}}}$ with respect to the physical inputs $m_{\pi,K}$ and $f_{\pi,K}$ at the isoQCD point, enabling a straightforward translation of our final results to alternative prescriptions for these quantities.
Finally, we have also explored scale setting using only $f_{\pi}^{\mathrm{isoQCD}}$ as physical input, obtaining the result
\begin{equation}
  \sqrt{t_0^{\mathrm{ph}}}=0.1439(10)(9)\;\mathrm{fm}\;\left[f_{\pi}\right]\,.
\end{equation}
The combined continuum-limit and chiral extrapolations proved to be less well controlled than those obtained for $f_{\pi K}$, highlighting the need to incorporate additional (near-)physical-point ensembles and increased statistics to achieve a comparable final precision.
As noted above, we foresee that our mixed-action approach will offer substantial potential for future applications.
A preliminary version of our scale setting underlies the study of charm quark physics described in Ref.~\cite{Bussone:2023kag}, which we plan to update with additional ensembles.
A precise determination of the light, strange, and charm quark masses, as well as applications to B-physics, also lie within our scope.
%



\section*{Acknowledgements}

We are grateful to our colleagues in the Coordinated Lattice Simulations (CLS) initiative for the generation of the gauge field configuration ensembles employed in this study.
We would like to express our gratitude to our colleagues
Mattia Bruno, Nikolai Husung, Tomasz Korzec, Simon Kuberski, Alberto Ramos, Stefan Schaefer, and Rainer Sommer
for discussions and for providing us with valuable input.
We acknowledge PRACE for awarding us access to MareNostrum at Barcelona Supercomputing Center (BSC), Spain and to HAWK at GCS@HLRS, Germany.
The authors thankfully acknowledge the computer resources at MareNostrum and the technical support provided by Barcelona Supercomputing Center (FI-2020-3-0026).
We thank CESGA for granting access to the FinisTerrae systems.
This work is partially supported by grants PGC2018-094857-B-I00, PID2021-127526NB-I00, and PID2024-160152NB-I00, funded by MCIN/AEI/10.13039/\allowbreak 501100011033, by ``ERDF A way of making Europe'', and by FEDER, UE,
as well as by the Spanish Research Agency (Agencia Estatal de Investigaci\'on) through grants IFT Centro de Excelencia Severo Ochoa SEV-2016-0597 and No.~CEX2020-001007-S, funded by MCIN/AEI/\allowbreak 10.13039/501100011033,
and by the European Commission – NextGenerationEU, through Momentum CSIC Programme: Develop Your Digital Talent.
We also acknowledge support from the project H2020-MSCAITN-2018-813942 (EuroPLEx), under grant agreement No.~813942,
and the EU Horizon 2020 research and innovation programme, STRONG-2020 project, under grant agreement No.~824093.
We acknowledge HPC support by Emilio Ambite, staff hired under the Generation D initiative, promoted by Red.es, an organisation attached to the Spanish Ministry for Digital Transformation and the Civil Service, for the attraction and retention of talent through grants and training contracts, financed by the Recovery, Transformation and Resilience Plan through the European Union’s Next Generation funds.
%


\cleardoublepage

\begin{appendix}

\section{Correspondence of currents in the twisted and physical bases}
\label{app:twist}

For completeness, in this Appendix we provide the correspondence between the currents in the twisted and physical bases.
Using the notation introduced in~\req{eq:current_notation}, the transformation properties under~\req{eq:tmrot} of all possible currents in our four-flavour valence sector at full twist read
\begingroup
\allowdisplaybreaks
\begin{align}
&\ba{l@{\hspace{2mm}}c@{\hspace{2mm}}l@{\hspace{10mm}}l@{\hspace{2mm}}c@{\hspace{2mm}}l@{\hspace{10mm}}l@{\hspace{2mm}}c@{\hspace{2mm}}l}
V_\mu^{uu} 	&\to& V_\mu^{uu}\,,&
A_\mu^{uu} 	&\to& A_\mu^{uu}\,,&
P^{uu}		&\to& -iS^{uu}\,,\\[0.5ex]
S^{uu}		&\to& -iP^{uu}\,,&
T_{0k}^{uu}	&\to& \ihalf\varepsilon_{kmn}T_{mn}^{uu}\,,&
T_{mn}^{uu}	&\to& i\varepsilon_{kmn}T_{0k}^{uu}\,,\\[2.0ex]
V_\mu^{ud} 	&\to& iA_\mu^{ud}\,,&
A_\mu^{ud} 	&\to& iV_\mu^{ud}\,,&
P^{ud}		&\to& P^{ud}\,,\\[0.5ex]
S^{ud}		&\to& S^{ud}\,,&
T_{0k}^{ud}	&\to& T_{0k}^{ud}\,,&
T_{mn}^{ud}	&\to& T_{mn}^{ud}\,,\\[2.0ex]
V_\mu^{us}	&\to& iA_\mu^{us}\,,&
A_\mu^{us}	&\to& iV_\mu^{us}\,,&
P^{us}		&\to& P^{us}\,,\\[0.5ex]
S^{us}		&\to& S^{us}\,,&
T_{0k}^{us}	&\to& T_{0k}^{us}\,,&
T_{mn}^{us}	&\to& T_{mn}^{us}\,,\\[2.0ex]
V_\mu^{uc}	&\to& V_\mu^{uc}\,,&
A_\mu^{uc}	&\to& A_\mu^{uc}\,,&
P^{uc}		&\to& -iS^{uc}\,,\\[0.5ex]
S^{uc}		&\to& -iP^{uc}\,,&
T_{0k}^{uc}	&\to& \ihalf\varepsilon_{kmn}T_{mn}^{uc}\,,&
T_{mn}^{uc}	&\to& i\varepsilon_{kmn}T_{0k}^{uc}\,,\\[2.0ex]
\ea\\[2.0ex]
&\ba{l@{\hspace{2mm}}c@{\hspace{2mm}}l@{\hspace{10mm}}l@{\hspace{2mm}}c@{\hspace{2mm}}l@{\hspace{10mm}}l@{\hspace{2mm}}c@{\hspace{2mm}}l}
V_\mu^{du}	&\to& -iA_\mu^{du}\,,&
A_\mu^{du}	&\to& -iV_\mu^{du}\,,&
P^{du}		&\to& P^{du}\,,\\[0.5ex]
S^{du}		&\to& S^{du}\,,&
T_{0k}^{du}	&\to& T_{0k}^{du}\,,&
T_{mn}^{du}	&\to& T_{mn}^{du}\,,\\[2.0ex]
V_\mu^{dd}	&\to& V_\mu^{dd}\,,&
A_\mu^{dd}	&\to& A_\mu^{dd}\,,&
P^{dd}		&\to& iS^{dd}\,,\\[0.5ex]
S^{dd}		&\to& iP^{dd}\,,&
T_{0k}^{dd}	&\to& -\ihalf\varepsilon_{kmn}T_{mn}^{dd}\,,&
T_{mn}^{dd}	&\to& -i\varepsilon_{kmn}T_{0k}^{dd}\,,\\[2.0ex]
V_\mu^{ds}	&\to& V_\mu^{ds}\,,&
A_\mu^{ds}	&\to& A_\mu^{ds}\,,&
P^{ds}		&\to& iS^{ds}\,,\\[0.5ex]
S^{ds}		&\to& iP^{ds}\,,&
T_{0k}^{ds}	&\to& -\ihalf\varepsilon_{kmn}T_{mn}^{ds}\,,&
T_{mn}^{ds}	&\to& -i\varepsilon_{kmn}T_{0k}^{ds}\,,\\[2.0ex]
V_\mu^{dc}	&\to& -iA_\mu^{dc}\,,&
A_\mu^{dc}	&\to& -iV_\mu^{dc}\,,&
P^{dc}		&\to& P^{dc}\,,\\[0.5ex]
S^{dc}		&\to& S^{dc}\,,&
T_{0k}^{dc}	&\to& T_{0k}^{dc}\,,&
T_{mn}^{dc}	&\to& T_{mn}^{dc}\,,\\[2.0ex]
\ea\\[2.0ex]
&\ba{l@{\hspace{2mm}}c@{\hspace{2mm}}l@{\hspace{10mm}}l@{\hspace{2mm}}c@{\hspace{2mm}}l@{\hspace{10mm}}l@{\hspace{2mm}}c@{\hspace{2mm}}l}
V_\mu^{su} 	&\to& -iA_\mu^{su}\,,&
A_\mu^{su} 	&\to& -iV_\mu^{su}\,,&
P^{su}		&\to& P^{su}\,,\\[0.5ex]
S^{su}		&\to& S^{su}\,,&
T_{0k}^{su}	&\to& T_{0k}^{su}\,,&
T_{mn}^{su}	&\to& T_{mn}^{su}\,,\\[2.0ex]
V_\mu^{sd} 	&\to& V_\mu^{sd}\,,&
A_\mu^{sd} 	&\to& A_\mu^{sd}\,,&
P^{sd}		&\to& iS^{sd}\,,\\[0.5ex]
S^{sd}		&\to& iP^{sd}\,,&
T_{0k}^{sd}	&\to& -\ihalf\varepsilon_{kmn}T_{mn}^{sd}\,,&
T_{mn}^{sd}	&\to& -i\varepsilon_{kmn}T_{0k}^{sd}\,,\\[2.0ex]
V_\mu^{ss}	&\to& V_\mu^{ss}\,,&
A_\mu^{ss}	&\to& A_\mu^{ss}\,,&
P^{ss}		&\to& iS^{ss}\,,\\[0.5ex]
S^{ss}		&\to& iP^{ss}\,,&
T_{0k}^{ss}	&\to& -\ihalf\varepsilon_{kmn}T_{mn}^{ss}\,,&
T_{mn}^{ss}	&\to& -i\varepsilon_{kmn}T_{0k}^{ss}\,,\\[2.0ex]
V_\mu^{sc}	&\to& -iA_\mu^{sc}\,,&
A_\mu^{sc}	&\to& -iV_\mu^{sc}\,,&
P^{sc}		&\to& P^{sc}\,,\\[0.5ex]
S^{sc}		&\to& S^{sc}\,,&
T_{0k}^{sc}	&\to& T_{0k}^{sc}\,,&
T_{mn}^{sc}	&\to& T_{mn}^{sc}\,,\\[2.0ex]
\ea\\[2.0ex]
&\ba{l@{\hspace{2mm}}c@{\hspace{2mm}}l@{\hspace{10mm}}l@{\hspace{2mm}}c@{\hspace{2mm}}l@{\hspace{10mm}}l@{\hspace{2mm}}c@{\hspace{2mm}}l}
V_\mu^{cu} 	&\to& V_\mu^{cu}\,,&
A_\mu^{cu} 	&\to& A_\mu^{cu}\,,&
P^{cu}		&\to& -iS^{cu}\,,\\[0.5ex]
S^{cu}		&\to& -iP^{cu}\,,&
T_{0k}^{cu}	&\to& \ihalf\varepsilon_{kmn}T_{mn}^{cu}\,,&
T_{mn}^{cu}	&\to& i\varepsilon_{kmn}T_{0k}^{cu}\,,\\[2.0ex]
V_\mu^{cd} 	&\to& iA_\mu^{cd}\,,&
A_\mu^{cd} 	&\to& iV_\mu^{cd}\,,&
P^{cd}		&\to& P^{cd}\,,\\[0.5ex]
S^{cd}		&\to& S^{cd}\,,&
T_{0k}^{cd}	&\to& T_{0k}^{cd}\,,&
T_{mn}^{cd}	&\to& T_{mn}^{cd}\,,\\[2.0ex]
V_\mu^{cs}	&\to& iA_\mu^{cs}\,,&
A_\mu^{cs}	&\to& iV_\mu^{cs}\,,&
P^{cs}		&\to& P^{cs}\,,\\[0.5ex]
S^{cs}		&\to& S^{cs}\,,&
T_{0k}^{cs}	&\to& T_{0k}^{cs}\,,&
T_{mn}^{cs}	&\to& T_{mn}^{cs}\,,\\[2.0ex]
V_\mu^{cc}	&\to& V_\mu^{cc}\,,&
A_\mu^{cc}	&\to& A_\mu^{cc}\,,&
P^{cc}		&\to& -iS^{cc}\,,\\[0.5ex]
S^{cc}		&\to& -iP^{cc}\,,&
T_{0k}^{cc}	&\to& \ihalf\varepsilon_{kmn}T_{mn}^{cc}\,,&
T_{mn}^{cc}	&\to& i\varepsilon_{kmn}T_{0k}^{cc}\,.\\[2.0ex]
\ea
\end{align}
\endgroup


\section{$\Oa$-improvement of tmQCD}
\label{app:Oa}

In this appendix, we review the $\Oa$-improvement structure of lattice QCD with Wilson quarks including a chirally rotated twisted mass term (tmQCD) at generic values of the masses.
We follow a strategy similar to that adopted in Ref.~\cite{Bhattacharya:2005rb} for the standard Wilson action, in order to extend the results for two light mass-degenerate quarks provided in Ref.~\cite{Frezzotti:2001ea}.
This framework will then serve as the basis for analysing $\Oa$-improvement in our partially quenched mixed-action setup, with the aim of determining the extent to which automatic $\Oa$-improvement at maximal twist~\cite{Frezzotti:2003ni} carries over to the mixed-action observables considered in this work.

\subsection{Symanzik effective field theory and $\Oa$-improvement}
\label{sec:SymEFT}

Symanzik's improvement programme~\cite{Symanzik:1983dc,Symanzik:1983gh,Luscher:1984xn} is a systematic way of removing the leading cutoff effects in observables computed with a lattice regularisation.
We begin by recalling some general considerations and refer to Ref.~\cite{Luscher:1996sc} for a detailed account of the implementation of the improvement programme in QCD.
Asymptotically close to the continuum limit, the lattice-regulated theory can be described in terms of a local Symanzik effective field theory (SymEFT) with a Lagrangian density given by
\begin{gather}
  \label{eq:Oa}
  \cL_{\mathrm{\scriptscriptstyle eff}} = \cL_0 + a \cL_1 + a^2 \cL_2 + \dots\,,
\end{gather}
where $\cL_0$ is the continuum Lagrangian of Euclidean QCD.
The generic term $\cL_k$, $k \geq 1$, is a linear combination of all local gauge-invariant composite operators $\cO_{k,l}$ of engineering dimension $4+k$ with vacuum quantum numbers allowed by the symmetries of the regularised theory, viz.
\begin{gather}
  \cL_{k} = \sum_l c_l\,\cO_{k,l}\,, \quad k\geq1\,.
\end{gather}
Discretisation effects also arise from local composite fields.
We consider a local gauge-invariant operator $\phi$ constructed from quark and gluon fields on the lattice.
The generalisation of the argument to various classes of operators poses no additional complications.
A renormalised, on-shell (i.e.\ $x_1 \neq x_2$), connected two-point function can then be written as
\begin{gather}
  G(x_1,x_2) = Z_\phi^2 \left\langle\phi(x_1)\phi(x_2)\right\rangle\,,
\end{gather}
where, for simplicity, we assume that $\phi$ renormalises multiplicatively with a renormalisation factor $Z_\phi$.
In the SymEFT, the renormalised field $Z_\phi \,\phi$ is represented by an effective field
\begin{gather}
  \label{eq:Oa_c}
  \phi_{\mathrm{\scriptscriptstyle eff}} = \phi_0 + a\sum_l c_{\phi,l}\, \phi_{1,l} + \dots\,,
\end{gather}
where $\phi_{k,l}$ are operators of dimension $\mathrm{dim}[\phi] + k$, which share the same transformation properties as $\phi_0$ under all symmetries preserved by the regularisation.
Considering the above expansions for $\mathcal{L}_{\mathrm{\scriptscriptstyle eff}}$ and $\phi_{\mathrm{\scriptscriptstyle eff}}$ in Eqs.~(\ref{eq:Oa}) and (\ref{eq:Oa_c}), respectively, and expanding
\begin{gather}
  e^{-S_{\mathrm{\scriptscriptstyle eff}}}=e^{-S_0}\left\{1-aS_1+\ldots\right\}\,,
\end{gather}
with $S_k=\int\dif^4x\cL_k(x)$, one obtains the following expression for the lattice correlation function
\begin{gather}
  \begin{split}
    \label{eq:G2SymEFT}
    G(x_1,x_2) =
    \left\langle\phi_0(x_1)\phi_0(x_2)\right\rangle_0
    &-a\sum_l c_l \left\langle\phi_0(x_1)\,\phi_0(x_2)\left\{\int\dif^4y\,\cO_{1,l}(y)\right\}\right\rangle_0\\
    &+a\sum_l c_{\phi,l} \left\{
    \left\langle \phi_{1,l}(x_1)\, \phi_0(x_2)\right\rangle_0+
    \left\langle\phi_0(x_1)\, \phi_{1,l}(x_2)\right\rangle_0
    \right\}\\
    &+\mbox{O}(a^2)\,,
  \end{split}
\end{gather}
where expectation values, denoted by $\left\langle \cdot \right\rangle_0$, are computed in the theory with continuum action $S_0$.
An analogous expression applies to a generic connected $n$-point function with arbitrary renormalised operator insertions.
Given this structure of the leading cutoff effects, the $\Oa$-improvement programme proceeds by adopting a regularised action and a set of composite operators in which counterterms proportional to $S_1$ and $\phi_{1,l}$, respectively, are introduced with suitably tuned coefficients, so as to cancel the $\Oa$ terms present in $\left\langle \phi(x_1)\,\phi(x_2) \right\rangle$.
It is important to emphasise that, despite the formal expansion in powers of $a$, quantum effects generally induce logarithmic corrections in the lattice spacing, driven by the anomalous dimensions of the operators $\mathcal{O}_{k,l}$ and $\phi_{k,l}$.
These corrections are known to significantly affect the continuum-limit scaling in, for example, the $\mathrm{O}(3)$ sigma model~\cite{Balog:2009np,Balog:2009yj}.
The study of these logarithmic corrections for Wilson-type regularisations of the QCD action has been undertaken in a series of works~\cite{Husung:2019ytz,Husung:2021tml,Husung:2021mfl,Husung:2022kvi,Husung:2024cgc}.

\subsection{Discrete symmetries of Wtm regularisations}
\label{sec:discrsym}

Our task is thus to identify the relevant sets of higher-dimensional operators that contribute to Eqs.~(\ref{eq:Oa})~and~(\ref{eq:Oa_c}).
To that purpose, let us first lay out the symmetries under which operator behaviour is classified.
We will start by writing a completely generic Wilson fermion action of the form
\begin{gather}
  \label{eq:genWtm}
  S_{\mathrm{\scriptscriptstyle F}} = a^4 \sum_{i=1}^{\NF} \sum_x
  \bar\psi_i(x)\left\{\frac{1}{2}\gamma_\mu\left(\nabla_\mu^\ast+\nabla_\mu\right)
  -\frac{ar_i}{2}\nabla_\mu^\ast\nabla_\mu + m_{0,i} + i\gamma_5\mu_i\right\}\psi_i(x)\,,
\end{gather}
where $i$ is a flavour index, and we are employing the twisted basis for the fermion fields.
Note that, for the sake of generality, we adopt a slightly different convention from that used for our $\NF = 4$ action in Eqs.~(\ref{eq:mval})--(\ref{eq:mval2}), in that we do not assign specific signs to the twisted masses $\mu_{i}$ reflecting the structure of the twist.
We are also allowing for a generic Wilson parameter $r_i$ for each flavour.
We assume that a mass-independent renormalisation scheme has been chosen.
Under renormalisation, a critical mass term of the form
\begin{gather}
  \label{eq:genWtmmcrit}
  - a^4 \sum_{i=1}^{\NF} \sum_x \mcrit\bar\psi_i(x)\psi_i(x)\,,
\end{gather}
has to be added, leading to the following expression for the bare subtracted quark mass $m_i$
\begin{gather}
  m_i = m_{0,i}-\mcrit\,.
\end{gather}
In compact notation, using the mass matrices $\mathbf{m}$ and $\boldsymbol{\mu}$ in flavour space, the mass term of the generic Wtm action can be written as
\begin{gather}
  \label{eq:gen_mass_term}
  \bar\psi(x)\{\mathbf{m}+i\gamma_5\boldsymbol{\mu}\}\psi(x)\,.
\end{gather}
The action based in Eqs.~(\ref{eq:genWtm})-(\ref{eq:genWtmmcrit}) exhibits the following discrete symmetries for each individual flavour (see Ref.~\cite{Frezzotti:2003ni} for a detailed discussion):
\begin{itemize}
\item $\mathcal{R}_5 \times D$, where
  \begin{gather}
    \label{eq:r5sym}
    \bar\psi_i(x) \xrightarrow{~\mathcal{R}_5~} -\bar\psi_i(x)\gamma_5\,, \quad
    \psi_i(x) \xrightarrow{~\mathcal{R}_5~} \gamma_5\psi_i(x)\,;\\
    \label{eq:dsym}
    \bar\psi_i(x) \xrightarrow{~D~} e^{i3\pi/2}\bar\psi_i(-x)\,, \quad
    \psi_i(x) \xrightarrow{~D~} e^{i3\pi/2}\psi_i(-x)\,, \quad
    U_\mu(x) \xrightarrow{~D~} U_\mu(-x-a\hat\mu)^\dagger\,.
  \end{gather}
\item $\mathcal{R}_5 \times [m_i \to -m_i] \times [\mu_i \to -\mu_i] \times [r_i \to -r_i]$.
\item $P \times [\mu_i \to -\mu_i]$, where $P$ is the parity transformation
  \begin{gather}
    \bar\psi_i(x) \xrightarrow{~P~} \bar\psi_i(x_P)\gamma_0\,, \quad
    \psi_i(x) \xrightarrow{~P~} \gamma_0\psi_i(x_P)\,;\\
    U_0(x) \xrightarrow{~P~} U_0(x_P)\,, \quad
    U_k(x) \xrightarrow{~P~} U_k(x_P-a\hat{k})^\dagger\,,
  \end{gather}
  and $x_P=(x_0;-\vx)$ for $x=(x_0,\vx)$.
\item $T \times [\mu_i \to -\mu_i]$, where $T$ is the time reversal transformation
  \begin{gather}
    \bar\psi_i(x) \xrightarrow{~T~} i\bar\psi_i(x_T)\gamma_0\gamma_5\,, \quad
    \psi_i(x) \xrightarrow{~T~} i\gamma_0\gamma_5\psi_i(x_T)\,;\\
    U_0(x) \xrightarrow{~T~} U_0(x_T-a\hat{0})^\dagger\,, \quad
    U_k(x) \xrightarrow{~T~} U_k(x_T)\,,
  \end{gather}
  and $x_T=(-x_0;\vx)$ for $x=(x_0,\vx)$.
\item $C$, the standard charge conjugation symmetry given by
  \begin{gather}
    \bar\psi_i(x) \xrightarrow{~C~} -i\psi_i(x)^T \gamma_0\gamma_2\,, \quad
    \psi_i(x) \xrightarrow{~C~} i\gamma_0\gamma_2\bar\psi(x)^T\,;\\
    U_\mu(x) \xrightarrow{~C~} U_\mu(x)^\ast\,,
  \end{gather}
  where the superscript $T$ indicates transposition.
\end{itemize}

\subsection{$\Oa$-improvement of the action}

To recall the structure of the $\Oa$ counterterms to the action, let us first consider, without loss of generality, the simple case of $\NF=2$ massive Wilson fermions.
Using partial integration and gauge invariance, the resulting list of operators with the correct discrete and flavour symmetries that enter $\cL_1$ reads~\cite{Luscher:1996sc}
\begin{align*}
  \cO_1^{(1)} & = \bar\psi\, i\sigma_{\mu\nu} F_{\mu\nu} \psi \,,\\
  \cO_1^{(2)} & = \bar\psi\, \overrightarrow{D}_\mu \overrightarrow{D}_\mu \psi \,
  +\, \bar\psi\, \overleftarrow{D}_\mu \overleftarrow{D}_\mu \psi \,.\\
  \cO_1^{(3)} & = m F_{\mu\nu}F_{\mu\nu} \,,\\
  \cO_1^{(4)} & = m\left[  \bar\psi\, \gamma_\mu \overrightarrow{D}_\mu \psi
    - \bar\psi\, \overleftarrow{D}_\mu \gamma_\mu \psi \right] \,,\\
  \cO_1^{(5)} & = m^2 \bar\psi\, \psi\,,
\end{align*}
where $\psi$ is a flavour doublet, and arrows indicate the field on which the covariant derivative acts.
Note that we have adopted a continuum notation, consistently with the structure of SymEFT.
It was shown in Ref.~\cite{Luscher:1996sc} that, when these operators are inserted into on-shell correlation functions, their contributions can always be reabsorbed into a redefinition of the composite operators, or eliminated altogether by using the equations of motion, except for those involving $\mathcal{O}_1^{(1)}$ --- the so-called Sheikholeslami–Wohlert term~\cite{Sheikholeslami:1985ij}.
For the purpose of the improvement programme, it will therefore be sufficient to add (a suitably regularised version of) this term to the lattice action, with an improvement coefficient $\icsw(g_0^2)$.
If completely generic mass terms, including twisted masses, are allowed, the list of mass-dependent operators similar to $\cO_1^{(3-5)}$ will become larger.
However, the conclusion still holds, and the Sheikholeslami–Wohlert term suffices to improve the action, up to a rescaling of the bare parameters by terms depending on the subtracted bare mass and the bare twisted mass parameters~\cite{Frezzotti:2000nk,Frezzotti:2003ni,Frezzotti:2003xj}.

\subsection{$\Oa$-improvement of non-singlet currents}

In order to determine the $\Oa$-improvement structure of quark bilinear operators for a generic Wilson twisted-mass (Wtm) action, we adopt a strategy similar to that in Ref.~\cite{Bhattacharya:2005rb}.
Following the convenient notation for -- non-diagonal and diagonal -- flavour non-singlet bilinears employed in Ref.~\cite{Bhattacharya:2005rb}, we consider composite fields of the form $\tr(\lambda\cO_{\scriptscriptstyle \Gamma})$, where $\lambda$ is a generator of $\mathrm{SU}(\NF)$.\,\footnote{
We primarily focus on the $\Oa$-improvement of flavour non-singlet currents (see Ref.~\cite{Bussone:2018ljj} for an initial account of this topic).
The complete analysis of the flavour-singlet currents, $\tr(\cO_{\scriptscriptstyle \Gamma})$, is a relatively straightforward extension, but it lies beyond the scope of this work.
}
More specifically, the composite field
\begin{gather}
  \cO_{\scriptscriptstyle \Gamma} = \bar\psi\,\Gamma\psi\,, \qquad
  \Gamma \in \{\mathbf{1},\gamma_\mu,i\sigma_{\mu\nu},\gamma_\mu\gamma_5,\gamma_5\}\,,
\end{gather}
is treated as a matrix with indices in flavour space, while spin and colour indices are traced over.
This is equivalent to
\begin{gather}
  \cO_{\scriptscriptstyle \Gamma} \in \{S,V_\mu,T_{\mu\nu},A_\mu,P\}\,,
\end{gather}
in terms of the currents defined in~\req{eq:current_notation}.
The contraction of the matrix  $\cO_{\scriptscriptstyle \Gamma}$ with the generator $\lambda$ is denoted by  $\tr(\lambda\cO_{\scriptscriptstyle \Gamma})$, where the trace $\tr$ acts on flavour indices.
The notation is trivially generalised to other composite fields.
Let us first consider the chiral limit, where the structure of mass terms is irrelevant.
Using a superscript $\mathrm{I}_\chi$ to denote improved currents in the chiral limit, one then has the well-known result~\cite{Jansen:1995ck, Luscher:1996sc, Bhattacharya:2005rb}
\begin{align}
  \label{eq:Simpc}
  \tr \left( \lambda S \right)^{\mathrm{\scriptscriptstyle I}_\chi} &= \tr \left( \lambda S \right)\,, \\
  \label{eq:Vimpc}
  \tr \left( \lambda V_\mu \right)^{\mathrm{\scriptscriptstyle I}_\chi} &= \tr \left( \lambda V_\mu \right) 
  + a \icV \tr\left(\lambda \tilde{\partial}_\nu T_{\mu\nu}\right)\,, \\
  \label{eq:Timpc}
  \tr \left( \lambda T_{\mu\nu} \right)^{\mathrm{\scriptscriptstyle I}_\chi} &= \tr \left( \lambda T_{\mu\nu} \right) 
  + a \icT \tr\left[\lambda \left(\tilde{\partial}_\mu V_\nu  - \tilde{\partial}_\nu V_\mu\right)\right]\,, \\
  \label{eq:Aimpc}
  \tr \left( \lambda A_\mu \right)^{\mathrm{\scriptscriptstyle I}_\chi} &= \tr \left( \lambda A_\mu \right) 
  + a \icA  \tr\left(\lambda \tilde{\partial}_\mu P\right)\,, \\
  \label{eq:Pimpc}
  \tr \left( \lambda P \right)^{\mathrm{\scriptscriptstyle I}_\chi} &= \tr \left( \lambda P \right)\,,
\end{align}
where terms that may contribute to off-shell correlation functions (see, e.g., Ref.~\cite{Capitani:2000xi}) have been dropped.
The difference between the standard Wilson case and the more general twisted-mass formulation arises when considering improvement terms involving quark masses.
Using the notation introduced in \req{eq:gen_mass_term}, current improvement in the standard Wilson case requires only the consideration of the combinations
\begin{gather}
  \tr(\lambda\mathbf{m}\cO_{\scriptscriptstyle \Gamma})\,, \quad
  \tr(\mathbf{m})\tr(\lambda\cO_{\scriptscriptstyle \Gamma})\,, \quad
  \tr(\lambda\mathbf{m})\tr(\cO_{\scriptscriptstyle \Gamma})\,,
\end{gather}
in the non-singlet case, with the additional combinations
\begin{gather}
  \tr(\mathbf{m}\cO_{\scriptscriptstyle \Gamma})\,, \quad
  \tr(\mathbf{m})\tr(\cO_{\scriptscriptstyle \Gamma})\,,
\end{gather}
relevant for singlet currents.
When $\boldsymbol{\mu} \neq 0$, similar combinations involving the twisted mass matrix are possible.
The non-singlet ones are
\begin{gather}
  i\tr(\lambda\boldsymbol{\mu}\cO_{\scriptscriptstyle \Gamma\gamma_{\scriptscriptstyle 5}})\,, \quad
  i\tr(\boldsymbol{\mu})\tr(\lambda\cO_{\scriptscriptstyle \Gamma\gamma_{\scriptscriptstyle 5}})\,, \quad
  i\tr(\lambda\boldsymbol{\mu})\tr(\cO_{\scriptscriptstyle \Gamma\gamma_{\scriptscriptstyle 5}})\,,
\end{gather}
while in the singlet case
\begin{gather}
  i\tr(\boldsymbol{\mu}\cO_{\scriptscriptstyle \Gamma\gamma_{\scriptscriptstyle 5}})\,, \quad
  i\tr(\boldsymbol{\mu})\tr(\cO_{\scriptscriptstyle \Gamma\gamma_{\scriptscriptstyle 5}})\,,
\end{gather}
appear.
In order to classify the possible correction terms for each non-singlet current, we use the Wtm discrete symmetries listed in Sect.~\ref{sec:discrsym}.
All currents have well-defined transformation properties under $\mathcal{R}_5$ --- vector and axial-vector currents are even, whereas scalar and pseudoscalar densities, as well as tensor currents, are odd --- while they acquire only an irrelevant global phase under the $D$ part of the symmetry, introduced in \req{eq:dsym}.
Concerning the dimension-four operators involving masses, they also have well-defined phases under $\mathcal{R}_5 \times [\mu_i \to -\mu_i]$ --- or, alternatively, $P \times [\mu_i \to -\mu_i]$.
Using the corresponding transformation properties, and verifying that no additional discrete symmetries allow a further reduction of the mixing, we obtain the following general expression for the improved (unrenormalised) currents in the $\mathcal{R}_5$-even channel
\begin{gather}
  \label{eq:Oaceven}
  \begin{split}
    \tr(\lambda\cO_{\scriptscriptstyle \Gamma})^{\mathrm{\scriptscriptstyle I}} =
    & \left[1+a\ibGb\tr(\mathbf{m})\right]\tr(\lambda\cO_{\scriptscriptstyle \Gamma})^{\mathrm{\scriptscriptstyle I}_\chi}
    + a \ibG \tr\left(\left\{\frac{\lambda}{2},\mathbf{m}\right\}\cO_{\scriptscriptstyle \Gamma}\right)
    + a f_{\scriptscriptstyle \Gamma}\tr(\lambda\mathbf{m})\tr(\cO_{\scriptscriptstyle \Gamma})\\
    & + a \ibGc\tr\left(\left[\frac{\lambda}{2},i\boldsymbol{\mu}\right]\cO_{\scriptscriptstyle \Gamma\gamma_{\scriptscriptstyle 5}}\right)\,,
  \end{split}
\end{gather}
and the corresponding one in the $\mathcal{R}_5$-odd channel,
\begin{gather}
  \label{eq:Oacodd}
  \begin{split}
    \tr(\lambda\cO_{\scriptscriptstyle \Gamma})^{\mathrm{\scriptscriptstyle I}} =
    & \left[1+a\ibGb\tr(\mathbf{m})\right]\tr(\lambda\cO_{\scriptscriptstyle \Gamma})^{\mathrm{\scriptscriptstyle I}_\chi}
    + a \ibG \tr\left(\left\{\frac{\lambda}{2},\mathbf{m}\right\}\cO_{\scriptscriptstyle \Gamma}\right)
    + a f_{\scriptscriptstyle \Gamma}\tr(\lambda\mathbf{m})\tr(\cO_{\scriptscriptstyle \Gamma})\\
    & + a \ibGc\tr\left(\left\{\frac{\lambda}{2},i\boldsymbol{\mu}\right\}\cO_{\scriptscriptstyle \Gamma\gamma_{\scriptscriptstyle 5}}\right)
    + a \ibGh\tr(i\boldsymbol{\mu})\tr(\lambda\cO_{\scriptscriptstyle \Gamma\gamma_{\scriptscriptstyle 5}})
    + a\check{f}_{\scriptscriptstyle \Gamma} \tr(i\lambda\boldsymbol{\mu})\tr(\cO_{\scriptscriptstyle \Gamma\gamma_{\scriptscriptstyle 5}})\,.
  \end{split}
\end{gather}
The first terms of each expression involve the $\Oa$-improved operators in the chiral limit introduced in Eqs.~(\ref{eq:Simpc})–(\ref{eq:Pimpc}).\,\footnote{
We note that the improvement coefficient $\ibGc$ is denoted by $\ibGt$ in Ref.~\cite{Frezzotti:2001ea}.
This change in notation is motivated by the fact that the symbol $\ibGt$ has since been adopted to denote another improvement term proportional to the current quark masses (see, e.g., Eqs.~(\ref{eqn:decay_const_W}) and (\ref{eqn:mPCACI})).}
To obtain ready-to-use expressions, we finally distinguish between off-diagonal and diagonal flavour structures --- that is, currents with flavour indices $\cO_{ij}$ for $i \neq j$, and currents that allow the construction of diagonal non-singlet combinations, for which $\cO_{ii}-\cO_{jj}$ can be used as building blocks.
Combining this with Eqs.~(\ref{eq:Simpc})–(\ref{eq:Pimpc}), we obtain the following expressions for the off-diagonal case
\begin{align}
  \label{eq:Simp}
  (S^{ij})^{\mathrm{\scriptscriptstyle I}} &= \left[1+a\ibSb\tr(\mathbf{m})+a\ibS m_{ij}^{\scriptscriptstyle (+)}\right]S^{ij}
  + ia \left[\ibSh\tr(\boldsymbol{\mu})+\ibSc\mu_{ij}^{\scriptscriptstyle (+)}\right]P^{ij}\,,\\[1.0ex]
  \label{eq:Vimp}
  (V_\mu^{ij})^{\mathrm{\scriptscriptstyle I}} &= \left[1+a\ibVb\tr(\mathbf{m})+a\ibV m_{ij}^{\scriptscriptstyle (+)}\right]\left(V_\mu^{ij}
  + a \icV \tilde{\partial}_\nu T_{\mu\nu}^{ij}\right)
  + ia \ibVc\mu_{ij}^{\scriptscriptstyle (-)} A_\mu^{ij}\,,\\[1.0ex]
  \nonumber
  (T_{\mu\nu}^{ij})^{\mathrm{\scriptscriptstyle I}} &= \left[1+a\ibTb\tr(\mathbf{m})+a\ibT m_{ij}^{\scriptscriptstyle (+)}\right]\left(T_{\mu\nu}^{ij}
  + a \icT \left[\tilde{\partial}_\mu V_\nu^{ij}  - \tilde{\partial}_\nu V_\mu^{ij}\right]\right)\\
  \label{eq:Timp}
  & \qquad + ia \left[\ibTh\tr(\boldsymbol{\mu})+\ibTc\mu_{ij}^{\scriptscriptstyle (+)}\right]\tilde{T}_{\mu\nu}^{ij}\,,\\[1.0ex]
  \label{eq:Aimp}
  (A_\mu^{ij})^{\mathrm{\scriptscriptstyle I}} &= \left[1+a\ibAb\tr(\mathbf{m})+a\ibA m_{ij}^{\scriptscriptstyle (+)}\right]\left(A_\mu^{ij}
  + a \icA \tilde{\partial}_\mu P^{ij}\right)
  + ia \ibAc\mu_{ij}^{\scriptscriptstyle (-)}V_\mu^{ij}\,,\\[1.0ex]
  \label{eq:Pimp}
  (P^{ij})^{\mathrm{\scriptscriptstyle I}} &= \left[1+a\ibPb\tr(\mathbf{m})+a\ibP m_{ij}^{\scriptscriptstyle (+)}\right]P^{ij}
  + ia \left[\ibPh\tr(\boldsymbol{\mu})+\ibPc\mu_{ij}^{\scriptscriptstyle (+)}\right]S^{ij}\,,
\end{align}
where we define the average subtracted quark mass $m_{ij}^{\scriptscriptstyle (+)}$ as follows
\begin{gather}
  \label{eq:avsubtm}
  m_{ij}^{\scriptscriptstyle (+)} = \frac{m_i+m_j}{2}\,,
\end{gather}
and the half-sum and half-difference $\mu_{ij}^{\scriptscriptstyle (\pm)}$ of the twisted-mass parameters as
\begin{gather}
  \mu_{ij}^{\scriptscriptstyle (\pm)} = \frac{\mu_i \pm \mu_j}{2}\,.
\end{gather}
The diagonal flavour non-singlet bilinears read
\begin{align}
  \nonumber
  (S^{ii}-S^{jj})^{\mathrm{\scriptscriptstyle I}} &= [1+a\ibSb\tr(\mathbf{m})](S^{ii}-S^{jj})\\
  \nonumber
  & \qquad
  + a \ibS(m_i S^{ii}-m_j S^{jj})
  + a f_{\mathrm{\scriptscriptstyle S}} (m_i-m_j)\tr(S)\\
  \label{eq:Simpd}
  & \qquad
  + ia \ibSh \tr(\boldsymbol{\mu})(P^{ii}-P^{jj})
  + ia \ibSc (\mu_i P^{ii}-\mu_j P^{jj})
  + ia \check{f}_{\mathrm{\scriptscriptstyle S}}(\mu_i-\mu_j)\tr(P)
  \,,\\[1.0ex]
  \nonumber
  (V_\mu^{ii}-V_\mu^{jj})^{\mathrm{\scriptscriptstyle I}} &= [1+a\ibVb\tr(\mathbf{m})]\left\{
  (V_\mu^{ii}-V_\mu^{jj})
  + a \icV \tilde{\partial}_\nu (T_{\mu\nu}^{ii}-T_{\mu\nu}^{jj})\right\}\\
  \label{eq:Vimpd}
  & \qquad
  + a \ibV(m_i V_\mu^{ii}-m_j V_{\mu}^{jj})
  + a f_{\mathrm{\scriptscriptstyle V}} (m_i-m_j)\tr(V_\mu)\,,
  \\[1.0ex]
  \nonumber
  (T_{\mu\nu}^{ii}-T_{\mu\nu}^{jj})^{\mathrm{\scriptscriptstyle I}} &= [1+a\ibTb\tr(\mathbf{m})]\left\{
  (T_{\mu\nu}^{ii}-T_{\mu\nu}^{jj})
  + a \icT \left[\tilde{\partial}_\mu (V_\nu^{ii}-V_\nu^{jj})  - \tilde{\partial}_\nu (V_\mu^{ii}-V_\mu^{jj})\right]
  \right\}\\
  \nonumber
  & \qquad
  + a \ibT(m_i T_{\mu\nu}^{ii}-m_j T_{\mu\nu}^{jj})
  + a f_{\mathrm{\scriptscriptstyle T}} (m_i-m_j)\tr(T_{\mu\nu})\\
  \label{eq:Timpd}
  & \qquad
  + ia \ibTh \tr(\boldsymbol{\mu})(\tilde T_{\mu\nu}^{ii}-\tilde T_{\mu\nu}^{jj})
  + ia \ibTc (\mu_i \tilde T_{\mu\nu}^{ii}-\mu_j \tilde T_{\mu\nu}^{jj})
  + ia \check{f}_{\mathrm{\scriptscriptstyle T}}(\mu_i-\mu_j)\tr(\tilde T_{\mu\nu})
  \,,\\[1.0ex]
  \nonumber
  (A_\mu^{ii}-A_\mu^{jj})^{\mathrm{\scriptscriptstyle I}} &= [1+a\ibAb\tr(\mathbf{m})]\left\{
  (A_\mu^{ii}-A_\mu^{jj})
  + a \icA \tilde{\partial}_\mu (P^{ii}-P^{jj})\right\}\\
  \label{eq:Aimpd}
  & \qquad
  + a \ibA(m_i A_\mu^{ii}-m_j A_{\mu}^{jj})
  + a \fA (m_i-m_j)\tr(A_\mu)\,,\\[1.0ex]
  \nonumber
  (P^{ii}-P^{jj})^{\mathrm{\scriptscriptstyle I}} &= [1+a\ibPb\tr(\mathbf{m})](P^{ii}-P^{jj})\\
  \nonumber
  & \qquad
  + a \ibP(m_i P^{ii}-m_j P^{jj})
  + a \fP (m_i-m_j)\tr(P)\\
  \label{eq:Pimpd}
  & \qquad
  + ia \ibPh \tr(\boldsymbol{\mu})(S^{ii}-S^{jj})
  + ia \ibPc (\mu_i S^{ii}-\mu_j S^{jj})
  + ia \check{f}_{\mathrm{\scriptscriptstyle P}}(\mu_i-\mu_j)\tr(S)
  \,.
\end{align}
We note that, in the case of the tensor currents in Eqs.~(\ref{eq:Timp}) and~(\ref{eq:Timpd}), we have flipped the signs of the terms associated with $\tilde{T}_{\mu\nu}$ with respect to the general expression in the $\mathcal{R}_5$-odd case in \req{eq:Oacodd}, owing to the identity $\sigma_{\mu\nu}\gamma_5 = -\tilde{\sigma}_{\mu\nu}$.
This sign flip ensures consistency with the sign structure of the terms in Eqs.~(\ref{eq:Simp})–(\ref{eq:Pimpd}).
It is also worth noting that, up to Eqs.~(\ref{eq:Oaceven})–(\ref{eq:Oacodd}), the expressions remain valid for non-diagonal mass matrices $\mathbf{m}$ and $\boldsymbol{\mu}$, provided they do not contain an imaginary part, in order to avoid complications with the anomaly.
To derive Eqs.~(\ref{eq:Simp})–(\ref{eq:Pimpd}), however, we have assumed the practical case of diagonal mass matrices.

\subsection{Automatic $\Oa$-improvement and the partially quenched mixed-action setup}
\label{sec:OaMA}

We start by considering the argument for automatic $\Oa$-improvement at maximal twist, first put forward in Ref.~\cite{Frezzotti:2003ni}.\,\footnote{See also Refs.~\cite{Frezzotti:2005gi,Shindler:2005vj,Sint:2005qz,Shindler:2007vp,Aoki:2006gh} for streamlined versions of the argument and further insights into the structure of automatic $\Oa$-improvement.}
In the fully twisted case, subtracted standard quark masses are set to zero, possibly up to $\Oa$ effects.
Consequently, all terms proportional to $\mathbf{m}$ in Eqs.~(\ref{eq:Oaceven})–(\ref{eq:Oacodd}) vanish.
Equivalently, the terms depending on (combinations of) $m_i$ vanish in Eqs.~(\ref{eq:Simp})–(\ref{eq:Pimp}) and (\ref{eq:Simpd})–(\ref{eq:Pimpd}).
Concerning the terms proportional to (combinations of) $\mu_i$, it is straightforward to verify that they systematically possess opposite physical parity to the dimension-three fields they correct.
While parity is not preserved when $\boldsymbol{\mu}\neq 0$, it is broken only at $\Oa$, which means that the matrix elements of parity-odd operators will be $\Oa$.
As a consequence, the insertion of any of these counterterms in on-shell correlation functions will give rise to terms of $\Oasq$, since the $\Oa$ amplitude will be multiplied by the explicit factor of $a$ that lowers the field dimension.
It follows that no Symanzik counterterms to the currents are needed in order to obtain physical observables that scale towards the continuum as $\Oasq$.
An equivalent conclusion, as discussed in Ref.~\cite{Bussone:2018ljj}, can also be obtained using the Wilson Averaging (WA) procedure introduced in Ref.~\cite{Frezzotti:2003ni}, which involves averaging correlation functions with opposite signs of the Wilson parameter $r_i$.
The discrete symmetries of the Wtm regularisation described in Sect.~\ref{sec:discrsym} allow the WA to be reformulated as a Mass Average (MA), where correlation functions with opposite signs of the subtracted mass $m_i$ are considered.
The automatic $\Oa$-improvement of physical observables at maximal twist follows from the limiting case in which the MA is performed with $m_i = 0$~\cite{Bussone:2018ljj}.
The argument remains valid when the Sheikholeslami-Wohlert term~\cite{Sheikholeslami:1985ij} is included, as its contribution to physical observables in the fully twisted case is of $\Oasq$.
Moreover, employing a non-perturbative determination of the improvement coefficient $\icsw$~\cite{Luscher:1996ug, Bulava:2013cta} eliminates double insertions of the Pauli term, also contributing at $\Oasq$, within the SymEFT framework.
We now turn to the considered mixed-action formulation, corresponding to a partially quenched setup in which the twisted-mass parameters appear exclusively in the valence sector.
A first observation can be made when considering quenching.
If contributions from sea quarks to the dynamics are neglected, then the terms in Eqs.~(\ref{eq:Oaceven})–(\ref{eq:Oacodd}) originating from sea-quark loops vanish.
These terms can be readily identified as those containing factors of $\tr(\mathbf{m})$ and $\tr(\boldsymbol{\mu})$, i.e., those associated with improvement coefficients $\ibGb$ and $\ibGh$, respectively.
For example, in the case of quenched standard Wilson fermions, only counterterms with coefficients $c_\Gamma$, $\ibG$, and $f_{\scriptscriptstyle \Gamma}$ need to be considered.
Conversely, when present, the terms generated solely by quark loops have coefficients that, in a perturbative expansion, start at $\mbox{O}(\alphas^2(1/a))$.
This applies to both coefficients $\ibGb$ and $\ibGh$.
The partially quenched setup under consideration employs the same massless Wilson-Dirac operator in both the sea and valence sectors -- including the Sheikholeslami–Wohlert term with a non-perturbatively tuned value of $\icsw$ -- while massive terms involving twisted-mass parameters are introduced only in the valence sector.
The analysis of symmetry properties performed above for the unitary setup can be readily extended by including sea and valence fields together with ghost fields, which cancel the contribution of valence fields to the fermion determinant.
By considering the Wilson average and the Mass average of the valence sector for $\mathcal{R}_{5}$-even and parity-even lattice correlation functions involving quark fields, one observes that the symmetries of the partially quenched setup eliminate all $\Oa$ lattice artefacts, except for those proportional to the trace of the sea-quark mass matrix $\trmsea$.
More specifically, $\Oa$ terms of the form
\begin{gather}
  \trmsea\bar\psi^{\ival}\lambda\Gamma\psi^{\ival}\,,
\end{gather}
are allowed by the symmetries even in the fully twisted valence sector, while no additional contributions -- involving, e.g., ghost fields -- modify the $\Oa$-improvement structure of the quark bilinears~\cite{Bussone:2018ljj}.
This is consistent with the expectation that automatic $\Oa$-improvement does not apply to the massive Wilson-Dirac operator used in the sea sector.
The conclusion is that full $\Oa$-improvement of physical observables, based on the considered mixed-action setup, can be achieved by employing valence currents of the form
\begin{align}
  \label{eq:Simpv-mixed}
  (S^{ij})^{\star} &= \left[1+a\ibSb\trmsea\right]S^{ij}\,,\\[1.0ex]
  \label{eq:Vimpv-mixed}
  (V_\mu^{ij})^{\star} &= \left[1+a\ibVb\trmsea\right]V_\mu^{ij}\,,\\[1.0ex]
  \label{eq:Timpv-mixed}
  (T_{\mu\nu}^{ij})^{\star} &= \left[1+a\ibTb\trmsea\right]T_{\mu\nu}^{ij}\,,\\[1.0ex]
  \label{eq:Aimpv-mixed}
  (A_\mu^{ij})^{\star} &= \left[1+a\ibAb\trmsea\right]A_\mu^{ij}\,,\\[1.0ex]
  \label{eq:Pimpv-mixed}
  (P^{ij})^{\star} &= \left[1+a\ibPb\trmsea\right]P^{ij}\,,
\end{align}
and
\begin{align}
  \label{eq:Simpd-mixed}
  (S^{ii}-S^{jj})^{\star} &= [1+a\ibSb\trmsea](S^{ii}-S^{jj})\,,\\[1.0ex]
  \label{eq:Vimpd-mixed}
  (V_\mu^{ii}-V_\mu^{jj})^{\star} &= [1+a\ibVb\trmsea](V_\mu^{ii}-V_\mu^{jj})\,,\\[1.0ex]
  \label{eq:Timpd-mixed}
  (T_{\mu\nu}^{ii}-T_{\mu\nu}^{jj})^{\star} &= [1+a\ibTb\trmsea]
  (T_{\mu\nu}^{ii}-T_{\mu\nu}^{jj})\,,\\[1.0ex]
  \label{eq:Aimpd-mixed}
  (A_\mu^{ii}-A_\mu^{jj})^{\star} &= [1+a\ibAb\trmsea]
  (A_\mu^{ii}-A_\mu^{jj})\,,\\[1.0ex]
  \label{eq:Pimpd-mixed}
  (P^{ii}-P^{jj})^{\star} &= [1+a\ibPb\trmsea](P^{ii}-P^{jj})\,.
\end{align}
As noted above, the $\Oa$ terms that are not eliminated depend on the improvement coefficients $\bar{b}_{\Gamma}$ which arise only at $\mbox{O}(\alphas^2)$.
Moreover, for the considered $\NF= 2 + 1$ CLS ensembles, the light and strange sea-quark masses satisfy  $a\trmsea \lesssim 10^{-2}$.
Overall, the $a\bar{b}_{\Gamma}\trmsea$ term is therefore expected to introduce only a very small correction.
In the present study, this type of residual contributions are neglected in both the unitary and mixed-action setups.
The dedicated numerical studies reported in Sec.~\ref{sec:results} indicate that the continuum-limit scaling of the physical quantities based on these two regularisations are consistent with leading lattice artefacts of $\Oasq$, as can be seen, for instance, in Fig.~\ref{ch_ss:fig:universality}.
%


\subsection{Continuum-limit scaling of the strange-quark twist angle}
\label{app:twistangles}

\begin{figure}[htbp!]
  \centering
  \includegraphics[width=0.7\linewidth]{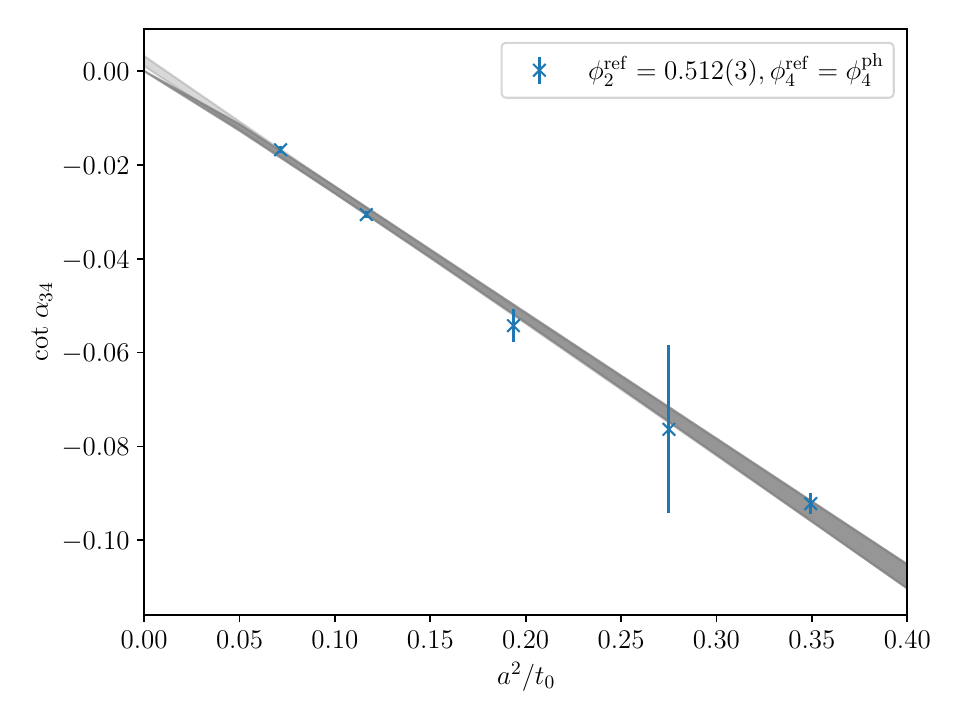}
  \caption{
    Continuum-limit scaling of $\cot\alpha_{34}$ -- the cotangent of the strange-quark twist angle defined in \req{eqn:twist_34} -- evaluated on the line of constant physics $\phi_2^{\mathrm{ref}}=0.512(3)$ and $\phi_4^{\mathrm{ref}}=\phi_4^{\mathrm{ph}}$.
    The light-grey band displays a linear fit in $a^2/t_0$, whose p-value is approximately $0.3$.
    The extrapolated continuum value is consistent with zero within two standard deviations.
    For comparison, the dark grey band shows a linear fit that is constrained to vanish in the continuum limit.
  }
  \label{fig:twist}
\end{figure}

We close this Appendix about the improvement of the mixed action setup by presenting numerical evidence for the continuum-limit scaling of the twist angle in the strange-quark sector.  
As explained in Sect.~\ref{subsec:matching}, the maximal-twist condition is imposed in the mass-degenerate light-quark sector via \req{eqn:maximal_twist}, where the mass-degenerate light-quark flavours are labelled by indices $1$ and $2$.  
In practice, the renormalised twist angle introduced in \req{eq:bare_twist_angle} is fixed so that
\begin{equation}
  \cot\alpha_{12} \;\equiv\; \frac{m_{12,\mathrm{R}}^{\mathrm{Wtm}}}{\mu_{12,\mathrm{R}}} \;=\; 0,
\end{equation}
and hence $\alpha_{12}=\tfrac{\pi}{2}$ on all ensembles.  
The matching procedure of Sect.~\ref{subsec:matching} determines the valence hopping parameter $\kappa^{\ival*}$ for which the maximal-twist condition in \req{eqn:maximal_twist} holds on a given ensemble.  
We note that the subtracted quark mass obtained using this determination of $\kappa^{\ival*}$ inherits $\Oa$ lattice artefacts from the employed maximal-twist condition.  
In general, since such artefacts may include terms of $\mbox{O}(a\mu_{i}^2)$, it is advantageous to impose the maximal-twist condition in the light-quark sector, $i=1,2$.  
The value of $\kappa^{\ival*}$ is then employed in the computation of physical quantities with light, strange or heavier valence masses on that ensemble.  
Following the discussion in Secs.~\ref{sec:setup} and~\ref{sec:OaMA}, the use of the maximal-twist condition in \req{eqn:maximal_twist} guarantees that physical observables computed in the mixed-action setup scale with cutoff effects of $\mbox{O}(a^2)$, up to possible residual contributions of $\mbox{O}\!\bigl(\alphas^2\,a\, \trmsea \bigr)$.
As discussed in Sec.~\ref{sec:results}, this expectation has been explicitly verified numerically; see, for instance, Fig.~\ref{ch_ss:fig:universality}.
On the other hand, matrix elements of parity-odd operators vanish in the continuum limit and, at finite lattice spacing, may receive $\Oa$ discretisation effects.
An example is given by the valence PCAC quark mass in the strange sector, $m_{34,\mathrm{R}}^{\mathrm{Wtm}}$, which in the physical-quark basis is related to a parity-odd two-point correlator of a vector current and a pseudoscalar density.
Consequently, a meaningful complementary test of the Wtm mixed action is to conduct a numerical investigation of the the continuum-limit scaling of this quantity.
More specifically, we consider $\alpha_{34}$ -- the renormalised twist angle in the strange sector -- as follows
\begin{equation}
  \label{eqn:twist_34}
  \cot\alpha_{34} \;=\; \frac{m_{34,\mathrm{R}}^{\mathrm{Wtm}}}{\mu_{34,\mathrm{R}}}=\frac{\ZA m_{34}^{\mathrm{Wtm}}}{\mu_{34}}\,,
\end{equation}
and study its approach to $\alpha_{34}=\tfrac{\pi}{2}$ in the continuum limit.
In a similar manner to the treatment of the pseudoscalar decay constants in Sec.~\ref{sec:match}, the Wtm strange-quark PCAC mass, $m_{34}^{\mathrm{Wtm}}$, was interpolated to the valence mass parameters $(\kappa^{\ival*},\,a\mu_1^{*},\,a\mu_3^{*})$ that satisfy the matching condition.
A subsequent interpolation is performed to reach the selected line of constant physics, defined by the reference values $\phi_2^{\mathrm{ref}} = 0.512(3)$ and $\phi_4^{\mathrm{ref}} = \phi_4^{\mathrm{ph}}$.
The choice of $\phi_2^{\mathrm{ref}}$ was made to minimise the interpolation range, based on the available data for $\phi_2$ at each value of $\beta$.
In practice, for each ensemble, a linear interpolation to $\phi_2^{\mathrm{ref}}$ is performed using the two nearest measurements in $\phi_2$.
To estimate a systematic uncertainty associated with this interpolation, we also perform a quadratic fit in $\phi_2$ over all available points and take the deviation between the quadratic and linear interpolations at $\phi_2^{\mathrm{ref}}$ as an additional source of error.
The continuum-limit scaling of $\cot\alpha_{34}$ at the reference point is shown in Fig.~\ref{fig:twist}.  
Although $\Oa$ lattice artefacts can in principle contribute to $\cot\alpha_{34}$, the data are well described by a linear dependence in $a^2$; the corresponding fit has a p-value of approximately $0.3$, and the continuum-extrapolated value is consistent with zero within two sigma.
We emphasise that, in the computation of $m_{12}^{\mathrm{Wtm}}$ used to impose the maximal-twist condition in \req{eqn:maximal_twist}, and in the determination of $\cot\alpha_{34}$ in \req{eqn:twist_34}, the mass-independent improvement coefficient $\icA$ has been included, while the effect of including the known mass-dependent coefficients is observed to be negligibly small.
The observed $a^2$-scaling in Fig.~\ref{fig:twist} therefore suggests that the residual $\Oa$ effects are numerically small for this observable.  
We remark that the tree-level values of the some of the $\Oa$ improvement coefficients associated with the twisted mass term are not expected to be independent among them (see, for instance, Ref.~\cite{Frezzotti:2001ea}), and a suitable prescription may be adopted to suppress those contributing to the computation of the twist angles from the PCAC quark mass at tree level.
%


\section{Reweighting}
\label{app:flagged_s}

\begin{figure}[htbp!]
  \centering
  \includegraphics[width=0.9\linewidth]{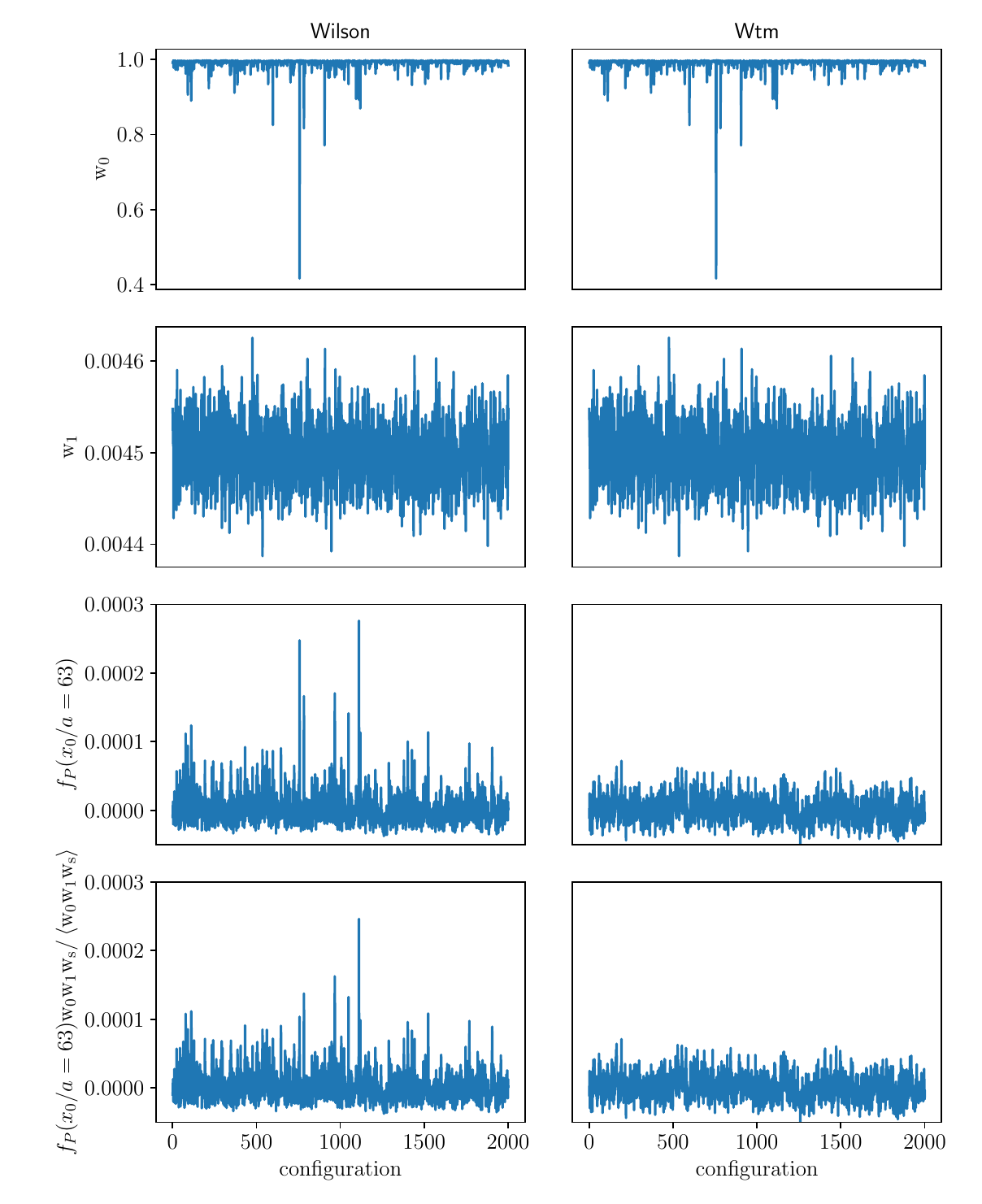}
  \caption{
    Illustration of the impact of reweighting factors for ensemble D200 in the Wilson unitary ({\it left} panels) and Wtm mixed-action ({\it right} panels) setups.
    The figure displays the Monte Carlo histories of the reweighting factors and the light pseudoscalar two-point correlator $\fP(x_0)$ in Eqs.~(\ref{eqn:fP}) and (\ref{eqn:sym}), with the source position fixed at Euclidean time $y_{0}/a = 1$ and the sink at $x_{0}/a = 63$, approximately at the midpoint of the temporal extent.
    The top row displays the light-sector twisted-mass reweighting factor ${\mathrm{w}}_0$ for each configuration.
    The second row shows the strange-quark reweighting factor ${\mathrm{w}}_1$, used to correct for the rational approximation in the RHMC algorithm.
    We note that the left and right panels are the same for the first and second rows since the same reweighting factor are applied in the Wilson and Wtm cases.
    The third row presents the Monte Carlo history of the unreweighted light pseudoscalar correlator.
    The bottom row shows the same correlator after applying the complete set of reweighting factors, ${\mathrm{w}}_0$, ${\mathrm{w}}_1$, and the factor ${\mathrm{w}}_{\mathrm{s}}$ accounting for the rare occurrence of a negative strange-quark determinant.
    In this ensemble, ${\mathrm{w}}_{\mathrm{s}}$ takes the value $+1$ for all configurations.
  }
  \label{fig:rwf}
\end{figure}

The target $\NF=2+1$ flavour theory is attained through the application of reweighting factors acting in the light mass-degenerate up-down sector and on the strange quark sector.
The first type is the reweighting factor, denoted ${\mathrm{w}}_0$, applied to remove the twisted-mass term introduced in the light-quark sector to improve the stability of the HMC~\cite{0810.0946, Luscher:2012av, Bruno:2014jqa}.
In the strange quark sector, a reweighting factor~\cite{Bruno:2014jqa} denoted ${\mathrm{w}}_1$, is employed to correct for the rational approximation induced by the use of the Rational Hybrid Monte Carlo (RHMC) algorithm~\cite{Kennedy:1998cu,Clark:2006fx}.
In addition, where present, configurations with a negative strange-determinant sign are corrected by a sign flip~\cite{Mohler:2020txx} in the ensemble averages.
In practice, these sign flips occur only on a small subset of configurations and the strange-quark reweighting factors only lead to a small increase of the statistical uncertainty.
As indicated by the column LMD in Table~\ref{tab:CLS_ens}, the reweighting factors ${\mathrm{w}}_0$ and ${\mathrm{w}}_1$ are computed either using the low-mode deflation (LMD) technique employed in~\cite{Kuberski:2023zky}, or via a more conventional stochastic estimation method~\cite{Bruno:2014jqa}.
The impact of the reweighting procedure is illustrated in Fig.~\ref{fig:rwf} for both the Wilson unitary and Wtm mixed-action setups, showing the Monte Carlo history of a light pseudoscalar correlator for ensemble D200 before and after applying the complete reweighting procedure.
We note that the same set of reweighting factors is applied in the two setups, since the reweighting induces an effect in the sea-quark action.
Overall, we observe that the application of reweighting induces only minor shifts in the central values and a limited increase in the statistical uncertainties in both the Wilson and Wtm cases.
In particular, the integrated autocorrelation time $\tau_{\mathrm{int}}$, determined following the procedure outlined in Appendix~\ref{app:analysis}, does not vary significantly after the application of reweighting.
%


\section{Boundary effects}
\label{app:obc}

\begin{figure}[htbp!]
  \centering
  \includegraphics[width=0.9\linewidth]{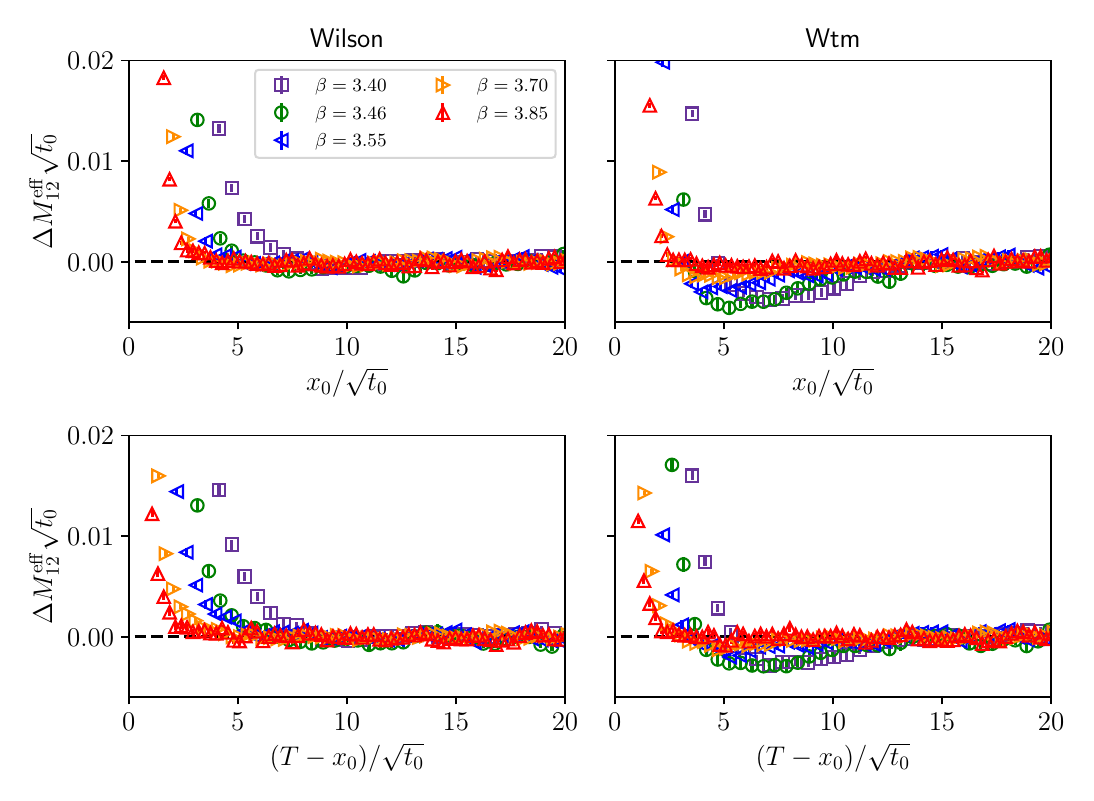}
  \caption{
    Boundary effects in the effective PCAC quark mass of the light sector for the unitary Wilson (\textit{left}) and mixed-action Wtm (\textit{right}) regularisations.
    On the vertical axis, $\Delta M_{12}^{\mathrm{eff}}(x_0) \sqrt{t_0}$ denotes the effective RGI quark mass, based on \req{eq:RGI}, expressed in units of $t_0$ and subtracted by the central value of the ground-state contribution determined from the plateau region.
    The correlation functions used to compute the effective PCAC quark mass via \req{eqn:mpcac} have a source located at $y_0/a = 1$, while the sink position $x_0$ is varied.
    The \textit{top} panel displays the region near the boundary at $x_0 = 0$, whereas the \textit{bottom} panel corresponds to the vicinity of the $x_0 = T$ boundary.
    Different colours and symbols represent various lattice spacings along a line of constant physics, defined by symmetric-point ensembles.
    It is observed that boundary effects propagating into the bulk are increasingly suppressed as the lattice spacing is reduced.
  }
  \label{fig:boundaries_m12_CL}
\end{figure}

\begin{figure}[htbp]
  \begin{center}
    \includegraphics[width=0.9\linewidth]{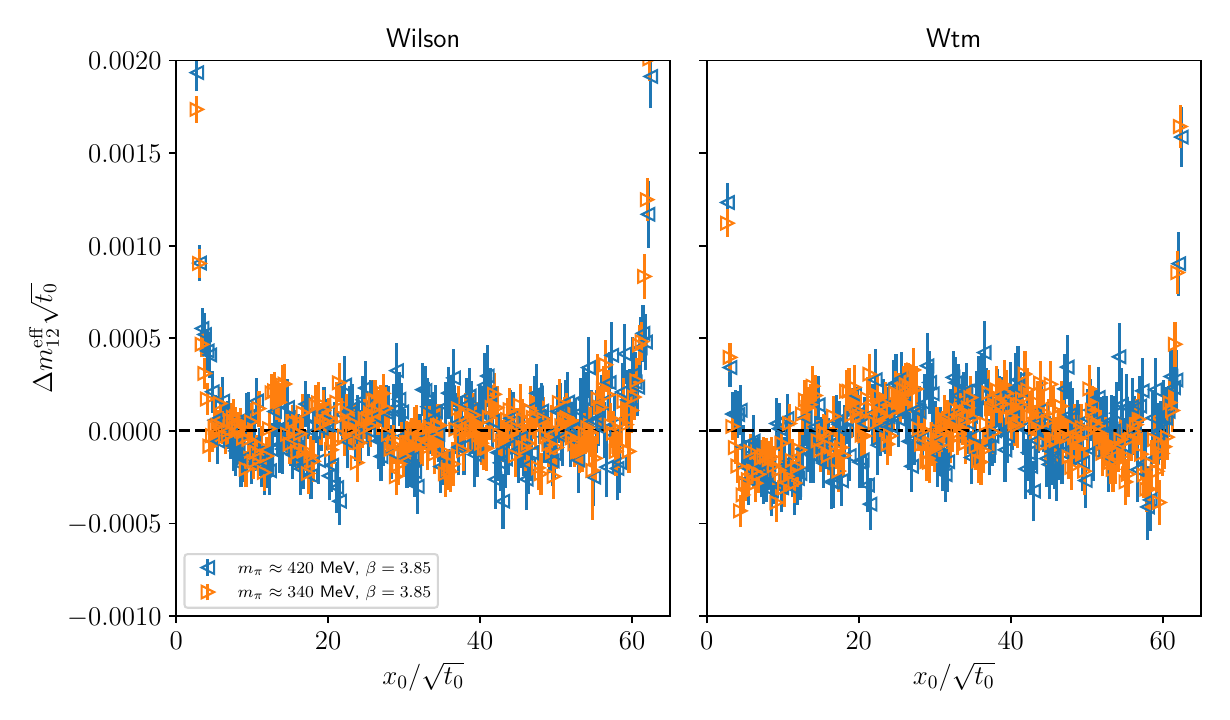}
  \end{center}
  \caption{
    Light-quark mass-dependence of boundary effects in the effective PCAC quark mass of the light sector for the unitary Wilson (\textit{left}) and mixed-action Wtm (\textit{right}) regularisations.
    On the vertical axis, $\Delta m_{12}^{\mathrm{eff}}(x_0) \sqrt{t_0}$ denotes the effective PCAC quark mass, based on \req{eqn:mpcac}, expressed in units of $t_0$ and subtracted by the central value of the ground-state contribution determined from the plateau region.
    The correlation functions used to compute the effective PCAC quark mass via \req{eqn:mpcac} have a source located at $y_0/a = 1$, while the sink position $x_0$ is varied.
    Different colours and symbols correspond to two values of the pion mass at the finest lattice spacing, $a \approx 0.039\,\mathrm{fm}$, corresponding to $\beta = 3.85$.
    Only a mild dependence on the quark mass is observed in the boundary effects affecting the PCAC quark mass for both regularisations.
  }
  \label{fig:boundaries_m12_mq}
\end{figure}

\begin{figure}[htbp]
  \centering
  \includegraphics[width=0.9\linewidth]{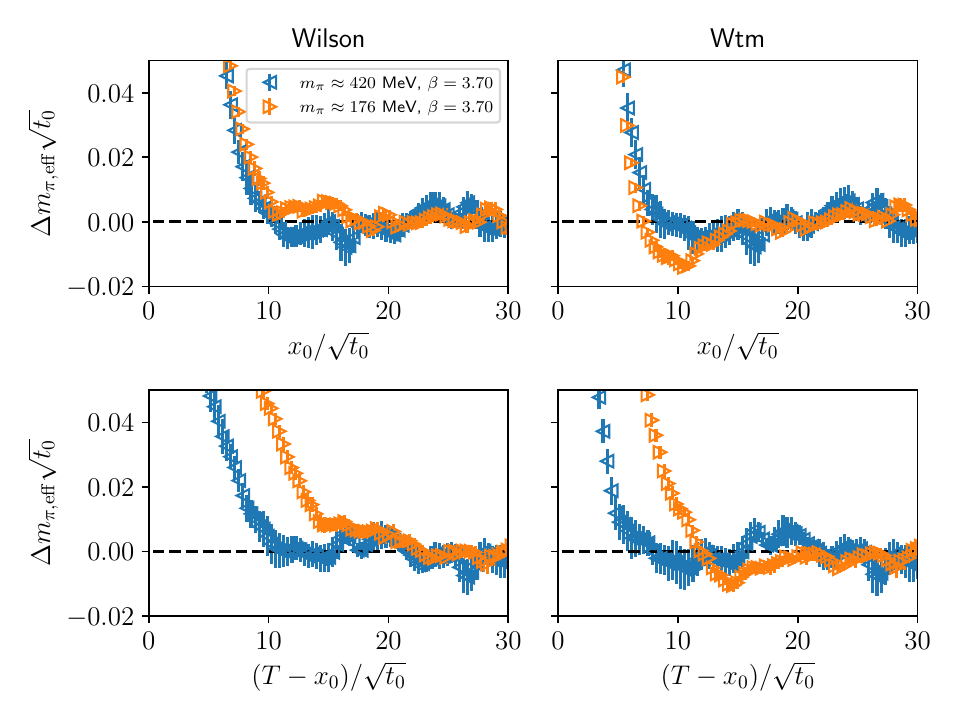}
  \caption{
    Pseudoscalar effective mass in the vicinity of the boundaries for the unitary Wilson (\textit{left}) and mixed-action Wtm (\textit{right}) regularisations.
    On the vertical axis, $\Delta m_{\pi,\mathrm{eff}}\,\sqrt{t_0}$ denotes the light pseudoscalar effective mass $m_{\mathrm{\scriptscriptstyle PS,eff}}^{12}(x_0)$, as defined in \req{eqn:meff}, expressed in units of $t_0$ and shifted by the central value of the ground-state contribution determined from the plateau region.
    The pseudoscalar correlation function used to compute the effective mass in \req{eqn:meff} has a source located at $y_0/a = 1$, while the sink position $x_0$ is varied.
    The \textit{top} panel displays the region near the boundary at $x_0 = 0$, whereas the \textit{bottom} panel corresponds to the vicinity of the $x_0 = T$ boundary.
    Different colours and symbols correspond to two values of the pion mass at a fixed value of the lattice spacing, $a \approx 0.049\,\mathrm{fm}$, corresponding to $\beta = 3.70$.
    The observed dependence on the quark mass near the boundaries is analysed in the context of ground-state extraction, as described in Appendix~\ref{app:groundstate}.
  }
  \label{fig:boundaries_mpi}
\end{figure}

In this Appendix, we analyse the effects associated with the use of open boundary conditions (OBC) in the temporal direction.
As noted in the caption of Table~\ref{tab:CLS_ens}, the majority of CLS ensembles employed in this work rely on OBC~\cite{Bruno:2014ova}.
More specifically, the gauge fields obey OBC~\cite{Luscher:2010we,Schaefer:2010hu,Luscher:2011kk}, implemented by setting the time component of the field-strength tensor to zero at the temporal boundaries, $x_0 = 0$ and $x_0 = T$.
Dirichlet boundary conditions are imposed on the fermion fields~\cite{hep-lat/0603029,hep-lat/9312079}.
Periodic boundary conditions are applied in all spatial directions.
As discussed in Sec.~\ref{subsec:obc}, the adoption of OBC facilitates smooth topological transitions and prevents the emergence of barriers as the continuum limit is approached~\cite{Luscher:2010we,Schaefer:2010hu,Luscher:2011kk,Luscher:2012av}.
Nevertheless, the presence of temporal boundaries breaks translational invariance in time, resulting in correlation functions that depend on the distance from the boundaries.
Sufficiently far from the temporal boundaries, local observables converge towards their vacuum expectation values, with corrections that are exponentially suppressed.
This enables the identification of plateau regions where boundary effects are negligible.
The spectral decomposition of correlation functions in the presence of temporal boundaries has been investigated in Refs.~\cite{hep-lat/9903040,Bruno:2015hfq}.
Boundary effects may also induce lattice artefacts near $x_0 = 0$ and $x_0 = T$, partly due to residual $\mbox{O}(a g_0^2)$ contributions stemming from the use of tree-level values for the boundary improvement coefficients.
It is therefore important to establish the minimum separation from the boundaries that can be considered safe, beyond which such effects can be disregarded.
We begin by examining the PCAC quark masses for both the unitary Wilson and mixed-action Wtm regularisations, as these are protected from excited-state contamination, allowing boundary effects to be directly analysed.
In particular, we focus on the light-sector RGI quark masses $M_{12}$, defined as follows
\begin{equation}
  \label{eq:RGI}
  M_{ij}=\frac{M}{m_{\mathrm{R}}(\mu_\mathrm{ren})}\times m_{ij}^{\mathrm{R}}(\mu_\mathrm{ren})\,,
\end{equation}
with indices $i=1$ and $j=2$.
The first factor on the right-hand side is the continuum, flavour-independent, non-perturbative running factor determined in Ref.~\cite{Campos:2018ahf}, while $m_{ij}^{\mathrm{R}}(\mu_\mathrm{ren})$ denotes the renormalised quark mass in the Schr\"odinger Functional scheme at a hadronic renormalisation scale, $\mu_\mathrm{ren} = 233(8)\,\mathrm{MeV}$.
The renormalised quark mass is obtained from Eqs.~(\ref{eqn:mpcac}) and (\ref{eqn:mPCACI}), in conjunction with the non-perturbative renormalisation factor $\ZP(\mu_\mathrm{ren})$ computed in Ref.~\cite{Campos:2018ahf}.
The gradient flow scale $t_0$ is employed to construct the dimensionless quantities $M_{12} \sqrt{t_0}$ and $x_0/\sqrt{t_0}$.
To enhance the visibility and facilitate comparison of deviations from plateau behaviour, the central value of the ground-state contribution is subtracted from the effective RGI quark mass $M_{12}^{\mathrm{eff}}(x_0) \sqrt{t_0}$.
This subtraction, denoted by $\Delta$ in $\Delta M_{12}^{\mathrm{eff}}(x_0) \sqrt{t_0}$, is illustrated in Fig.~\ref{fig:boundaries_m12_CL} near the boundaries at $x_0 = 0$ and $x_0 = T$.
Significant lattice artefacts are observed close to the boundaries for both the unitary Wilson and mixed-action Wtm setups on symmetric point ensembles.
At the coarsest lattice spacing, these effects are substantially suppressed at a separation of $14\,\sqrt{t_0} \approx 2\,\mathrm{fm}$.
Boundary effects diminish rapidly in the bulk region as the lattice spacing is reduced.
The light-quark mass-dependence of boundary effects is illustrated in Fig.~\ref{fig:boundaries_m12_mq}, which shows the effective PCAC quark mass $m_{12}^{\mathrm{eff}}(x_0)$ -- subtracted by the central value of the ground-state contribution determined from the plateau region -- for two pion mass values at fixed lattice spacing.
Only a mild dependence on the quark mass is observed in the boundary effects affecting the PCAC quark mass for the two regularisations.
We now examine the pion effective mass $m_{\pi,\mathrm{eff}}$, which corresponds to $m_{\mathrm{\scriptscriptstyle PS, eff}}^{12}(x_0)$ as defined in \req{eqn:meff}, in the vicinity of the boundaries.
In this case, the Euclidean time dependence in $x_0$ receives additional contributions from excited states near the source position at $y_0 = a$.
Conversely, the region close to the boundary at $x_0 = T$ is primarily influenced by boundary effects.
Figure~\ref{fig:boundaries_mpi} illustrates the Euclidean time dependence of the light pseudoscalar effective mass for two pion mass values at fixed coupling $\beta$, for both the Wilson and Wtm regularisations.
A significant dependence on the quark mass can be observed near the boundaries.
The occurrence of these effects is analysed in the context of ground-state extraction, as described in Appendix~\ref{app:groundstate}.
%


\section{Error analysis methods}
\label{app:analysis}

\subsection{Statistical error analysis}
\label{app:statanalysis}

\begin{figure}[htbp!]
  \centering
  \includegraphics[width=1.\linewidth]{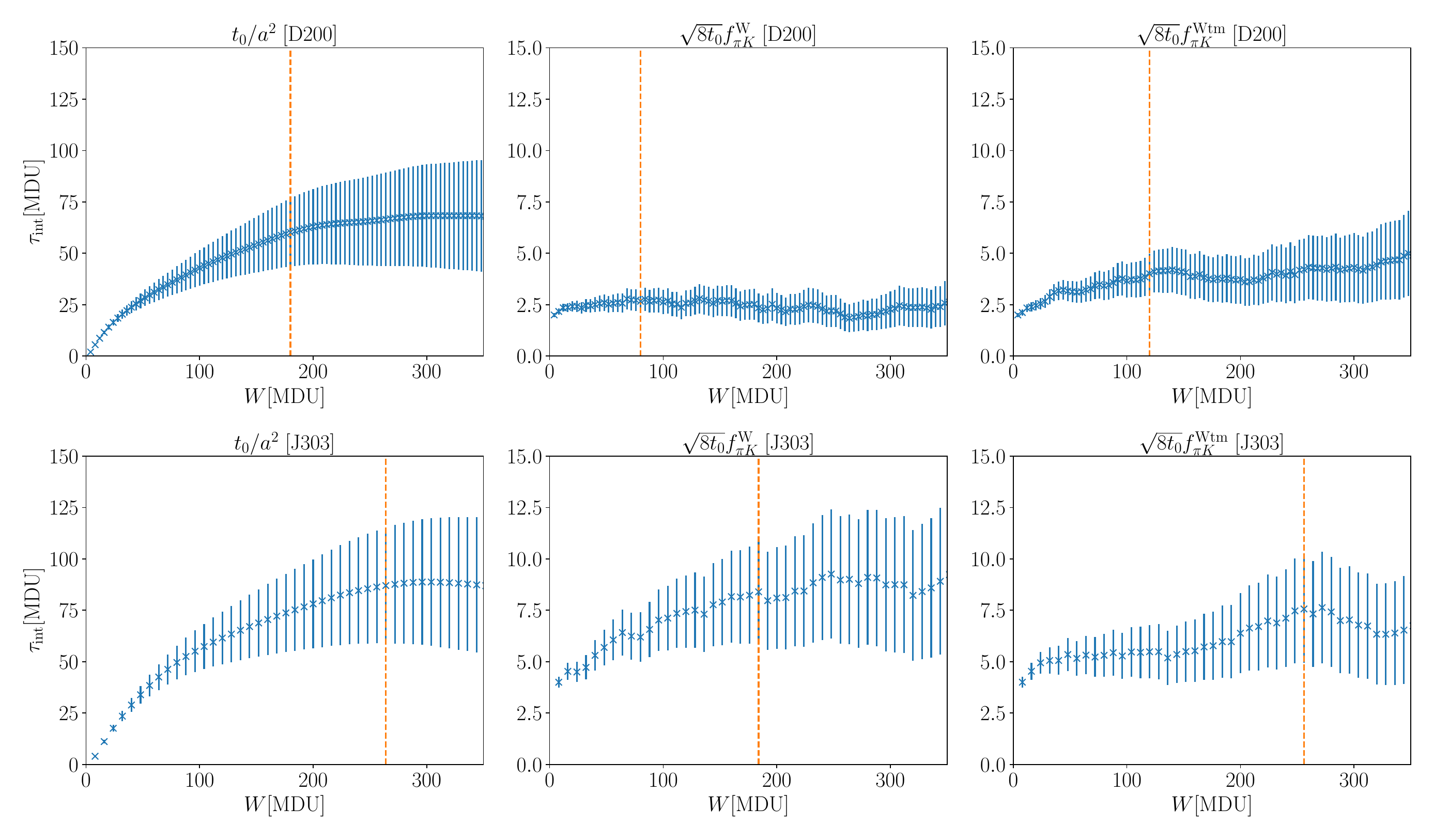}
  \caption{
    From {\it left} to {\it right}: dependence on the summation window $W$ of the integrated autocorrelation time $\tau_{\mathrm{int}}^{\alpha}(F,W)$, as defined in \req{eqn:tint}, for the observables $F$ corresponding to $t_0/a^2$, $\sqrt{8t_0}\,f_{\pi K}^{\mathrm{W}}$ (Wilson regularisation), and $\sqrt{8t_0}\,f_{\pi K}^{\mathrm{Wtm}}$ (Wtm mixed action), respectively, after applying the mass-shifting procedure of \req{eqn:derphi4}.
    The {\it top} row presents results for ensemble D200, corresponding to a lattice spacing $a \approx 0.063\,\mathrm{fm}$, while the {\it bottom} row displays results for ensemble J303, with $a \approx 0.049\,\mathrm{fm}$.
    Both ensembles employ open boundary conditions in the temporal direction.
    The dashed orange vertical line indicates the selected value of $\overline{W}$.
    While $t_0/a^2$ exhibits relatively large integrated autocorrelation times, the combination $\sqrt{8t_0}\,f_{\pi K}$ yields smaller values of $\tau_{\mathrm{int}}$, which are comparable across the two regularisations.
  }
  \label{fig:autocorr}
\end{figure}

\begin{figure}[htbp!]
  \centering
  \includegraphics[width=1.\linewidth]{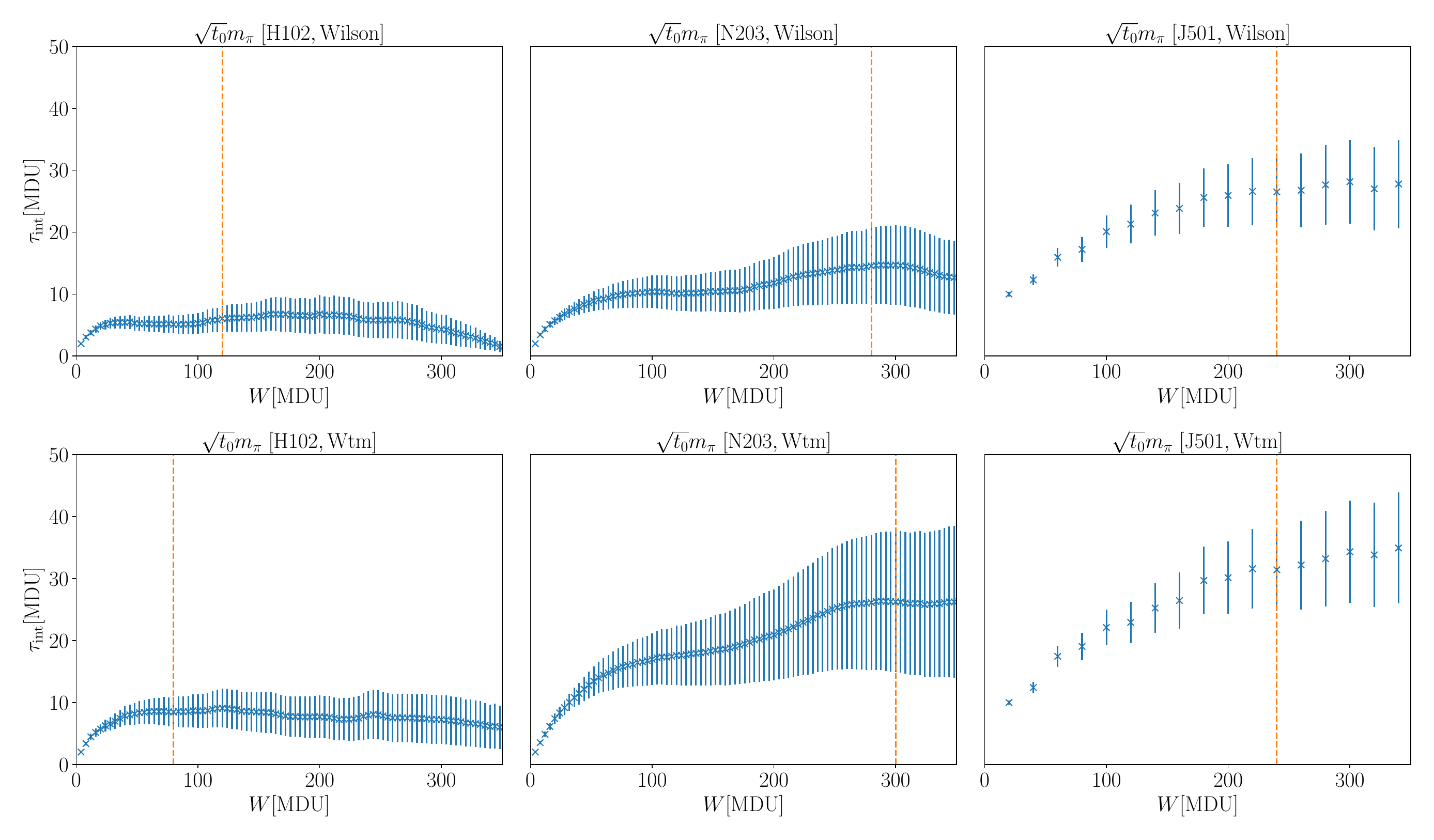}
  \caption{
    Dependence of the integrated autocorrelation time, defined in \req{eqn:tint}, of $\sqrt{t_0}\,m_{\pi}$ on the summation window $W$ for the Wilson regularisation ({\it top} row) and the Wtm mixed action ({\it bottom} row).
    From {\it left} to {\it right}, the panels correspond to the ensembles H102 ($a \approx 0.085\,\mathrm{fm}$), N203 ($a \approx 0.065\,\mathrm{fm}$), and J501 ($a \approx 0.039\,\mathrm{fm}$).
    All ensembles employ open boundary conditions in the temporal direction and correspond to approximately the same value of the pion mass (cf. Table~\ref{tab:CLS_ens}).
    The dashed orange vertical line indicates the selected value of $\overline{W}$.
    As expected, an increase in the integrated autocorrelation time is observed as the lattice spacing is reduced.
  }
  \label{fig:tau_mpi}
\end{figure}

The statistical uncertainties of the Monte Carlo (MC) data are computed using the $\Gamma$-method \cite{Wolff:2003sm, 1009.5228}, as implemented in the \texttt{ADerrors} package~\cite{Ramos:2018vgu}, which employs automatic differentiation techniques to compute and propagate uncertainties.
For completeness, we summarise below some generic expressions defining the autocorrelation functions at the core of the $\Gamma$-method.
A more detailed account can be found in Refs.~\cite{Wolff:2003sm, Luscher:2010ae, 1009.5228}.
Given a set of statistically independent ensembles indexed by $\alpha$, we consider a primary observable with expectation value $P_i^{\alpha}$, labelled by $i = 1, \dots, n_{\mathrm{obs}}$.
A Markov Chain Monte Carlo process is used to generate $n_\alpha$ gauge-field configurations for ensemble $\alpha$.
Evaluating the observable $P_i^{\alpha}$ on these configurations yields a series of measurements
\begin{equation}
  p_i^{\alpha}(t)\,,\quad t = 1, \dots, n_{\alpha}\,.
\end{equation}
The stochastic estimator of $P_i^{\alpha}$ is given by the sample mean
\begin{equation}
  \label{eqn:pbar}
  \bar{p}_i^{\,\alpha}=\frac{1}{n_{\alpha}}\sum_{t=1}^{n_{\alpha}}p_i^{\alpha}(t)\,,
\end{equation} 
which is used to define the deviation from the true expectation value
\begin{equation}
  \bar{\delta} p_i^{\alpha} = \bar{p}_i^{\,\alpha} - P_i^{\alpha}\,.
\end{equation}
The deviations of the individual measurements given by
\begin{equation}
  \delta p_i^{\alpha}(t)=p_i^{\alpha}(t)-\bar{p}_i^{\,\alpha}\,,
\end{equation}
are used to construct the estimator of the autocorrelation function
\begin{equation}
  \label{eq:GammaP}
  \overline{\Gamma}^{\,\alpha}_{ij}(t) = \frac{1}{n_{\alpha} - t} \sum_{t'=1}^{n_{\alpha} - t} \delta p_i^{\alpha}(t')\, \delta p_j^{\alpha}(t' + t)\,.
\end{equation}
\noindent For a derived observable $F$, defined as a function $f$ of the expectation values $P_i^{\alpha}$, the estimator $\bar{F}$ reads
\begin{equation}
  \bar{F} = f(\bar{p}_i^{\,\alpha})\,.
\end{equation}
A Taylor expansion of $f$ in the deviations $\bar{\delta} p_i^{\,\alpha}$ allows to construct an estimator of the autocorrelation function of $F$
\begin{equation}
  \label{eq:GammaG}
  \overline{\Gamma}^{\alpha}_{F}(t)=\sum_{ij}\bar{f}_i^{\alpha}\bar{f}_j^{\alpha}\frac{1}{n_{\alpha}-t}\sum_{t'=1}^{n_{\alpha}-t}\delta p_i^{\alpha}(t')\, \delta p_j^{\alpha}(t'+t) = \sum_{ij}\bar{f}_i^{\alpha}\bar{f}_j^{\alpha}\overline{\Gamma}_{ij}^{\alpha}(t)\,,
\end{equation}
where the coefficients $\bar{f}_i^{\alpha}$ are the derivatives of $f$ with respect to $P_i^{\alpha}$, evaluated at $\bar{p}_i^{\,\alpha}$
\begin{equation}
  \bar{f}_i^{\alpha} = \left. \frac{\partial f}{\partial P_i^{\alpha}} \right|_{\bar{p}_i^{\,\alpha}}\,.
\end{equation}
The second equality in \req{eq:GammaG}, obtained using \req{eq:GammaP}, shows that the computation of $\overline{\Gamma}^{\alpha}_{F}$ requires only the projected autocorrelation function.
The statistical error $\sigma_F$ of $F$ is given by
\begin{equation}
  \sigma_F^2 = \sum_{\alpha} 2 \tau_{\mathrm{int}}^{\alpha}(F) \frac{\overline{\Gamma}_F^{\alpha}(0)}{n_{\alpha}}\,,
\end{equation}
where the sum runs over all ensembles contributing to $F$.
The integrated autocorrelation time of $F$ is defined as follows
\begin{equation}
  \label{eqn:tintformal}
  \tau_{\mathrm{int}}^{\alpha}(F) = \frac{1}{2} + \sum_{t=1}^{\infty} \frac{\overline{\Gamma}_F^{\alpha}(t)}{\overline{\Gamma}_F^{\alpha}(0)}\,.
\end{equation}
However, a practical estimator of $\tau_{\mathrm{int}}^{\alpha}(F)$ requires a truncation of the values of $t$ in \req{eqn:tintformal} inside a summation window controlled by $W$,
\begin{equation}
  \label{eqn:tint}
  \tau_{\mathrm{int}}^{\alpha}(F,W) = \frac{1}{2} + \sum_{t=1}^{W} \frac{\overline\Gamma_F^{\alpha}(t)}{\overline\Gamma_F^{\alpha}(0)}\,.
\end{equation}
The choice of $W$ necessitates careful consideration~\cite{Wolff:2003sm,1009.5228}.
Values that are too small risk introducing systematic uncertainties through truncation of the tail of the autocorrelation function, whereas excessively large values can lead to inflated statistical fluctuations by incorporating data from regions in which the signal is lost.
The exponential decay of $\overline{\Gamma}_F^{\alpha}(t)$ at asymptotically large $t$ is governed by the exponential autocorrelation time $\tau_{\mathrm{exp}}^{\alpha}$, corresponding to the slowest mode of the Markov process.
The corresponding truncation error is therefore of $\mbox{O}(\exp(-W/\tau_{\mathrm{exp}}^{\alpha}))$.
Following the proposal of Ref.~\cite{Wolff:2003sm}, we select the value $W = \overline{W}$ that minimises the combined statistical and systematic error, under the assumption $\tau_{\mathrm{exp}}^{\alpha} \approx S_{\tau} \tau_{\mathrm{int}}^{\alpha}$.
The default value of the parameter $S_{\tau}$ is set to $S_{\tau} = 4$.
Direct inspection of the stability of $\tau_{\mathrm{int}}^{\alpha}(F)$ as a function of $W$ is used to adjust the value of $S_{\tau}$ until $\overline{W}$ is observed to lie within a plateau regime.
Figure~\ref{fig:autocorr} displays the $W$-dependence of the integrated autocorrelation time defined in \req{eqn:tint} for key observables considered in this work: the gradient flow scale $t_0/a^2$ and $\sqrt{8t_0}\,f_{\pi K}$, in both the Wilson and Wtm regularisations, after the application of the mass-shifting procedure, for ensembles D200 and J303.
Similarly, Fig.~\ref{fig:tau_mpi} illustrates the dependence on $W$ of the integrated autocorrelation time of the pion mass, expressed in units of $t_0$, for the ensembles H102, N203, and J501. 
%


\subsection{Estimation of systematic uncertainties from model variation}
\label{app:MA}

A common practice for exploring systematic uncertainties in a target observable, based on a given set of data, is to consider various well-motivated functional forms that model the dependence on relevant parameters, or to apply cuts to the data in order to favour a regime where the chosen model is expected to describe the data most accurately.
In the following, variations in the functional forms and the application of data cuts will be collectively referred to as model variations.
The estimation of the quoted systematic uncertainty is frequently challenging, necessitating a comprehensive exploration of the relevant model space and a careful examination of the impact of the physically motivated variations.
When assessing the magnitude of systematic effects, a conservative approach -- where the largest deviation among a set of physically motivated models that provide a satisfactory description of the data is quoted as the systematic uncertainty -- can be complemented by a model-averaging procedure.
In the latter, a probability-weighted combination of the model variations yields an alternative estimate of the systematic uncertainties.
In this Appendix, we provide a concise account of the model-averaging procedures employed in this work to assess the magnitude of systematic effects.
More specifically, model variations were applied to the isolation of the ground-state signal in correlation functions and to the combined continuum and chiral extrapolations of $\sqrt{8t_0}f_{\pi K}$.

\subsubsection{Formulation of the $\chi^2$ statistic}

For a given dataset $D = \{y_i\}_{i=1,\dots,n_{\mathrm{dat}}}$ and a set of models $\mathcal{M} = \{M\}$, each characterised by a different number of parameters $\boldsymbol{p} = (p_1, \ldots, p_{n_{\mathrm{param}}})$, we aim to determine the probability $P\left(M|D\right)$ that a given model $M$ represents the true underlying distribution of the data $D$.
To this end, we fit the dataset $D$ to each model $M$ using the generalised least-squares method.
This procedure determines the free parameters of model $M$ by minimising the negative log-likelihood, which is also referred to as the Information Criterion (IC),
\begin{equation}
  \label{eqn:logchi}
  \mathrm{IC} = -2\log P\left(D|M\right)= \chi^2 + {\mathrm{cnst}}\,,
\end{equation}
depending on the chi-squared statistic.
We consider the following general form of the chi-squared function
\begin{equation}
  \label{eqn:chisq}
  \chi_{\mathcal{W}}^2 = \sum_{i,j=1}^{n_{\mathrm{dat}}}\left(y_i-f_M(x_i,\boldsymbol{p})\right)\, \mathcal{W}_{ij}\, \left(y_j-f_M(x_j,\boldsymbol{p})\right)\,,
\end{equation}
where $\mathcal{W}$ represents a weight matrix and $f_M(x_i, \boldsymbol{p})$ denotes the model function, parameterised by the set of free parameters $\boldsymbol{p}$, and dependent on the elements of the set of independent variables $\{x_i\}$.
In the specific case where $\mathcal{W} = C^{-1}$, with $C$ being the covariance matrix of the data set $\{y_i\}$, the procedure corresponds to a correlated fit.
In cases where the independent variables $\{x_i\}$ carry uncertainties -- such as in the combined continuum and chiral fits discussed in Sec.~\ref{sec:results}, where $\phi_2$ and $a^2/t_0$ serve as instances of $x_i$ -- the $\chi_{\mathcal{W}}^2$ function in \req{eqn:chisq} may be generalised to incorporate both the correlations among the $\{x_i\}$ data and those between $\{x_i\}$ and $\{y_j\}$.
To achieve this, one may construct an augmented data vector
\begin{equation}
  Y=(y_1,...,y_{n_{\mathrm{dat}}},x_1,...,x_{N})^T,
\end{equation}
where the dimension $N$ of the data set $\{x_i\}$ is assumed to be an integer multiple of $n_{\mathrm{dat}}$.
By considering the function $F_i$, depending on the additional parameters $\{\tilde{p}_j\}_{j=n_{\mathrm{dat}}+1,\dots,n_{\mathrm{dat}}+N}$
\begin{equation}
  F_i=\left\{\begin{matrix}
  f_M(x_i,\boldsymbol{p}) & \text{if}\ i\leq n_{\mathrm{dat}}\,, \\ 
  \tilde{p}_i & \text{if}\ i >n_{\mathrm{dat}}\,,
  \end{matrix}\right.
\end{equation}
the following definition of the $\chi^2$ function can be considered
\begin{equation}
  \label{eqn:chisq_full}
  \chi_{\mathcal{W}}^2=\sum_{i,j=1}^{n_{\mathrm{dat}}+N}\left(Y-F\right)^T_i \, \mathcal{W}_{ij}\, \left(Y-F\right)_j\,.
\end{equation}
In the case of a correlated fit, $\mathcal{W} = C^{-1}$, the matrix $C$ denotes the covariance matrix of the augmented data vector $Y$. 
When analysing highly correlated data sets in which the number of samples is not substantially larger than the number of data points, spuriously small eigenvalues in the correlation matrix may impair the ability to accurately invert the covariance matrix $C$~\cite{Michael:1994sz}, thereby compromising the reliability of correlated fits.
Following Ref.~\cite{Bruno:2022mfy}, the goodness-of-fit -- quantified by the p-value -- can still be estimated in cases where the standard correlated fit is replaced by a modification of the weight matrix $\mathcal{W}$.
We will, for instance, exploit this property to perform uncorrelated fits, in which the weight matrix satisfies $\mathcal{W}^{-1} = \mathrm{diag}(C)$, for the extraction of the ground-state contribution from fits to data on a range of Euclidean times.
In the combined continuum and chiral fits described in Sec.~\ref{sec:results}, the correlations among the $\{y_i\}$ data are retained, while those involving the $\{x_i\}$ data are neglected.
This approximation enables a stable inversion of the weight matrix $\mathcal{W}$ while also capturing the dominant correlations, as evidenced by the fact that the expectation value of the chi-squared lies close to the number of degrees of freedom, as required for a correlated fit~\cite{Bruno:2022mfy}.
Furthermore, as discussed in Sec.~\ref{subsec:t0}, we have also considered more general forms of the weight matrix $\mathcal{W}$, in which contributions modelling systematic effects are incorporated alongside the correlation matrix.

\subsubsection{Model averaging procedures}

Minimising the chi-squared function provides both an estimate of the expectation value $\langle \chi^2 \rangle$ and the goodness-of-fit, expressed through the p-value, for a given model.  
From the same fit, one can compute information criteria based on the Akaike Information Criterion (AIC)~\cite{akaike1976canonical} or the Takeuchi Information Criterion (TIC)~\cite{Takeuchi76}.
In this context, model variation seeks to assess the distance of a chosen, physically motivated model, from the true underlying distribution that generated the data.
Various procedures for performing model averaging have been proposed and applied in the literature, that consider a number of information-criterion approaches to Bayesian and frequentist weighting schemes~\cite{Kullback:1951zyt,akaike1976canonical,Akaike:1998zah,Takeuchi76,Watanabe2007,BMW:2013fzj,Borsanyi:2020mff,Jay:2020jkz,Neil:2022joj,Neil:2023pgt,Frison:2023lwb}.
A concrete implementation of the model averaging (MA) procedure consists in assigning a probability weight $W^{(i)}$ to each model $M_i$, based on the model selection criterion.
A weighted average of an observable $\mathcal{O}$ over the considered model variations,
\begin{align}
  \label{eq:OWMA}
  \langle \mathcal{O} \rangle_{\mathrm{MA}} &= \sum_i \mathcal{O}^{(i)} W^{(i)}\,,
\end{align}
provides the central value and statistical uncertainty of $\mathcal{O}$. The associated systematic uncertainty is then estimated from the weighted variance across the model ensemble
\begin{align}
  \label{eq:WMsyst}
  \sigma^2_{\mathrm{syst}}&=\left< \mathcal{O}^2\right>_{\mathrm{MA}} - \left<\mathcal{O}\right>^2_{\mathrm{MA}}\,.
\end{align}
As reported in Appendix~\ref{app_variations}, several model selection criteria are employed in our analysis to estimate the magnitude of systematic effects.
The approaches considered include the application of information criteria, as well as the evaluation of the maximal deviation observed across models with statistically acceptable p-values.
When considering the chi-squared function defined in \req{eqn:chisq_full}, which depends on the weight matrix $\mathcal{W}$, it is convenient to adopt a formulation of the Takeuchi Information Criterion (TIC)~\cite{Frison:2023lwb} expressed in terms of the expectation value of the chi-squared
\begin{equation}
  \label{eqn:TICchiexp}
  \mathrm{TIC}_{\mathcal{W}} = \chi_{\mathcal{W}}^2 - 2\langle \chi_{\mathcal{W}}^2 \rangle\,.
\end{equation}
The corresponding probability weight of the $i$-th model can be written as follows
\begin{equation}
  \label{eqn:weight}
  W^{(i)} = \mathcal{N} \exp\left(-\frac{1}{2} \mathrm{TIC}^{(i)}_{\mathcal{W}} \right)\,,
\end{equation}
where the normalisation $\mathcal{N}$ is fixed so that $\sum_i W^{(i)} =1$.
In the case of correlated fits, where $\mathcal{W} = C^{-1}$, the expected value of the chi-squared coincides with the number of degrees of freedom, i.e., $\langle \chi^2 \rangle = n_{\mathrm{dat}} - n_{\mathrm{param}}$, where $\chi^2$ denotes the correlated chi-squared.
For a fixed value of $n_{\mathrm{dat}}$ -- corresponding to the absence of data cuts and thus to a term contributing only an additive constant to the information criterion -- the expression in \req{eqn:TICchiexp} reduces to the Akaike Information Criterion, $\mathrm{AIC}= \chi^2 + 2 n_{\mathrm{param}}$.

\subsubsection{Model variation with data cuts}

The truncation of the data set enables the selection of regimes of variables wherein the model is anticipated to perform better.
The implementation of data cuts therefore serves as a diagnostic tool to investigate potential systematic effects.
We consider a fixed number of data points, denoted by $n_{\mathrm{dat}}$, such that the application of $n_{\mathrm{cut}}$ cuts results in the retention of $n_{\mathrm{dat}} - n_{\mathrm{cut}}$ data points in the model under consideration.

Formally, this corresponds to the selection of a model in which the data excluded by the cuts is described by a trivial model, parameterised by auxiliary parameters $\boldsymbol{p}_{\mathrm{aux}}$ that do not contribute to the minimum of the $\chi^2$
\begin{equation}
  f_M(x_i,\boldsymbol{p},\boldsymbol{p}_{\mathrm{aux}}) = p_{{\mathrm{aux}},i}\,,\quad \mathrm{if}\;i\in I_{\mathrm{cut}}\,,
\end{equation}
where $I_{\mathrm{cut}}$ denotes the set of indices, of dimension $n_{\mathrm{cut}}$, corresponding to the data points that have been excluded by the cuts.
When computing information criteria, these auxiliary parameters are treated in the same manner as all other model parameters.
It is worth noting that the normalisation factor $\det(C)^{-1/2}$ of the likelihood function remains unchanged between models with and without data cuts, and can therefore be safely ignored.
The computation of the effective number of degrees of freedom within the context of the TIC~\cite{Frison:2023lwb} continues to be given by \req{eqn:weight}, included the limiting cases of correlated and uncorrelated fits.
In the case of correlated fits, where data cuts are applied such that $n_{\mathrm{dat}} - n_{\mathrm{cut}}$ data points are retained, the resulting information criterion derived from \req{eqn:TICchiexp} can be expressed, up to an additive constant, as follows
\begin{equation}
  \label{eqn:TICC}
  \mathrm{TIC}_{C^{-1}} = \chi^2 + 2n_{\mathrm{param}} + 2n_{\mathrm{cut}}\,.
\end{equation}
This formulation of the information criterion coincides with that advocated in Ref.~\cite{Jay:2020jkz}; see also Refs.~\cite{Neil:2022joj,Neil:2023pgt} for comparisons with alternative criteria.
In practice, we exploit the fact that the probability weight in \req{eqn:weight} may also be assigned to models incorporating data cuts, in order to probe systematic effects associated with variations in lattice spacing, pion mass, and volume within the combined continuum and chiral extrapolations employed in the determination of $t_0^{\mathrm{ph}}$.
Tables~\ref{apex_ma:tab:w1}--\ref{apex_ma:tab:comb4} and \ref{apex_ma:tab:comb_fpi1}--\ref{apex_ma:tab:comb_fpi2} report the p-values and weights $W$ associated with these fits.
Analogously, cuts are applied to the Euclidean time intervals used in the extraction of the gradient flow scale $t_0/a^2$, meson masses and decay constants, as well as PCAC quark masses.

\subsubsection{Application of model variation to ground-state isolation}
\label{app:groundstate}

\begin{figure}[htbp!]
  \centering
  \includegraphics[width=1.\linewidth]{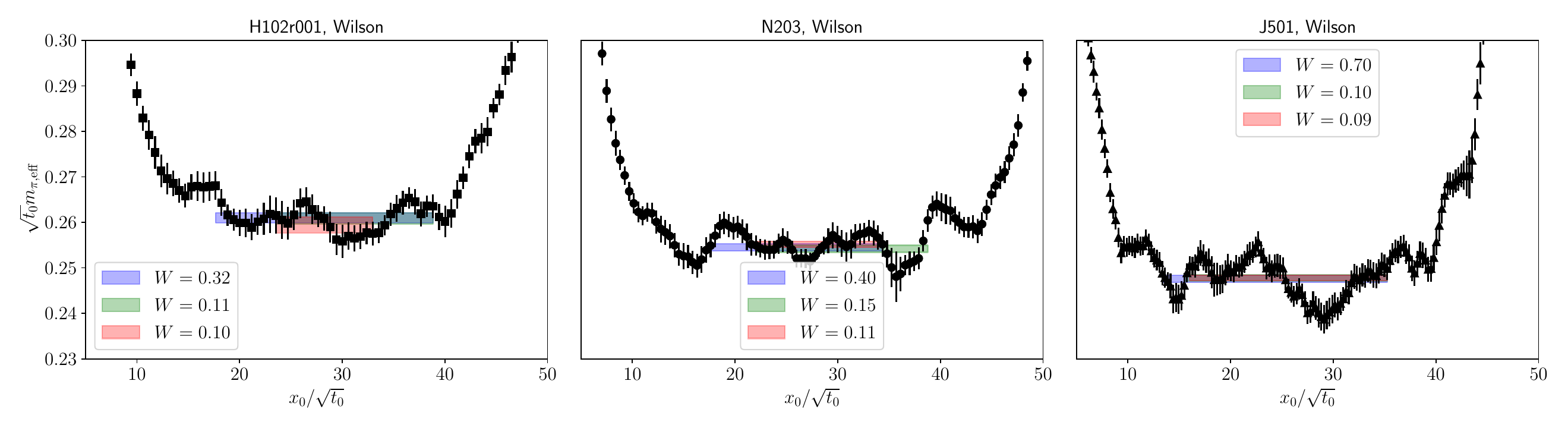}
  \includegraphics[width=1.\linewidth]{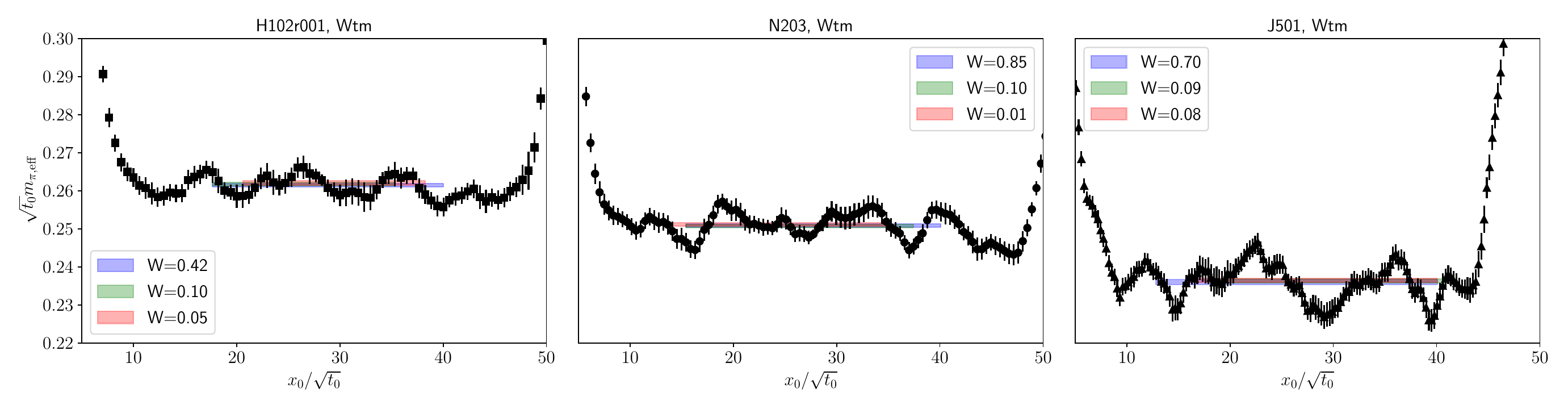}
  \caption{
    Euclidean time dependence of the pion effective mass in units of $\sqrt{t_0}$ for the unitary Wilson ({\it top}) and mixed-action Wtm ({\it bottom}) regularisations.
    On the vertical axis, $m_{\pi,\mathrm{eff}}$ denotes the light-sector effective pseudoscalar mass $m_{\mathrm{\scriptscriptstyle PS, eff}}^{12}(x_0)$, as defined in \req{eqn:meff}.
    From {\it left} to {\it right}, the panels correspond to the ensembles H102 ($a \approx 0.085\,\mathrm{fm}$), N203 ($a \approx 0.065\,\mathrm{fm}$), and J501 ($a \approx 0.039\,\mathrm{fm}$).
    All ensembles employ open boundary conditions in the temporal direction.
    The pseudoscalar correlation function used to compute the effective mass has a source located at $y_0/a = 1$, while the sink position $x_0$ is varied.
    The sea pion masses of these ensembles are approximately equal (cf.~Table~\ref{tab:CLS_ens}).
    The valence Wtm pion mass corresponds to a choice of mass parameters within the grid of simulated points (cf.~Sec.~\ref{sec:match}), prior to the matching procedure between sea and valence quark masses.
    For improved visibility, only the three models with the largest normalised weights $W$, computed according to \req{eqn:weight}, are shown as horizontal bands spanning the Euclidean time intervals used in the fit.
    All displayed models have a p-value greater than $0.1$.
    The occurrence of long-range fluctuations in $x_0$ within the bulk region signals strong correlations, originating from the finite statistical precision of the sampling~\cite{Luscher:2012av,Bruno:2014ova}.
    Given that both regularisations share the same sea sector, a common pattern in the fluctuations is observed for the Wilson and Wtm regularisations.
    The integrated autocorrelation times of these quantities are shown in Fig.~\ref{fig:tau_mpi}.
  }
  \label{fig:mpi_effective}
\end{figure}

We now illustrate the application of model variation to the extraction of the ground-state contribution from an effective observable $\mathcal{O}^{\mathrm{eff}}(x_0)$ depending on the Euclidean time $x_0$ -- for example, the effective mass of a light pseudoscalar meson extracted from \req{eqn:meff} -- on ensembles with open boundary conditions in the temporal direction.
We consider the following fit ansatz
\begin{equation}
  \label{eqn:f}
  f(x_0) = p_1 + p_2 \exp(-p_3 x_0) + p_4 \exp(-p_5 (T - x_0))\,,
\end{equation}
where $p_i$ are fit parameters, and $p_1$ represents the desired ground-state contribution.
The remaining parameters account for deviations from plateau behaviour due to excited-state contamination and boundary effects.
Data cuts are implemented by selecting different Euclidean time fit intervals.
When broad fit intervals are considered, the strong correlations in $x_0$ present in the data may lead to ill-conditioned covariance matrices.
To avoid this issue, uncorrelated fits are performed by setting $\mathcal{W}= \mathrm{diag}(C)^{-1}$ in \req{eqn:chisq}, and the goodness-of-fit is assessed using the p-value, as described in Ref.~\cite{Bruno:2022mfy}.
For each fit interval, we compute the value of $\chi_{\mathcal{W}}^2$ from \req{eqn:chisq}, along with the corresponding $\langle \chi_{\mathcal{W}}^2 \rangle$ and fit parameters, in particular $p_1$, which is the quantity of interest.
This enables the computation of the associated model weight $W$ according to \req{eqn:weight}.
The model averaging of the observable $\mathcal{O} = p_1$, as defined \req{eq:OWMA}, yields the central value and statistical uncertainty, while the systematic uncertainty $\sigma_{\mathrm{syst}}$ is estimated using \req{eq:WMsyst}.
The application of this method in the  extraction the ground-state of the pion effective mass for ensembles H102, N203 and J501 is illustrated in Fig.~\ref{fig:mpi_effective}, where only a representative subset of fit ranges is shown as an illustration of the method.
The extraction of the ground-state contribution can be challenging due to the presence of boundary effects, excited-state contamination, and strong correlations in $x_0$.
A robust definition of the goodness-of-fit is essential for selecting fits with sufficiently large p-values.
To monitor the stability of the ground-state extraction based on model averaging, we also consider a two-stage procedure~\cite{Bruno:2014ova}.
The first step involves fitting $\mathcal{O}^{\mathrm{eff}}(x_0)$ using the ansatz $f(x_0)$ defined in \req{eqn:f}, selecting the largest time interval for which the p-value exceeds $0.1$.
The parameters obtained from this fit are then used to define a fit range $\left[ x_{0,\mathrm{min}}, x_{0,\mathrm{max}} \right]$, within which the statistical uncertainty of the effective observable exceeds the systematic deviations from a plateau behaviour.
More specifically, $x_{0,\mathrm{min}}$ and $x_{0,\mathrm{max}}$ are determined by the following conditions
\begin{align}
  \label{eqn:plateau}
  \sigma\!\bigl(\mathcal{O}^{\mathrm{eff}}(x_{0,\mathrm{min}})\bigr) &= K p_2 \exp(-p_3 x_{0,\mathrm{min}}), \\
  \sigma\!\bigl(\mathcal{O}^{\mathrm{eff}}(x_{0,\mathrm{max}})\bigr) &= K p_4 \exp(-p_5 (T - x_{0,\mathrm{max}})),
\end{align}
where $\sigma\!\bigl(\mathcal{O}^{\mathrm{eff}}(x_0)\bigr)$ denotes the statistical uncertainty of the effective observable, and the parameter $K$ is set to $K = 3$ in practice.
The second step consists in determining the ground-state contribution by performing a constant fit to $\mathcal{O}^{\mathrm{eff}}(x_0)$ within the interval $\left[ x_{0,\mathrm{min}}, x_{0,\mathrm{max}} \right]$.
The application of this two-stage procedure to the determination of the effective mass for the J501 ensemble -- in the Wilson regularisation before the mass shift -- gives
\begin{equation}
  am_{\pi}=0.06605(16)\,.
\end{equation}
The following consistent determination is obtained from the model-averaging procedure
\begin{equation}
  am_{\pi}=0.06613(15)\,.
\end{equation}

\subsubsection{Decomposing submodel systematics}
\label{sec:decompsys}

Let us assume that the set of models $\mathcal{M}$ can be represented as a product of sets of submodels $\mathcal{M}_i$
\begin{equation}
  \label{eqn:MA-submodels}
  \mathcal{M} = \mathcal{M}_1 \otimes \mathcal{M}_2 \otimes \dots\,
\end{equation}
For instance, in the analysis of the combined continuum-chiral extrapolation $\mathcal{M}_1$ may contain the parametrisations of the lattice-spacing dependence in Eqs.~(\ref{ch_ss:eq:a2})--(\ref{ch_ss:eq:a2phi2}), while $\mathcal{M}_2$ may correspond to the continuum light-quark mass dependence given by Eqs.~(\ref{ch_ss:eq:SU3ChPT})~and~(\ref{ch_ss:eq:SU2pik}) for the $\chi$PT expressions, or by Eqs.(\ref{ch_ss:eq:Tay2})--(\ref{ch_ss:eq:Tay4}) for Taylor fits.
The model weights in \req{eqn:weight} can thus be labelled as $W_{M_1M_2\cdots}$, where $M_i\in\mathcal{M}_i$ is a specific choice of submodel.
Similarly to Eqs.~(\ref{eq:OWMA})~and~(\ref{eq:WMsyst}), the following partial weighted model average, conditioned on fixing submodel $M_i$, can be considered
\begin{align}
  \langle \mathcal{O}\rangle_{\mathrm{MA}\mid M_i} &= \sum_{M_{j\not=i}} \mathcal{O}^{M_1M_2\cdots} \frac{W_{M_1M_2\cdots}}{\overline{W}_{M_i}}\,,
\end{align}
where the marginal weight is given by
\begin{align}
  \overline{W}_{M_i} &= \sum_{M_{j\not=i}} W_{M_1M_2\cdots}\,.
\end{align}
The partial variance $\sigma_{\mathrm{syst},i}^{2}$ -- corresponding to the contribution to $\sigma_{\mathrm{syst}}^{2}$ originating from the model dependence along the component $\mathcal{M}_{i}$ --  is given by
\begin{align}
  \sigma_{\mathrm{syst},i}^2 &= \sum_{M_i} \left( \langle \mathcal{O}\rangle_{\mathrm{MA}\mid M_i}^2   \overline{W}_{M_i}\right) - \left(\sum_{M_i}  \langle \mathcal{O}\rangle_{\mathrm{MA}\mid M_i}\overline{W}_{M_i}\right)^2 \,.
  \label{eq:varsysti}
\end{align}
Using associativity in \req{eqn:MA-submodels}, one can extend the definition of the partial variance to the joint variation of along $\mathcal{M}_i$ and $\mathcal{M}_j$ as follows
\begin{equation}
  \sigma_{\mathrm{syst},ij}^2 = \sum_{M_i,M_j} \left( \langle \mathcal{O}\rangle_{\mathrm{MA}\mid M_iM_j}^2 \overline{W}_{M_iM_j}\right) - \left(\sum_{M_i,M_j}  \langle \mathcal{O}\rangle_{\mathrm{MA}\mid M_iM_j}\overline{W}_{M_iM_j}\right)^2\,,
\end{equation}
where the marginal weights read
\begin{equation}
  \overline{W}_{M_iM_j} = \sum_{M_{k\neq i,j}} W_{M_1M_2\cdots}\,.
\end{equation}
If the components are sufficiently independent and the sets of models are large, the covariance between the model-dependence arising from $\mathcal{M}_i$ and $\mathcal{M}_j$
\begin{equation}
  \Gamma_{ij} = \frac{1}{2} \left( \sigma_{\mathrm{syst},ij}^2  - \sigma_{\mathrm{syst},i}^2 - \sigma_{\mathrm{syst},j}^2 \right)\,,
\end{equation}
is expected to satisfy $\Gamma_{ij}\ll \sigma_{\mathrm{syst},ij}^2$, so that the partial systematics $\sigma_{\mathrm{syst},i}^2$ approximately sum up to $\sigma^2_{\mathrm{syst}}$.

\subsubsection{Final remarks on model variation and model average}

We close this Appendix by emphasising that the analysis of systematic effects is a challenging task that requires meticulous and thorough attention, particularly in cases involving extrapolations.
Model variation is essential for probing the model space, but the interpretation of model-averaged predictions demands direct scrutiny.
The reliability of the model average is closely linked to the availability of large, statistically robust datasets.
Moreover, the set of candidate models should be sufficiently diverse and mutually independent, so that the model average is not dominated by minor variations within a narrow region of model space.
Information criteria that incorporate data cuts tend to disfavour smaller datasets, even when their p-values indicate a good fit.
However, these criteria do not, by default, encode the physical basis behind such cuts.
Data exclusion is often employed to suppress regions where models are expected to be affected by larger systematic uncertainties.
While model-selection criteria may penalise data reduction exponentially, they do not directly account for the fact that imposing a data cut introduces physical information based on the known regime of validity of the models.
To address the challenges associated with estimating systematic effects through model variation, several refinements have been incorporated into the analysis.
In particular, a modification of the weight in the $\chi^2$ function is introduced in Sec.~\ref{subsec:t0} (see also Appendix~\ref{app_variations}) to encode physical information about the regimes where Symanzik EFT and $\chi$PT are expected to be most reliable in the combined continuum–chiral extrapolation.
This is complemented by a robust definition of the goodness-of-fit for generalised $\chi^2$ weights~\cite{Bruno:2022mfy}, and by the exploration of various model-selection criteria (see Appendix~\ref{app_variations}) to assess the sensitivity of the results to the choice of selection method.
%


\section{Combined continuum and chiral fits of mass derivatives}
\label{app:derivatives}

\begin{figure}[htbp!]
  \centering
  \includegraphics[width=.7\textwidth]{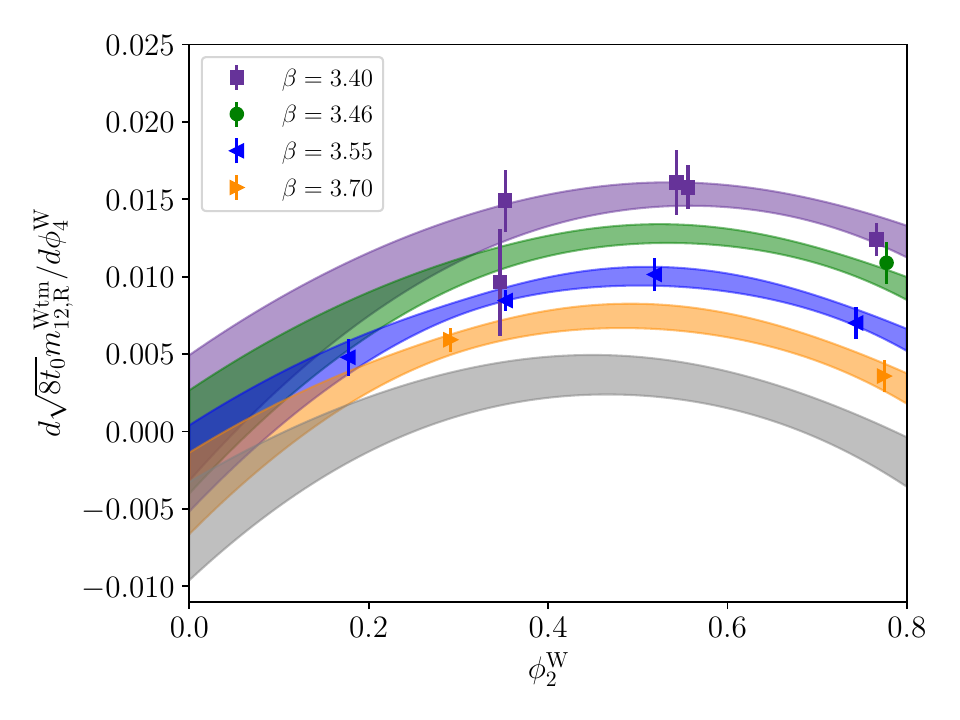}
  \caption{
    Mass derivative $d\left(\sqrt{8t_0} \, m_{12, \mathrm{R}}^{\mathrm{Wtm}}\right)/d\phi_4^{\mathrm{W}}$ defined in \req{eqn:dOdphi4} for the mixed-action Wtm setup as a function of $\phi_2^{\mathrm{W}}$.
    The bands represent the outcome of the combined continuum and chiral fit in \req{eqn:md_2}, projected either onto each value of $\beta$ or onto the continuum (lower grey band).
    The corresponding values of the fit parameters are collected in \req{eqn:dm12_wtm}. 
    The p-value of the fit is approximately $0.4$.
  }
  \label{fig:dm12_wtm}
\end{figure}

In Sec.~\ref{sec:match}, we introduced the mass-shifting procedure required to impose the condition $\phi_4 = \phi_4^{\mathrm{ph}}$ across all ensembles.
The procedure involves computing the derivative of an observable $\mathcal{O}$ with respect to $\phi_4^{\mathrm{W}}$, as defined in \req{eqn:dOdphi4}, followed by a combined continuum and chiral extrapolation.
The fit parameters obtained from the fit ansatz in \req{eqn:md_1} are collected in Table~\ref{tab:md}. 
\sisetup{
  table-number-alignment = center,
  table-figures-integer = 1, 
  table-figures-decimal = 3, 
  table-figures-uncertainty = 2, 
  separate-uncertainty = true,
  scientific-notation = fixed,   
  exponent-to-prefix = false,    
  table-align-uncertainty = true
}
\begin{table}[htbp!]
  \renewcommand{\arraystretch}{1.3}
  \begin{center}
    \begin{tabular}{c | @{\hspace{2em}} S[table-format=1.3(2)] @{\hspace{2em}} S[table-format=1.3(2)] @{\hspace{2em}} S[table-format=1.3(2)]}
      \toprule
      $\mathcal{O}$ & $A$ & $B$ & $C$ \\
      \toprule
      $\sqrt{8t_0}f_{\pi K}^{\mathrm{W}}$ & 0.006(8) & 0.002(12) & 0.046(40) \\ 
      $\sqrt{8t_0}f_{\pi}^{\mathrm{W}}$ & 0.002(8) & 0.013(13) & 0.027(41) \\ 
      $\sqrt{8t_0}f_{K}^{\mathrm{W}}$ & 0.007(9) & -0.003(14) & 0.062(42) \\ 
      $\phi_2^{\mathrm{W}}$ & 0.004(33) & 0.114(96) & 0.920(154) \\ 
      \midrule
      $\sqrt{8t_0}f_{\pi K}^{\mathrm{Wtm}}$ & -0.005(6) & 0.009(8) & -0.012(24) \\ 
      $\sqrt{8t_0}f_{\pi}^{\mathrm{Wtm}}$ & -0.007(7) & 0.017(8) & -0.030(28) \\ 
      $\sqrt{8t_0}f_{K}^{\mathrm{Wtm}}$ & -0.003(7) & 0.003(8) & 0.005(24) \\ 
      $\phi_2^{\mathrm{Wtm}}$ & 0.011(16) & -0.070(26) & 0.057(79) \\ 
      $\phi_4^{\mathrm{Wtm}}$ & -0.024(36) & -0.003(49) & -0.065(147) \\ 
      \bottomrule
    \end{tabular}
  \end{center}
  \renewcommand{\arraystretch}{1.0}
  \caption{
    Determination of the fit parameters in \req{eqn:md_1} for the mass derivatives $d\mathcal{O}/d\phi_4^{\mathrm{W}}$, defined in \req{eqn:dOdphi4}, of the observables $\mathcal{O}$ listed in the first column.
    The superscript ``$\mathrm{W}$'' refers to observables computed in the unitary Wilson setup, while ``$\mathrm{Wtm}$'' denotes those obtained using the Wtm mixed-action regularisation.
  }
  \label{tab:md}
\end{table}
For the particular case of $\mathcal{O} = \sqrt{8t_0} \, m_{12}^{\mathrm{Wtm}, \mathrm{R}}$, a fit to \req{eqn:md_2} is considered.
The corresponding values of the fit parameters read
\begin{align}
  \label{eqn:dm12_wtm}
  A=-0.0006(3), \quad B=0.044(12), \quad C=-0.049(11), \quad D=0.016(14), \quad E=0.031(22)\,.
\end{align}
The associated fit is illustrated in Fig.~\ref{fig:dm12_wtm}.
%


\section{Finite size effects}
\label{app:FVE}

Simulations of QCD in a finite spatial volume inherently introduce finite-size effects (FSE), which constitute a source of systematic uncertainty.
For observables that receive only exponentially suppressed finite-size corrections controlled by $m_{\pi}L$, it is customary to adopt the condition $m_{\pi}L \gtrsim 4$, while simultaneously employing spatial lattice extents satisfying $L \gtrsim 2\,\mathrm{fm}$.
The lattice sizes of the ensembles considered in this work are larger than $L > 2.3\,\mathrm{fm}$, with $m_{\pi}L > 3.9$, as reported in Table~\ref{tab:CLS_ens}.
In particular, in the physical-point ensemble E250, the lattice spatial extent is $L \approx 6\,\mathrm{fm}$, with $m_\pi L \approx 4.0$.
These constraints may be complemented by explicit tests of residual finite-size effects, either through simulations at multiple lattice volumes or by applying corrections derived from effective field theory.

Among the observables considered in this work, the pseudoscalar meson masses and decay constants are the most sensitive to FSE.
Chiral perturbation theory ($\chi$PT) provides a systematic framework for studying these effects.
For convenience, we reproduce below the $\chi$PT expressions for the FSE corrections employed in this work, and refer to the original derivations in Refs.~\cite{Gasser:1986vb, Colangelo:2003hf, Colangelo:2005gd} for further details.
We consider the observable $\mathcal{O} \in \{m_{\pi}, m_K, f_{\pi}, f_K\}$, and denote by $\mathcal{O}^{(\infty)}$ its value in the infinite-volume limit and by $\mathcal{O}^{(L)}$ its value at finite spatial volume $L^3$.
The finite-size correction $R_\mathcal{O}$ is then defined via
\begin{equation}
  \label{eq:ROFSE}
  \mathcal{O}^{(\infty)} = \mathcal{O}^{(L)} \frac{1}{1 + R_\mathcal{O}}\,.
\end{equation}
The NLO $\chi$PT expressions for $R_\mathcal{O}$ read
\begin{align}
  \label{eq:Rmpi}
  R_{m_{\pi}}&=\frac{1}{4}\xi_{\pi}\tilde{g}_1(\lambda_{\pi})-\frac{1}{12}\xi_{\eta}\tilde{g}_1(\lambda_{\eta})\,, \\
  \label{eq:RmK}
  R_{m_K}&=\frac{1}{6}\xi_{\eta}\tilde{g}_1(\lambda_{\eta})\,, \\
  \label{eq:RfK}
  R_{f_K}&=-\xi_{\pi}\tilde{g}_1(\lambda_{\pi})-\frac{1}{2}\xi_{K}\tilde{g}_1(\lambda_{K})\,, \\
  \label{eq:Rfpi}
  R_{f_{\pi}}&=-\frac{3}{8}\xi_{\pi}\tilde{g}_1(\lambda_{\pi})-\frac{3}{4}\xi_{K}\tilde{g}_1(\lambda_{K})-\frac{3}{8}\xi_{\eta}\tilde{g}_1(\lambda_{\eta})\,.
\end{align}
These expressions depend on the following definitions
\begin{align}
  \label{eq:xips}
  \xi_{\mathrm{\scriptscriptstyle PS}}&=\frac{m_{\mathrm{\scriptscriptstyle PS}}^2}{(4\pi f_{\pi})^2}\,, \\
  \label{eq:lambdaps}
  \lambda_{\mathrm{\scriptscriptstyle PS}}&=m_{\mathrm{\scriptscriptstyle PS}}L\,, \\
  \tilde{g}_1(x)&=\sum_{n=1}^{\infty}\frac{4m(n)}{\sqrt{n}x}K_1\left(\sqrt{n}x\right)\,,
\end{align}
where the pseudoscalar meson masses $\mathrm{PS} = \pi, K, \eta$ can be related via the leading order expression $m_{\eta}^2 = \frac{4}{3} m_K^2 - \frac{1}{3} m_{\pi}^2$, and $K_1(x)$ denotes the modified Bessel function of the second kind.
For convenience, the multiplicities $m(n)$~\cite{Colangelo:2003hf} are listed in Table~\ref{apex_fv:tab:mn}.
\begin{table}[htbp!]
  \begin{center}
    \begin{tabular}{c c c c c c c c c c c}
      \toprule
      $n$ & 1 & 2 & 3 & 4 & 5 & 6 & 7 & 8 & 9 & 10 \\
      $m(n)$ & 6 & 12 & 8 & 6 & 24 & 24 & 0 & 12 & 30 & 24 \\
      \midrule
      $n$ & 11 & 12 & 13 & 14 & 15 & 16 & 17 & 18 & 19 & 20 \\
      $m(n)$ & 24 & 8 & 24 & 48 & 0 & 6 & 48 & 36 & 24 & 24 \\
      \bottomrule
    \end{tabular}
  \end{center}
  \caption{
    Multiplicities $m(n)$ calculated in~\cite{Colangelo:2003hf} for $n\leq20$.
  }
  \label{apex_fv:tab:mn}
\end{table}

\begin{figure}[ht!]
  \centering
  \includegraphics[width=.49\textwidth]{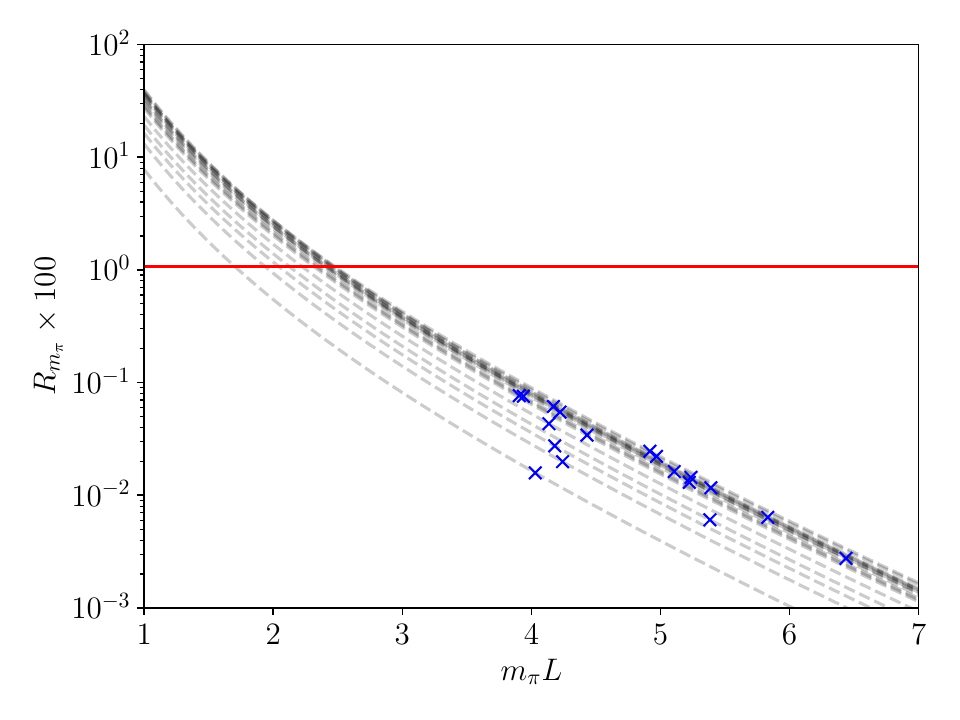}
  \includegraphics[width=.49\textwidth]{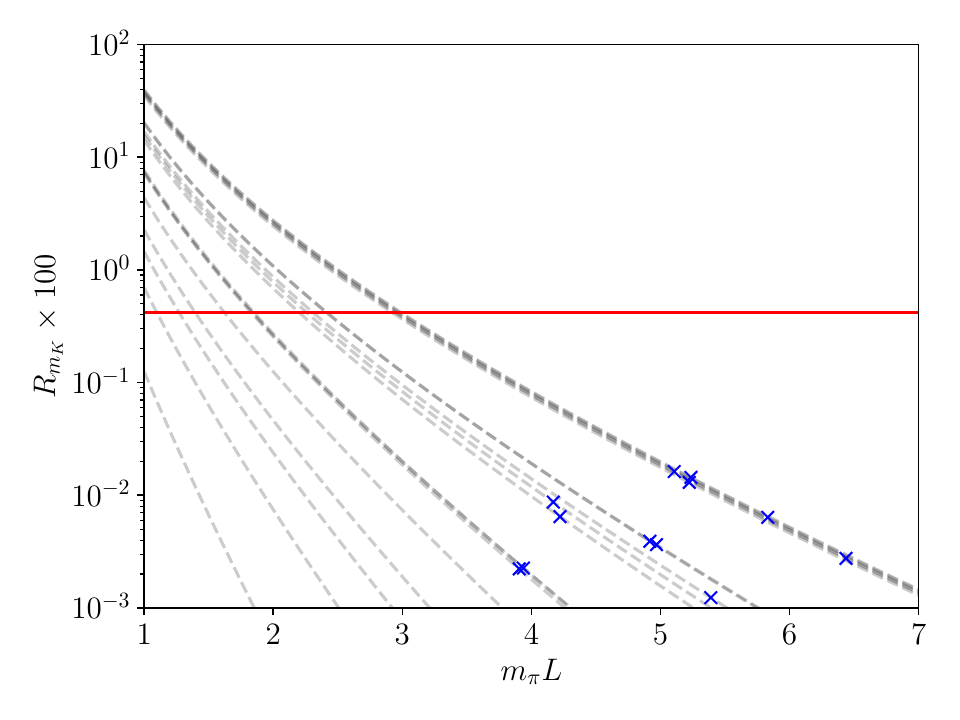}
  \includegraphics[width=.49\textwidth]{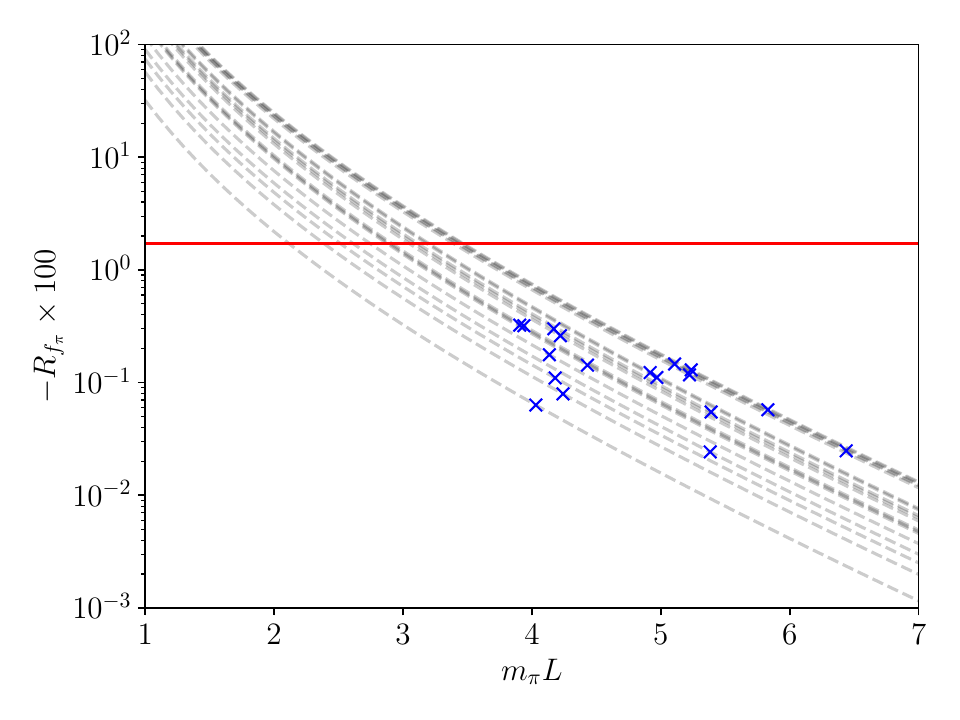}
  \includegraphics[width=.49\textwidth]{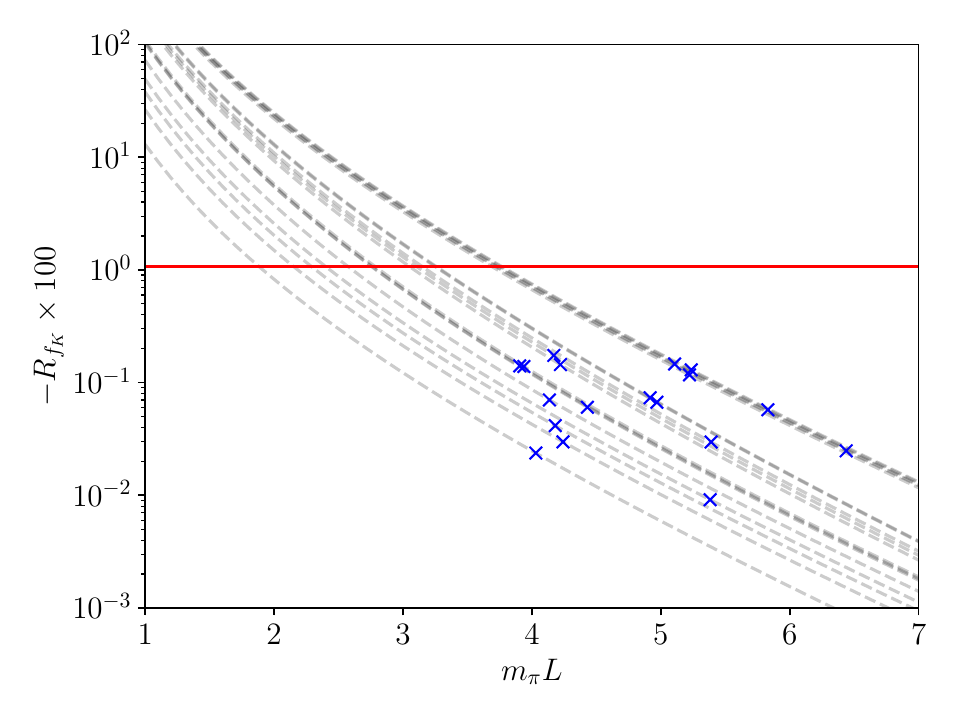}
  \caption{
    Finite-size correction $R_{\mathcal{O}}$, introduced in \req{eq:ROFSE}, for $\mathcal{O} \in \{m_{\pi}, m_{K}, f_{\pi}, f_{K}\}$ as given by the NLO $\chi$PT expressions in Eqs.~(\ref{eq:Rmpi})--(\ref{eq:Rfpi}), shown as a function of $m_{\pi}L$ (grey dashed lines).
    The blue crosses, located at the values of $m_{\pi}L$ corresponding to the gauge-field ensembles considered in this work, represent $R_{\mathcal{O}}$ computed using the pion and kaon masses and the pion decay constant from the Wilson regularisation in Eqs.~(\ref{eq:xips}) and (\ref{eq:lambdaps}).
    Although the finite-size corrections are observed to be at most at the few per-mille level, and below the statistical precision of the lattice data (the red horizontal line illustrates a representative example of the relative statistical precision), they are still included in the analysis.
  }
  \label{fig:fse}
\end{figure}

For the ensembles with the smallest lattice volumes and lightest pion masses -- i.e. those most affected by FSE -- we observe that the magnitude of the corrections remains well below half a standard deviation, as illustrated in Fig. \ref{fig:fse}.
Nevertheless, these corrections are applied across all ensembles.
In the case of the gradient flow scale $t_0$, FSE from $\chi$PT arise only at NNLO and are therefore expected to be very small~\cite{Bar:2013ora}.
Furthermore, since the PCAC quark mass is a short-distance quantity, it exhibits reduced sensitivity to FSE.
%


\section{Independent continuum and chiral fits of $\sqrt{8t_{0}}f_{\pi K}$}
\label{app:chiral_w_wtm}

\begin{figure}[htbp!]
  \centering
  \includegraphics[width=0.8\textwidth]{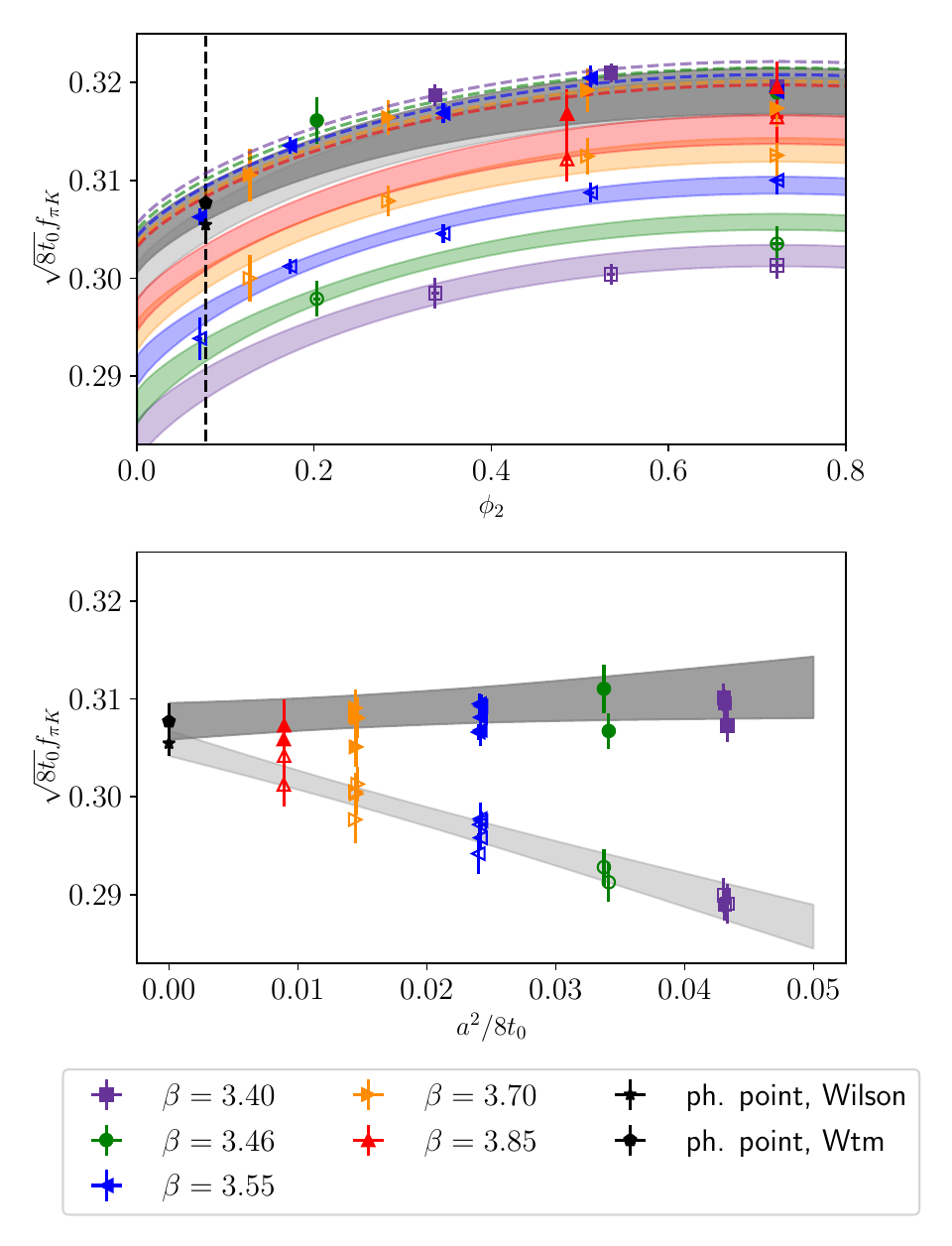}
  \caption{
    Light-quark mass-dependence of $\sqrt{8t_0}f_{\pi K}$ for the  $\mathrm{SU}(3)$ $\chi$PT model, incorporating pure $\Oasq$ cutoff effects and excluding data cuts, corresponding to the label $[\chi\mathrm{SU}(3)][a^2][-]$, associated to Eqs.~(\ref{ch_ss:eq:SU3ChPT}) and (\ref{ch_ss:eq:a2}).
    Independent fits are performed for the Wilson (empty symbols) and Wtm mixed-action (filled symbols) datasets.
    The modified weight in the $\chi^2$ function, introduced in \req{ch_ss:eq:Wpenal}, is employed.
    Coloured bands represent the pion-mass dependence at each lattice spacing for the Wilson data.
    For improved visual clarity, dashed lines indicate the central values of the mass-dependence in the Wtm mixed-action case.
    The light-grey band denotes the continuum-limit mass-dependence derived from the Wilson data, while a darker-grey band corresponds to the mixed-action results.
    The location of the physical mass $\phi_2^{\mathrm{ph}}$ is marked by a vertical dashed line.
    \textit{Bottom}:
    The lattice-spacing dependence of $\sqrt{8t_0}f_{\pi K}$ is shown for the same model.
    Data points are projected to the physical pion mass $\phi_2^{\mathrm{ph}}$ using the corresponding fit results for the continuum mass-dependence $\Phi_{\pi K}^{\mathrm{cont}}(\phi_2)$.
    The values of $\sqrt{8t_0}f_{\pi K}$ at the physical point are indicated by black filled symbols, which agree within one sigma.
    The p-value of the fit to the Wilson dataset is $0.83$, while that for the Wtm mixed-action data is $0.79$.
  }
  \label{fig:continuum-chiralindep}
\end{figure}

In this appendix, we provide a concise illustration of the results from the combined continuum and chiral extrapolations of $\sqrt{8t_0}f_{\pi K}$, obtained through independent fits to the Wilson unitary data and the Wtm mixed-action regularisation. 
This approach enables a universality test, as a common continuum-limit result is not enforced for the two regularisations.
We focus on the model $[\chi \mathrm{SU(3)}][a^2]$ (cf. Appendix~\ref{apex_model_av_t0} for a complete specification), with no cuts applied to the dataset.
The results are displayed in Fig.~\ref{fig:continuum-chiralindep}.
We observe that the continuum-limit results obtained from the Wilson and Wtm mixed-action analyses agree within one standard deviation.
%


\section{Variation in chi-squared weights and model-selection criteria}
\label{app_variations}

\begin{figure}[htbt!]
  \centering
  \includegraphics[width=1.\textwidth]{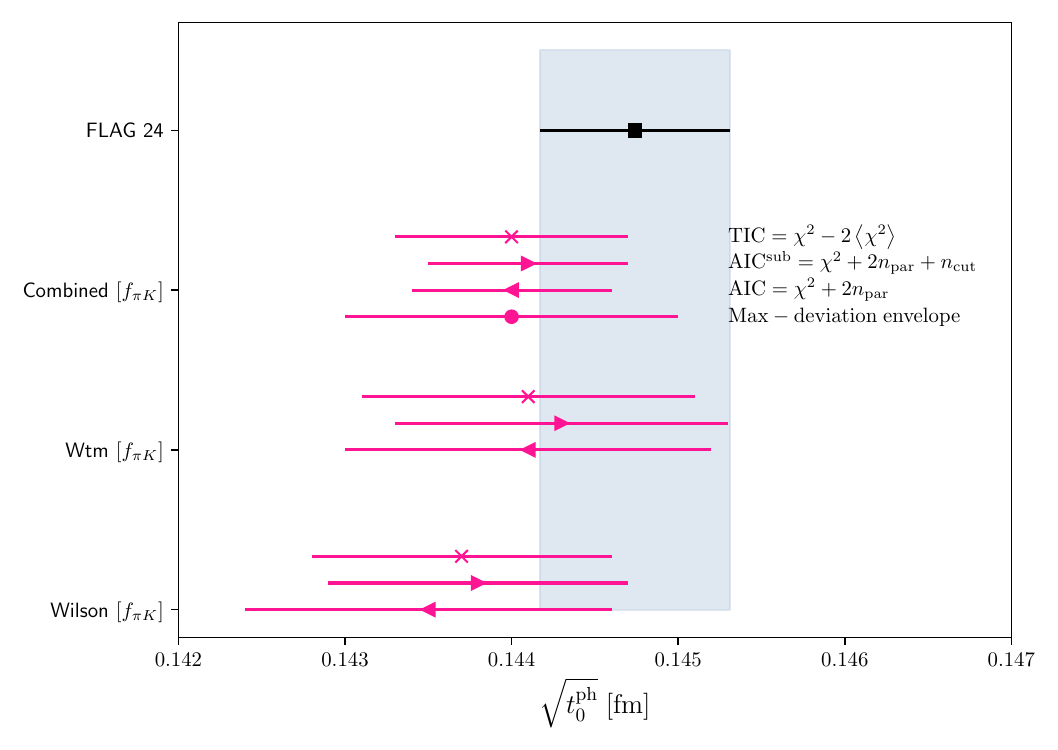}
  \caption{
    Comparison of determinations of $\sqrt{t_0^{\mathrm{ph}}}$ obtained using different model-selection criteria -- the TIC, see \req{eqn:TICchiexp}, the Akaike Information Criterion (AIC), see \req{eq:AIC}, and the modified criterion $\mathrm{AIC}^{\mathrm{sub}}$, see \req{eq:AIC_sub} -- for the three analysis variants: the unitary Wilson regularisation, the Wtm mixed-action setup, and the combination of both.
    The $\NF=2+1$ FLAG~24 average result~\cite{FlavourLatticeAveragingGroupFLAG:2024oxs} is denoted by the black symbol and the vertical band.
  }
  \label{fig:variationsIC}
\end{figure}

In this Appendix, we summarise the exploration of the stability of the determination of $\sqrt{t_0^{\mathrm{ph}}}$ under variations of the weight matrix used in the definition of the chi-squared function, as introduced in Sec.~\ref{subsec:t0}.
We also examine the impact of different model-selection criteria employed in the model variation procedure described in Appendix~\ref{app:MA}.
The weight matrix $\mathcal{W}$, defined in \req{ch_ss:eq:Wpenal}, enters the definition of the chi-squared function as given in \req{eqn:chisq_full}, and depends on the additional factor $z_i$.
The expression for $z_i$, provided in \req{ch_ss:eq:penal}, is a function of the parameters $p_{\beta}$ and $p_{\phi_2}$, which model higher-order effects in the lattice-spacing and quark-mass dependence, respectively.
Setting $p_{\beta} = p_{\phi_2} = 0$ corresponds to the standard approach, in which the weight of the chi-squared function is given by the inverse of the covariance matrix.
Considering non-zero values for these parameters constitutes a conservative approach, leading to larger uncertainties in the fitted parameters.
As discussed in Sec.~\ref{subsec:t0}, setting either $p_{\beta}$ or $p_{\phi_2}$ to infinity is equivalent to completely removing the contribution of ensembles at the coarsest lattice spacing or at the symmetric point, respectively.
Given that the inclusion of these ensembles yields good p-values, we opt for finite values of $p_{\beta}$ and $p_{\phi_2}$ in the main analysis.
We have explored various strategies for selecting the values of $p_{\beta}$ and $p_{\phi_{2}}$, including direct assessment of the possible impact of $\mbox{O}(a^{4})$ effects for each regularisation, and of $\mbox{O}(m_{\pi}^{4})$ effects in the lattice data.
This is combined with the requirement that the statistical precision of ensembles at the coarsest lattice spacing or at the symmetric point does not overly constrain the fit.
We observe that the same values of $p_{\beta}$ can be employed for both the Wilson and Wtm analyses, given that the statistical precision of the two regularisations is comparable.
Specifically, we choose $p_{\beta} = 1.5\times10^{-2}$ and $p_{\phi_2} = 5.8\times10^{-3}$, as this choice allows for a broader exploration of the model space by assigning greater weight in the model averaging to fits that include cuts, compared to the standard case with $p_{\beta} = p_{\phi_2} = 0$.
The selected value of $p_{\beta}$ effectively equates the weight $\mathcal{W}$ in \req{ch_ss:eq:Wpenal} of the ensemble H101 at the coarsest lattice spacing to that of J500 at the finest lattice spacing.
In this way, the inverse weight of H101 in the chi-squared is roughly twice its statistical error.
A similar procedure is followed to select the value of $p_{\phi_2}$, such that the weight $\mathcal{W}$ of ensembles at the symmetric point equates to that of ensemble E250, lying at the physical point.
The impact of varying $p_{\beta}$ and $p_{\phi_2}$ is summarised in Table~\ref{tab:variations_chi2W}.
We observe clear signs of stability in the results for $t_0^{\mathrm{ph}}$ under these variations.
\begin{table}[htbp!]
  \centering
  \begin{tabular}{l | l}
    \toprule
    & TIC, $p_{\beta}=p_{\phi_2}=0$ \\
    \midrule
    $\sqrt{t_0^{\mathrm{ph}}}\;[\mathrm{fm}]\; \mathrm{Wilson}$ & 0.1441(7)(4) \\ 
    $\sqrt{t_0^{\mathrm{ph}}}\;[\mathrm{fm}]\; \mathrm{Wtm}$ & 0.1442(8)(4) \\
    $\sqrt{t_0^{\mathrm{ph}}}\;[\mathrm{fm}]\; \mathrm{Combined}$ & 0.1441(5)(4) \\ 
    \midrule
    & TIC, $p_{\beta}=1.5\times10^{-2},\;p_{\phi_2}=5.8\times10^{-3}$ \\
    \midrule
    $\sqrt{t_0^{\mathrm{ph}}}\;[\mathrm{fm}]\; \mathrm{Wilson}$ & 0.1437(8)(4) \\ 
    $\sqrt{t_0^{\mathrm{ph}}}\;[\mathrm{fm}]\; \mathrm{Wtm}$ & 0.1441(8)(4) \\
    $\sqrt{t_0^{\mathrm{ph}}}\;[\mathrm{fm}]\; \mathrm{Combined}$ & 0.1440(6)(4) \\
    \midrule
    & TIC, $p_{\beta}=p_{\phi_2}\to\infty$ \\
    \midrule
    $\sqrt{t_0^{\mathrm{ph}}}\;[\mathrm{fm}]\; \mathrm{Wilson}$ & 0.1437(10)(3) \\ 
    $\sqrt{t_0^{\mathrm{ph}}}\;[\mathrm{fm}]\; \mathrm{Wtm}$ & 0.1442(10)(4) \\
    $\sqrt{t_0^{\mathrm{ph}}}\;[\mathrm{fm}]\; \mathrm{Combined}$ & 0.1439(8)(3) \\ 
    \toprule
    & $\mathrm{AIC}^{\mathrm{sub}}$, $p_{\beta}=p_{\phi_2}=0$ \\
    \midrule
    $\sqrt{t_0^{\mathrm{ph}}}\;[\mathrm{fm}]\; \mathrm{Wilson}$ & 0.1440(7)(5) \\ 
    $\sqrt{t_0^{\mathrm{ph}}}\;[\mathrm{fm}]\; \mathrm{Wtm}$ & 0.1442(7)(5) \\
    $\sqrt{t_0^{\mathrm{ph}}}\;[\mathrm{fm}]\; \mathrm{Combined}$ & 0.1441(5)(5) \\ 
    \midrule
    & $\mathrm{AIC}^{\mathrm{sub}}$, $p_{\beta}=1.5\times10^{-2},\;p_{\phi_2}=5.8\times10^{-3}$ \\
    \midrule
    $\sqrt{t_0^{\mathrm{ph}}}\;[\mathrm{fm}]\; \mathrm{Wilson}$ & 0.1438(8)(4) \\ 
    $\sqrt{t_0^{\mathrm{ph}}}\;[\mathrm{fm}]\; \mathrm{Wtm}$ & 0.1443(9)(4) \\
    $\sqrt{t_0^{\mathrm{ph}}}\;[\mathrm{fm}]\; \mathrm{Combined}$ & 0.1441(5)(4) \\
    \midrule
    & $\mathrm{AIC}^{\mathrm{sub}}$, $p_{\beta}=p_{\phi_2}\to\infty$ \\
    \midrule
    $\sqrt{t_0^{\mathrm{ph}}}\;[\mathrm{fm}]\; \mathrm{Wilson}$ & 0.1437(10)(3) \\ 
    $\sqrt{t_0^{\mathrm{ph}}}\;[\mathrm{fm}]\; \mathrm{Wtm}$ & 0.1442(10)(4) \\
    $\sqrt{t_0^{\mathrm{ph}}}\;[\mathrm{fm}]\; \mathrm{Combined}$ & 0.1440(9)(3) \\
    \bottomrule
  \end{tabular}
  \caption{
    Summary of the study of the dependence of $\sqrt{t_0^{\mathrm{ph}}}$ on the variation of the parameters $p_{\mathrm{\beta}}$ and $p_{\mathrm{\phi_2}}$ in \req{ch_ss:eq:penal}.
    Two information criteria are included in this comparison: the TIC, as defined in \req{eqn:TICchiexp}, and the $\mathrm{AIC}^{\mathrm{sub}}$, given in \req{eq:AIC_sub}.
  }
  \label{tab:variations_chi2W}
\end{table}
We have employed various model selection criteria to investigate the estimates of the systematic effects on $t_0^{\mathrm{ph}}$ arising from model variations.
These variations are associated with the set of fit functions and data cuts employed in the combined continuum and chiral extrapolations of the decay constants, expressed in units of $t_0$.
In particular, we examine the impact of varying the IC employed in the model averaging procedure.
The general methodology is described in Appendix~\ref{app:MA}.
We consider the IC based on the formulation of the Takeuchi Information Criterion (TIC), as defined in \req{eqn:TICchiexp}.
As noted in Appendix~\ref{app:MA}, this IC reduces to the expression given in \req{eqn:TICC} for correlated fits.
Additionally, we employ the Akaike Information Criterion (AIC),
\begin{align}
  \label{eq:AIC}
  \mathrm{AIC} &= \chi^2 + 2n_{\mathrm{param}}\,.
\end{align}
We also test an alternative IC that uses an different weighting term associated with the presence of data cuts,
\begin{align}
  \label{eq:AIC_sub}
  \mathrm{AIC}^{\mathrm{sub}} &= \chi^2 + 2n_{\mathrm{param}} + n_{\mathrm{cut}}\,.
\end{align}
The notation employed in this IC is adopted from Ref.~\cite{Neil:2023pgt}, which also provides a comparative discussion with the IC specified in \req{eqn:TICC}.
Furthermore, we consider a procedure in which the quoted total uncertainty is defined as half of the largest deviation between any two models within the subset whose p-value exceeds $0.1$.
In this case, the central value is obtained via a standard weighted average over the selected models.
This procedure is referred to as ``Max-deviation envelope (p-value $> 0.1$)'' in Table~\ref{tab:variationsIC}, where we summarise the impact of the various model selection criteria on the extraction of $t_0^{\mathrm{ph}}$.
The corresponding graphical representation is shown in Fig.~\ref{fig:variationsIC}.
The results obtained using the various model-selection criteria exhibit good consistency.
The TIC yields slightly larger uncertainties in the combined analysis of the Wilson and Wtm regularisations than the AIC and $\mathrm{AIC}^{\mathrm{sub}}$.
As expected, the ``Max-deviation envelope (p-value $> 0.1$)'' produces the largest systematic uncertainty, as it adopts the maximal spread among all accepted fits.
Figure~\ref{ch_ss:fig:MA_comb} compares model variation based on the TIC in two ways: the upper panel shows the effect of data cuts integrated over all fit ans\"atze, while the lower panel shows the complementary case where the effect of fit ans\"atze is integrated over all the applied data cuts.
The vertical band, indicating the TIC-based model-average result, adequately covers the spread of model categories in both panels.
It is on the basis of this suitable behaviour that the TIC is adopted as the primary model-selection criterion for the main analysis.
\begin{table}[htbp!]
  \centering
  \begin{tabular}{l | l}
    \toprule
    & TIC \\
    \midrule
    $\sqrt{t_0^{\mathrm{ph}}}\;[\mathrm{fm}]\; \mathrm{Wilson}$ & 0.1437(8)(4)[9] \\
    $\sqrt{t_0^{\mathrm{ph}}}\;[\mathrm{fm}]\; \mathrm{Wtm}$ & 0.1441(8)(4)[9] \\
    $\sqrt{t_0^{\mathrm{ph}}}\;[\mathrm{fm}]\; \mathrm{Combined}$ & 0.1440(6)(4)[7] \\
    \midrule
    & AIC \\
    \midrule
    $\sqrt{t_0^{\mathrm{ph}}}\;[\mathrm{fm}]\; \mathrm{Wilson}$ & 0.1435(8)(7)[11] \\
    $\sqrt{t_0^{\mathrm{ph}}}\;[\mathrm{fm}]\; \mathrm{Wtm}$ & 0.1441(8)(7)[11] \\
    $\sqrt{t_0^{\mathrm{ph}}}\;[\mathrm{fm}]\; \mathrm{Combined}$ & 0.1440(6)(3)[7] \\
    \midrule
    & $\mathrm{AIC}^{\mathrm{sub}}$ \\
    \midrule
    $\sqrt{t_0^{\mathrm{ph}}}\;[\mathrm{fm}]\; \mathrm{Wilson}$ & 0.1438(8)(4)[9] \\
    $\sqrt{t_0^{\mathrm{ph}}}\;[\mathrm{fm}]\; \mathrm{Wtm}$ & 0.1443(9)(4)[10] \\
    $\sqrt{t_0^{\mathrm{ph}}}\;[\mathrm{fm}]\; \mathrm{Combined}$ & 0.1441(5)(4)[6] \\
    \midrule
    & Max-deviation envelope (p-value $> 0.1$)\\
    \midrule
    $\sqrt{t_0^{\mathrm{ph}}}\;[\mathrm{fm}]\; \mathrm{Combined}$ & 0.1440[10]\\
    \bottomrule
  \end{tabular}
  \caption{
    Summary of the study of the dependence of $\sqrt{t_0^{\mathrm{ph}}}$ on the variation of the model selection criteria, including the TIC, see \req{eqn:TICchiexp}, the Akaike Information Criterion (AIC), see \req{eq:AIC}, the modified criterion $\mathrm{AIC}^{\mathrm{sub}}$, see \req{eq:AIC_sub}, as well as the ``Max-deviation envelope (p-value $> 0.1$)'' procedure described in the text.
    The first uncertainty is the statistical error and the second the systematic error, while the uncertainty between squared brackets is the total error.
  }
  \label{tab:variationsIC}
\end{table}
%


\section{Dependence of $t_0^{\mathrm{ph}}$ on the isoQCD definition}
\label{app:prescription}

In this Appendix, we examine the impact on the value of $t_0^{\mathrm{ph}}$ resulting from changing the prescription used to define the isoQCD limit.
In Eqs.~(\ref{ch_ss:eq:t0ph_w})–(\ref{ch_ss:eq:t0ph_c}) we provided results for $\sqrt{t_0^{\mathrm{ph}}}$ using as physical input the values in Eqs.~(\ref{ch_ss:eq:isoQCD_mpi})--(\ref{ch_ss:eq:isoQCD_fpi}), based on the Edinburgh consensus advocated in the 2024 edition of the FLAG review~\cite{FlavourLatticeAveragingGroupFLAG:2024oxs}, supplemented by the value of $f_K^{\mathrm{isoQCD}}$ in \req{ch_ss:eq:isoQCD_fk}.
An alternative scheme for defining the isoQCD limit, based on the FLAG~21 report~\cite{FlavourLatticeAveragingGroupFLAG:2021npn}, has also been considered
\begin{align}
  \label{ch_ss:eq:FLAG21}
  m_{\pi}^{\mathrm{isoQCD}}&=134.9768(5)\,\mathrm{MeV}\,, \notag\\
  m_K^{\mathrm{isoQCD}}&=497.611(13)\,\mathrm{MeV}\,, \notag\\
  f_{\pi}^{\mathrm{isoQCD}}&=130.56(2)_{\mathrm{exp}}(13)_{\mathrm{QED}}(2)_{|V_{ud}|}\,\mathrm{MeV}\,, \notag\\
  f_K^{\mathrm{isoQCD}}&=157.2(2)_{\mathrm{exp}}(2)_{\mathrm{QED}}(4)_{|V_{us}|}\,{\mathrm{MeV}}\,.
\end{align}
Compared to Eqs.~(\ref{ch_ss:eq:isoQCD_mpi})--(\ref{ch_ss:eq:isoQCD_fk}), this prescription differs primarily in the value of $m_K^{\mathrm{isoQCD}}$.
Applying the complete scale setting analysis described in Sec.~\ref{subsec:t0}, with the input values in \req{ch_ss:eq:FLAG21}, leads to the following determinations
\begin{align}
  \label{ch_ss:eq:t0ph_w_21}
  \sqrt{t_0^{\mathrm{ph}}}&=0.1435(8)(4)\,\mathrm{fm}\,, & \mathrm{Wilson}\,, \\
  \label{ch_ss:eq:t0ph_tm_21}
  \sqrt{t_0^{\mathrm{ph}}}&=0.1439(8)(4)\,\mathrm{fm}\,, & \mathrm{Wtm}\,, \\
  \label{ch_ss:eq:t0ph_c_21}
  \sqrt{t_0^{\mathrm{ph}}}&=0.1438(6)(4)\,\mathrm{fm}\,, & \mathrm{Combined}\,.
\end{align}

\begin{figure}[htbp!]
  \centering
  \includegraphics[width=1.\textwidth]{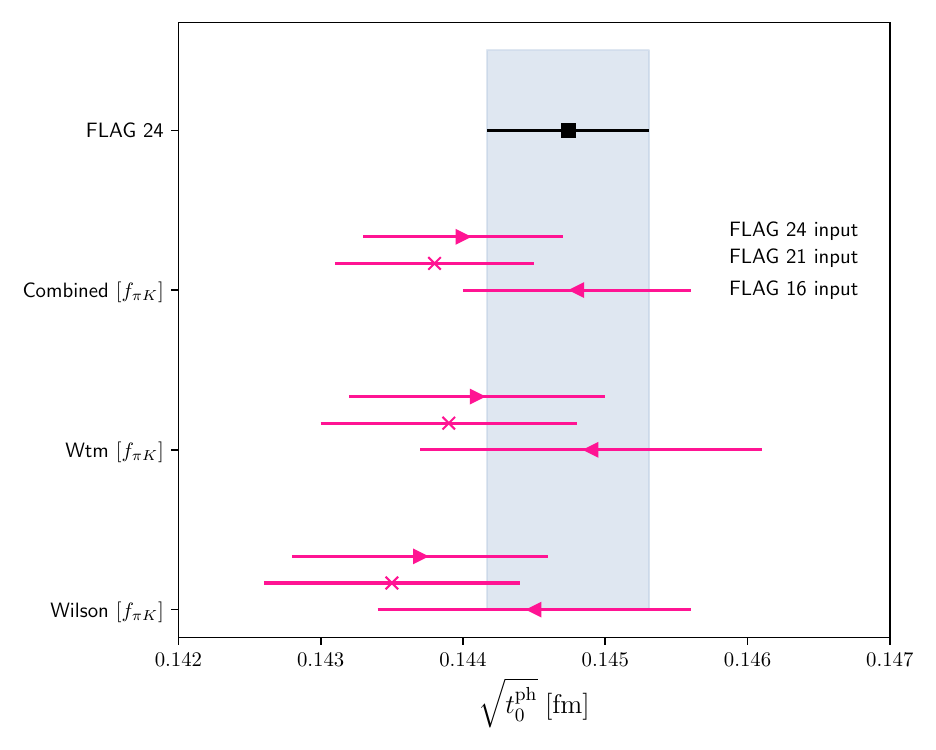}
  \caption{
    Dependence of the value of $\sqrt{t_0^{\mathrm{ph}}}$ on the prescription adopted to define the isoQCD limit for the three analysis variants: the unitary Wilson regularisation, the Wtm mixed-action setup, and the combination of both.
    The prescription adopted in this work follows the Edinburgh consensus, as defined in the FLAG~24 report~\cite{FlavourLatticeAveragingGroupFLAG:2024oxs}, and presented in Eqs.~(\ref{ch_ss:eq:isoQCD_mpi})--(\ref{ch_ss:eq:isoQCD_fpi}), supplemented by the isoQCD value of $f_K^{\mathrm{isoQCD}}$~\cite{FlavourLatticeAveragingGroupFLAG:2024oxs}, as given in \req{ch_ss:eq:isoQCD_fk}.
    The alternative schemes in Eqs.~(\ref{ch_ss:eq:FLAG21}) and (\ref{ch_ss:eq:FLAG16}) are based on the FLAG~21~\cite{FlavourLatticeAveragingGroupFLAG:2021npn} and FLAG~16~\cite{Aoki:2016frl} reports, respectively.
    The $\NF=2+1$ FLAG~24 average result~\cite{FlavourLatticeAveragingGroupFLAG:2024oxs} is denoted by the black filled square symbol and the vertical band.
  }
  \label{fig:prescription}
\end{figure}

A third prescription, based on the FLAG~16 report~\cite{Aoki:2016frl}, reads
\begin{align}
  \label{ch_ss:eq:FLAG16}
  m_{\pi}^{\mathrm{isoQCD}}&=134.8(3)\,\mathrm{MeV}\,, \notag\\
  m_K^{\mathrm{isoQCD}}&=494.2(3)\,\mathrm{MeV}\,, \notag\\
  f_{\pi}^{\mathrm{isoQCD}}&=130.4(2)\,\mathrm{MeV}\,, \notag\\
  f_K^{\mathrm{isoQCD}}&=156.2(7)\,\mathrm{MeV}\,.
\end{align}
Compared to Eqs.~(\ref{ch_ss:eq:isoQCD_mpi})--(\ref{ch_ss:eq:isoQCD_fk}), this scheme differs mostly in the value of $f_K^{\mathrm{isoQCD}}$.
Repeating the scale setting analysis with the input values in \req{ch_ss:eq:FLAG16} leads to the following determinations
\begin{align}
  \label{ch_ss:eq:t0ph_w_16}
  \sqrt{t_0^{\mathrm{ph}}}&=0.1445(9)(6)\,\mathrm{fm}\,, &\mathrm{Wilson}\,, \\
  \label{ch_ss:eq:t0ph_tm_16}
  \sqrt{t_0^{\mathrm{ph}}}&=0.1449(9)(5)\,\mathrm{fm}\,, &\mathrm{Wtm}\,, \\
  \label{ch_ss:eq:t0ph_c_16}
  \sqrt{t_0^{\mathrm{ph}}}&=0.1448(6)(4)\,\mathrm{fm}\,, &\mathrm{Combined}\,.
\end{align}
The results of this comparison are also illustrated in Fig.~\ref{fig:prescription}.
The results obtained with different isoQCD prescriptions can be mapped onto one another by using the derivatives of the relevant observables with respect to the scheme-defining quantities.
Table~\ref{tab:ders} lists the partial derivatives of $\sqrt{t_0^{\mathrm{ph}}}$ with respect to $m_{\pi,K}$ and $f_{\pi,K}$, evaluated at their isoQCD values specified in Eqs.~(\ref{ch_ss:eq:isoQCD_mpi})--(\ref{ch_ss:eq:isoQCD_fk}).
We have explicitly verified that the shifts obtained from the derivatives in Table~\ref{tab:ders} reproduce the values quoted in Eqs.~(\ref{ch_ss:eq:t0ph_w_21}) and (\ref{ch_ss:eq:t0ph_c_16}) -- i.e. the results for the FLAG~21 and FLAG~16 isoQCD prescriptions, respectively -- when starting from \req{ch_ss:eq:t0ph_c} which is based on the FLAG~24 input values in Eqs.~(\ref{ch_ss:eq:isoQCD_mpi})--(\ref{ch_ss:eq:isoQCD_fk}).
For related analyses of the scheme dependence in defining the isoQCD limit --- for quantities relevant to studies of the hadronic vacuum polarisation contribution to the muon anomalous magnetic moment --- we refer to Ref.~\cite{Aliberti:2025beg}.
\begin{table}[t!]
  \begin{center}
    \renewcommand{\arraystretch}{1.4}
    \begin{tabular}{c c c c}
      \toprule
      $\left.\frac{\partial\sqrt{t_0^{\mathrm{ph}}}}{\partial
        m_{\pi}}\frac{m_{\pi}}{\sqrt{t_0^{\mathrm{ph}}}}\right|_{\mathrm{isoQCD}}$ & $\left.\frac{\partial\sqrt{t_0^{\mathrm{ph}}}}{\partial m_{K}}\frac{m_{K}}{\sqrt{t_0^{\mathrm{ph}}}}\right|_{\mathrm{isoQCD}}$ & $\left.\frac{\partial\sqrt{t_0^{\mathrm{ph}}}}{\partial f_{\pi}}\frac{f_{\pi}}{\sqrt{t_0^{\mathrm{ph}}}}\right|_{\mathrm{isoQCD}}$ & $\left.\frac{\partial\sqrt{t_0^{\mathrm{ph}}}}{\partial f_{K}}\frac{f_{K}}{\sqrt{t_0^{\mathrm{ph}}}}\right|_{\mathrm{isoQCD}}$ \\
      \toprule
      $0.0128$ & $-0.3277$ & $-0.2931$ & $-0.7069$ \\
      \bottomrule
    \end{tabular}
    \renewcommand{\arraystretch}{1.0}
  \end{center}
  \caption{
    Partial derivatives of $\sqrt{t_0^{\mathrm{ph}}}$ with respect to $m_{\pi,K}$ and $f_{\pi,K}$, evaluated at their isoQCD values specified in Eqs.~(\ref{ch_ss:eq:isoQCD_mpi})--(\ref{ch_ss:eq:isoQCD_fk}).
    These derivatives can be used to convert the values of $\sqrt{t_0^{\mathrm{ph}}}$ to other isoQCD schemes characterised by different values of $m_{\pi,K}^{\mathrm{isoQCD}}$ and $f_{\pi,K}^{\mathrm{isoQCD}}$.
  }
  \label{tab:ders}
\end{table}
%
%
%


\clearpage

\section{Model-variation results for $t_0^{\mathrm{ph}}$ based on $f_{\pi K}^{\mathrm{isoQCD}}$}
\label{apex_model_av_t0}

This Appendix presents the results of the model variations considered in the determination of $t_0^{\mathrm{ph}}$, as described in Sec.~\ref{subsec:t0}, based on physical input from $f_{\pi K}^{\mathrm{isoQCD}}$ derived from the values given in Eqs.~(\ref{ch_ss:eq:isoQCD_fpi})--(\ref{ch_ss:eq:isoQCD_fk}).
Table~\ref{apex_ma:tab:labels} defines the labels used to identify each model, along with their corresponding functional forms and data cuts.
Tables~\ref{apex_ma:tab:w1}--\ref{apex_ma:tab:w2} report the determinations of $t_0^{\mathrm{ph}}$ from an independent analysis of the Wilson unitary setup, including the associated p-values~\cite{Bruno:2022mfy} and weights $W$ computed via Eqs.~(\ref{eqn:TICchiexp})--(\ref{eqn:weight}).
Tables~\ref{apex_ma:tab:tm1}--\ref{apex_ma:tab:tm2} and \ref{apex_ma:tab:comb1}--\ref{apex_ma:tab:comb4} present analogous results for the Wtm mixed-action setup and the combined analysis of Wilson and Wtm regularisations, respectively.
\begin{table}[htbp!]
  \begin{center}
    \renewcommand{\arraystretch}{1.3}
    \begin{tabular}{l | l}
      \toprule
      Model label & Description \\
      \toprule
      $[\chi \mathrm{SU(3)}]$ & SU(3) $\chi$PT in \req{ch_ss:eq:SU3ChPT} \\
      $[\chi \mathrm{SU(2)}]$ & SU(2) $\chi$PT in \req{ch_ss:eq:SU2pik} \\
      $[\mathrm{Tay2}]$ & Second order Taylor fit in \req{ch_ss:eq:Tay2} \\
      $[\mathrm{Tay4}]$ & Fourth order Taylor fit in \req{ch_ss:eq:Tay4} \\
      \midrule
      $[a^2]$ & $\Oasq$ effects in \req{ch_ss:eq:a2} \\
      $[a^2\alphas^{\hat{\Gamma}}]$ & $\Oasq$ effects in \req{ch_ss:eq:aas} \\
      $[a^2+a^2\phi_2]$ & $\Oasq$ effects in \req{ch_ss:eq:a2phi2} \\
      \midrule
      $[-]$ & No cut in data \\
      $[\beta>3.40]$ & Remove $\beta=3.40$ ensembles [\req{ch_ss:eq:cuts1}]\\
      $[\beta>3.46]$ & Remove $\beta=3.40$ and $\beta=3.46$ ensembles [\req{ch_ss:eq:cuts2}]\\
      $[m_{\pi}<420\,\mathrm{MeV}]$ & Remove symmetric point ensembles [\req{ch_ss:eq:cuts3}]\\
      $[m_{\pi}<350\,\mathrm{MeV}]$ & Remove $\phi_2>0.4$ ensembles [\req{ch_ss:eq:cuts4}]\\
      $[\beta>3.40\;\&\;m_{\pi}<420\,\mathrm{MeV}]$ & Remove symmetric point and $\beta=3.40$ ensembles [\req{ch_ss:eq:cuts5}]\\
      $[m_{\pi}L>4.1]$ & Remove ensembles with volumes $m_{\pi}L\leq4.1$ [\req{ch_ss:eq:cuts6}]\\
      \bottomrule
    \end{tabular}
    \renewcommand{\arraystretch}{1.0}
  \end{center}
  \caption{
    Correspondence between each fit model used in the combined continuum and chiral fits of $\sqrt{8t_0}f_{\pi K}$ -- with physical input from $f_{\pi K}^{\mathrm{isoQCD}}$ in Eqs.~(\ref{ch_ss:eq:isoQCD_fpi})--(\ref{ch_ss:eq:isoQCD_fk}) -- and the labels employed in Tables~\ref{apex_ma:tab:w1}--\ref{apex_ma:tab:comb4} and Figs.~\ref{ch_ss:fig:MA_w}--\ref{ch_ss:fig:MA_comb}.
    In the combined analysis of the Wilson and Wtm setups, independent cutoff effects are considered for each regularisation.
    In such cases, two successive labels are used: the first denotes the lattice artefacts for the Wilson setup, and the second for the Wtm mixed-action setup.
    If only one label is shown, it indicates that the same functional form for lattice artefacts is assumed for both regularisations, albeit with independent parameters.
  }
  \label{apex_ma:tab:labels}
\end{table}

\clearpage

\begin{table}[htbp!]
  \begin{center}
    \begin{tabular}{ l | l | l | l }
      \toprule
      Model & p-value & $W$ & $\sqrt{t_0^{\mathrm{ph}}}\,[\mathrm{fm}]$ \\
      \toprule
$[\chi \mathrm{SU(3)}][a^2][-]$ & 0.60 & 0.0645 & 0.1437(6) \\
$[\chi \mathrm{SU(3)}][a^2][\beta>3.40]$ & 0.52 & 0.0243 & 0.1435(8) \\
$[\chi \mathrm{SU(3)}][a^2][\beta>3.46]$ & 0.42 & 0.0113 & 0.1432(9) \\
$[\chi \mathrm{SU(3)}][a^2][m_{\pi}<420\;\rm{MeV}]$ & 0.51 & 0.0201 & 0.1437(6) \\
$[\chi \mathrm{SU(3)}][a^2][\beta>3.40\;\&\;m_{\pi}<420\;\rm{MeV}]$ & 0.32 & 0.0042 & 0.1431(10) \\
$[\chi \mathrm{SU(3)}][a^2][m_{\pi}<350\;\rm{MeV}]$ & 0.33 & 0.0025 & 0.1438(7) \\
$[\chi \mathrm{SU(3)}][a^2][m_{\pi}L>4.1]$ & 0.44 & 0.0107 & 0.1439(7) \\
$[\chi \mathrm{SU(3)}][a^2\alpha_s^{\hat{\Gamma}}][-]$ & 0.58 & 0.0662 & 0.1438(6) \\
$[\chi \mathrm{SU(3)}][a^2\alpha_s^{\hat{\Gamma}}][\beta>3.40]$ & 0.52 & 0.0237 & 0.1435(8) \\
$[\chi \mathrm{SU(3)}][a^2\alpha_s^{\hat{\Gamma}}][\beta>3.46]$ & 0.42 & 0.0113 & 0.1433(9) \\
$[\chi \mathrm{SU(3)}][a^2\alpha_s^{\hat{\Gamma}}][m_{\pi}<420\;\rm{MeV}]$ & 0.50 & 0.0191 & 0.1438(6) \\
$[\chi \mathrm{SU(3)}][a^2\alpha_s^{\hat{\Gamma}}][\beta>3.40\;\&\;m_{\pi}<420\;\rm{MeV}]$ & 0.34 & 0.0042 & 0.1432(10) \\
$[\chi \mathrm{SU(3)}][a^2\alpha_s^{\hat{\Gamma}}][m_{\pi}<350\;\rm{MeV}]$ & 0.34 & 0.0025 & 0.1439(7) \\
$[\chi \mathrm{SU(3)}][a^2\alpha_s^{\hat{\Gamma}}][m_{\pi}L>4.1]$ & 0.44 & 0.0103 & 0.1440(7) \\
$[\chi \mathrm{SU(3)}][a^2+a^2\phi_2][-]$ & 0.57 & 0.0430 & 0.1433(10) \\
$[\chi \mathrm{SU(3)}][a^2+a^2\phi_2][\beta>3.40]$ & 0.45 & 0.0161 & 0.1432(13) \\
$[\chi \mathrm{SU(3)}][a^2+a^2\phi_2][\beta>3.46]$ & 0.33 & 0.0051 & 0.1433(17) \\
$[\chi \mathrm{SU(3)}][a^2+a^2\phi_2][m_{\pi}<420\;\rm{MeV}]$ & 0.44 & 0.0103 & 0.1433(13) \\
$[\chi \mathrm{SU(3)}][a^2+a^2\phi_2][m_{\pi}L>4.1]$ & 0.37 & 0.0061 & 0.1437(13) \\
      \midrule
$[\chi \mathrm{SU(2)}][a^2][-]$ & 0.56 & 0.0383 & 0.1438(9) \\
$[\chi \mathrm{SU(2)}][a^2][\beta>3.40]$ & 0.47 & 0.0156 & 0.1436(11) \\
$[\chi \mathrm{SU(2)}][a^2][\beta>3.46]$ & 0.39 & 0.0069 & 0.1434(11) \\
$[\chi \mathrm{SU(2)}][a^2][m_{\pi}<420\;\rm{MeV}]$ & 0.45 & 0.0110 & 0.1438(10) \\
$[\chi \mathrm{SU(2)}][a^2][\beta>3.40\;\&\;m_{\pi}<420\;\rm{MeV}]$ & 0.25 & 0.0021 & 0.1434(13) \\
$[\chi \mathrm{SU(2)}][a^2][m_{\pi}<350\;\rm{MeV}]$ & 0.68 & 0.0043 & 0.1423(13) \\
$[\chi \mathrm{SU(2)}][a^2][m_{\pi}L>4.1]$ & 0.42 & 0.0070 & 0.1442(10) \\
$[\chi \mathrm{SU(2)}][a^2\alpha_s^{\hat{\Gamma}}][-]$ & 0.55 & 0.0365 & 0.1439(9) \\
$[\chi \mathrm{SU(2)}][a^2\alpha_s^{\hat{\Gamma}}][\beta>3.40]$ & 0.45 & 0.0173 & 0.1437(11) \\
$[\chi \mathrm{SU(2)}][a^2\alpha_s^{\hat{\Gamma}}][\beta>3.46]$ & 0.39 & 0.0065 & 0.1435(12) \\
$[\chi \mathrm{SU(2)}][a^2\alpha_s^{\hat{\Gamma}}][m_{\pi}<420\;\rm{MeV}]$ & 0.43 & 0.0097 & 0.1439(10) \\
$[\chi \mathrm{SU(2)}][a^2\alpha_s^{\hat{\Gamma}}][\beta>3.40\;\&\;m_{\pi}<420\;\rm{MeV}]$ & 0.25 & 0.0021 & 0.1435(13) \\
$[\chi \mathrm{SU(2)}][a^2\alpha_s^{\hat{\Gamma}}][m_{\pi}<350\;\rm{MeV}]$ & 0.69 & 0.0044 & 0.1424(13) \\
$[\chi \mathrm{SU(2)}][a^2\alpha_s^{\hat{\Gamma}}][m_{\pi}L>4.1]$ & 0.41 & 0.0062 & 0.1443(10) \\
$[\chi \mathrm{SU(2)}][a^2+a^2\phi_2][-]$ & 0.54 & 0.0275 & 0.1430(12) \\
$[\chi \mathrm{SU(2)}][a^2+a^2\phi_2][\beta>3.40]$ & 0.42 & 0.0108 & 0.1429(16) \\
$[\chi \mathrm{SU(2)}][a^2+a^2\phi_2][\beta>3.46]$ & 0.29 & 0.0030 & 0.1428(20) \\
$[\chi \mathrm{SU(2)}][a^2+a^2\phi_2][m_{\pi}<420\;\rm{MeV}]$ & 0.41 & 0.0065 & 0.1429(15) \\
$[\chi \mathrm{SU(2)}][a^2+a^2\phi_2][m_{\pi}L>4.1]$ & 0.41 & 0.0047 & 0.1430(17) \\
      \bottomrule
    \end{tabular}
  \end{center}
  \caption{
    Model variations used to determine $\sqrt{t_0^{\mathrm{ph}}}$, as described in Sec.~\ref{subsec:t0}, based on $f_{\pi K}^{\mathrm{isoQCD}}$, employing only the unitary Wilson regularisation.
    The first column identifies the model via Table~\ref{apex_ma:tab:labels}; 
    the second reports the p-value~\cite{Bruno:2022mfy}, 
    the third gives the model weight $W$ from Eqs.~(\ref{eqn:TICchiexp})--(\ref{eqn:weight}),
    and the fourth lists the resulting $\sqrt{t_0^{\mathrm{ph}}}$.
    All models use the weight matrix $\mathcal{W}$ from \req{ch_ss:eq:Wpenal} in the chi-squared function of \req{eqn:chisq}, with coefficients $p_{\beta}$ and $p_{\phi_2}$ in \req{ch_ss:eq:penal} fixed as detailed in Sec.~\ref{subsec:t0} and Appendix~\ref{app_variations}.
  }
  \label{apex_ma:tab:w1}
\end{table}

\clearpage

\begin{table}[htbp!]
  \begin{center}
    \begin{tabular}{ l | l | l | l }
      \toprule
      Model & p-value & $W$ & $\sqrt{t_0^{\mathrm{ph}}}\,[\mathrm{fm}]$ \\
      \toprule
$[\mathrm{Tay2}][a^2][-]$ & 0.56 & 0.0572 & 0.1441(7) \\
$[\mathrm{Tay2}][a^2][\beta>3.40]$ & 0.48 & 0.0241 & 0.1439(9) \\
$[\mathrm{Tay2}][a^2][\beta>3.46]$ & 0.39 & 0.0080 & 0.1436(10) \\
$[\mathrm{Tay2}][a^2][m_{\pi}<420\;\rm{MeV}]$ & 0.47 & 0.0170 & 0.1440(7) \\
$[\mathrm{Tay2}][a^2][\beta>3.40\;\&\;m_{\pi}<420\;\rm{MeV}]$ & 0.30 & 0.0037 & 0.1436(11) \\
$[\mathrm{Tay2}][a^2][m_{\pi}<350\;\rm{MeV}]$ & 0.22 & 0.0014 & 0.1441(9) \\
$[\mathrm{Tay2}][a^2][m_{\pi}L>4.1]$ & 0.51 & 0.0150 & 0.1443(8) \\
$[\mathrm{Tay2}][a^2\alpha_s^{\hat{\Gamma}}][-]$ & 0.55 & 0.0567 & 0.1442(8) \\
$[\mathrm{Tay2}][a^2\alpha_s^{\hat{\Gamma}}][\beta>3.40]$ & 0.48 & 0.0245 & 0.1440(9) \\
$[\mathrm{Tay2}][a^2\alpha_s^{\hat{\Gamma}}][\beta>3.46]$ & 0.39 & 0.0080 & 0.1437(10) \\
$[\mathrm{Tay2}][a^2\alpha_s^{\hat{\Gamma}}][m_{\pi}<420\;\rm{MeV}]$ & 0.48 & 0.0163 & 0.1442(8) \\
$[\mathrm{Tay2}][a^2\alpha_s^{\hat{\Gamma}}][\beta>3.40\;\&\;m_{\pi}<420\;\rm{MeV}]$ & 0.30 & 0.0037 & 0.1437(11) \\
$[\mathrm{Tay2}][a^2\alpha_s^{\hat{\Gamma}}][m_{\pi}<350\;\rm{MeV}]$ & 0.22 & 0.0014 & 0.1442(9) \\
$[\mathrm{Tay2}][a^2\alpha_s^{\hat{\Gamma}}][m_{\pi}L>4.1]$ & 0.49 & 0.0142 & 0.1444(8) \\
$[\mathrm{Tay2}][a^2+a^2\phi_2][-]$ & 0.55 & 0.0444 & 0.1434(11) \\
$[\mathrm{Tay2}][a^2+a^2\phi_2][\beta>3.40]$ & 0.44 & 0.0158 & 0.1433(14) \\
$[\mathrm{Tay2}][a^2+a^2\phi_2][\beta>3.46]$ & 0.31 & 0.0051 & 0.1434(18) \\
$[\mathrm{Tay2}][a^2+a^2\phi_2][m_{\pi}<420\;\rm{MeV}]$ & 0.42 & 0.0112 & 0.1434(14) \\
$[\mathrm{Tay2}][a^2+a^2\phi_2][m_{\pi}L>4.1]$ & 0.46 & 0.0089 & 0.1436(14) \\
      \midrule
$[\mathrm{Tay4}][a^2][-]$ & 0.46 & 0.0223 & 0.1439(9) \\
$[\mathrm{Tay4}][a^2][\beta>3.40]$ & 0.37 & 0.0088 & 0.1437(11) \\
$[\mathrm{Tay4}][a^2][\beta>3.46]$ & 0.28 & 0.0031 & 0.1435(12) \\
$[\mathrm{Tay4}][a^2][m_{\pi}<420\;\rm{MeV}]$ & 0.37 & 0.0066 & 0.1439(9) \\
$[\mathrm{Tay4}][a^2][\beta>3.40\;\&\;m_{\pi}<420\;\rm{MeV}]$ & 0.20 & 0.0014 & 0.1434(12) \\
$[\mathrm{Tay4}][a^2][m_{\pi}<350\;\rm{MeV}]$ & 0.70 & 0.0043 & 0.1423(12) \\
$[\mathrm{Tay4}][a^2][m_{\pi}L>4.1]$ & 0.43 & 0.0068 & 0.1449(12) \\
$[\mathrm{Tay4}][a^2+a^2\phi_2][-]$ & 0.45 & 0.0148 & 0.1434(11) \\
$[\mathrm{Tay4}][a^2+a^2\phi_2][\beta>3.40]$ & 0.34 & 0.0051 & 0.1433(14) \\
$[\mathrm{Tay4}][a^2+a^2\phi_2][\beta>3.46]$ & 0.19 & 0.0018 & 0.1434(18) \\
$[\mathrm{Tay4}][a^2+a^2\phi_2][m_{\pi}<420\;\rm{MeV}]$ & 0.31 & 0.0043 & 0.1434(14) \\
$[\mathrm{Tay4}][a^2+a^2\phi_2][m_{\pi}L>4.1]$ & 0.55 & 0.0081 & 0.1436(15) \\
      \bottomrule
    \end{tabular}
  \end{center}
  \caption{
    Continued from Table~\ref{apex_ma:tab:w1}.
  }
  \label{apex_ma:tab:w2}
\end{table}

\clearpage

\begin{table}[htbp!]
  \begin{center}
    \begin{tabular}{ l | l | l | l }
      \toprule
      Model & p-value & $W$ & $\sqrt{t_0^{\mathrm{ph}}}\,[\mathrm{fm}]$ \\
      \toprule
      $[\chi \mathrm{SU(3)}][a^2][-]$ & 0.74 & 0.0601 & 0.1443(5) \\
      $[\chi \mathrm{SU(3)}][a^2][\beta>3.40]$ & 0.64 & 0.0343 & 0.1441(7) \\
      $[\chi \mathrm{SU(3)}][a^2][\beta>3.46]$ & 0.54 & 0.0098 & 0.1440(9) \\
      $[\chi \mathrm{SU(3)}][a^2][m_{\pi}<420\;\rm{MeV}]$ & 0.69 & 0.0287 & 0.1441(5) \\
      $[\chi \mathrm{SU(3)}][a^2][\beta>3.40\;\&\;m_{\pi}<420\;\rm{MeV}]$ & 0.46 & 0.0074 & 0.1438(9) \\
      $[\chi \mathrm{SU(3)}][a^2][m_{\pi}<350\;\rm{MeV}]$ & 0.50 & 0.0026 & 0.1444(6) \\
      $[\chi \mathrm{SU(3)}][a^2][m_{\pi}L>4.1]$ & 0.76 & 0.0123 & 0.1443(6) \\
      $[\chi \mathrm{SU(3)}][a^2\alpha_s^{\hat{\Gamma}}][-]$ & 0.74 & 0.0570 & 0.1443(5) \\
      $[\chi \mathrm{SU(3)}][a^2\alpha_s^{\hat{\Gamma}}][\beta>3.40]$ & 0.65 & 0.0344 & 0.1441(7) \\
      $[\chi \mathrm{SU(3)}][a^2\alpha_s^{\hat{\Gamma}}][\beta>3.46]$ & 0.53 & 0.0099 & 0.1440(9) \\
      $[\chi \mathrm{SU(3)}][a^2\alpha_s^{\hat{\Gamma}}][m_{\pi}<420\;\rm{MeV}]$ & 0.70 & 0.0288 & 0.1441(5) \\
      $[\chi \mathrm{SU(3)}][a^2\alpha_s^{\hat{\Gamma}}][\beta>3.40\;\&\;m_{\pi}<420\;\rm{MeV}]$ & 0.44 & 0.0075 & 0.1438(9) \\
      $[\chi \mathrm{SU(3)}][a^2\alpha_s^{\hat{\Gamma}}][m_{\pi}<350\;\rm{MeV}]$ & 0.49 & 0.0026 & 0.1444(6) \\
      $[\chi \mathrm{SU(3)}][a^2\alpha_s^{\hat{\Gamma}}][m_{\pi}L>4.1]$ & 0.77 & 0.0122 & 0.1443(6) \\
      $[\chi \mathrm{SU(3)}][a^2+a^2\phi_2][-]$ & 0.74 & 0.0495 & 0.1436(9) \\
      $[\chi \mathrm{SU(3)}][a^2+a^2\phi_2][\beta>3.40]$ & 0.63 & 0.0159 & 0.1434(13) \\
      $[\chi \mathrm{SU(3)}][a^2+a^2\phi_2][\beta>3.46]$ & 0.44 & 0.0061 & 0.1444(18) \\
      $[\chi \mathrm{SU(3)}][a^2+a^2\phi_2][m_{\pi}<420\;\rm{MeV}]$ & 0.63 & 0.0147 & 0.1441(11) \\
      $[\chi \mathrm{SU(3)}][a^2+a^2\phi_2][m_{\pi}L>4.1]$ & 0.75 & 0.0101 & 0.1433(14) \\
      \midrule
      $[\chi \mathrm{SU(2)}][a^2][-]$ & 0.74 & 0.0345 & 0.1440(9) \\
      $[\chi \mathrm{SU(2)}][a^2][\beta>3.40]$ & 0.63 & 0.0168 & 0.1439(11) \\
      $[\chi \mathrm{SU(2)}][a^2][\beta>3.46]$ & 0.48 & 0.0068 & 0.1439(12) \\
      $[\chi \mathrm{SU(2)}][a^2][m_{\pi}<420\;\rm{MeV}]$ & 0.63 & 0.0102 & 0.1440(10) \\
      $[\chi \mathrm{SU(2)}][a^2][\beta>3.40\;\&\;m_{\pi}<420\;\rm{MeV}]$ & 0.32 & 0.0021 & 0.1438(13) \\
      $[\chi \mathrm{SU(2)}][a^2][m_{\pi}<350\;\rm{MeV}]$ & 0.99 & 0.0046 & 0.1430(12) \\
      $[\chi \mathrm{SU(2)}][a^2][m_{\pi}L>4.1]$ & 0.68 & 0.0073 & 0.1444(10) \\
      $[\chi \mathrm{SU(2)}][a^2\alpha_s^{\hat{\Gamma}}][-]$ & 0.73 & 0.0323 & 0.1440(9) \\
      $[\chi \mathrm{SU(2)}][a^2\alpha_s^{\hat{\Gamma}}][\beta>3.40]$ & 0.62 & 0.0169 & 0.1439(11) \\
      $[\chi \mathrm{SU(2)}][a^2\alpha_s^{\hat{\Gamma}}][\beta>3.46]$ & 0.48 & 0.0070 & 0.1439(12) \\
      $[\chi \mathrm{SU(2)}][a^2\alpha_s^{\hat{\Gamma}}][m_{\pi}<420\;\rm{MeV}]$ & 0.64 & 0.0105 & 0.1440(10) \\
      $[\chi \mathrm{SU(2)}][a^2\alpha_s^{\hat{\Gamma}}][\beta>3.40\;\&\;m_{\pi}<420\;\rm{MeV}]$ & 0.32 & 0.0021 & 0.1438(13) \\
      $[\chi \mathrm{SU(2)}][a^2\alpha_s^{\hat{\Gamma}}][m_{\pi}<350\;\rm{MeV}]$ & 0.99 & 0.0046 & 0.1430(12) \\
      $[\chi \mathrm{SU(2)}][a^2\alpha_s^{\hat{\Gamma}}][m_{\pi}L>4.1]$ & 0.68 & 0.0073 & 0.1443(11) \\
      $[\chi \mathrm{SU(2)}][a^2+a^2\phi_2][-]$ & 0.70 & 0.0242 & 0.1435(12) \\
      $[\chi \mathrm{SU(2)}][a^2+a^2\phi_2][\beta>3.40]$ & 0.54 & 0.0086 & 0.1433(16) \\
      $[\chi \mathrm{SU(2)}][a^2+a^2\phi_2][\beta>3.46]$ & 0.38 & 0.0032 & 0.1445(23) \\
      $[\chi \mathrm{SU(2)}][a^2+a^2\phi_2][m_{\pi}<420\;\rm{MeV}]$ & 0.53 & 0.0056 & 0.1440(14) \\
      $[\chi \mathrm{SU(2)}][a^2+a^2\phi_2][m_{\pi}L>4.1]$ & 0.71 & 0.0064 & 0.1428(18) \\
      \bottomrule
    \end{tabular}
  \end{center}
  \caption{
    Model variations used to determine $\sqrt{t_0^{\mathrm{ph}}}$, as described in Sec.~\ref{subsec:t0}, based on $f_{\pi K}^{\mathrm{isoQCD}}$, employing only the mixed action Wtm regularisation.
    The first column identifies the model via Table~\ref{apex_ma:tab:labels}; 
    the second reports the p-value~\cite{Bruno:2022mfy}, 
    the third gives the model weight $W$ from Eqs.~(\ref{eqn:TICchiexp})--(\ref{eqn:weight}),
    and the fourth lists the resulting $\sqrt{t_0^{\mathrm{ph}}}$.
    All models use the weight matrix $\mathcal{W}$ from \req{ch_ss:eq:Wpenal} in the chi-squared function of \req{eqn:chisq}, with coefficients $p_{\beta}$ and $p_{\phi_2}$ in \req{ch_ss:eq:penal} fixed as detailed in Sec.~\ref{subsec:t0} and Appendix~\ref{app_variations}.
  }
  \label{apex_ma:tab:tm1}
\end{table}

\clearpage

\begin{table}[htbp!]
  \begin{center}
    \begin{tabular}{ l | l | l | l }
      \toprule
      Model & p-value & $W$ & $\sqrt{t_0^{\mathrm{ph}}}\,[\mathrm{fm}]$ \\
      \toprule
      $[\mathrm{Tay2}][a^2][-]$ & 0.65 & 0.0377 & 0.1447(7) \\
      $[\mathrm{Tay2}][a^2][\beta>3.40]$ & 0.57 & 0.0214 & 0.1445(9) \\
      $[\mathrm{Tay2}][a^2][\beta>3.46]$ & 0.46 & 0.0079 & 0.1444(10) \\
      $[\mathrm{Tay2}][a^2][m_{\pi}<420\;\rm{MeV}]$ & 0.62 & 0.0198 & 0.1445(7) \\
      $[\mathrm{Tay2}][a^2][\beta>3.40\;\&\;m_{\pi}<420\;\rm{MeV}]$ & 0.38 & 0.0050 & 0.1442(11) \\
      $[\mathrm{Tay2}][a^2][m_{\pi}<350\;\rm{MeV}]$ & 0.36 & 0.0016 & 0.1447(8) \\
      $[\mathrm{Tay2}][a^2][m_{\pi}L>4.1]$ & 0.79 & 0.0133 & 0.1447(7) \\
      $[\mathrm{Tay2}][a^2\alpha_s^{\hat{\Gamma}}][-]$ & 0.66 & 0.0376 & 0.1447(7) \\
      $[\mathrm{Tay2}][a^2\alpha_s^{\hat{\Gamma}}][\beta>3.40]$ & 0.57 & 0.0215 & 0.1445(9) \\
      $[\mathrm{Tay2}][a^2\alpha_s^{\hat{\Gamma}}][\beta>3.46]$ & 0.46 & 0.0073 & 0.1444(10) \\
      $[\mathrm{Tay2}][a^2\alpha_s^{\hat{\Gamma}}][m_{\pi}<420\;\rm{MeV}]$ & 0.61 & 0.0194 & 0.1444(7) \\
      $[\mathrm{Tay2}][a^2\alpha_s^{\hat{\Gamma}}][\beta>3.40\;\&\;m_{\pi}<420\;\rm{MeV}]$ & 0.38 & 0.0051 & 0.1442(11) \\
      $[\mathrm{Tay2}][a^2\alpha_s^{\hat{\Gamma}}][m_{\pi}<350\;\rm{MeV}]$ & 0.34 & 0.0016 & 0.1446(8) \\
      $[\mathrm{Tay2}][a^2\alpha_s^{\hat{\Gamma}}][m_{\pi}L>4.1]$ & 0.79 & 0.0138 & 0.1447(7) \\
      $[\mathrm{Tay2}][a^2+a^2\phi_2][-]$ & 0.68 & 0.0316 & 0.1438(10) \\
      $[\mathrm{Tay2}][a^2+a^2\phi_2][\beta>3.40]$ & 0.54 & 0.0122 & 0.1436(14) \\
      $[\mathrm{Tay2}][a^2+a^2\phi_2][\beta>3.46]$ & 0.36 & 0.0045 & 0.1446(20) \\
      $[\mathrm{Tay2}][a^2+a^2\phi_2][m_{\pi}<420\;\rm{MeV}]$ & 0.54 & 0.0107 & 0.1442(13) \\
      $[\mathrm{Tay2}][a^2+a^2\phi_2][m_{\pi}L>4.1]$ & 0.82 & 0.0152 & 0.1432(15) \\
      \midrule
      $[\mathrm{Tay4}][a^2][-]$ & 0.68 & 0.0205 & 0.1441(9) \\
      $[\mathrm{Tay4}][a^2][\beta>3.40]$ & 0.55 & 0.0102 & 0.1440(11) \\
      $[\mathrm{Tay4}][a^2][\beta>3.46]$ & 0.40 & 0.0036 & 0.1440(12) \\
      $[\mathrm{Tay4}][a^2][m_{\pi}<420\;\rm{MeV}]$ & 0.59 & 0.0087 & 0.1440(9) \\
      $[\mathrm{Tay4}][a^2][\beta>3.40\;\&\;m_{\pi}<420\;\rm{MeV}]$ & 0.29 & 0.0018 & 0.1438(13) \\
      $[\mathrm{Tay4}][a^2][m_{\pi}<350\;\rm{MeV}]$ & 0.99 & 0.0047 & 0.1430(11) \\
      $[\mathrm{Tay4}][a^2][m_{\pi}L>4.1]$ & 0.71 & 0.0062 & 0.1451(13) \\
      $[\mathrm{Tay4}][a^2+a^2\phi_2][-]$ & 0.65 & 0.0170 & 0.1437(11) \\
      $[\mathrm{Tay4}][a^2+a^2\phi_2][\beta>3.40]$ & 0.49 & 0.0054 & 0.1435(14) \\
      $[\mathrm{Tay4}][a^2+a^2\phi_2][\beta>3.46]$ & 0.33 & 0.0023 & 0.1448(19) \\
      $[\mathrm{Tay4}][a^2+a^2\phi_2][m_{\pi}<420\;\rm{MeV}]$ & 0.51 & 0.0051 & 0.1442(13) \\
      $[\mathrm{Tay4}][a^2+a^2\phi_2][m_{\pi}L>4.1]$ & 0.95 & 0.0122 & 0.1433(16) \\
      \bottomrule
    \end{tabular}
  \end{center}
  \caption{
    Continued from Table~\ref{apex_ma:tab:tm1}.
  }
  \label{apex_ma:tab:tm2}
\end{table}

\clearpage

\begin{table}[htbp!]
  \begin{center}
    \begin{tabular}{ l | l | l | l }
      \toprule
      Model & p-value & $W$ & $\sqrt{t_0^{\mathrm{ph}}}$ [fm] \\
      \toprule
      $[\chi \mathrm{SU(3)}][a^2][-]$ & 0.80 & 0.0484 & 0.1441(3) \\
$[\chi \mathrm{SU(3)}][a^2][\beta>3.40]$ & 0.74 & 0.0149 & 0.1439(5) \\
$[\chi \mathrm{SU(3)}][a^2][\beta>3.46]$ & 0.74 & 0.0028 & 0.1436(7) \\
$[\chi \mathrm{SU(3)}][a^2][m_{\pi}<420\;\rm{MeV}]$ & 0.81 & 0.0123 & 0.1439(3) \\
$[\chi \mathrm{SU(3)}][a^2][\beta>3.40\;\&\;m_{\pi}<420\;\rm{MeV}]$ & 0.71 & 0.0010 & 0.1434(8) \\
$[\chi \mathrm{SU(3)}][a^2][m_{\pi}<350\;\rm{MeV}]$ & 0.69 & 0.0001 & 0.1441(4) \\
$[\chi \mathrm{SU(3)}][a^2][m_{\pi}L>4.1]$ & 0.66 & 0.0012 & 0.1441(4) \\
$[\chi \mathrm{SU(3)}][a^2\alpha_s^{\hat{\Gamma}}][-]$ & 0.77 & 0.0423 & 0.1441(3) \\
$[\chi \mathrm{SU(3)}][a^2\alpha_s^{\hat{\Gamma}}][\beta>3.40]$ & 0.73 & 0.0124 & 0.1439(5) \\
$[\chi \mathrm{SU(3)}][a^2\alpha_s^{\hat{\Gamma}}][\beta>3.46]$ & 0.73 & 0.0027 & 0.1436(7) \\
$[\chi \mathrm{SU(3)}][a^2\alpha_s^{\hat{\Gamma}}][m_{\pi}<420\;\rm{MeV}]$ & 0.79 & 0.0113 & 0.1439(4) \\
$[\chi \mathrm{SU(3)}][a^2\alpha_s^{\hat{\Gamma}}][\beta>3.40\;\&\;m_{\pi}<420\;\rm{MeV}]$ & 0.71 & 0.0011 & 0.1434(8) \\
$[\chi \mathrm{SU(3)}][a^2\alpha_s^{\hat{\Gamma}}][m_{\pi}<350\;\rm{MeV}]$ & 0.70 & 0.0001 & 0.1442(4) \\
$[\chi \mathrm{SU(3)}][a^2\alpha_s^{\hat{\Gamma}}][m_{\pi}L>4.1]$ & 0.64 & 0.0010 & 0.1441(4) \\
$[\chi \mathrm{SU(3)}][a^2][a^2+a^2\phi_2][-]$ & 0.89 & 0.0703 & 0.1437(4) \\
$[\chi \mathrm{SU(3)}][a^2][a^2+a^2\phi_2][\beta>3.40]$ & 0.83 & 0.0187 & 0.1434(6) \\
$[\chi \mathrm{SU(3)}][a^2][a^2+a^2\phi_2][\beta>3.46]$ & 0.75 & 0.0026 & 0.1433(7) \\
$[\chi \mathrm{SU(3)}][a^2][a^2+a^2\phi_2][m_{\pi}<420\;\rm{MeV}]$ & 0.83 & 0.0094 & 0.1437(4) \\
$[\chi \mathrm{SU(3)}][a^2][a^2+a^2\phi_2][\beta>3.40\;\&\;m_{\pi}<420\;\rm{MeV}]$ & 0.67 & 0.0007 & 0.1433(8) \\
$[\chi \mathrm{SU(3)}][a^2][a^2+a^2\phi_2][m_{\pi}<350\;\rm{MeV}]$ & 0.66 & 0.0001 & 0.1439(5) \\
$[\chi \mathrm{SU(3)}][a^2][a^2+a^2\phi_2][m_{\pi}L>4.1]$ & 0.80 & 0.0017 & 0.1436(5) \\
$[\chi \mathrm{SU(3)}][a^2+a^2\phi_2][a^2][[-]$ & 0.85 & 0.0404 & 0.1444(4) \\
$[\chi \mathrm{SU(3)}][a^2+a^2\phi_2][a^2][\beta>3.40]$ & 0.79 & 0.0108 & 0.1442(5) \\
$[\chi \mathrm{SU(3)}][a^2+a^2\phi_2][a^2][\beta>3.46]$ & 0.75 & 0.0022 & 0.1439(7) \\
$[\chi \mathrm{SU(3)}][a^2+a^2\phi_2][a^2][m_{\pi}<420\;\rm{MeV}]$ & 0.82 & 0.0103 & 0.1442(4) \\
$[\chi \mathrm{SU(3)}][a^2+a^2\phi_2][a^2][\beta>3.40\;\&\;m_{\pi}<420\;\rm{MeV}]$ & 0.70 & 0.0004 & 0.1437(8) \\
$[\chi \mathrm{SU(3)}][a^2+a^2\phi_2][a^2][m_{\pi}<350\;\rm{MeV}]$ & 0.61 & 0.000 & 0.1442(5) \\
$[\chi \mathrm{SU(3)}][a^2+a^2\phi_2][a^2][m_{\pi}L>4.1]$ & 0.75 & 0.0011 & 0.1446(5) \\
$[\chi \mathrm{SU(3)}][a^2+a^2\phi_2][-]$ & 0.87 & 0.0523 & 0.1435(7) \\
$[\chi \mathrm{SU(3)}][a^2+a^2\phi_2][\beta>3.40]$ & 0.80 & 0.0139 & 0.1433(10) \\
$[\chi \mathrm{SU(3)}][a^2+a^2\phi_2][\beta>3.46]$ & 0.72 & 0.0008 & 0.1438(14) \\
$[\chi \mathrm{SU(3)}][a^2+a^2\phi_2][m_{\pi}<420\;\rm{MeV}]$ & 0.79 & 0.0048 & 0.1438(9) \\
$[\chi \mathrm{SU(3)}][a^2+a^2\phi_2][m_{\pi}L>4.1]$ & 0.77 & 0.0009 & 0.1435(10) \\
        \bottomrule
    \end{tabular}
  \end{center}
  \caption{
    Model variations used to determine $\sqrt{t_0^{\mathrm{ph}}}$, as described in Sec.~\ref{subsec:t0}, based on $f_{\pi K}^{\mathrm{isoQCD}}$, employing the combined analysis of the Wilson unitary and the Wtm mixed action regularisations.
    The first column identifies the model via Table~\ref{apex_ma:tab:labels}; 
    the second reports the p-value~\cite{Bruno:2022mfy}, 
    the third gives the model weight $W$ from Eqs.~(\ref{eqn:TICchiexp})--(\ref{eqn:weight}),
    and the fourth lists the resulting $\sqrt{t_0^{\mathrm{ph}}}$.
    All models use the weight matrix $\mathcal{W}$ from \req{ch_ss:eq:Wpenal} in the chi-squared function of \req{eqn:chisq}, with coefficients $p_{\beta}$ and $p_{\phi_2}$ in \req{ch_ss:eq:penal} fixed as detailed in Sec.~\ref{subsec:t0} and Appendix~\ref{app_variations}.
  }
  \label{apex_ma:tab:comb1}
\end{table}

\clearpage

\begin{table}[htbp!]
  \begin{center}
    \begin{tabular}{ l | l | l | l }
      \toprule
      Model & p-value & $W$ & $\sqrt{t_0^{\mathrm{ph}}}\,[\mathrm{fm}]$ \\
      \toprule
      $[\chi \mathrm{SU(2)}][a^2][-]$ & 0.77 & 0.0255 & 0.1440(8) \\
$[\chi \mathrm{SU(2)}][a^2][\beta>3.40]$ & 0.70 & 0.0080 & 0.1438(9) \\
$[\chi \mathrm{SU(2)}][a^2][\beta>3.46]$ & 0.72 & 0.0021 & 0.1437(10) \\
$[\chi \mathrm{SU(2)}][a^2][m_{\pi}<420\;\rm{MeV}]$ & 0.78 & 0.0061 & 0.1440(8) \\
$[\chi \mathrm{SU(2)}][a^2][m_{\pi}L>4.1]$ & 0.62 & 0.0007 & 0.1442(9) \\
$[\chi \mathrm{SU(2)}][a^2\alpha_s^{\hat{\Gamma}}][-]$ & 0.75 & 0.0218 & 0.1440(8) \\
$[\chi \mathrm{SU(2)}][a^2\alpha_s^{\hat{\Gamma}}][\beta>3.40]$ & 0.69 & 0.0067 & 0.1438(10) \\
$[\chi \mathrm{SU(2)}][a^2\alpha_s^{\hat{\Gamma}}][\beta>3.46]$ & 0.71 & 0.0019 & 0.1437(10) \\
$[\chi \mathrm{SU(2)}][a^2\alpha_s^{\hat{\Gamma}}][m_{\pi}<420\;\rm{MeV}]$ & 0.77 & 0.0057 & 0.1440(8) \\
$[\chi \mathrm{SU(2)}][a^2\alpha_s^{\hat{\Gamma}}][m_{\pi}L>4.1]$ & 0.60 & 0.0005 & 0.1442(9) \\
$[\chi \mathrm{SU(2)}][a^2][a^2+a^2\phi_2][-]$ & 0.88 & 0.0463 & 0.1437(8) \\
$[\chi \mathrm{SU(2)}][a^2][a^2+a^2\phi_2][\beta>3.40]$ & 0.81 & 0.0117 & 0.1435(10) \\
$[\chi \mathrm{SU(2)}][a^2][a^2+a^2\phi_2][\beta>3.46]$ & 0.74 & 0.0012 & 0.1434(10) \\
$[\chi \mathrm{SU(2)}][a^2][a^2+a^2\phi_2][m_{\pi}<420\;\rm{MeV}]$ & 0.82 & 0.0042 & 0.1438(8) \\
$[\chi \mathrm{SU(2)}][a^2][a^2+a^2\phi_2][m_{\pi}L>4.1]$ & 0.80 & 0.0011 & 0.1439(9) \\
$[\chi \mathrm{SU(2)}][a^2+a^2\phi_2][a^2][-]$ & 0.82 & 0.0198 & 0.1442(8) \\
$[\chi \mathrm{SU(2)}][a^2+a^2\phi_2][a^2][\beta>3.40]$ & 0.76 & 0.0093 & 0.1441(10) \\
$[\chi \mathrm{SU(2)}][a^2+a^2\phi_2][a^2][\beta>3.46]$ & 0.73 & 0.0013 & 0.1440(10) \\
$[\chi \mathrm{SU(2)}][a^2+a^2\phi_2][a^2][m_{\pi}<420\;\rm{MeV}]$ & 0.80 & 0.0038 & 0.1442(8) \\
$[\chi \mathrm{SU(2)}][a^2+a^2\phi_2][a^2][m_{\pi}L>4.1]$ & 0.70 & 0.0005 & 0.1446(9) \\
$[\chi \mathrm{SU(2)}][a^2+a^2\phi_2][-]$ & 0.86 & 0.0261 & 0.1434(10) \\
$[\chi \mathrm{SU(2)}][a^2+a^2\phi_2][\beta>3.40]$ & 0.78 & 0.0074 & 0.1431(13) \\
$[\chi \mathrm{SU(2)}][a^2+a^2\phi_2][\beta>3.46]$ & 0.68 & 0.0005 & 0.1436(18) \\
$[\chi \mathrm{SU(2)}][a^2+a^2\phi_2][m_{\pi}<420\;\rm{MeV}]$ & 0.79 & 0.0028 & 0.1435(12) \\
$[\chi \mathrm{SU(2)}][a^2+a^2\phi_2][m_{\pi}L>4.1]$ & 0.80 & 0.0010 & 0.1429(14) \\
      \bottomrule
    \end{tabular}
  \end{center}
  \caption{
    Continued from Table~\ref{apex_ma:tab:comb1}.
  }
  \label{apex_ma:tab:comb2}
\end{table}

\clearpage

\begin{table}[htbp!]
  \begin{center}
    \begin{tabular}{ l | l | l | l }
      \toprule
      Model & p-value & $W$ & $\sqrt{t_0^{\mathrm{ph}}}\,[\mathrm{fm}]$ \\
      \toprule
      $[\mathrm{Tay2}][a^2][-]$ & 0.74 & 0.0380 & 0.1445(6) \\
$[\mathrm{Tay2}][a^2][\beta>3.40]$ & 0.71 & 0.011 & 0.1443(7) \\
$[\mathrm{Tay2}][a^2][\beta>3.46]$ & 0.70 & 0.0018 & 0.1440(9) \\
$[\mathrm{Tay2}][a^2][m_{\pi}<420\;\rm{MeV}]$ & 0.76 & 0.0097 & 0.1443(6) \\
$[\mathrm{Tay2}][a^2][\beta>3.40\;\&\;m_{\pi}<420\;\rm{MeV}]$ & 0.67 & 0.0009 & 0.1439(9) \\
$[\mathrm{Tay2}][a^2][m_{\pi}<350\;\rm{MeV}]$ & 0.57 & 0.0001 & 0.1444(7) \\
$[\mathrm{Tay2}][a^2][m_{\pi}L>4.1]$ & 0.70 & 0.0016 & 0.1445(6) \\
$[\mathrm{Tay2}][a^2\alpha_s^{\hat{\Gamma}}][-]$ & 0.72 & 0.0291 & 0.1445(6) \\
$[\mathrm{Tay2}][a^2\alpha_s^{\hat{\Gamma}}][\beta>3.40]$ & 0.68 & 0.0090 & 0.1443(7) \\
$[\mathrm{Tay2}][a^2\alpha_s^{\hat{\Gamma}}][\beta>3.46]$ & 0.69 & 0.0024 & 0.1440(9) \\
$[\mathrm{Tay2}][a^2\alpha_s^{\hat{\Gamma}}][m_{\pi}<420\;\rm{MeV}]$ & 0.75 & 0.0079 & 0.1443(6) \\
$[\mathrm{Tay2}][a^2\alpha_s^{\hat{\Gamma}}][\beta>3.40\;\&\;m_{\pi}<420\;\rm{MeV}]$ & 0.68 & 0.0008 & 0.1439(9) \\
$[\mathrm{Tay2}][a^2\alpha_s^{\hat{\Gamma}}][m_{\pi}<350\;\rm{MeV}]$ & 0.57 & 0.0001 & 0.1444(7) \\
$[\mathrm{Tay2}][a^2\alpha_s^{\hat{\Gamma}}][m_{\pi}L>4.1]$ & 0.68 & 0.0013 & 0.1445(7) \\
$[\mathrm{Tay2}][a^2][a^2+a^2\phi_2][-]$ & 0.86 & 0.0531 & 0.1441(6) \\
$[\mathrm{Tay2}][a^2][a^2+a^2\phi_2][\beta>3.40]$ & 0.81 & 0.0129 & 0.1439(8) \\
$[\mathrm{Tay2}][a^2][a^2+a^2\phi_2][\beta>3.46]$ & 0.72 & 0.0016 & 0.1438(9) \\
$[\mathrm{Tay2}][a^2][a^2+a^2\phi_2][m_{\pi}<420\;\rm{MeV}]$ & 0.81 & 0.0079 & 0.1441(6) \\
$[\mathrm{Tay2}][a^2][a^2+a^2\phi_2][\beta>3.40\;\&\;m_{\pi}<420\;\rm{MeV}]$ & 0.62 & 0.0005 & 0.1437(9) \\
$[\mathrm{Tay2}][a^2][a^2+a^2\phi_2][m_{\pi}<350\;\rm{MeV}]$ & 0.52 & 0.0000 & 0.1442(7) \\
$[\mathrm{Tay2}][a^2][a^2+a^2\phi_2][m_{\pi}L>4.1]$ & 0.85 & 0.0027 & 0.1440(7) \\
$[\mathrm{Tay2}][a^2+a^2\phi_2][a^2][-]$ & 0.79 & 0.0268 & 0.1448(6) \\
$[\mathrm{Tay2}][a^2+a^2\phi_2][a^2][\beta>3.40]$ & 0.75 & 0.0080 & 0.1446(8) \\
$[\mathrm{Tay2}][a^2+a^2\phi_2][a^2][\beta>3.46]$ & 0.71 & 0.0014 & 0.1443(9) \\
$[\mathrm{Tay2}][a^2+a^2\phi_2][a^2][m_{\pi}<420\;\rm{MeV}]$ & 0.78 & 0.0075 & 0.1445(6) \\
$[\mathrm{Tay2}][a^2+a^2\phi_2][a^2][\beta>3.40\;\&\;m_{\pi}<420\;\rm{MeV}]$ & 0.63 & 0.0004 & 0.1441(10) \\
$[\mathrm{Tay2}][a^2+a^2\phi_2][a^2][m_{\pi}<350\;\rm{MeV}]$ & 0.50 & 0.0000 & 0.1445(7) \\
$[\mathrm{Tay2}][a^2+a^2\phi_2][a^2][m_{\pi}L>4.1]$ & 0.77 & 0.0011 & 0.1449(7) \\
$[\mathrm{Tay2}][a^2+a^2\phi_2][-]$ & 0.87 & 0.0413 & 0.1437(8) \\
$[\mathrm{Tay2}][a^2+a^2\phi_2][\beta>3.40]$ & 0.78 & 0.0084 & 0.1434(12) \\
$[\mathrm{Tay2}][a^2+a^2\phi_2][\beta>3.46]$ & 0.67 & 0.0008 & 0.1439(15) \\
$[\mathrm{Tay2}][a^2+a^2\phi_2][m_{\pi}<420\;\rm{MeV}]$ & 0.77 & 0.0039 & 0.1438(11) \\
$[\mathrm{Tay2}][a^2+a^2\phi_2][m_{\pi}L>4.1]$ & 0.84 & 0.0016 & 0.1435(12) \\
      \bottomrule
    \end{tabular}
  \end{center}
  \caption{
    Continued from Table~\ref{apex_ma:tab:comb2}.
  }
  \label{apex_ma:tab:comb3}
\end{table}

\clearpage

\begin{table}[htbp!]
  \begin{center}
    \begin{tabular}{ l | l | l | l }
      \toprule
      Model & p-value & $W$ & $\sqrt{t_0^{\mathrm{ph}}}\,[\mathrm{fm}]$ \\
      \toprule
      $[\mathrm{Tay4}][a^2][-]$ & 0.72 & 0.0157 & 0.1442(8) \\
$[\mathrm{Tay4}][a^2][\beta>3.40]$ & 0.68 & 0.0047 & 0.1440(9) \\
$[\mathrm{Tay4}][a^2][\beta>3.46]$ & 0.66 & 0.0008 & 0.1438(10) \\
$[\mathrm{Tay4}][a^2][m_{\pi}<420\;\rm{MeV}]$ & 0.72 & 0.0041 & 0.1440(8) \\
$[\mathrm{Tay4}][a^2][m_{\pi}L>4.1]$ & 0.68 & 0.0007 & 0.1450(10) \\
$[\mathrm{Tay4}][a^2][a^2+a^2\phi_2][-]$ & 0.84 & 0.0199 & 0.1439(8) \\
$[\mathrm{Tay4}][a^2][a^2+a^2\phi_2][\beta>3.40]$ & 0.76 & 0.0054 & 0.1437(9) \\
$[\mathrm{Tay4}][a^2][a^2+a^2\phi_2][\beta>3.46]$ & 0.66 & 0.0006 & 0.1436(10) \\
$[\mathrm{Tay4}][a^2][a^2+a^2\phi_2][m_{\pi}<420\;\rm{MeV}]$ & 0.77 & 0.0026 & 0.1439(8) \\
$[\mathrm{Tay4}][a^2][a^2+a^2\phi_2][m_{\pi}L>4.1]$ & 0.87 & 0.0012 & 0.1447(10) \\
$[\mathrm{Tay4}][a^2+a^2\phi_2][a^2][-]$ & 0.79 & 0.0159 & 0.1444(8) \\
$[\mathrm{Tay4}][a^2+a^2\phi_2][a^2][\beta>3.40]$ & 0.72 & 0.0037 & 0.1442(9) \\
$[\mathrm{Tay4}][a^2+a^2\phi_2][a^2][\beta>3.46]$ & 0.68 & 0.0006 & 0.1440(10) \\
$[\mathrm{Tay4}][a^2+a^2\phi_2][a^2][m_{\pi}<420\;\rm{MeV}]$ & 0.75 & 0.0023 & 0.1442(8) \\
$[\mathrm{Tay4}][a^2+a^2\phi_2][a^2][m_{\pi}L>4.1]$ & 0.73 & 0.0005 & 0.1452(10) \\
$[\mathrm{Tay4}][a^2+a^2\phi_2][-]$ & 0.83 & 0.0116 & 0.1437(9) \\
$[\mathrm{Tay4}][a^2+a^2\phi_2][\beta>3.40]$ & 0.73 & 0.0032 & 0.1434(12) \\
$[\mathrm{Tay4}][a^2+a^2\phi_2][\beta>3.46]$ & 0.62 & 0.0003 & 0.1440(15) \\
$[\mathrm{Tay4}][a^2+a^2\phi_2][m_{\pi}<420\;\rm{MeV}]$ & 0.71 & 0.0017 & 0.1439(11) \\
$[\mathrm{Tay4}][a^2+a^2\phi_2][m_{\pi}L>4.1]$ & 0.94 & 0.0021 & 0.1434(13) \\
      \bottomrule
    \end{tabular}
  \end{center}
  \caption{
    Continued from Table~\ref{apex_ma:tab:comb3}.
  }
  \label{apex_ma:tab:comb4}
\end{table}
%



\clearpage

\section{Model-variation results for $t_0^{\mathrm{ph}}$ based on $f_{\pi}^{\mathrm{isoQCD}}$}
\label{apex_model_av_t0_fpi}

This Appendix presents the results of the model variations considered in the determination of $t_0^{\mathrm{ph}}$, as described in Sec.~\ref{subsec:t0fpi}, based on physical input from $f_{\pi}^{\mathrm{isoQCD}}$ in \req{ch_ss:eq:isoQCD_fpi}.
Table~\ref{apex_ma:tab:labels_fpi} defines the labels used to identify each model, along with their corresponding functional forms and data cuts.
Tables~\ref{apex_ma:tab:comb_fpi1}--\ref{apex_ma:tab:comb_fpi2} report the determinations of $t_0^{\mathrm{ph}}$ from the combined analysis of Wilson and Wtm regularisations, including the associated p-values~\cite{Bruno:2022mfy} and weights $W$ computed via Eqs.~(\ref{eqn:TICchiexp})--(\ref{eqn:weight}).
\begin{table}[htbp!]
  \begin{center}
    \renewcommand{\arraystretch}{1.3}
    \begin{tabular}{l | l}
      \toprule
      Model label & Description \\
      \toprule
      $[\chi \mathrm{SU(3)}]$ & SU(3) $\chi$PT in Eqs.~(\ref{eq:SU3_fpi})~and~(\ref{eq:SU3_fk}) \\
      $[\chi \mathrm{SU(2)}]$ & SU(2) $\chi$PT in Eqs.~(\ref{eq:SU2_fpi})~and~(\ref{eq:SU2_fk}) \\
      \midrule
      $[a^2]$ & $\Oasq$ effects in \req{ch_ss:eq:a2} \\
      $[a^2\alphas^{\hat{\Gamma}}]$ & $\Oasq$ effects in \req{ch_ss:eq:aas} \\
      $[a^2+a^2\phi_2]$ & $\Oasq$ effects in \req{ch_ss:eq:a2phi2} \\
      \midrule
      $[-]$ & No cut in data \\
      $[\beta>3.40]$ & Remove $\beta=3.40$ ensembles [\req{ch_ss:eq:cuts1}]\\
      $[\beta>3.46]$ & Remove $\beta=3.40$ and $\beta=3.46$ ensembles [\req{ch_ss:eq:cuts2}]\\
      $[m_{\pi}<420\,\mathrm{MeV}]$ & Remove symmetric point ensembles [\req{ch_ss:eq:cuts3}]\\
      $[m_{\pi}L>4.1]$ & Remove ensembles with volumes $m_{\pi}L\leq4.1$ [\req{ch_ss:eq:cuts6}]\\
      \bottomrule
    \end{tabular}
    \renewcommand{\arraystretch}{1.0}
  \end{center}
  \caption{
    Correspondence between each fit model used in the combined continuum and chiral extrapolations of $\sqrt{8t_0}f_{\pi}$ and $\sqrt{8t_0}f_{K}$ -- with physical input from $f_{\pi}^{\mathrm{isoQCD}}$ in \req{ch_ss:eq:isoQCD_fpi} -- and the labels employed in Tables~\ref{apex_ma:tab:comb_fpi1}--\ref{apex_ma:tab:comb_fpi2} and Fig.~\ref{fig:MA_fpi}.
    The combined analysis of the Wilson and Wtm setups considers independent cutoff effects for each regularisation.
    Therefore, two successive labels are used: the first denotes the lattice artefacts for the Wilson setup, and the second for the Wtm mixed-action setup.
    If only one label is shown, it indicates that the same functional form for lattice artefacts is assumed for both regularisations, albeit with independent parameters.
  }
  \label{apex_ma:tab:labels_fpi}
\end{table}

\clearpage

\begin{table}[htbp!]
  \begin{center}
    \begin{tabular}{ l | l | l | l }
      \toprule
      Model & p-value & $W$ & $\sqrt{t_0^{\mathrm{ph}}}\,[\mathrm{fm}]$ \\
      \toprule

$[\chi {\rm SU(3)}][a^2][-]$ & 0.10 & 0.0054 & 0.1455(7) \\
$[\chi {\rm SU(3)}][a^2][\beta>3.40]$ & 0.23 & 0.0237 & 0.1445(9) \\
$[\chi {\rm SU(3)}][a^2][\beta>3.46]$ & 0.19 & 0.0050 & 0.1437(11) \\
$[\chi {\rm SU(3)}][a^2][m_{\pi}<420\;\rm{MeV}]$ & 0.14 & 0.0016 & 0.1454(7) \\
$[\chi {\rm SU(3)}][a^2][m_{\pi}L>4.1]$ & 0.11 & 0.0002 & 0.1456(7) \\
$[\chi {\rm SU(3)}][a^2\alpha_s^{\hat{\Gamma}}][-]$ & 0.11 & 0.0061 & 0.1456(7) \\
$[\chi {\rm SU(3)}][a^2\alpha_s^{\hat{\Gamma}}][\beta>3.40]$ & 0.21 & 0.0231 & 0.1445(9) \\
$[\chi {\rm SU(3)}][a^2\alpha_s^{\hat{\Gamma}}][\beta>3.46]$ & 0.19 & 0.0049 & 0.1437(11) \\
$[\chi {\rm SU(3)}][a^2\alpha_s^{\hat{\Gamma}}][m_{\pi}<420\;\rm{MeV}]$ & 0.15 & 0.0021 & 0.1454(7) \\
$[\chi {\rm SU(3)}][a^2\alpha_s^{\hat{\Gamma}}][m_{\pi}L>4.1]$ & 0.11 & 0.0002 & 0.1456(7) \\
$[\chi {\rm SU(3)}][a^2][a^2+a^2\phi_2][-]$ & 0.19 & 0.0625 & 0.1447(7) \\
$[\chi {\rm SU(3)}][a^2][a^2+a^2\phi_2][\beta>3.40]$ & 0.35 & 0.3014 & 0.1438(9) \\
$[\chi {\rm SU(3)}][a^2][a^2+a^2\phi_2][\beta>3.46]$ & 0.40 & 0.0481 & 0.1429(11) \\
$[\chi {\rm SU(3)}][a^2][a^2+a^2\phi_2][m_{\pi}<420\;\rm{MeV}]$ & 0.19 & 0.0041 & 0.1446(8) \\
$[\chi {\rm SU(3)}][a^2][a^2+a^2\phi_2][m_{\pi}L>4.1]$ & 0.26 & 0.0028 & 0.1446(8) \\
$[\chi {\rm SU(3)}][a^2+a^2\phi_2][a^2][[-]$ & 0.17 & 0.0288 & 0.1460(7) \\
$[\chi {\rm SU(3)}][a^2+a^2\phi_2][a^2][\beta>3.40]$ & 0.23 & 0.0401 & 0.1449(9) \\
$[\chi {\rm SU(3)}][a^2+a^2\phi_2][a^2][\beta>3.46]$ & 0.21 & 0.0063 & 0.1443(10) \\
$[\chi {\rm SU(3)}][a^2+a^2\phi_2][a^2][m_{\pi}<420\;\rm{MeV}]$ & 0.19 & 0.0013 & 0.1458(7) \\
$[\chi {\rm SU(3)}][a^2+a^2\phi_2][a^2][m_{\pi}L>4.1]$ & 0.16 & 0.0011 & 0.1462(7) \\
$[\chi {\rm SU(3)}][a^2+a^2\phi_2][-]$ & 0.18 & 0.0291 & 0.1451(9) \\
$[\chi {\rm SU(3)}][a^2+a^2\phi_2][\beta>3.40]$ & 0.32 & 0.0745 & 0.1436(11) \\
$[\chi {\rm SU(3)}][a^2+a^2\phi_2][\beta>3.46]$ & 0.41 & 0.0614 & 0.1421(13) \\
$[\chi {\rm SU(3)}][a^2+a^2\phi_2][m_{\pi}<420\;\rm{MeV}]$ & 0.19 & 0.0085 & 0.1451(9) \\
$[\chi {\rm SU(3)}][a^2+a^2\phi_2][m_{\pi}L>4.1]$ & 0.22 & 0.0025 & 0.1445(10) \\

        \bottomrule
    \end{tabular}
  \end{center}
  \caption{
    Model variations used to determine $\sqrt{t_0^{\mathrm{ph}}}$, as described in Sec.~\ref{subsec:t0fpi}, based on $f_{\pi}^{\mathrm{isoQCD}}$, employing the combined analysis of the Wilson unitary and the Wtm mixed action regularisations.
    The first column identifies the model via Table~\ref{apex_ma:tab:labels_fpi}; 
    the second reports the p-value~\cite{Bruno:2022mfy}, 
    the third gives the model weight $W$ from Eqs.~(\ref{eqn:TICchiexp})--(\ref{eqn:weight}),
    and the fourth lists the resulting $\sqrt{t_0^{\mathrm{ph}}}$.
    All models use the weight matrix $\mathcal{W}$ from \req{ch_ss:eq:Wpenal} in the chi-squared function of \req{eqn:chisq}, with coefficients $p_{\beta}$ and $p_{\phi_2}$ in \req{ch_ss:eq:penal} fixed as detailed in Sec.~\ref{subsec:t0} and Appendix~\ref{app_variations}.
  }
  \label{apex_ma:tab:comb_fpi1}
\end{table}

\clearpage

\begin{table}[htbp!]
  \begin{center}
    \begin{tabular}{ l | l | l | l }
      \toprule
      Model & p-value & $W$ & $\sqrt{t_0^{\mathrm{ph}}}\,[\mathrm{fm}]$ \\
      \toprule
$[\chi {\rm SU(2)}][a^2][-]$ & 0.12 & 0.0047 & 0.1457(7) \\
$[\chi {\rm SU(2)}][a^2][\beta>3.40]$ & 0.20 & 0.0096 & 0.1446(9) \\
$[\chi {\rm SU(2)}][a^2][\beta>3.46]$ & 0.18 & 0.0015 & 0.1439(10) \\
$[\chi {\rm SU(2)}][a^2][m_{\pi}<420\;\rm{MeV}]$ & 0.10 & 0.0002 & 0.1453(8) \\
$[\chi {\rm SU(2)}][a^2][m_{\pi}L>4.1]$ & 0.11 & 0.0001 & 0.1459(8) \\
$[\chi {\rm SU(2)}][a^2\alpha_s^{\hat{\Gamma}}][-]$ & 0.13 & 0.0054 & 0.1458(7) \\
$[\chi {\rm SU(2)}][a^2\alpha_s^{\hat{\Gamma}}][\beta>3.40]$ & 0.21 & 0.0087 & 0.1446(9) \\
$[\chi {\rm SU(2)}][a^2\alpha_s^{\hat{\Gamma}}][\beta>3.46]$ & 0.17 & 0.0014 & 0.1439(10) \\
$[\chi {\rm SU(2)}][a^2\alpha_s^{\hat{\Gamma}}][m_{\pi}<420\;\rm{MeV}]$ & 0.10 & 0.0002 & 0.1453(8) \\
$[\chi {\rm SU(2)}][a^2\alpha_s^{\hat{\Gamma}}][m_{\pi}L>4.1]$ & 0.12 & 0.0001 & 0.1460(8) \\
$[\chi {\rm SU(2)}][a^2][a^2+a^2\phi_2][-]$ & 0.21 & 0.0429 & 0.1449(8) \\
$[\chi {\rm SU(2)}][a^2][a^2+a^2\phi_2][\beta>3.40]$ & 0.34 & 0.0574 & 0.1439(10) \\
$[\chi {\rm SU(2)}][a^2][a^2+a^2\phi_2][\beta>3.46]$ & 0.36 & 0.0156 & 0.1431(11) \\
$[\chi {\rm SU(2)}][a^2][a^2+a^2\phi_2][m_{\pi}<420\;\rm{MeV}]$ & 0.15 & 0.0006 & 0.1447(9) \\
$[\chi {\rm SU(2)}][a^2][a^2+a^2\phi_2][m_{\pi}L>4.1]$ & 0.26 & 0.0023 & 0.1449(8) \\
$[\chi {\rm SU(2)}][a^2+a^2\phi_2][a^2][[-]$ & 0.19 & 0.0180 & 0.1461(7) \\
$[\chi {\rm SU(2)}][a^2+a^2\phi_2][a^2][\beta>3.40]$ & 0.22 & 0.0070 & 0.1451(9) \\
$[\chi {\rm SU(2)}][a^2+a^2\phi_2][a^2][\beta>3.46]$ & 0.19 & 0.0014 & 0.1446(10) \\
$[\chi {\rm SU(2)}][a^2+a^2\phi_2][a^2][m_{\pi}<420\;\rm{MeV}]$ & 0.14 & 0.0004 & 0.1457(8) \\
$[\chi {\rm SU(2)}][a^2+a^2\phi_2][a^2][m_{\pi}L>4.1]$ & 0.15 & 0.0004 & 0.1464(8) \\
$[\chi {\rm SU(2)}][a^2+a^2\phi_2][-]$ & 0.20 & 0.0121 & 0.1452(9) \\
$[\chi {\rm SU(2)}][a^2+a^2\phi_2][\beta>3.40]$ & 0.32 & 0.0326 & 0.1437(11) \\
$[\chi {\rm SU(2)}][a^2+a^2\phi_2][\beta>3.46]$ & 0.36 & 0.0315 & 0.1424(13) \\
$[\chi {\rm SU(2)}][a^2+a^2\phi_2][m_{\pi}<420\;\rm{MeV}]$ & 0.14 & 0.0004 & 0.1450(9) \\
$[\chi {\rm SU(2)}][a^2+a^2\phi_2][m_{\pi}L>4.1]$ & 0.22 & 0.0008 & 0.1447(11) \\
        \bottomrule
    \end{tabular}
  \end{center}
  \caption{
    Continued from Table~\ref{apex_ma:tab:comb_fpi1}.
  }
  \label{apex_ma:tab:comb_fpi2}
\end{table}
%



\section{Tables of observables entering the scale setting analysis}
\label{app:Tables}

In this Appendix, we compile the determinations of the observables entering the scale setting analysis.
Table~\ref{apex_ensembles:tab:obs_w} reports the unshifted values of the observables for the Wilson unitary setup.
Table~\ref{apex_ensembles:tab:obs_w_ms} presents the corresponding Wilson observables after applying the mass shift described in \req{eqn:derphi4} to enforce $\phi_4 = \phi_4^{\mathrm{ph}}$.
Table~\ref{apex_ensembles:tab:obs_comb} provides the Wtm mixed-action observables for which the condition $\phi_4 = \phi_4^{\mathrm{ph}}$ is enforced in both the sea and valence sectors.
\begin{sidewaystable}[htbp!]
  \begin{center}
    \begin{tabular}{c c c c c c c c}
      id & $t_0/a^2$ & $\phi_2^{\mathrm{W}}$ & $\phi_4^{\mathrm{W}}$ & $am_{12}^{\mathrm{W}}$ & $am_{13}^{\mathrm{W}}$ & $af_{\pi}^{\mathrm{W}}$ & $af_K^{\mathrm{W}}$ \\
      \toprule
      H101 & 2.8619(84) & 0.7664(30) & 1.1496(44) & 0.009233(39) & 0.009233(39) & 0.06289(26) & 0.06289(26) \\
      H102 & 2.8836(63) & 0.5508(32) & 1.1226(54) & 0.006522(38) & 0.010211(38) & 0.06037(22) & 0.06341(18) \\
      H105 & 2.8875(59) & 0.3487(32) & 1.1160(45) & 0.004015(33) & 0.011428(46) & 0.05701(40) & 0.06394(23) \\
      \midrule
      H400 & 3.6356(83) & 0.7775(40) & 1.1662(61) & 0.008302(39) & 0.008302(39) & 0.05635(22) & 0.05635(22) \\
      D450 & 3.6942(51) & 0.2091(25) & 1.1045(39) & 0.002134(22) & 0.010812(19) & 0.04974(34) & 0.05679(27) \\
      \midrule
      N202 & 5.1662(112) & 0.7434(24) & 1.1151(36) & 0.006863(13) & 0.006863(13) & 0.04805(13) & 0.04805(13) \\
      N203 & 5.1517(56) & 0.5189(23) & 1.1105(40) & 0.004750(13) & 0.007920(14) & 0.04622(12) & 0.04869(10) \\
      N200 & 5.1600(68) & 0.3525(17) & 1.1123(30) & 0.003158(11) & 0.008663(10) & 0.04418(10) & 0.04864(10) \\
      D200 & 5.1789(41) & 0.1772(10) & 1.1043(14) & 0.001552(6) & 0.009403(5) & 0.04223(10) & 0.04866(9) \\
      E250 & 5.2034(39) & 0.0734(16) & 1.0945(33) & 0.000640(15) & 0.009773(11) & 0.04003(68) & 0.04784(41) \\
      \midrule
      N300 & 8.5872(173) & 0.7749(40) & 1.1623(60) & 0.005513(6) & 0.005513(6) & 0.03778(11) & 0.03778(11) \\
      N302 & 8.5212(172) & 0.5184(34) & 1.1372(57) & 0.003720(7) & 0.006412(10) & 0.03635(11) & 0.03838(13) \\
      J303 & 8.6147(104) & 0.2912(13) & 1.1325(32) & 0.002050(5) & 0.007202(4) & 0.03404(12) & 0.03841(13) \\
      E300 & 8.6282(116) & 0.1334(10) & 1.1289(20) & 0.000934(5) & 0.007731(6) & 0.03229(20) & 0.03783(33) \\
      \midrule
      J500 & 13.9802(300) & 0.7431(35) & 1.1146(52) & 0.004220(5) & 0.004220(5) & 0.02986(21) & 0.02986(21) \\
      J501 & 14.0241(472) & 0.4907(21) & 1.1194(31) & 0.002741(3) & 0.004960(3) & 0.02821(22) & 0.02994(21) \\
      \bottomrule
    \end{tabular}
  \end{center}
  \caption{
    Determination of the unshifted observables from the Wilson unitary setup that enter the scale setting analysis.
    Improved and renormalised decay constants are reported, whereas for the PCAC quark masses, the bare and unimproved values are quoted.
    The various replicas of a given ensemble listed in Table~\ref{tab:CLS_ens} have been combined.
    The ensembles H102r001 and H102r002 are not replicas: although they share the same physical parameters, they differ in algorithmic parameters, and the corresponding observables are therefore averaged in the analysis.
    Similarly, the combination of the two replicas H105r001 and H105r002 has been averaged with the non-replica ensemble H105r005.
  }
  \label{apex_ensembles:tab:obs_w}
\end{sidewaystable}

\clearpage

\begin{sidewaystable}
  \begin{center}
    \begin{tabular}{c c c c c c c c}
      id & $t_0/a^2$ & $\phi_2^{\mathrm{W}}$ & $\phi_4^{\mathrm{W}}$ & $am_{12}^{\mathrm{W}}$ & $am_{13}^{\mathrm{W}}$ & $af_{\pi}^{\mathrm{W}}$ & $af_K^{\mathrm{W}}$ \\
      \toprule
H101 & 2.8865(91) & 0.7207(62) & 1.0811(101) & 0.008656(79) & 0.008656(79) & 0.06267(37) & 0.06267(37)\\
H102 & 2.9001(68) & 0.5341(41) & 1.0811(101) & 0.006307(50) & 0.009758(97) & 0.06028(26) & 0.06332(25)\\
H105 & 2.9050(65) & 0.3360(43) & 1.0811(101) & 0.003864(45) & 0.010976(100) & 0.05710(45) & 0.06427(30)\\
\midrule
H400 & 3.6647(82) & 0.7207(62) & 1.0811(101) & 0.007645(86) & 0.007645(86) & 0.05604(34) & 0.05604(34)\\
D450 & 3.7059(83) & 0.2026(32) & 1.0811(101) & 0.002060(37) & 0.010556(93) & 0.04972(34) & 0.05719(34)\\
\midrule
N202 & 5.1777(122) & 0.7207(62) & 1.0811(101) & 0.006653(68) & 0.006653(68) & 0.04816(21) & 0.04816(21)\\
N203 & 5.1635(72) & 0.5117(31) & 1.0811(101) & 0.004659(31) & 0.007668(77) & 0.04630(15) & 0.04891(17)\\
N200 & 5.1739(80) & 0.3454(26) & 1.0811(101) & 0.003096(20) & 0.008380(78) & 0.04423(14) & 0.04889(16)\\
D200 & 5.1903(65) & 0.1725(22) & 1.0811(101) & 0.001505(19) & 0.009190(74) & 0.04225(14) & 0.04900(15)\\
E250 & 5.2104(62) & 0.0708(23) & 1.0811(101) & 0.000606(28) & 0.009647(77) & 0.04003(69) & 0.04825(51)\\
\midrule
N300 & 8.6127(191) & 0.7207(62) & 1.0811(101) & 0.005096(56) & 0.005096(56) & 0.03763(23) & 0.03763(23)\\
N302 & 8.5427(197) & 0.5083(38) & 1.0811(101) & 0.003600(26) & 0.006047(75) & 0.03637(21) & 0.03848(26)\\
J303 & 8.6378(125) & 0.2836(24) & 1.0811(101) & 0.001999(14) & 0.006842(71) & 0.03402(24) & 0.03852(19)\\
E300 & 8.6517(142) & 0.1274(18) & 1.0811(101) & 0.000885(13) & 0.007405(60) & 0.03223(22) & 0.03794(39)\\
\midrule
J500 & 13.9907(312) & 0.7207(62) & 1.0811(101) & 0.004097(39) & 0.004097(39) & 0.02991(23) & 0.02991(23)\\
J501 & 14.0387(484) & 0.4854(30) & 1.0811(101) & 0.002687(14) & 0.004761(49) & 0.02826(24) & 0.03005(23)\\
      \bottomrule
    \end{tabular}
  \end{center}
  \caption{
    Determination of the observables from the Wilson unitary setup that enter the scale setting analysis, after applying the mass shift described in \req{eqn:derphi4} to enforce $\phi_4 = \phi_4^{\mathrm{ph}}$, as specified in \req{ch_ss:eq:phi4ph}.
    Further details are provided in the caption of Table~\ref{apex_ensembles:tab:obs_w}.
  }
  \label{apex_ensembles:tab:obs_w_ms}
\end{sidewaystable}

\clearpage

\begin{sidewaystable}
  \begin{center}
    \begin{tabular}{c c c c c c c c}
      id & $t_0/a^2$ & $\phi_2^{\mathrm{Wtm}}$ & $\phi_4^{\mathrm{Wtm}}$ & $a\mu_{12}$ & $a\mu_{13}$ & $af_{\pi}^{\mathrm{Wtm}}$ & $af_K^{\mathrm{Wtm}}$ \\
      \toprule
H101 & 2.8865(91) & 0.7207(62) & 1.0811(101) & 0.006321(60) & 0.006321(60) & 0.06647(29) & 0.06647(29)\\
H102 & 2.9001(68) & 0.5341(41) & 1.0811(101) & 0.004627(36) & 0.007263(58) & 0.06425(22) & 0.06789(17)\\
H105 & 2.9050(65) & 0.3360(43) & 1.0811(101) & 0.002861(34) & 0.008031(70) & 0.06163(23) & 0.06825(20)\\
\midrule
H400 & 3.6647(82) & 0.7207(62) & 1.0811(101) & 0.005724(57) & 0.005724(57) & 0.05889(27) & 0.05889(27)\\
D450 & 3.7059(83) & 0.2026(32) & 1.0811(101) & 0.001523(25) & 0.007834(87) & 0.05299(31) & 0.06047(30)\\
\midrule
N202 & 5.1777(122) & 0.7207(62) & 1.0811(101) & 0.005042(50) & 0.005042(50) & 0.04956(19) & 0.04956(19)\\
N203 & 5.1635(72) & 0.5117(31) & 1.0811(101) & 0.003545(23) & 0.005801(62) & 0.04809(20) & 0.05068(20)\\
N200 & 5.1739(80) & 0.3454(26) & 1.0811(101) & 0.002357(17) & 0.006389(59) & 0.04610(27) & 0.05070(14)\\
D200 & 5.1903(65) & 0.1725(22) & 1.0811(101) & 0.001187(13) & 0.007019(79) & 0.04463(11) & 0.05060(17)\\
E250 & 5.2104(62) & 0.0708(23) & 1.0811(101) & 0.000462(15) & 0.007352(88) & 0.04200(37) & 0.05012(45)\\
\midrule
N300 & 8.6127(191) & 0.7207(62) & 1.0811(101) & 0.004044(40) & 0.004044(40) & 0.03851(19) & 0.03851(19)\\
N302 & 8.5427(197) & 0.5083(38) & 1.0811(101) & 0.002848(17) & 0.004693(62) & 0.03725(19) & 0.03921(21)\\
J303 & 8.6378(125) & 0.2836(24) & 1.0811(101) & 0.001571(11) & 0.005324(56) & 0.03527(15) & 0.03929(20)\\
E300 & 8.6517(142) & 0.1274(18) & 1.0811(101) & 0.000707(9) & 0.005803(63) & 0.03339(16) & 0.03907(45)\\
\midrule
J500 & 13.9907(312) & 0.7207(62) & 1.0811(101) & 0.003228(33) & 0.003228(33) & 0.03020(23) & 0.03020(23)\\
J501 & 14.0387(484) & 0.4854(30) & 1.0811(101) & 0.002166(11) & 0.003768(43) & 0.02868(19) & 0.03031(22)\\
      \bottomrule
    \end{tabular}
  \end{center}
  \caption{
    Determination of the observables from the Wtm mixed-action setup that enter the scale setting analysis, after applying the mass shift described in \req{eqn:derphi4} to enforce $\phi_4 = \phi_4^{\mathrm{ph}}$, as specified in \req{ch_ss:eq:phi4ph} in the sea sector.
    The matching of the sea and valence quark masses, and the tuning of the Wtm regularisation to maximal twist, are carried out as explained in Sec.~\ref{sec:match}.
    The values of $t_0/a^2$, $\phi_2$, and $\phi_4$ are, by construction, identical to those in Table~\ref{apex_ensembles:tab:obs_w_ms}.
    The bare up, down, and strange twisted quark masses are used in the combination $\mu_{ij} = \frac{1}{2}\left( \mu_i + \mu_j \right)$.
    Further details are provided in the caption of Table~\ref{apex_ensembles:tab:obs_w}.
  }
  \label{apex_ensembles:tab:obs_comb}
\end{sidewaystable}

\end{appendix}
%


\cleardoublepage

\bibliographystyle{JHEP}
\bibliography{biblio}

@article{1009.5228,
    author = "Schaefer, Stefan and Sommer, Rainer and Virotta, Francesco",
    collaboration = "ALPHA",
    title = "{Critical slowing down and error analysis in lattice QCD simulations}",
    eprint = "1009.5228",
    archivePrefix = "arXiv",
    primaryClass = "hep-lat",
    reportNumber = "DESY-10-151, SFB-CPP-10-81, HU-EP-10-55",
    doi = "10.1016/j.nuclphysb.2010.11.020",
    journal = "Nucl. Phys. B",
    volume = "845",
    pages = "93--119",
    year = "2011"
}

@article{Kennedy:1998cu,
    author = "Kennedy, A. D. and Horv{\'a}th, Ivan and Sint, Stefan",
    editor = "DeGrand, Thomas A. and DeTar, Carleton E. and Sugar, R. and Toussaint, D.",
    title = "{A New exact method for dynamical fermion computations with nonlocal actions}",
    eprint = "hep-lat/9809092",
    archivePrefix = "arXiv",
    doi = "10.1016/S0920-5632(99)85217-7",
    journal = "Nucl. Phys. B Proc. Suppl.",
    volume = "73",
    pages = "834--836",
    year = "1999"
}

@article{Mohler:2020txx,
    author = "Mohler, Daniel and Schaefer, Stefan",
    title = "{Remarks on strange-quark simulations with Wilson fermions}",
    eprint = "2003.13359",
    archivePrefix = "arXiv",
    primaryClass = "hep-lat",
    reportNumber = "DESY-20-041, MITP/20-010, DESY 20-041 ; MITP/20-010",
    doi = "10.1103/PhysRevD.102.074506",
    journal = "Phys. Rev. D",
    volume = "102",
    number = "7",
    pages = "074506",
    year = "2020"
}

@article{0810.0946,
    author = "L{\"u}scher, Martin and Palombi, Filippo",
    editor = "Aubin, Christopher and Cohen, Saul and Dawson, Chris and Dudek, Jozef and Edwards, Robert and Joo, Balint and Lin, Huey-Wen and Orginos, Kostas and Richards, David and Thacker, Hank",
    title = "{Fluctuations and reweighting of the quark determinant on large lattices}",
    eprint = "0810.0946",
    archivePrefix = "arXiv",
    primaryClass = "hep-lat",
    reportNumber = "CERN-PH-TH-2008-205",
    doi = "10.22323/1.066.0049",
    journal = "PoS",
    volume = "LATTICE2008",
    pages = "049",
    year = "2008"
}

@article{Bhattacharya:2005rb,
    author = "Bhattacharya, Tanmoy and Gupta, Rajan and Lee, Weonjong and Sharpe, Stephen R. and Wu, Jackson M. S.",
    title = "{Improved bilinears in lattice QCD with non-degenerate quarks}",
    eprint = "hep-lat/0511014",
    archivePrefix = "arXiv",
    reportNumber = "LA-UR-05-8131",
    doi = "10.1103/PhysRevD.73.034504",
    journal = "Phys. Rev. D",
    volume = "73",
    pages = "034504",
    year = "2006"
}

@article{Frezzotti:2001ea,
    author = "Frezzotti, Roberto and Sint, Stefan and Weisz, Peter",
    collaboration = "ALPHA",
    title = "{O(a) improved twisted mass lattice QCD}",
    eprint = "hep-lat/0104014",
    archivePrefix = "arXiv",
    reportNumber = "CERN-TH-2001-014, MPI-PHT-2000-52, BICOCCA-FT-01-02",
    doi = "10.1088/1126-6708/2001/07/048",
    journal = "JHEP",
    volume = "07",
    pages = "048",
    year = "2001"
}

@article{Frezzotti:2003ni,
    author = "Frezzotti, R. and Rossi, G. C.",
    title = "{Chirally improving Wilson fermions. 1. O(a) improvement}",
    eprint = "hep-lat/0306014",
    archivePrefix = "arXiv",
    reportNumber = "BICOCCA-FT-03-15, ROM2F-2003-15",
    doi = "10.1088/1126-6708/2004/08/007",
    journal = "JHEP",
    volume = "08",
    pages = "007",
    year = "2004"
}

@article{Symanzik:1983dc,
    author = "Symanzik, K.",
    title = "{Continuum Limit and Improved Action in Lattice Theories. 1. Principles and $\varphi^4$ Theory}",
    reportNumber = "DESY-83-016",
    doi = "10.1016/0550-3213(83)90468-6",
    journal = "Nucl. Phys. B",
    volume = "226",
    pages = "187--204",
    year = "1983"
}

@article{Symanzik:1983gh,
    author = "Symanzik, K.",
    title = "{Continuum Limit and Improved Action in Lattice Theories. 2. O(N) Nonlinear Sigma Model in Perturbation Theory}",
    reportNumber = "DESY-83-026",
    doi = "10.1016/0550-3213(83)90469-8",
    journal = "Nucl. Phys. B",
    volume = "226",
    pages = "205--227",
    year = "1983"
}

@article{Luscher:1984xn,
    author = "L{\"u}scher, M. and Weisz, P.",
    title = "{On-shell improved lattice gauge theories}",
    reportNumber = "DESY-84-030",
    doi = "10.1007/BF01205792",
    journal = "Commun. Math. Phys.",
    volume = "98",
    number = "3",
    pages = "433",
    year = "1985",
    note = "[Erratum: Commun.Math.Phys. 98, 433 (1985)]"
}

@article{Balog:2009np,
    author = "Balog, Janos and Niedermayer, Ferenc and Weisz, Peter",
    title = "{The Puzzle of apparent linear lattice artifacts in the 2d non-linear sigma-model and Symanzik's solution}",
    eprint = "0905.1730",
    archivePrefix = "arXiv",
    primaryClass = "hep-lat",
    reportNumber = "MPP-2009-40",
    doi = "10.1016/j.nuclphysb.2009.09.007",
    journal = "Nucl. Phys. B",
    volume = "824",
    pages = "563--615",
    year = "2010"
}

@article{Balog:2009yj,
    author = "Balog, Janos and Niedermayer, Ferenc and Weisz, Peter",
    title = "{Logarithmic corrections to O(a**2) lattice artifacts}",
    eprint = "0901.4033",
    archivePrefix = "arXiv",
    primaryClass = "hep-lat",
    reportNumber = "MPP-2009-10",
    doi = "10.1016/j.physletb.2009.04.082",
    journal = "Phys. Lett. B",
    volume = "676",
    pages = "188--192",
    year = "2009"
}

@article{Husung:2019ytz,
    author = "Husung, Nikolai and Marquard, Peter and Sommer, Rainer",
    title = "{Asymptotic behavior of cutoff effects in Yang\textendash{}Mills theory and in Wilson\textquoteright{}s lattice QCD}",
    eprint = "1912.08498",
    archivePrefix = "arXiv",
    primaryClass = "hep-lat",
    reportNumber = "DESY-19-188, DESY 19-188",
    doi = "10.1140/epjc/s10052-020-7685-4",
    journal = "Eur. Phys. J. C",
    volume = "80",
    number = "3",
    pages = "200",
    year = "2020"
}

@phdthesis{Husung:2021tml,
    author = "Husung, Nikolai Andr\'e",
    title = "{Logarithmic corrections in Symanzik\textquoteright{}s effective theory of lattice QCD}",
    doi = "10.18452/22944",
    school = "Humboldt U., Berlin, Humboldt U., Berlin",
    month = "8",
    year = "2021"
}

@article{Husung:2021mfl,
    author = "Husung, Nikolai and Marquard, Peter and Sommer, Rainer",
    title = "{The asymptotic approach to the continuum of lattice QCD spectral observables}",
    eprint = "2111.02347",
    archivePrefix = "arXiv",
    primaryClass = "hep-lat",
    reportNumber = "DESY-21-177, HU-EP-21/45",
    doi = "10.1016/j.physletb.2022.137069",
    journal = "Phys. Lett. B",
    volume = "829",
    pages = "137069",
    year = "2022"
}

@article{Husung:2022kvi,
    author = "Husung, Nikolai",
    title = "{Logarithmic corrections to O(a) and O($a^2$) effects in lattice QCD with Wilson or Ginsparg\textendash{}Wilson quarks}",
    eprint = "2206.03536",
    archivePrefix = "arXiv",
    primaryClass = "hep-lat",
    doi = "10.1140/epjc/s10052-023-11258-8",
    journal = "Eur. Phys. J. C",
    volume = "83",
    number = "2",
    pages = "142",
    year = "2023",
    note = "[Erratum: Eur.Phys.J.C 83, 144 (2023)]"
}

@article{Sheikholeslami:1985ij,
    author = "Sheikholeslami, B. and Wohlert, R.",
    title = "{Improved Continuum Limit Lattice Action for QCD with Wilson Fermions}",
    reportNumber = "DESY-85-024",
    doi = "10.1016/0550-3213(85)90002-1",
    journal = "Nucl. Phys. B",
    volume = "259",
    pages = "572",
    year = "1985"
}

@article{Frezzotti:2000nk,
    author = "Frezzotti, Roberto and Grassi, Pietro Antonio and Sint, Stefan and Weisz, Peter",
    collaboration = "Alpha",
    title = "{Lattice QCD with a chirally twisted mass term}",
    eprint = "hep-lat/0101001",
    archivePrefix = "arXiv",
    reportNumber = "CERN-TH-2000-384, MPI-PHT-2000-51, BICOCCA-FT-0027, NYU-TH-00-09-13",
    doi = "10.1088/1126-6708/2001/08/058",
    journal = "JHEP",
    volume = "08",
    pages = "058",
    year = "2001"
}

@article{Frezzotti:2003xj,
    author = "Frezzotti, R. and Rossi, G. C.",
    editor = "Kalloniatis, A. C. and Leinweber, D. B. and Williams, A. G.",
    title = "{Twisted mass lattice QCD with mass nondegenerate quarks}",
    eprint = "hep-lat/0311008",
    archivePrefix = "arXiv",
    reportNumber = "BICOCCA-FT-03-31, ROM2F-2003-32",
    doi = "10.1016/S0920-5632(03)02477-0",
    journal = "Nucl. Phys. B Proc. Suppl.",
    volume = "128",
    pages = "193--202",
    year = "2004"
}

@article{Capitani:2000xi,
    author = "Capitani, S. and G{\"o}ckeler, M. and Horsley, R. and Perlt, H. and Rakow, Paul E. L. and Schierholz, G. and Schiller, A.",
    title = "{Renormalization and off-shell improvement in lattice perturbation theory}",
    eprint = "hep-lat/0007004",
    archivePrefix = "arXiv",
    reportNumber = "DESY-00-049, TPR-00-07, LU-ITP-2000-002, MIT-CTP-2981, HUB-EP-00-26",
    doi = "10.1016/S0550-3213(00)00590-3",
    journal = "Nucl. Phys. B",
    volume = "593",
    pages = "183--228",
    year = "2001"
}

@article{Frezzotti:2005gi,
    author = "Frezzotti, R. and Martinelli, G. and Papinutto, M. and Rossi, G. C.",
    title = "{Reducing cutoff effects in maximally twisted lattice QCD close to the chiral limit}",
    eprint = "hep-lat/0503034",
    archivePrefix = "arXiv",
    reportNumber = "DESY-04-249, ROM2F-2005-15, BICOCCA-FT-05-6",
    doi = "10.1088/1126-6708/2006/04/038",
    journal = "JHEP",
    volume = "04",
    pages = "038",
    year = "2006"
}

@article{Shindler:2005vj,
    author = "Shindler, Andrea",
    editor = "Michael, Christopher",
    title = "{Twisted mass lattice QCD: Recent developments and results}",
    eprint = "hep-lat/0511002",
    archivePrefix = "arXiv",
    reportNumber = "DESY-05-216, SFB-CPP-05-73",
    doi = "10.22323/1.020.0014",
    journal = "PoS",
    volume = "LAT2005",
    pages = "014",
    year = "2006"
}

@article{Sint:2005qz,
    author = "Sint, Stefan",
    editor = "Michael, Christopher",
    title = "{The Schr{\"o}dinger functional with chirally rotated boundary conditions}",
    eprint = "hep-lat/0511034",
    archivePrefix = "arXiv",
    reportNumber = "FTUAM-05-13, IFT-UAM-CSIC-05-39",
    doi = "10.22323/1.020.0235",
    journal = "PoS",
    volume = "LAT2005",
    pages = "235",
    year = "2006"
}

@article{Shindler:2007vp,
    author = "Shindler, Andrea",
    title = "{Twisted mass lattice QCD}",
    eprint = "0707.4093",
    archivePrefix = "arXiv",
    primaryClass = "hep-lat",
    reportNumber = "DESY-07-080, SFB-CPP-07-25",
    doi = "10.1016/j.physrep.2008.03.001",
    journal = "Phys. Rept.",
    volume = "461",
    pages = "37--110",
    year = "2008"
}

@article{Bussone:2018ljj,
    author = "Bussone, Andrea and Chaves, Sergio and Herdo\'\i{}za, Gregorio and Pena, Carlos and Preti, David and Romero, Jos\'e \'Angel and Ugarrio, Javier",
    collaboration = "ALPHA",
    title = "{Heavy-quark physics with a tmQCD valence action}",
    eprint = "1812.01474",
    archivePrefix = "arXiv",
    primaryClass = "hep-lat",
    reportNumber = "IFT-UAM/CSIC-18-100; FTUAM-18-22",
    doi = "10.22323/1.334.0270",
    journal = "PoS",
    volume = "LATTICE2018",
    pages = "270",
    year = "2019"
}

@article{hep-lat/0603029,
    author = "L{\"u}scher, Martin",
    title = "{The Schr{\"o}dinger functional in lattice QCD with exact chiral symmetry}",
    eprint = "hep-lat/0603029",
    archivePrefix = "arXiv",
    reportNumber = "CERN-PH-TH-2006-052",
    doi = "10.1088/1126-6708/2006/05/042",
    journal = "JHEP",
    volume = "05",
    pages = "042",
    year = "2006"
}

@article{hep-lat/9312079,
    author = "Sint, Stefan",
    title = "{On the Schr{\"o}dinger functional in QCD}",
    eprint = "hep-lat/9312079",
    archivePrefix = "arXiv",
    reportNumber = "DESY-93-165, DESY-93--165",
    doi = "10.1016/0550-3213(94)90228-3",
    journal = "Nucl. Phys. B",
    volume = "421",
    pages = "135--158",
    year = "1994"
}

@article{hep-lat/9903040,
    author = "Guagnelli, Marco and Heitger, Jochen and Sommer, Rainer and Wittig, Hartmut",
    collaboration = "ALPHA",
    title = "{Hadron masses and matrix elements from the QCD Schr{\"o}dinger functional}",
    eprint = "hep-lat/9903040",
    archivePrefix = "arXiv",
    reportNumber = "DESY-99-023, OUTP-99-07-P, OUTP-99-07P",
    doi = "10.1016/S0550-3213(99)00466-6",
    journal = "Nucl. Phys. B",
    volume = "560",
    pages = "465--481",
    year = "1999"
}

@phdthesis{Bruno:2015hfq,
    author = "Bruno, Mattia",
    title = "{The energy scale of the 3-flavour Lambda parameter}",
    doi = "10.18452/17516",
    school = "Humboldt U., Berlin",
    year = "2015"
}

@article{Aoki:2016frl,
    author = "Aoki, S. and others",
    title = "{Review of lattice results concerning low-energy particle physics}",
    eprint = "1607.00299",
    archivePrefix = "arXiv",
    primaryClass = "hep-lat",
    reportNumber = "CP3-Origins-2016-023, DESY-16-111, DIAS-2016-23, Edinburgh-2016-11, FTUAM-16-23, HIM-2016-02, IFT-UAM-CSIC-16-057, LPT-Orsay-16-47, MITP-16-059, RM3-TH-16-7, ROM2F-2016-05, YITP-16-77",
    doi = "10.1140/epjc/s10052-016-4509-7",
    journal = "Eur. Phys. J. C",
    volume = "77",
    number = "2",
    pages = "112",
    year = "2017"
}

@article{FlavourLatticeAveragingGroupFLAG:2021npn,
    author = "Aoki, Y. and others",
    collaboration = "Flavour Lattice Averaging Group (FLAG)",
    title = "{FLAG Review 2021}",
    eprint = "2111.09849",
    archivePrefix = "arXiv",
    primaryClass = "hep-lat",
    reportNumber = "CERN-TH-2021-191, JLAB-THY-21-3528, FERMILAB-PUB-21-620-SCD-T",
    doi = "10.1140/epjc/s10052-022-10536-1",
    journal = "Eur. Phys. J. C",
    volume = "82",
    number = "10",
    pages = "869",
    year = "2022"
}

@article{Bruno:2014jqa,
    author = "Bruno, Mattia and others",
    title = "{Simulation of QCD with N$_{f} =$ 2 $+$ 1 flavors of non-perturbatively improved Wilson fermions}",
    eprint = "1411.3982",
    archivePrefix = "arXiv",
    primaryClass = "hep-lat",
    reportNumber = "DESY-14-216, FTUAM-14-48, HIM-2014-01, HU-EP-14-51, MITP-14-091, SFB-CPP-14-89, IFT-UAM-CSIC-14-117",
    doi = "10.1007/JHEP02(2015)043",
    journal = "JHEP",
    volume = "02",
    pages = "043",
    year = "2015"
}

@article{Bruno:2016plf,
    author = "Bruno, Mattia and Korzec, Tomasz and Schaefer, Stefan",
    title = "{Setting the scale for the CLS $2 + 1$ flavor ensembles}",
    eprint = "1608.08900",
    archivePrefix = "arXiv",
    primaryClass = "hep-lat",
    reportNumber = "DESY-16-162, WUB-16-05",
    doi = "10.1103/PhysRevD.95.074504",
    journal = "Phys. Rev. D",
    volume = "95",
    number = "7",
    pages = "074504",
    year = "2017"
}

@article{Mohler:2017wnb,
    author = "Mohler, Daniel and Schaefer, Stefan and Simeth, Jakob",
    editor = "Della Morte, M. and Fritzsch, P. and G\'amiz S\'anchez, E. and Pena Ruano, C.",
    title = "{CLS 2+1 flavor simulations at physical light- and strange-quark masses}",
    eprint = "1712.04884",
    archivePrefix = "arXiv",
    primaryClass = "hep-lat",
    reportNumber = "DESY-17-166, HIM-2017-08, MITP-17-093",
    doi = "10.1051/epjconf/201817502010",
    journal = "EPJ Web Conf.",
    volume = "175",
    pages = "02010",
    year = "2018"
}

@article{Luscher:1985zq,
    author = "L{\"u}scher, M. and Weisz, P.",
    title = "{Computation of the Action for On-Shell Improved Lattice Gauge Theories at Weak Coupling}",
    reportNumber = "DESY-85-035",
    doi = "10.1016/0370-2693(85)90966-9",
    journal = "Phys. Lett. B",
    volume = "158",
    pages = "250--254",
    year = "1985"
}

@article{Wilson:1974sk,
    author = "Wilson, Kenneth G.",
    editor = "Taylor, J. C.",
    title = "{Confinement of Quarks}",
    reportNumber = "CLNS-262",
    doi = "10.1103/PhysRevD.10.2445",
    journal = "Phys. Rev. D",
    volume = "10",
    pages = "2445--2459",
    year = "1974"
}

@article{Bulava:2013cta,
    author = "Bulava, John and Schaefer, Stefan",
    title = "{Improvement of $N_f$ = 3 lattice QCD with Wilson fermions and tree-level improved gauge action}",
    eprint = "1304.7093",
    archivePrefix = "arXiv",
    primaryClass = "hep-lat",
    reportNumber = "CERN-PH-TH-2013-082, TCD-MATH-13-06",
    doi = "10.1016/j.nuclphysb.2013.05.019",
    journal = "Nucl. Phys. B",
    volume = "874",
    pages = "188--197",
    year = "2013"
}

@article{Pena:2004gb,
    author = "Pena, Carlos and Sint, Stefan and Vladikas, Anastassios",
    title = "{Twisted mass QCD and lattice approaches to the Delta I = 1/2 rule}",
    eprint = "hep-lat/0405028",
    archivePrefix = "arXiv",
    reportNumber = "DESY-04-079, FTUAM-03-22, IFT-UAM-CSIC-03-31, ROM2F-2004-10",
    doi = "10.1088/1126-6708/2004/09/069",
    journal = "JHEP",
    volume = "09",
    pages = "069",
    year = "2004"
}

@article{Frezzotti:2004wz,
    author = "Frezzotti, R. and Rossi, G. C.",
    title = "{Chirally improving Wilson fermions. II. Four-quark operators}",
    eprint = "hep-lat/0407002",
    archivePrefix = "arXiv",
    reportNumber = "BICOCCA-FT-04-7, ROM2F-2004-12, DESY-04-069",
    doi = "10.1088/1126-6708/2004/10/070",
    journal = "JHEP",
    volume = "10",
    pages = "070",
    year = "2004"
}

@article{Bruno:2014ova,
    author = "Bruno, Mattia and Schaefer, Stefan and Sommer, Rainer",
    collaboration = "ALPHA",
    title = "{Topological susceptibility and the sampling of field space in N$_{f}$ = 2 lattice QCD simulations}",
    eprint = "1406.5363",
    archivePrefix = "arXiv",
    primaryClass = "hep-lat",
    reportNumber = "DESY-14-101, SFB-CPP-14-29",
    doi = "10.1007/JHEP08(2014)150",
    journal = "JHEP",
    volume = "08",
    pages = "150",
    year = "2014"
}

@article{Luscher:2010we,
    author = "L{\"u}scher, Martin",
    editor = "Rossi, Giancarlo",
    title = "{Topology, the Wilson flow and the HMC algorithm}",
    eprint = "1009.5877",
    archivePrefix = "arXiv",
    primaryClass = "hep-lat",
    reportNumber = "CERN-PH-TH-2010-207",
    doi = "10.22323/1.105.0015",
    journal = "PoS",
    volume = "LATTICE2010",
    pages = "015",
    year = "2010"
}

@article{Luscher:2012av,
    author = "L{\"u}scher, Martin and Schaefer, Stefan",
    title = "{Lattice QCD with open boundary conditions and twisted-mass reweighting}",
    eprint = "1206.2809",
    archivePrefix = "arXiv",
    primaryClass = "hep-lat",
    reportNumber = "CERN-PH-TH-2012-161",
    doi = "10.1016/j.cpc.2012.10.003",
    journal = "Comput. Phys. Commun.",
    volume = "184",
    pages = "519--528",
    year = "2013"
}

@article{Luscher:2011kk,
    author = "L{\"u}scher, Martin and Schaefer, Stefan",
    title = "{Lattice QCD without topology barriers}",
    eprint = "1105.4749",
    archivePrefix = "arXiv",
    primaryClass = "hep-lat",
    reportNumber = "CERN-PH-TH-2011-116",
    doi = "10.1007/JHEP07(2011)036",
    journal = "JHEP",
    volume = "07",
    pages = "036",
    year = "2011"
}

@article{DallaBrida:2018tpn,
    author = "Dalla Brida, Mattia and Korzec, Tomasz and Sint, Stefan and Vilaseca, Pol",
    title = "{High precision renormalization of the flavour non-singlet Noether currents in lattice QCD with Wilson quarks}",
    eprint = "1808.09236",
    archivePrefix = "arXiv",
    primaryClass = "hep-lat",
    doi = "10.1140/epjc/s10052-018-6514-5",
    journal = "Eur. Phys. J. C",
    volume = "79",
    number = "1",
    pages = "23",
    year = "2019"
}

@article{Taniguchi:1998pf,
    author = "Taniguchi, Yusuke and Ukawa, Akira",
    title = "{Perturbative calculation of improvement coefficients to $\mbox{O}(g^2a)$ for bilinear quark operators in lattice QCD}",
    eprint = "hep-lat/9806015",
    archivePrefix = "arXiv",
    reportNumber = "UTCCP-P-40, UTHEP-384",
    doi = "10.1103/PhysRevD.58.114503",
    journal = "Phys. Rev. D",
    volume = "58",
    pages = "114503",
    year = "1998"
}

@article{Bar:2013ora,
    author = "B{\"a}r, Oliver and Golterman, Maarten",
    title = "{Chiral perturbation theory for gradient flow observables}",
    eprint = "1312.4999",
    archivePrefix = "arXiv",
    primaryClass = "hep-lat",
    doi = "10.1103/PhysRevD.89.034505",
    journal = "Phys. Rev. D",
    volume = "89",
    number = "3",
    pages = "034505",
    year = "2014",
    note = "[Erratum: Phys.Rev.D 89, 099905 (2014)]"
}

@article{Bietenholz:2011qq,
    author = "Bietenholz, W. and others",
    title = "{Flavour blindness and patterns of flavour symmetry breaking in lattice simulations of up, down and strange quarks}",
    eprint = "1102.5300",
    archivePrefix = "arXiv",
    primaryClass = "hep-lat",
    reportNumber = "DESY-11-030, EDINBURGH-2011-09, LTH-909",
    doi = "10.1103/PhysRevD.84.054509",
    journal = "Phys. Rev. D",
    volume = "84",
    pages = "054509",
    year = "2011"
}

@article{DallaBrida:2016kgh,
    author = "Dalla Brida, Mattia and Fritzsch, Patrick and Korzec, Tomasz and Ramos, Alberto and Sint, Stefan and Sommer, Rainer",
    collaboration = "ALPHA",
    title = "{Slow running of the Gradient Flow coupling from 200 MeV to 4 GeV in $N_{\rm f}=3$ QCD}",
    eprint = "1607.06423",
    archivePrefix = "arXiv",
    primaryClass = "hep-lat",
    reportNumber = "CERN-TH-2016-160, DESY-16-133, IFT-UAM-CSIC-16-067, WUB-16-02, YITP-16-89",
    doi = "10.1103/PhysRevD.95.014507",
    journal = "Phys. Rev. D",
    volume = "95",
    number = "1",
    pages = "014507",
    year = "2017"
}

@phdthesis{Strassberger:2023xnj,
    author = "Stra\ss{}berger, Ben",
    title = "{Towards Higher Precision Lattice QCD Results: Improved Scale Setting and Domain Decomposition Solvers}",
    doi = "10.18452/26517",
    school = "Humboldt U., Berlin",
    year = "2023"
}

@article{Bussone:2023kag,
    author = "Bussone, Andrea and Conigli, Alessandro and Frison, Julien and Herdo\'\i{}za, Gregorio and Pena, Carlos and Preti, David and S\'aez, Alejandro and Ugarrio, Javier",
    collaboration = "Alpha",
    title = "{Hadronic physics from a Wilson fermion mixed-action approach: charm quark mass and $D_{(s)}$ meson decay constants}",
    eprint = "2309.14154",
    archivePrefix = "arXiv",
    primaryClass = "hep-lat",
    reportNumber = "IFT-UAM/CSIC-23-114",
    doi = "10.1140/epjc/s10052-024-12816-4",
    journal = "Eur. Phys. J. C",
    volume = "84",
    number = "5",
    pages = "506",
    year = "2024"
}

@article{DellaMorte:2017dyu,
    author = {Della Morte, M. and Francis, A. and G\"ulpers, V. and Herdo\'\i{}za, G. and von Hippel, G. and Horch, H. and J\"ager, B. and Meyer, H. B. and Nyffeler, A. and Wittig, H.},
    title = "{The hadronic vacuum polarization contribution to the muon $g-2$ from lattice QCD}",
    eprint = "1705.01775",
    archivePrefix = "arXiv",
    primaryClass = "hep-lat",
    reportNumber = "CP3-ORIGINS-2017-015, HIM-2017-02, IFT-UAM-CSIC-17-039, MITP-17-030",
    doi = "10.1007/JHEP10(2017)020",
    journal = "JHEP",
    volume = "10",
    pages = "020",
    year = "2017"
}

@article{Aoyama:2020ynm,
    author = "Aoyama, T. and others",
    title = "{The anomalous magnetic moment of the muon in the Standard Model}",
    eprint = "2006.04822",
    archivePrefix = "arXiv",
    primaryClass = "hep-ph",
    reportNumber = "FERMILAB-PUB-20-207-T, INT-PUB-20-021, KEK Preprint 2020-5,
  MITP/20-028, KEK Preprint 2020-5, MITP/20-028, CERN-TH-2020-075, IFT-UAM/CSIC-20-74, LMU-ASC 18/20, LTH 1234,
  LU TP 20-20, LTH 1234, LU TP 20-20, MAN/HEP/2020/003, PSI-PR-20-06, UWThPh 2020-14, ZU-TH 18/20",
    doi = "10.1016/j.physrep.2020.07.006",
    journal = "Phys. Rept.",
    volume = "887",
    pages = "1--166",
    year = "2020"
}

@article{Luscher:2010iy,
    author = "L{\"u}scher, Martin",
    title = "{Properties and uses of the Wilson flow in lattice QCD}",
    eprint = "1006.4518",
    archivePrefix = "arXiv",
    primaryClass = "hep-lat",
    reportNumber = "CERN-PH-TH-2010-143",
    doi = "10.1007/JHEP08(2010)071",
    journal = "JHEP",
    volume = "08",
    pages = "071",
    year = "2010",
    note = "[Erratum: JHEP 03, 092 (2014)]"
}

@article{Schaefer:2010hu,
    author = "Schaefer, Stefan and Sommer, Rainer and Virotta, Francesco",
    collaboration = "ALPHA",
    title = "{Critical slowing down and error analysis in lattice QCD simulations}",
    eprint = "1009.5228",
    archivePrefix = "arXiv",
    primaryClass = "hep-lat",
    reportNumber = "DESY-10-151, SFB-CPP-10-81, HU-EP-10-55",
    doi = "10.1016/j.nuclphysb.2010.11.020",
    journal = "Nucl. Phys. B",
    volume = "845",
    pages = "93--119",
    year = "2011"
}

@article{Bali:2016umi,
    author = {Bali, Gunnar S. and Scholz, Enno E. and Simeth, Jakob and S\"oldner, Wolfgang},
    collaboration = "RQCD",
    title = "{Lattice simulations with $N_f=2+1$ improved Wilson fermions at a fixed strange quark mass}",
    eprint = "1606.09039",
    archivePrefix = "arXiv",
    primaryClass = "hep-lat",
    doi = "10.1103/PhysRevD.94.074501",
    journal = "Phys. Rev. D",
    volume = "94",
    number = "7",
    pages = "074501",
    year = "2016"
}

@article{Tantalo:2023onv,
    author = "Tantalo, Nazario",
    title = "{Matching lattice QC+ED to Nature}",
    eprint = "2301.02097",
    archivePrefix = "arXiv",
    primaryClass = "hep-lat",
    doi = "10.22323/1.430.0249",
    journal = "PoS",
    volume = "LATTICE2022",
    pages = "249",
    year = "2023"
}

@article{Gasser:1984gg,
    author = "Gasser, J. and Leutwyler, H.",
    title = "{Chiral Perturbation Theory: Expansions in the Mass of the Strange Quark}",
    reportNumber = "CERN-TH-3798",
    doi = "10.1016/0550-3213(85)90492-4",
    journal = "Nucl. Phys. B",
    volume = "250",
    pages = "465--516",
    year = "1985"
}

@article{Luscher:1996sc,
    author = "L{\"u}scher, Martin and Sint, Stefan and Sommer, Rainer and Weisz, Peter",
    title = "{Chiral symmetry and O(a) improvement in lattice QCD}",
    eprint = "hep-lat/9605038",
    archivePrefix = "arXiv",
    reportNumber = "DESY-96-086, CERN-TH-96-138, MPI-PHT-96-38",
    doi = "10.1016/0550-3213(96)00378-1",
    journal = "Nucl. Phys. B",
    volume = "478",
    pages = "365--400",
    year = "1996"
}

@article{Jansen:1995ck,
    author = "Jansen, Karl and Liu, Chuan and L{\"u}scher, Martin and Simma, Hubert and Sint, Stefan and Sommer, Rainer and Weisz, Peter and Wolff, Ulli",
    title = "{Nonperturbative renormalization of lattice QCD at all scales}",
    eprint = "hep-lat/9512009",
    archivePrefix = "arXiv",
    reportNumber = "DESY-95-230, CERN-TH-95-327, MPI-PHT-95-124",
    doi = "10.1016/0370-2693(96)00075-5",
    journal = "Phys. Lett. B",
    volume = "372",
    pages = "275--282",
    year = "1996"
}

@inproceedings{Sint:2007ug,
    author = "Sint, Stefan",
    title = "{Lattice QCD with a chiral twist}",
    booktitle = "{Workshop on Perspectives in Lattice QCD}",
    eprint = "hep-lat/0702008",
    archivePrefix = "arXiv",
    reportNumber = "TRINLAT-07-01",
    doi = "10.1142/9789812790927-0004",
    month = "2",
    year = "2007"
}

@article{Sharpe:1997by,
    author = "Sharpe, Stephen R.",
    title = "{Enhanced chiral logarithms in partially quenched QCD}",
    eprint = "hep-lat/9707018",
    archivePrefix = "arXiv",
    reportNumber = "UW-PT-97-18",
    doi = "10.1103/PhysRevD.62.099901",
    journal = "Phys. Rev. D",
    volume = "56",
    pages = "7052--7058",
    year = "1997",
    note = "[Erratum: Phys.Rev.D 62, 099901 (2000)]"
}

@article{Cichy:2010ta,
    author = "Cichy, Krzysztof and Herdo{\'i}za, Gregorio and Jansen, Karl",
    title = "{Continuum Limit of Overlap Valence Quarks on a Twisted Mass Sea}",
    eprint = "1012.4412",
    archivePrefix = "arXiv",
    primaryClass = "hep-lat",
    reportNumber = "DESY-10-240, FTUAM-10-35, SFB-CPP-10-130",
    doi = "10.1016/j.nuclphysb.2011.01.021",
    journal = "Nucl. Phys. B",
    volume = "847",
    pages = "179--196",
    year = "2011"
}

@article{EuropeanTwistedMass:2014osg,
    author = "Carrasco, N. and others",
    collaboration = "European Twisted Mass",
    title = "{Up, down, strange and charm quark masses with N$_f$ = 2+1+1 twisted mass lattice QCD}",
    eprint = "1403.4504",
    archivePrefix = "arXiv",
    primaryClass = "hep-lat",
    reportNumber = "PREPRINT-MITP-14-020, ROM2F-2014-01, RM3-TH-14-4",
    doi = "10.1016/j.nuclphysb.2014.07.025",
    journal = "Nucl. Phys. B",
    volume = "887",
    pages = "19--68",
    year = "2014"
}

@article{Clark:2006fx,
    author = "Clark, M. A. and Kennedy, A. D.",
    title = "{Accelerating dynamical fermion computations using the rational hybrid Monte Carlo (RHMC) algorithm with multiple pseudofermion fields}",
    eprint = "hep-lat/0608015",
    archivePrefix = "arXiv",
    doi = "10.1103/PhysRevLett.98.051601",
    journal = "Phys. Rev. Lett.",
    volume = "98",
    pages = "051601",
    year = "2007"
}

@article{Kuberski:2023zky,
    author = "Kuberski, Simon",
    title = "{Low-mode deflation for twisted-mass and RHMC reweighting in lattice QCD}",
    eprint = "2306.02385",
    archivePrefix = "arXiv",
    primaryClass = "hep-lat",
    reportNumber = "MITP-23-021",
    doi = "10.1016/j.cpc.2024.109173",
    journal = "Comput. Phys. Commun.",
    volume = "300",
    pages = "109173",
    year = "2024"
}

@article{deDivitiis:2019xla,
    author = {de Divitiis, Giulia Maria and Fritzsch, Patrick and Heitger, Jochen and K\"oster, Carl Christian and Kuberski, Simon and Vladikas, Anastassios},
    collaboration = "ALPHA",
    title = "{Non-perturbative determination of improvement coefficients $b_\mathrm{m}$ and $b_\mathrm{A}-b_\mathrm{P}$ and normalisation factor $Z_\mathrm{m}Z_\mathrm{P}/Z_\mathrm{A}$ with $N_\mathrm{f}= 3$ Wilson fermions}",
    eprint = "1906.03445",
    archivePrefix = "arXiv",
    primaryClass = "hep-lat",
    reportNumber = "CERN-TH-2019-085, MS-TP-19-13",
    doi = "10.1140/epjc/s10052-019-7287-1",
    journal = "Eur. Phys. J. C",
    volume = "79",
    number = "9",
    pages = "797",
    year = "2019"
}

@article{Colangelo:2005gd,
    author = "Colangelo, Gilberto and D{\"u}rr, Stephan and Haefeli, Christoph",
    title = "{Finite volume effects for meson masses and decay constants}",
    eprint = "hep-lat/0503014",
    archivePrefix = "arXiv",
    doi = "10.1016/j.nuclphysb.2005.05.015",
    journal = "Nucl. Phys. B",
    volume = "721",
    pages = "136--174",
    year = "2005"
}

@article{Gasser:1983yg,
    author = "Gasser, J. and Leutwyler, H.",
    title = "{Chiral Perturbation Theory to One Loop}",
    reportNumber = "CERN-TH-3689",
    doi = "10.1016/0003-4916(84)90242-2",
    journal = "Annals Phys.",
    volume = "158",
    pages = "142",
    year = "1984"
}

@article{ParticleDataGroup:2024cfk,
    author = "Navas, S. and others",
    collaboration = "Particle Data Group",
    title = "{Review of particle physics}",
    doi = "10.1103/PhysRevD.110.030001",
    journal = "Phys. Rev. D",
    volume = "110",
    number = "3",
    pages = "030001",
    year = "2024"
}

@article{ExtendedTwistedMass:2021qui,
    author = "Alexandrou, C. and others",
    collaboration = "Extended Twisted Mass",
    title = "{Ratio of kaon and pion leptonic decay constants with Nf=2+1+1 Wilson-clover twisted-mass fermions}",
    eprint = "2104.06747",
    archivePrefix = "arXiv",
    primaryClass = "hep-lat",
    doi = "10.1103/PhysRevD.104.074520",
    journal = "Phys. Rev. D",
    volume = "104",
    number = "7",
    pages = "074520",
    year = "2021"
}

@article{Miller:2020evg,
    author = "Miller, Nolan and others",
    title = {{Scale setting the M\"obius domain wall fermion on gradient-flowed HISQ action using the omega baryon mass and the gradient-flow scales $t_0$ and $w_0$}},
    eprint = "2011.12166",
    archivePrefix = "arXiv",
    primaryClass = "hep-lat",
    reportNumber = "LLNL-JRNL-816949, RIKEN-iTHEMS-Report-20, JLAB-THY-20-3290",
    doi = "10.1103/PhysRevD.103.054511",
    journal = "Phys. Rev. D",
    volume = "103",
    number = "5",
    pages = "054511",
    year = "2021"
}

@article{MILC:2015tqx,
    author = "Bazavov, A. and others",
    collaboration = "MILC",
    title = "{Gradient flow and scale setting on MILC HISQ ensembles}",
    eprint = "1503.02769",
    archivePrefix = "arXiv",
    primaryClass = "hep-lat",
    reportNumber = "FERMILAB-PUB-15-284-T",
    doi = "10.1103/PhysRevD.93.094510",
    journal = "Phys. Rev. D",
    volume = "93",
    number = "9",
    pages = "094510",
    year = "2016"
}

@article{Dowdall:2013rya,
    author = "Dowdall, R. J. and Davies, C. T. H. and Lepage, G. P. and McNeile, C.",
    title = "{Vus from pi and K decay constants in full lattice QCD with physical u, d, s and c quarks}",
    eprint = "1303.1670",
    archivePrefix = "arXiv",
    primaryClass = "hep-lat",
    doi = "10.1103/PhysRevD.88.074504",
    journal = "Phys. Rev. D",
    volume = "88",
    pages = "074504",
    year = "2013"
}

@article{RQCD:2022xux,
    author = {Bali, Gunnar S. and Collins, Sara and Georg, Peter and Jenkins, Daniel and Korcyl, Piotr and Sch\"afer, Andreas and Scholz, Enno E. and Simeth, Jakob and S\"oldner, Wolfgang and Weish\"aupl, Simon},
    collaboration = "RQCD",
    title = "{Scale setting and the light baryon spectrum in N$_{f}$ = 2 + 1 QCD with Wilson fermions}",
    eprint = "2211.03744",
    archivePrefix = "arXiv",
    primaryClass = "hep-lat",
    doi = "10.1007/JHEP05(2023)035",
    journal = "JHEP",
    volume = "05",
    pages = "035",
    year = "2023"
}

@article{RBC:2014ntl,
    author = "Blum, T. and others",
    collaboration = "RBC, UKQCD",
    title = "{Domain wall QCD with physical quark masses}",
    eprint = "1411.7017",
    archivePrefix = "arXiv",
    primaryClass = "hep-lat",
    reportNumber = "KEK-TH-1769, RBRC-1095, DAMTP-2014-86",
    doi = "10.1103/PhysRevD.93.074505",
    journal = "Phys. Rev. D",
    volume = "93",
    number = "7",
    pages = "074505",
    year = "2016"
}

@article{BMW:2012hcm,
    author = {Bors\'anyi, Szabolcs and D\"urr, Stephan and Fodor, Zolt\'an and Hoelbling, Christian and Katz, S\'andor D. and Krieg, Stefan and Kurth, Thorsten and Lellouch, Laurent and Lippert, Thomas and McNeile, Craig},
    collaboration = "BMW",
    title = "{High-precision scale setting in lattice QCD}",
    eprint = "1203.4469",
    archivePrefix = "arXiv",
    primaryClass = "hep-lat",
    reportNumber = "ITP-BUDAPEST-657, CPT-P004-2012, WUB-12-02",
    doi = "10.1007/JHEP09(2012)010",
    journal = "JHEP",
    volume = "09",
    pages = "010",
    year = "2012"
}

@article{Strassberger:2021tsu,
    author = "Strassberger, Ben and others",
    title = "{Scale setting for CLS 2+1 simulations}",
    eprint = "2112.06696",
    archivePrefix = "arXiv",
    primaryClass = "hep-lat",
    reportNumber = "DESY-21-175, WUB/21-05",
    doi = "10.22323/1.396.0135",
    journal = "PoS",
    volume = "LATTICE2021",
    pages = "135",
    year = "2022"
}

@article{Hudspith:2024kzk,
    author = "Hudspith, Renwick J. and Lutz, Matthias F. M. and Mohler, Daniel",
    title = "{Precise Omega baryons from lattice QCD}",
    eprint = "2404.02769",
    archivePrefix = "arXiv",
    primaryClass = "hep-lat",
    month = "4",
    year = "2024"
}

@article{Bornyakov:2015eaa,
    author = "Bornyakov, V. G. and others",
    title = "{Wilson flow and scale setting from lattice QCD}",
    eprint = "1508.05916",
    archivePrefix = "arXiv",
    primaryClass = "hep-lat",
    reportNumber = "ADP-15-25-T927, DESY-15-154, EDINBURGH-2015-19, LIVERPOOL-LTH-1054",
    month = "8",
    year = "2015"
}

@article{Brida:2025gii,
    author = {Brida, Mattia Dalla and H\"ollwieser, Roman and Knechtli, Francesco and Korzec, Tomasz and Ramos, Alberto and Sint, Stefan and Sommer, Rainer},
    title = "{The strength of the interaction between quarks and gluons}",
    eprint = "2501.06633",
    archivePrefix = "arXiv",
    primaryClass = "hep-ph",
    month = "1",
    year = "2025"
}

@article{FlavourLatticeAveragingGroupFLAG:2024oxs,
    author = "Aoki, Y. and others",
    collaboration = "Flavour Lattice Averaging Group (FLAG)",
    title = "{FLAG Review 2024}",
    eprint = "2411.04268",
    archivePrefix = "arXiv",
    primaryClass = "hep-lat",
    reportNumber = "CERN-TH-2024-192, FERMILAB-PUB-24-0785-T",
    month = "11",
    year = "2024"
}

@article{Bulava:2015bxa,
    author = "Bulava, John and Della Morte, Michele and Heitger, Jochen and Wittemeier, Christian",
    collaboration = "ALPHA",
    title = "{Non-perturbative improvement of the axial current in $N_f$=3 lattice QCD with Wilson fermions and tree-level improved gauge action}",
    eprint = "1502.04999",
    archivePrefix = "arXiv",
    primaryClass = "hep-lat",
    reportNumber = "TCDMATH-15-01, MS-TP-15-02, CP3-ORIGINS-2015-001, DIAS-2015-1",
    doi = "10.1016/j.nuclphysb.2015.05.003",
    journal = "Nucl. Phys. B",
    volume = "896",
    pages = "555--568",
    year = "2015"
}

@article{Frison:2023lwb,
    author = "Frison, Julien",
    title = "{Towards fully bayesian analyses in Lattice QCD}",
    eprint = "2302.06550",
    archivePrefix = "arXiv",
    primaryClass = "hep-lat",
    month = "2",
    year = "2023"
}

@article{Bruno:2022mfy,
    author = "Bruno, Mattia and Sommer, Rainer",
    title = "{On fits to correlated and auto-correlated data}",
    eprint = "2209.14188",
    archivePrefix = "arXiv",
    primaryClass = "hep-lat",
    doi = "10.1016/j.cpc.2022.108643",
    journal = "Comput. Phys. Commun.",
    volume = "285",
    pages = "108643",
    year = "2023"
}

@article{Bar:2002nr,
    author = "B{\"a}r, Oliver and Rupak, Gautam and Shoresh, Noam",
    title = "{Simulations with different lattice Dirac operators for valence and sea quarks}",
    eprint = "hep-lat/0210050",
    archivePrefix = "arXiv",
    reportNumber = "MIT-CTP-3319, LBNL-51684, BUHEP-02-36",
    doi = "10.1103/PhysRevD.67.114505",
    journal = "Phys. Rev. D",
    volume = "67",
    pages = "114505",
    year = "2003"
}

@article{Bar:2003mh,
    author = "B{\"a}r, Oliver and Rupak, Gautam and Shoresh, Noam",
    title = "{Chiral perturbation theory at O(a**2) for lattice QCD}",
    eprint = "hep-lat/0306021",
    archivePrefix = "arXiv",
    reportNumber = "UTHEP-469, LBNL-52989, BUHEP-03-13",
    doi = "10.1103/PhysRevD.70.034508",
    journal = "Phys. Rev. D",
    volume = "70",
    pages = "034508",
    year = "2004"
}

@article{Golterman:2005xa,
    author = "Golterman, Maarten and Izubuchi, Taku and Shamir, Yigal",
    title = "{The Role of the double pole in lattice QCD with mixed actions}",
    eprint = "hep-lat/0504013",
    archivePrefix = "arXiv",
    reportNumber = "KANAZAWA-05-05, RBRC-489",
    doi = "10.1103/PhysRevD.71.114508",
    journal = "Phys. Rev. D",
    volume = "71",
    pages = "114508",
    year = "2005"
}

@article{Hasenfratz:2006bq,
    author = "Hasenfratz, Anna and Hoffmann, Roland",
    title = "{Mixed action simulations on staggered background: Interpretation and result for the 2-flavor QCD chiral condensate}",
    eprint = "hep-lat/0609067",
    archivePrefix = "arXiv",
    reportNumber = "COLO-HEP-519",
    doi = "10.1103/PhysRevD.74.114509",
    journal = "Phys. Rev. D",
    volume = "74",
    pages = "114509",
    year = "2006"
}

@article{Aubin:2008wk,
    author = "Aubin, C. and Laiho, Jack and Van de Water, Ruth S.",
    title = "{Discretization effects and the scalar meson correlator in mixed-action lattice simulations}",
    eprint = "0803.0129",
    archivePrefix = "arXiv",
    primaryClass = "hep-lat",
    reportNumber = "FERMILAB-PUB-08-053-T",
    doi = "10.1103/PhysRevD.77.114501",
    journal = "Phys. Rev. D",
    volume = "77",
    pages = "114501",
    year = "2008"
}

@article{Bernardoni:2010nf,
    author = "Bernardoni, Fabio and Hern{\'a}ndez, Pilar and Garron, Nicolas and Necco, Silvia and Pena, Carlos",
    title = "{Probing the chiral regime of $N_{f}$= 2 QCD with mixed actions}",
    eprint = "1008.1870",
    archivePrefix = "arXiv",
    primaryClass = "hep-lat",
    reportNumber = "IFIC-10-27, FTUV-10-0810, FTUAM-10-10, IFT-UAM-CSIC-10-47, CERN-PH-TH-2010-178, EDINBURGH-2010-22",
    doi = "10.1103/PhysRevD.83.054503",
    journal = "Phys. Rev. D",
    volume = "83",
    pages = "054503",
    year = "2011"
}

@article{Cichy:2012vg,
    author = "Cichy, Krzysztof and Drach, Vincent and Garc{\'i}a-Ramos, Elena and Herdo{\'i}za, Gregorio and Jansen, Karl",
    title = "{Overlap valence quarks on a twisted mass sea: a case study for mixed action Lattice QCD}",
    eprint = "1211.1605",
    archivePrefix = "arXiv",
    primaryClass = "hep-lat",
    reportNumber = "DESY-12-192, HU-EP-12-32, SFB-CPP-12-80, FTUAM-12-99, IFT-UAM-CSIC-12-67",
    doi = "10.1016/j.nuclphysb.2012.12.011",
    journal = "Nucl. Phys. B",
    volume = "869",
    pages = "131--163",
    year = "2013"
}

@article{Colangelo:2003hf,
    author = "Colangelo, Gilberto and D{\"u}rr, Stephan",
    title = "{The Pion mass in finite volume}",
    eprint = "hep-lat/0311023",
    archivePrefix = "arXiv",
    reportNumber = "DESY-03-177",
    doi = "10.1140/epjc/s2004-01593-y",
    journal = "Eur. Phys. J. C",
    volume = "33",
    pages = "543--553",
    year = "2004"
}

@article{Colangelo:2010ba,
    author = "Colangelo, Gilberto and Fuhrer, Andreas and Lanz, Stefan",
    title = "{Finite volume effects for nucleon and heavy meson masses}",
    eprint = "1005.1485",
    archivePrefix = "arXiv",
    primaryClass = "hep-lat",
    reportNumber = "LU-TP-10-12",
    doi = "10.1103/PhysRevD.82.034506",
    journal = "Phys. Rev. D",
    volume = "82",
    pages = "034506",
    year = "2010"
}

@article{Gasser:1986vb,
    author = "Gasser, J. and Leutwyler, H.",
    title = "{Light Quarks at Low Temperatures}",
    reportNumber = "BUTP-86/19-BERN",
    doi = "10.1016/0370-2693(87)90492-8",
    journal = "Phys. Lett. B",
    volume = "184",
    pages = "83--88",
    year = "1987"
}

@article{Ramos:2018vgu,
    author = "Ramos, Alberto",
    title = "{Automatic differentiation for error analysis of Monte Carlo data}",
    eprint = "1809.01289",
    archivePrefix = "arXiv",
    primaryClass = "hep-lat",
    doi = "10.1016/j.cpc.2018.12.020",
    journal = "Comput. Phys. Commun.",
    volume = "238",
    pages = "19--35",
    year = "2019"
}

@article{Wolff:2003sm,
    author = "Wolff, Ulli",
    collaboration = "ALPHA",
    title = "{Monte Carlo errors with less errors}",
    eprint = "hep-lat/0306017",
    archivePrefix = "arXiv",
    reportNumber = "HU-EP-03-32, SFB-CPP-03-12",
    doi = "10.1016/S0010-4655(03)00467-3",
    journal = "Comput. Phys. Commun.",
    volume = "156",
    pages = "143--153",
    year = "2004",
    note = "[Erratum: Comput.Phys.Commun. 176, 383 (2007)]"
}

@article{MA,
    author = "Jay, William I. and Neil, Ethan T.",
    title = "{Bayesian model averaging for analysis of lattice field theory results}",
    eprint = "2008.01069",
    archivePrefix = "arXiv",
    primaryClass = "stat.ME",
    reportNumber = "FERMILAB-PUB-20-374-T",
    doi = "10.1103/PhysRevD.103.114502",
    journal = "Phys. Rev. D",
    volume = "103",
    pages = "114502",
    year = "2021"
}

@article{Michael:1998sg,
    author = "Michael, Christopher and Peisa, J.",
    collaboration = "UKQCD",
    title = "{Maximal variance reduction for stochastic propagators with applications to the static quark spectrum}",
    eprint = "hep-lat/9802015",
    archivePrefix = "arXiv",
    reportNumber = "LTH-420",
    doi = "10.1103/PhysRevD.58.034506",
    journal = "Phys. Rev. D",
    volume = "58",
    pages = "034506",
    year = "1998"
}

@article{Campos:2018ahf,
    author = "Campos, Isabel and Fritzsch, Patrick and Pena, Carlos and Preti, David and Ramos, Alberto and Vladikas, Anastassios",
    collaboration = "ALPHA",
    title = "{Non-perturbative quark mass renormalisation and running in $N_f=3$ QCD}",
    eprint = "1802.05243",
    archivePrefix = "arXiv",
    primaryClass = "hep-lat",
    reportNumber = "CERN-TH-2018-029, IFT-UAM/CSIC-18-011, FTUAM-18-4, IFT-UAM-CSIC-18-011",
    doi = "10.1140/epjc/s10052-018-5870-5",
    journal = "Eur. Phys. J. C",
    volume = "78",
    number = "5",
    pages = "387",
    year = "2018"
}

@inproceedings{Luscher:2010ae,
    author = "L{\"u}scher, Martin",
    title = "{Computational Strategies in Lattice QCD}",
    booktitle = "{Les Houches Summer School: Session 93: Modern perspectives in lattice QCD: Quantum field theory and high performance computing}",
    eprint = "1002.4232",
    archivePrefix = "arXiv",
    primaryClass = "hep-lat",
    reportNumber = "CERN-PH-TH-2010-047",
    pages = "331--399",
    month = "2",
    year = "2010"
}

@inbook{Akaike:1998zah,
    author = "Akaike, Hirotogu",
    title = "{Information Theory and an Extension of the Maximum Likelihood Principle}",
    doi = "10.1007/978-1-4612-1694-0-15",
    publisher = "Springer Science+Business Media",
    address = "New York",
    year = "1998"
}

@article{Jay:2020jkz,
    author = "Jay, William I. and Neil, Ethan T.",
    title = "{Bayesian model averaging for analysis of lattice field theory results}",
    eprint = "2008.01069",
    archivePrefix = "arXiv",
    primaryClass = "stat.ME",
    reportNumber = "FERMILAB-PUB-20-374-T",
    doi = "10.1103/PhysRevD.103.114502",
    journal = "Phys. Rev. D",
    volume = "103",
    pages = "114502",
    year = "2021"
}

@article{BMW:2013fzj,
    author = {D{\"u}rr, Stephan and others},
    collaboration = "BMW",
    title = "{Lattice QCD at the physical point meets SU(2) chiral perturbation theory}",
    eprint = "1310.3626",
    archivePrefix = "arXiv",
    primaryClass = "hep-lat",
    reportNumber = "CPT-P005-2013, WUB-13-14",
    doi = "10.1103/PhysRevD.90.114504",
    journal = "Phys. Rev. D",
    volume = "90",
    number = "11",
    pages = "114504",
    year = "2014"
}

@article{Borsanyi:2020mff,
    author = "Bors{\'a}nyi, Sz. and others",
    title = "{Leading hadronic contribution to the muon magnetic moment from lattice QCD}",
    eprint = "2002.12347",
    archivePrefix = "arXiv",
    primaryClass = "hep-lat",
    doi = "10.1038/s41586-021-03418-1",
    journal = "Nature",
    volume = "593",
    number = "7857",
    pages = "51--55",
    year = "2021"
}

@article{Takeuchi76,
author="Takeuchi, K.",
title="Distribution of information statistics and validity criteria of models",
journal="Mathematical Science",
year="1976",
volume="153",
pages="12-18",
URL="https://cir.nii.ac.jp/crid/1571135650620877312"
}

@article{Michael:1994sz,
    author = "Michael, Christopher and McKerrell, A.",
    title = "{Fitting correlated hadron mass spectrum data}",
    eprint = "hep-lat/9412087",
    archivePrefix = "arXiv",
    reportNumber = "LTH-342",
    doi = "10.1103/PhysRevD.51.3745",
    journal = "Phys. Rev. D",
    volume = "51",
    pages = "3745--3750",
    year = "1995"
}

@article{Neil:2023pgt,
    author = "Neil, Ethan T. and Sitison, Jacob W.",
    title = "{Model averaging approaches to data subset selection}",
    eprint = "2305.19417",
    archivePrefix = "arXiv",
    primaryClass = "stat.ME",
    doi = "10.1103/PhysRevE.108.045308",
    journal = "Phys. Rev. E",
    volume = "108",
    number = "4",
    pages = "045308",
    year = "2023"
}

@article{Neil:2022joj,
    author = "Neil, Ethan T. and Sitison, Jacob W.",
    title = "{Improved information criteria for Bayesian model averaging in lattice field theory}",
    eprint = "2208.14983",
    archivePrefix = "arXiv",
    primaryClass = "stat.ME",
    doi = "10.1103/PhysRevD.109.014510",
    journal = "Phys. Rev. D",
    volume = "109",
    number = "1",
    pages = "014510",
    year = "2024"
}

@article{Kullback:1951zyt,
    author = "Kullback, S. and Leibler, R. A.",
    title = "{On Information and Sufficiency}",
    doi = "10.1214/aoms/1177729694",
    journal = "The Annals of Mathematical Statistics",
    volume = "22",
    number = "1",
    pages = "79--86",
    year = "1951"
}

@incollection{akaike1976canonical,
  title={Canonical correlation analysis of time series and the use of an information criterion},
  author={Akaike, Hirotugu},
  booktitle={Mathematics in science and engineering},
  volume={126},
  pages={27--96},
  year={1976},
  publisher={Elsevier}
}

@misc{Watanabe2007,
  doi = {10.48550/ARXIV.0712.0653},
  
  url = {https://arxiv.org/abs/0712.0653},
  
  author = {Watanabe, Sumio},
  
  title = {Equations of States in Singular Statistical Estimation},
  
  publisher = {arXiv},
  
  year = {2007},
  
  copyright = {arXiv.org perpetual, non-exclusive license}
}

@article{Husung:2024cgc,
    author = "Husung, Nikolai",
    title = "{Lattice artifacts of local fermion bilinears up to $\text {O}(\text {a}^2)$}",
    eprint = "2409.00776",
    archivePrefix = "arXiv",
    primaryClass = "hep-lat",
    reportNumber = "IFT-UAM/CSIC-24-125",
    doi = "10.1140/epjc/s10052-025-13825-7",
    journal = "Eur. Phys. J. C",
    volume = "85",
    number = "4",
    pages = "427",
    year = "2025"
}

@article{Aoki:2006gh,
    author = "Aoki, Sinya and B{\"a}r, Oliver",
    editor = "Blum, Tom and Creutz, Michael and DeTar, Carleton and Karsch, Frithjof and Kronfeld, Andreas and Morningstar, Colin and Richards, David and Shigemitsu, Junko and Toussaint, Doug",
    title = "{Automatic O(a) improvement for twisted-mass QCD}",
    eprint = "hep-lat/0610098",
    archivePrefix = "arXiv",
    reportNumber = "UTHEP-532",
    doi = "10.22323/1.032.0165",
    journal = "PoS",
    volume = "LAT2006",
    pages = "165",
    year = "2006"
}

@article{Luscher:1996ug,
    author = "L{\"u}scher, Martin and Sint, Stefan and Sommer, Rainer and Weisz, Peter and Wolff, Ulli",
    title = "{Nonperturbative O(a) improvement of lattice QCD}",
    eprint = "hep-lat/9609035",
    archivePrefix = "arXiv",
    reportNumber = "CERN-TH-96-218, DESY-96-180, HUB-EP-96-46, MPI-PHT-96-75, FSU-SCRI-96-101",
    doi = "10.1016/S0550-3213(97)00080-1",
    journal = "Nucl. Phys. B",
    volume = "491",
    pages = "323--343",
    year = "1997"
}

@article{Aliberti:2025beg,
    author = "Aliberti, R. and others",
    title = "{The anomalous magnetic moment of the muon in the Standard Model: an update}",
    eprint = "2505.21476",
    archivePrefix = "arXiv",
    primaryClass = "hep-ph",
    reportNumber = "CERN-TH-2025-101, FERMILAB-PUB-25-0344-T, INT-PUB-25-015, IPARCOS-UCM-25-029, KEK Preprint 2025-22, LTH 1403, MITP-25-037, UWThPh 2025-15, UWThPh
  2025-15, ZU-TH 37/25, IPARCOS-UCM-25-029",
    doi = "10.1016/j.physrep.2025.08.002",
    journal = "Phys. Rept.",
    volume = "1143",
    pages = "1--158",
    year = "2025"
}

@article{Chen:2007ug,
    author = "Chen, Jiunn-Wei and O'Connell, Donal and Walker-Loud, Andre",
    title = "{Universality of mixed action extrapolation formulae}",
    eprint = "0706.0035",
    archivePrefix = "arXiv",
    primaryClass = "hep-lat",
    reportNumber = "CALT-68-2650, UMD-40762-389",
    doi = "10.1088/1126-6708/2009/04/090",
    journal = "JHEP",
    volume = "04",
    pages = "090",
    year = "2009"
}

@article{Chen:2009su,
    author = "Chen, Jiunn-Wei and Golterman, Maarten and O'Connell, Donal and Walker-Loud, Andre",
    title = "{Mixed Action Effective Field Theory: An Addendum}",
    eprint = "0905.2566",
    archivePrefix = "arXiv",
    primaryClass = "hep-lat",
    doi = "10.1103/PhysRevD.79.117502",
    journal = "Phys. Rev. D",
    volume = "79",
    pages = "117502",
    year = "2009"
}

@article{Larsen:2025wvg,
    author = "Larsen, Rasmus and Mukherjee, Swagato and Petreczky, Peter and Shu, Hai-Tao and Weber, Johannes Heinrich",
    title = "{Scale Setting and Strong Coupling Determination in the Gradient Flow Scheme for 2+1 Flavor Lattice QCD}",
    eprint = "2502.08061",
    archivePrefix = "arXiv",
    primaryClass = "hep-lat",
    month = "2",
    year = "2025"
}

@article{Aoki:1998ar,
    author = "Aoki, Sinya and Nagai, Kei-ichi and Taniguchi, Yusuke and Ukawa, Akira",
    title = "{Perturbative renormalization factors of bilinear quark operators for improved gluon and quark actions in lattice QCD}",
    eprint = "hep-lat/9802034",
    archivePrefix = "arXiv",
    reportNumber = "UTCCP-P-33, MPI-PHT-98-15, UTHEP-380",
    doi = "10.1103/PhysRevD.58.074505",
    journal = "Phys. Rev. D",
    volume = "58",
    pages = "074505",
    year = "1998"
}

@article{Korcyl:2016ugy,
    author = "Korcyl, Piotr and Bali, Gunnar S.",
    title = "{Non-perturbative determination of improvement coefficients using coordinate space correlators in $N_f=2+1$ lattice QCD}",
    eprint = "1607.07090",
    archivePrefix = "arXiv",
    primaryClass = "hep-lat",
    doi = "10.1103/PhysRevD.95.014505",
    journal = "Phys. Rev. D",
    volume = "95",
    number = "1",
    pages = "014505",
    year = "2017"
}

@article{Korcyl:2016cmx,
    author = "Korcyl, Piotr and Bali, Gunnar S.",
    title = "{Non-perturbative determination of improvement coefficients using coordinate space correlators in $N_f=2+1$ lattice QCD}",
    eprint = "1609.09477",
    archivePrefix = "arXiv",
    primaryClass = "hep-lat",
    doi = "10.22323/1.256.0190",
    journal = "PoS",
    volume = "LATTICE2016",
    pages = "190",
    year = "2016"
}

@article{Costa:2025xej,
    author = "Costa, Marios and Gavriel, Demetrianos and Panagopoulos, Haralambos and Spanoudes, Gregoris",
    title = "{Perturbative determination of $\mathcal{O}(a\,m)$ improvement on the QCD running coupling}",
    eprint = "2503.00463",
    archivePrefix = "arXiv",
    primaryClass = "hep-lat",
    month = "3",
    year = "2025"
}

@article{DallaBrida:2023fpl,
    author = {Dalla Brida, Mattia and H{\"o}llwieser, Roman and Knechtli, Francesco and Korzec, Tomasz and Sint, Stefan and Sommer, Rainer},
    collaboration = "ALPHA",
    title = "{Heavy Wilson quarks and O(a) improvement: nonperturbative results for b$_{g}$}",
    eprint = "2401.00216",
    archivePrefix = "arXiv",
    primaryClass = "hep-lat",
    reportNumber = "CERN-TH-2023-230",
    doi = "10.1007/JHEP01(2024)188",
    journal = "JHEP",
    volume = "2024",
    number = "01",
    pages = "188",
    year = "2024"
}

@article{Sint:1995ch,
    author = "Sint, Stefan and Sommer, Rainer",
    title = "{The Running coupling from the QCD Schrodinger functional: A One loop analysis}",
    eprint = "hep-lat/9508012",
    archivePrefix = "arXiv",
    reportNumber = "MPI-PHT-95-69, CERN-TH-95-208",
    doi = "10.1016/0550-3213(96)00020-X",
    journal = "Nucl. Phys. B",
    volume = "465",
    pages = "71--98",
    year = "1996"
}

@article{Narayanan:2006rf,
    author = "Narayanan, R. and Neuberger, H.",
    title = "{Infinite N phase transitions in continuum Wilson loop operators}",
    eprint = "hep-th/0601210",
    archivePrefix = "arXiv",
    doi = "10.1088/1126-6708/2006/03/064",
    journal = "JHEP",
    volume = "03",
    pages = "064",
    year = "2006"
}

\end{document}